# THE ARCHITECTURE OF AN AUTONOMIC, RESOURCE-AWARE, WORKSTATION-BASED DISTRIBUTED DATABASE SYSTEM

Angus Macdonald

PhD Thesis

February 2012

University of
St Andrews

## Abstract

Distributed software systems that are designed to run over workstation machines within organisations are termed *workstation-based*. Workstation-based systems are characterised by dynamically changing sets of machines that are used primarily for other, user-centric tasks. They must be able to adapt to and utilize spare capacity when and where it is available, and ensure that the non-availability of an individual machine does not affect the availability of the system.

This thesis focuses on the requirements and design of a workstation-based database system, which is motivated by an analysis of existing database architectures that are typically run over static, specially provisioned sets of machines.

A typical clustered database system — one that is run over a number of specially provisioned machines — executes queries interactively, returning a synchronous response to applications, with its data made durable and resilient to the failure of machines. There are no existing workstation-based databases. Furthermore, other workstation-based systems do not attempt to achieve the requirements of interactivity and durability, because they are typically used to execute asynchronous batch processing jobs that tolerate data loss — results can be re-computed. These systems use external servers to store the final results of computations rather than workstation machines.

This thesis describes the design and implementation of a workstation-based database system and investigates its viability by evaluating its performance against existing clustered database systems and testing its availability during machine failures.

## ACKNOWLEDGEMENTS

I'd like to thank my supervisors, Professor Alan Dearle and Dr Graham Kirby, for the opportunities, the support, and the education that they have given me. Words cannot express how grateful I am for this.

To Greg, Davie, Stuart, Alex, and the many others who provided support and entertainment away from work, and who pushed me (mostly) in the right direction, thank you.

Thanks to my many officemates through the years, particularly Rob and Andy for helping me start, and Masih for helping me finish.

Many thanks to the countless students I've had the pleasure to teach, for providing challenging discussions and inspiration in equal measure. I learned more from you, than you did from me.

Finally, the biggest of thanks to my sister Julie, and to my parents, Anne and John Macdonald, for their love and support, and for making me who I am today. This is for them.

## DECLARATIONS

I, Angus Macdonald, hereby certify that this thesis, which is approximately 55,000 words in length, has been written by me, that it is the record of work carried out by me and that it has not been submitted in any previous application for a higher degree.

I was admitted as a research student in September 2007 and as a candidate for the degree of Doctor of Philosophy in September 2007; the higher study for which this is a record was carried out in the University of St Andrews between 2007 and 2012.

Date: Signature of candidate:

I hereby certify that the candidate has fulfilled the conditions of the Resolution and Regulations appropriate for the degree of Doctor of Philosophy in the University of St Andrews and that the candidate is qualified to submit this thesis in application for that degree.

Date: Signature of supervisor:

In submitting this thesis to the University of St Andrews I understand that I am giving permission for it to be made available for use in accordance with the regulations of the University Library for the time being in force, subject to any copyright vested in the work not being affected thereby. I also understand that the title and the abstract will be published, and that a copy of the work may be made and supplied to any bona fide library or research worker, that my thesis will be electronically accessible for personal or research use unless exempt by award of an embargo as requested below, and that the library has the right to migrate my thesis into new electronic forms as required to ensure continued access to the thesis. I have obtained any third-party copyright permissions that may be required in order to allow such access and migration, or have requested the appropriate embargo below.

The following is an agreed request by candidate and supervisor regarding the electronic publication of this thesis: Access to printed copy and electronic publication of thesis through the University of St Andrews.

Date: Signature of candidate:

Date: Signature of supervisor:

## TABLE OF CONTENTS

## TABLE OF FIGURES

# Chapter 1: Introduction

# 1 INTRODUCTION

Distributed software systems that are designed to run over workstation machines within organisations are called *workstation-based systems*. The goal of these systems is to harness under-utilized computational and storage resources of existing machines instead of provisioning new machines specifically for a given task.

Existing workstation-based systems focus on computationally-intensive batch processing tasks, typically using non-workstation-based machines to store the results of computations. This is in part due to the perception that workstation machines are unreliable, since they are primarily used for other tasks which constrain their resources and limit their availability. This thesis extends existing work with the design and implementation of a system that makes use of the storage capacity of workstation machines in addition to their computational capacity.

***More specifically, this thesis investigates the viability of systems that use the un-utilized capacity of workstations to provide services, such as databases, that are typically run in server clusters (groups of co-located, specially provisioned machines)***. These services are termed as interactive, workstation-based systems because they require synchronous responses to application queries. To provide this, they must utilise storage capacity on workstations rather than relying on specially provisioned servers or clusters.

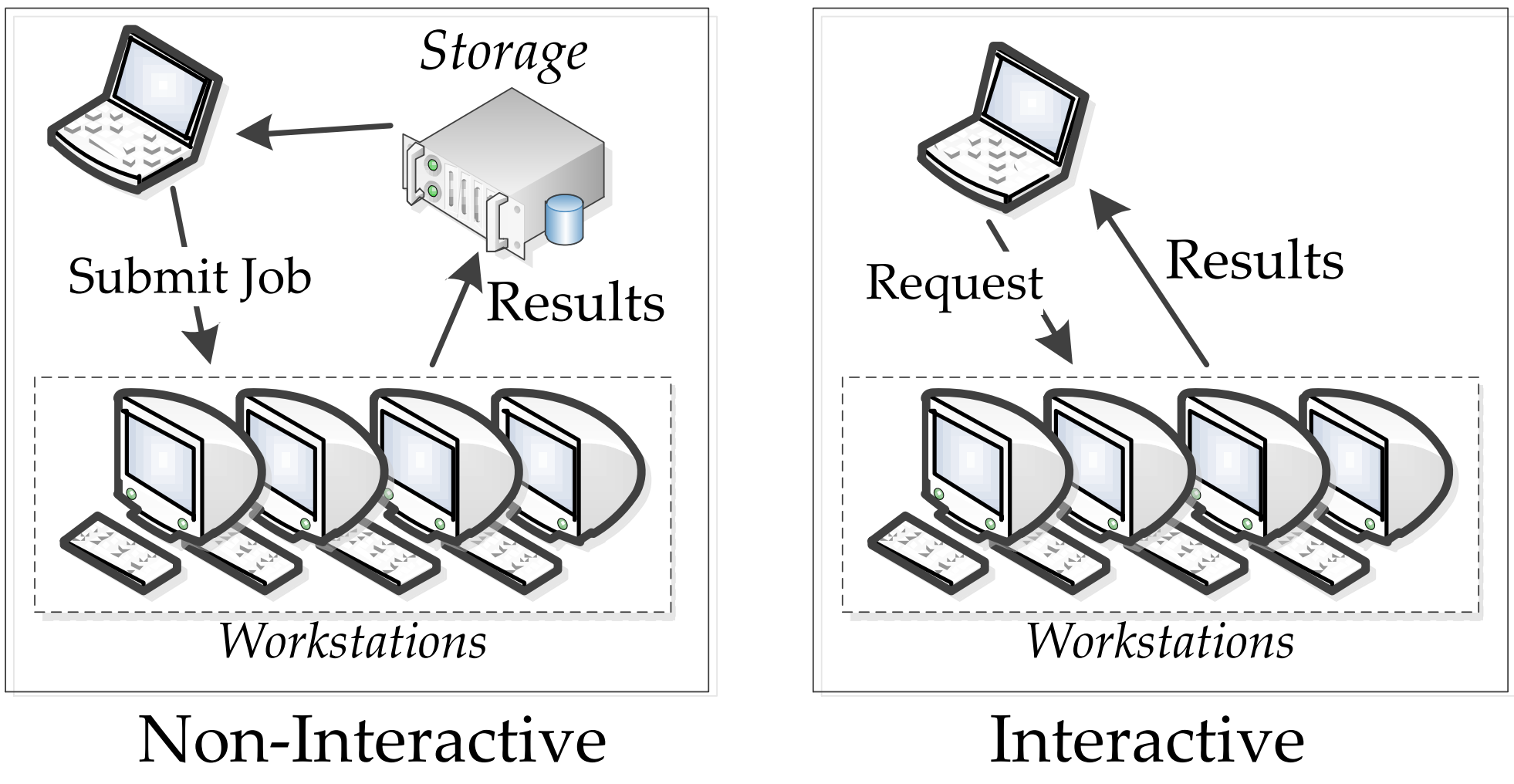

**Figure 1: (*left*) A non-interactive system: jobs are submitted by a user or application to workstations, and results are stored on a server when computations have completed. (*right*) An interactive system: a request is made to the workstation-based system and a response is synchronously returned.**

To test the viability of this approach, this thesis focuses on the creation of a database designed to make use of un-utilized resources within organizations.

## 1.1 DATABASE SYSTEMS

Database systems are a particularly challenging test of workstation-based systems, because they must be able to store large quantities of highly structured data and they are judged on the speed of their responses to queries. For a workstation-based system to be considered viable, it must be a competitive alternative to existing non-workstation-based systems, and it must be capable of operating in an environment where machines are unreliable and often fail.

The focus on database systems is motivated by their typical use within enterprises. A database system is commonly run over servers in a server room and replicated, which requires a number of machines to be provisioned. It is then used by many users or applications within an enterprise.

Database systems typically require specially provisioned resources to operate, in contrast to workstation-based systems which make use of existing, under-utilized resources within organizations. This is important because idle machines still use a substantial portion of their peak power consumption[1], which is costly to organizations that do not effectively utilize existing infrastructure. If workstation machines are able to run a database system in place of server room machines, the overall cost to an organisation can be reduced because fewer machines need to be provisioned.

There are many database systems that are currently run in server rooms, but whose workload potentially allows them to be used over workstation machines. These databases are used for development or other non-production purposes, or are production systems that are not typically under heavy load but still require strong transactional integrity.

However, existing database systems are not designed to run over workstation machines, which have different properties to clustered machines —a greater expectation of machine failure, dynamic membership, and dynamic resource availability. The primary motivation of this thesis is to investigate the viability of a workstation-based database system by analysing the architectures that are suited to such a system.

[1] Servers with near 0% utilization still use around 50% of the power used at peak utilization [94].

## 1.2 EXISTING APPROACHES

In contrast to the proposed workstation-based approach, existing approaches to creating replicated database systems typically use clustered database systems running over small numbers of specially provisioned machines. Copies of data are created across these machines to improve resilience to failure, and the database is able to provide strict transactional guarantees.

Larger companies and cloud service providers use cloud datastores to store much larger volumes of data than clustered databases (*petabytes* rather than *terabytes*) across hundreds or thousands of machines, typically at the expense of strict transactional guarantees, meaning clients may see stale data. Machine failure is more common in this environment, so these systems are designed to handle a wider variety of failures automatically, compared to clustered systems that typically rely on manual intervention.

### 1.2.1 Limitations of Existing Approaches

Existing work on clustered distributed databases has focused on systems that run on relatively static sets of machines, while work on datastores typically sacrifices transactional guarantees to run on a larger, more dynamic set of machines. This thesis analyses the needs and requirements of a hybrid system, which provides strong transactional guarantees and must be resilient enough to run on an unreliable and dynamically changing set of machines, automatically and without administrator intervention.

## 1.3 THESIS CONTRIBUTION

This thesis evaluates the viability of interactive workstation-based systems through the design of an autonomic, resource-aware interactive workstation-based database system, and the implementation of a subset of this design. It makes a number of contributions, including a taxonomy of distributed database systems design, a set of requirements of an interactive workstation-based system, a design that meets these requirements, and an evaluation of the resulting system. This system was implemented for this thesis, but is partially based on an existing database system named H2 [87].

The taxonomy of modern database systems architectures highlights the heterogeneity of existing solutions and illustrates the trade-offs involved made in their design. Modern distributed database systems are reviewed and analysed in the context of this taxonomy, along with other relevant work on distributed systems. In addition, the concept of *interactive* workstation-based systems is introduced along with an analysis of existing *non-interactive* workstation-based systems.

Having introduced the concept of interactive, workstation-based systems, the thesis presents the requirements for such systems. These requirements are based on the perceived needs of workstation-based systems and of workstation-based databases more specifically.

The primary contribution of this thesis is the knowledge gained in the design of an autonomic, resource-aware, interactive workstation-based database system named D2O. A part of this design is implemented and evaluated to test the viability of workstation-based systems in terms of query performance and fault tolerance.

The designed system, D2O, is a workstation-based system, which distinguishes it from existing databases, which typically fall into a number of discrete categories: *clustered systems* that run inside a single machine room and are optimized for high throughput ACID transactions, and *cloud databases* and *datastores* that run in and across data centres, providing high availability, but offering fewer transactional guarantees. D2O provides the same transactional guarantees as clustered database systems, but is designed to run over a more dynamically changing set of machines than traditional clustered databases.

Existing approaches to handling node failure in clustered environments typically require downtime or manual intervention[1–3]. Some systems are able to continue to operate after the failure of a small number of nodes, but are designed to run in relatively static environments where the set of nodes in the system rarely changes and failed nodes are quickly replaced [4-5]. D2O is designed to meet a further challenge, where the set of machines running the system is dynamic, and the system cannot rely on manual intervention to restart failed machines. It is designed to adapt as the set of available machines and usage patterns change over time.

An implementation of D2O — H2O — which implements the database and fault tolerant components of D2O (it is not autonomic or resource-aware), is created for the purpose of evaluating part of D2O's design. H2O is partially based on an existing non-distributed database system named H2 — H2 is extended to support distributed transactions and replication.

Two evaluations are presented. The first evaluates the performance of H2O in comparison to two existing clustered database systems. The second evaluates H2O's response to node failure in a series of tests. These tests are run on a machine cluster, but they are designed to simulate the types of failure that may often occur in a workstation-based system. Both sets of evaluations are used to evaluate the viability of workstation-based systems.

## 1.4 Thesis Structure

This chapter, *Chapter 1*, has introduced the motivations behind this work and the area it covers. The remainder of the thesis is structured as follows.

*Chapter 2 (Background Concepts)* presents a summary of distributed database systems architectures, and explains how these architectures determine the scale and effectiveness of these systems under different workloads.

*Chapter 3 (Related Work)* presents a survey of modern distributed database systems, which are analysed based on the terminology and concepts introduced in the previous chapter. It concludes with an analysis of the limitations of existing systems, and motivations for looking further at interactive, workstation-based systems.

*Chapter 4 (Requirements)* lists the requirements for an interactive workstation-based database system.

*Chapter 5 (Design)* presents the design of D2O, the database system created to meet the requirements of the previous chapter. The motivations behind this design are explained through comparisons with the systems discussed in *Chapter 3* and the requirements outlined in *Chapter 4*.

*Chapter 6 (Implementation)* discusses some of the detail behind H2O, the implementation of D2O, with particular focus on aspects that are relevant to the evaluation in the next chapter.

*Chapter 7 (Evaluation)* evaluates the effectiveness of the H2O implementation by measuring transaction throughput when running a database benchmark, and when the system is subjected to node failure. These experiments are used to evaluate the viability of H2O as an interactive workstation-based system.

*Chapter 8 (Conclusion)* analyses the initial proposition of this thesis, that interactive, workstation-based systems are viable. It also presents future work which may follow the work covered in this thesis.

# Chapter 2: Background Concepts

# 2 BACKGROUND CONCEPTS

This chapter defines the terminology and concepts used throughout this thesis to describe distributed database systems and other related work.

## 2.1 TERMINOLOGY

Authors often use different terminology to describe the same concepts. To avoid ambiguity the terminology introduced here is used consistently through the rest of this work.

A *database* is broadly defined as a software program that manages the storage, organization, and retrieval of data [29]. The majority of databases described in the following chapters are relational and provide an SQL interface, though other data models are also discussed where relevant.

A *distributed database management system (DDBMS)* is defined to be database software that is designed to be run over multiple physical machines [29]. To an application it is conceptually a single entity, though there are multiple database processes running on multiple machines. These processes are referred to as *database instances*[2].

The machines on which databases instances execute are referred to as *nodes* (consistent with common distributed systems terminology, including *DeCandia et al.* [6]). In some cases the term *workstation* is used to indicate a specific type of node which is being used primarily as a user's machine.

Each relational database instance can manage access to multiple *tables*. Where there are copies of a table stored on a number of instances each individual copy is called a *replica*.

The term *database system* is used to refer to a collection of database instances that are connected together with a single schema[3]. Each database system has a *schema*, which stores information on the tables that are currently stored across the database instances in the system.

---

[2] In database literature *database instances* are typically referred to as *database sites or database nodes* [29], however this terminology is not used here as the word *site/node* implies that the entire machine running the database is being referred to, rather than just the database process.
[3] Most database literature does not differentiate between the specific instance of a DDBMS (called a *database system* in this thesis) and the more general concept of the DDBMS itself.

These concepts are illustrated in Figure 2, below, which shows a database system containing three database instances, each on a different node. There are also two tables, *X* and *Y*, each with two replicas, and a schema with a single replica.

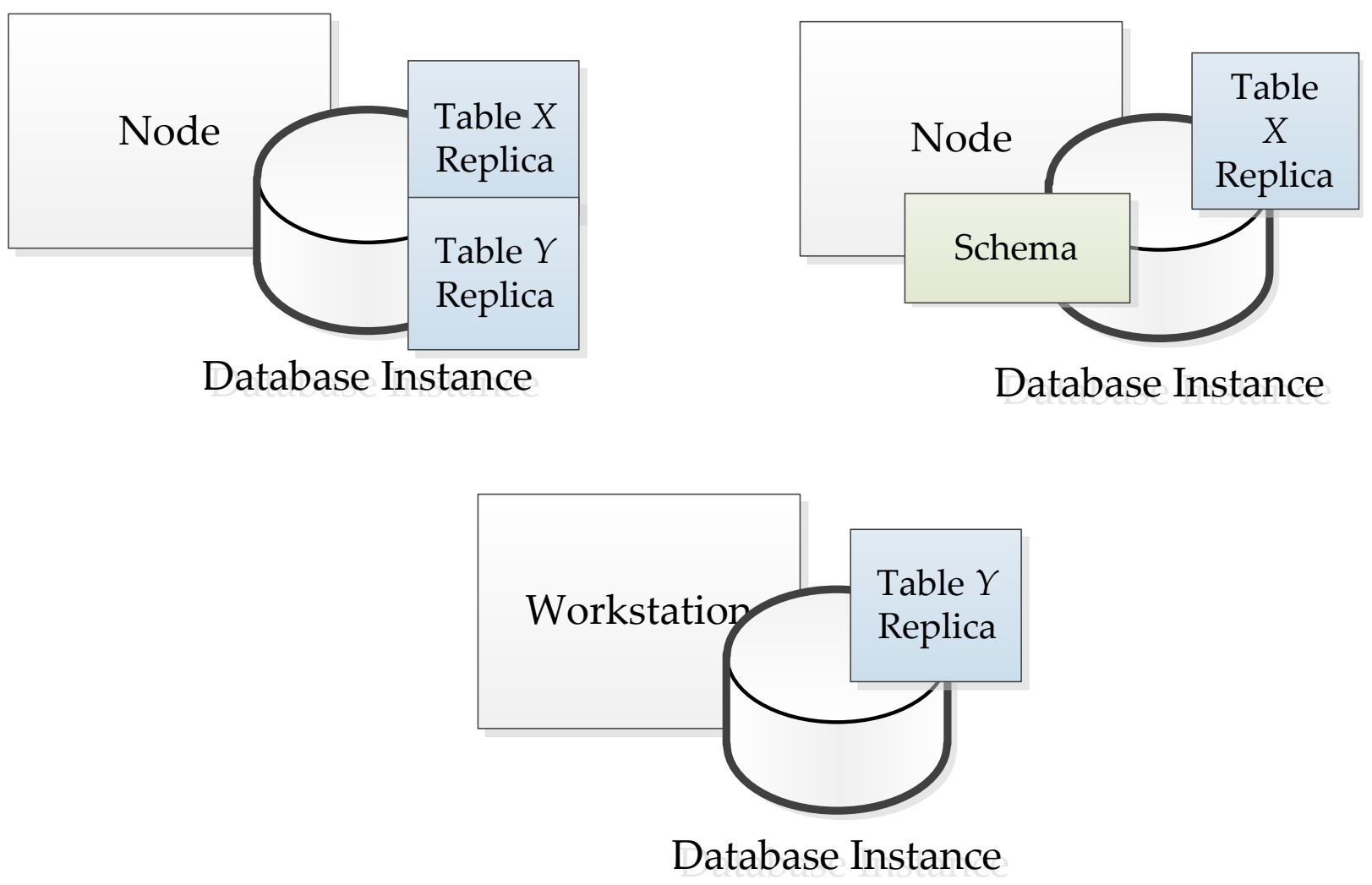

**Figure 2: Illustration of a database system with the terminology used in this thesis.**

This work focuses on the creation of a database designed to make use of *un-utilized resources* within organizations. These resources include CPU, RAM, and disk capacity on workstation machines.

## 2.2 Aspects of Distributed Database Architectures

### 2.2.1 Heterogeneity

Distributed databases can be created from many loosely coupled instances that are connected to allow applications to query over disparate data sources, or from many tightly coupled instances that are used to backup data and to scale.

A distributed database is *heterogeneous* with respect to autonomy if it connects disparate DDBMSs together, where each of these databases has its own schema[4]. Consequently, a heterogeneous system is usually middleware software — software that connects multiple distinct DDBMSs and their schemas with each other, and enables queries to be sent between them. Each database within the system has complete autonomy, and may even be running a different database implementation. Distribution is used as a mechanism for connecting disparate data sources.

In contrast, a *homogeneous* distributed database system (with respect to autonomy) comprises multiple database instances that share a single central schema. Each instance runs the same software and is able to connect to other instances through functionality provided by the database software itself, rather than external middleware. A homogeneous database is tightly integrated, so query planning and data placement decisions can be made globally. Distribution is used to spread data across many machines, to balance load and to maintain backups in case of failure.

This thesis focuses on *homogeneous distributed databases*, so all databases referred to here are homogeneous, unless otherwise stated.

### 2.2.2 Hardware

Most modern distributed databases, whether heterogeneous or homogeneous, are designed to run on off-the-shelf computers connected through a network. These types of database are described as "*shared-nothing*" [7], because each instance runs on separate nodes which share nothing — memory, processor, or disk — with each other.

[4] In database literature, heterogeneity can refer to a differing data model or access mechanism provided by a set of databases. In this thesis heterogeneity is used to refer to the autonomy of a set of databases — a distributed database is heterogeneous if it connects a number of autonomous (self-contained) databases [29].

Other databases use *shared-disk* architectures meaning multiple processors share the same disk. Shared-disk architectures are beneficial because data does not have to be sent between nodes for processing or replication [8], but they require specialized hardware which is more expensive to provision. Most modern distributed DDBMSs are *shared-nothing*.

These approaches are said to either *scale out* or *scale up*. Shared-nothing systems *scale out*, meaning they support larger workloads by adding more machines, whereas shared-disk and single machine architectures *scale up*, meaning they support larger workloads by adding more capacity and power to existing machines, or by replacing them entirely [9].

All databases described in this work are assumed to be *shared-nothing*, unless stated otherwise.

### 2.2.3 Locale

A database's architecture is partly determined by how it is intended to be deployed — the locale over which it is expected to run. For instance, in a ***clustered DDBMS*** every node is located within a local-area network. This locality makes it less costly to send data between nodes and less likely network partitions[5] will occur, compared to a wide-area distribution. Wider distributions normally connect many disparate database instances to form a wide-area heterogeneous system. Clustered databases, including *MySQL* [1] and *PostgreSQL* [2], are discussed later in 3.1.1.

Various cloud-based database architectures have become popular in recent years. ***Cloud datastores*** such as *Amazon Dynamo* [6] run across large data centres and are designed to scale with the demands of large web companies. ***Cloud-based databases*** such as *Xeround* [10] (also described as *database-as-a-service* [11]) are more similar to clustered databases in scale and operation, but they run on virtual machines in remote cloud data centres instead of locally provisioned hardware. *Amazon EC2* [12], which allows customers to create virtual machine instances on demand, is the most widely used virtual machine hosting service.

[5] A network partition exists if two or more nodes are unable to communicate with each other.

This work describes another type of system, a ***workstation-based database***, which is similar in scale to a cluster, but the database is run over workstation machines within an enterprise, rather than in a dedicated machine cluster. The network connections between workstations are often slower than clustered systems.

These architectures are summarized below in Table 1.

| Type | Distribution | Scale (# of instances) | Latency between instances |
|---|---|---|---|
| Clustered | LAN | 100's | Low |
| Workstation-based | LAN | 100's | Low |
| Wide-area | WAN | 100's | High |
| Cloud-based database | LAN (of Data Centre) | 100's | Low |
| Cloud datastore | WAN (between Data Centres) | 1000's | High |

**Table 1: Summary of databases by distribution**

This broadly categorises the architectures of modern distributed databases. The next section looks at how these systems are implemented.

## 2.3 DATABASE COMPONENTS

This section discusses the components of a typical DDBMS and outlines how they differ, based on the intended use of the database.

### 2.3.1 Fault Tolerance

The intended scale and locale of a database affects how it is designed to handle machine or network failure. In a clustered DDBMS, the chance of failure is low because machine and network failure is rare and there are a small number of machines. These systems are said to have *low membership churn*, because the set of machines that are active in the system rarely changes.

In contrast, DDBMSs spread widely over thousands of machines are more likely to experience failure[6], because there are many more machines that could fail, and the network connections between them are more unreliable. These systems are said to have *high membership churn* if there is a high rate of change in the set of machines active in the system.

The expectation of churn is important because it determines how a DDBMS handles machine and network failure. In clusters, where failure is rare, it is often acceptable to run at a slightly reduced capacity before an administrator brings instances back online. In wider-scale systems such as cloud datastores, where failure is more common, new instances are often automatically integrated into the system with no downtime.

### 2.3.2 Transactional Properties

The transactional properties of a database specify what guarantees an application has about the data it is querying or updating. Many databases are termed *ACID compliant*, meaning they guarantee that transactions will be *atomic*, *consistent*, *isolated*, and *durable*[7]. This strict set of transactional properties is required for many applications, but can be costly to implement. For example, to maintain atomicity, co-ordination is needed between instances executing distributed transactions (discussed in the next section). Other databases provide looser guarantees which speed up transactions, but which are not suitable for many applications.

---

[6] The chance that an individual machine fails may be the same as in a cluster, but since there are many more machines, the system observes many more failures.

[7] To be *atomic*, a transaction must either complete in full or fail in full, leaving database state unchanged. A transaction must take the database from one valid state to another to be *consistent*, cannot see or modify data that exists as part of another uncommitted transaction to be *isolated*, and must not lose these changes once it has committed to be *durable*.

The most common alternative to ACID is *eventual consistency*, in contrast to the *immediate consistency* provided by ACID. Eventual consistency guarantees that for each replica of data an update will either eventually reach the replica or the replica will be removed [13]. However, it is possible that a query may read old data — data that is inconsistent with the current state of the database. This sacrifice is made to improve the speed of updates and the availability of data, and is discussed in section 2.3.6.

***Distributed Atomic Commit***

Databases supporting ACID transactions must be able to update data atomically, which can be particularly costly in distributed transactions where multiple replicas are involved. This section discusses *consensus protocols*, which are used to support atomicity by ensuring that every instance reaches consensus on the command to be executed, either *commit* or *rollback*, on replica sites. The most common of these — *two-phase commit*, *three-phase commit*, and *Paxos* — are discussed here.

Two- and three-phase commit are called centralized consensus protocols because they use a central transaction co-ordinator to manage the commit process.

Two-Phase Commit

Two-phase commit has a *prepare* phase and a *commit* phase [14]. It works as follows:

1. The co-ordinator sends a *PREPARE* message to participants in a transaction.
2. Each participant responds by indicating their readiness to commit. To provide durability, the update must be written to disk before this message is sent.
3. On receipt of a positive response to the *PREPARE* message from each participant, the co-ordinator sends a *COMMIT* message to all participants.
4. Each participant commits the update locally.

If an update fails prior to the *PREPARE* message being sent, one or more participants may respond with a *ROLLBACK* message, or they may never respond in the case of network or node failures. When the co-ordinator receives a *ROLLBACK* message or 'times out' waiting for a response, it sends *ROLLBACK* to all participants in place of the *COMMIT* message.

In the event of node failure after acknowledgement of the *PREPARE* message, an update can proceed because it has been written to disk. If the failed node restarts, it communicates with the co-ordinator to check that a *COMMIT* message was sent, and then commits the update. In the event of a catastrophic failure, where the node will never restart[8], the co-ordinator eventually removes the failed node from the set of active nodes, meaning it does not try to send further updates to the failed node.

Two-phase commit is a blocking protocol, meaning participants block resources while waiting for messages from the co-ordinator. This ensures that instances always reach the same decision, even when the co-ordinator fails midway through a transaction, as illustrated below.

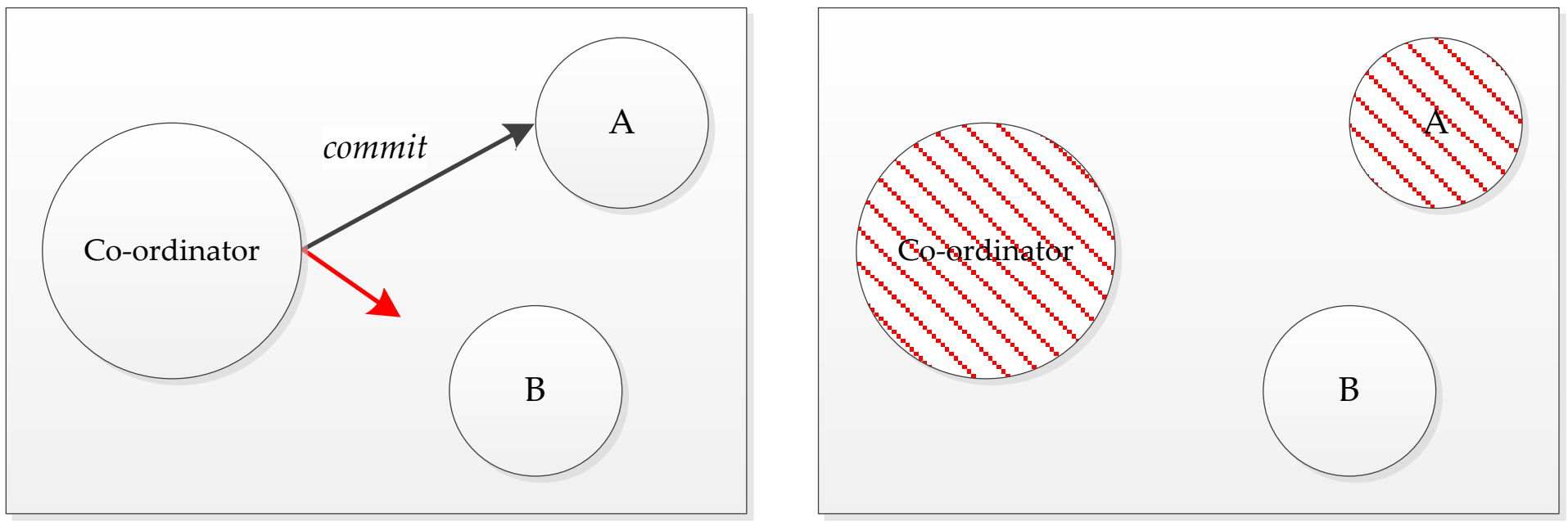

(i) Co-ordinator sends commit to *A*, then fails before it can be sent to *B*.

(ii) *A* commits transaction, then fails. *B* blocks because it doesn't know whether to commit or rollback

**Figure 3: An example of the two-phase commit blocking after node failure.**

In this illustration the co-ordinator has begun sending *COMMIT* messages to each machine involved in a transaction. *Machine A* has received the *COMMIT* message, but the co-ordinator has failed before sending the *COMMIT* message to *B*. Instance *B* is not able to commit, because it has not received the message to do so, but it is also not able to rollback, because it is possible that *A* has already committed the transaction. Instance *A* may have failed after committing the transaction, so *B* cannot contact it to determine the state of the transaction. This means that *B* must block until a new co-ordinator is started and has contacted all participating nodes, otherwise it is possible for some nodes to commit while others abort.

[8] A node may never restart if it is damaged (e.g. the hard disk fails), or destroyed. In contrast, a node which suffers a non-catastrophic failure may be able to restart at a later point.

### Three-Phase Commit

An alternative is the three-phase commit protocol, which is a non-blocking approach [14]. A third phase, the *pre-commit*, is added between the original *prepare* and *commit*. As with two-phase commit, participants respond to a *PREPARE* message indicating whether they intend to commit or abort. When they receive a *PRE-COMMIT* message they know that all participants have agreed to commit (the abort case is the same as two-phase commit) and they respond to acknowledge that the message was received. When all participants have responded to the *PRE-COMMIT* message, a *COMMIT* message is sent just as with two-phase commit.

The addition of a pre-commit phase means that every participant is aware of the global decision to commit prior to the first participant committing. This means that all participants can independently commit if the co-ordinator fails, rather than blocking [15].

### Paxos

*Paxos* is a decentralized consensus algorithm, meaning it has multiple co-ordinators (unlike two- and three-phase commit), and needs only a majority of these to be operational to reach consensus [16–18]. In the standard *Paxos* algorithm, designed for distributed consensus rather than atomic commits, *Paxos* allows a set of nodes to agree upon a single value. *Paxos* guarantees *safety*, meaning only a single value can ever be chosen, but does not guarantee *progress*, since a value may not be chosen if a majority of nodes are unavailable.

A *Paxos* node can take on any or all of three roles: *proposer*, *acceptor*, and *learner*. A *proposer* proposes a value that it wants agreement upon. It does this by sending a proposal containing a value to the set of all *acceptors*, which decide whether to accept the value. Each acceptor chooses a value independently — it may receive multiple proposals, each from a different *proposer* — and sends its decision to *learners*, which determine whether any value has been accepted. For a value to be accepted by *Paxos*, a majority of acceptors must choose the same value. In practice, a single node may take on many or all of these roles, but in the examples in this section each role is run on a separate node, as illustrated below.

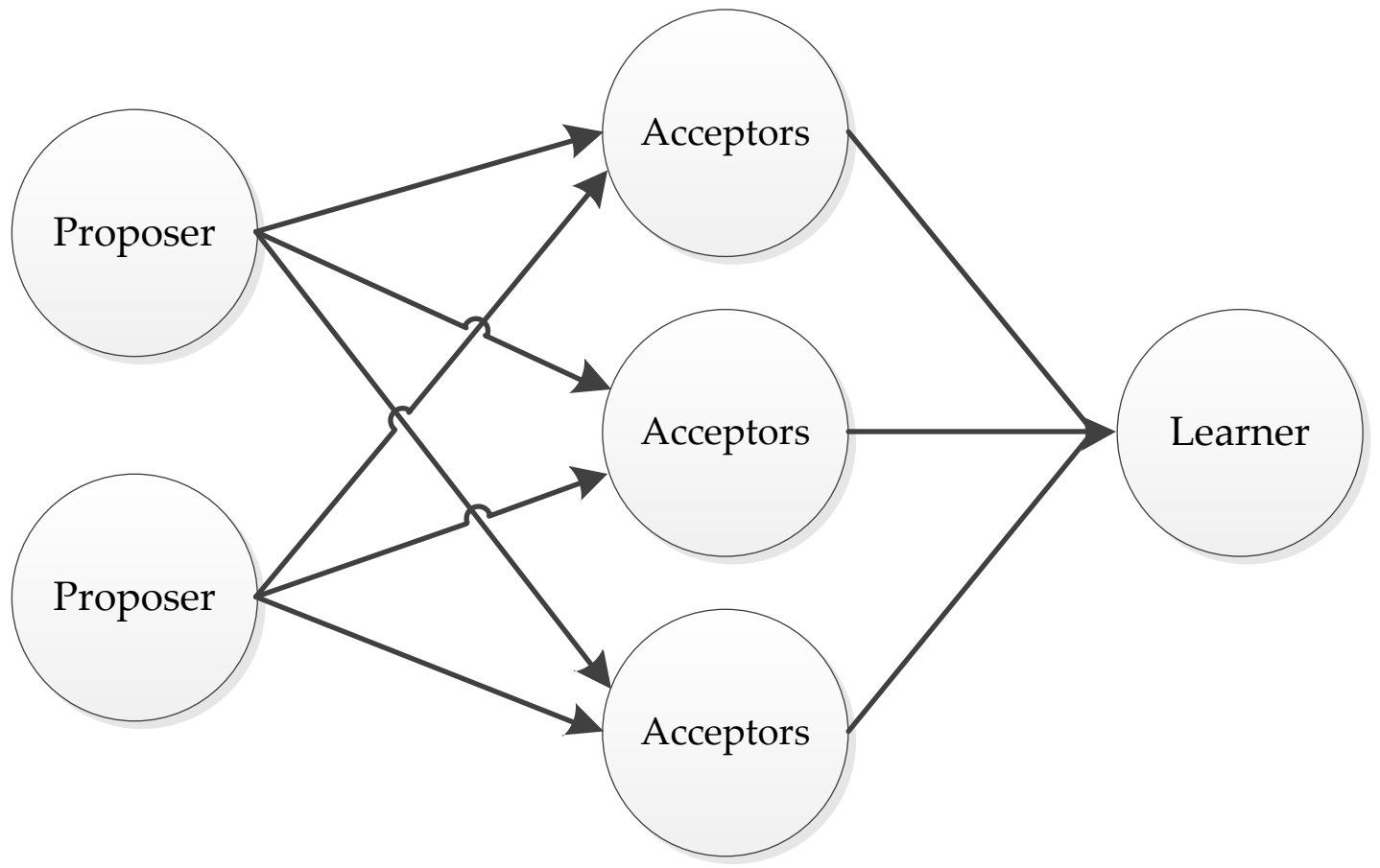

**Figure 4: Basic *Paxos* architecture. A number of *proposers* make proposals to *acceptors*. When an *acceptor* accepts a value it sends the result to learner nodes.**

*Standard Paxos Algorithm*

In the standard *Paxos* algorithm proposers send two types of messages to acceptors: *prepare* and *accept* requests. In the first stage of this algorithm a proposer sends a *prepare request* to each acceptor containing a proposed value, *v*, and a proposal number, *n*. Each proposer's proposal number must be a positive, monotonically increasing, unique, natural number, with respect to other proposers' proposal numbers[9].

In the example illustrated below, there are two proposers, both making prepare requests. The request from *proposer A* reaches *acceptors X* and *Y* before the request from *proposer B*, but the request from *proposer B* reaches *acceptor Z* first.

[9] The method of ensuring the uniqueness of proposal numbers when there are multiple proposers is not specified in the Paxos algorithm itself.

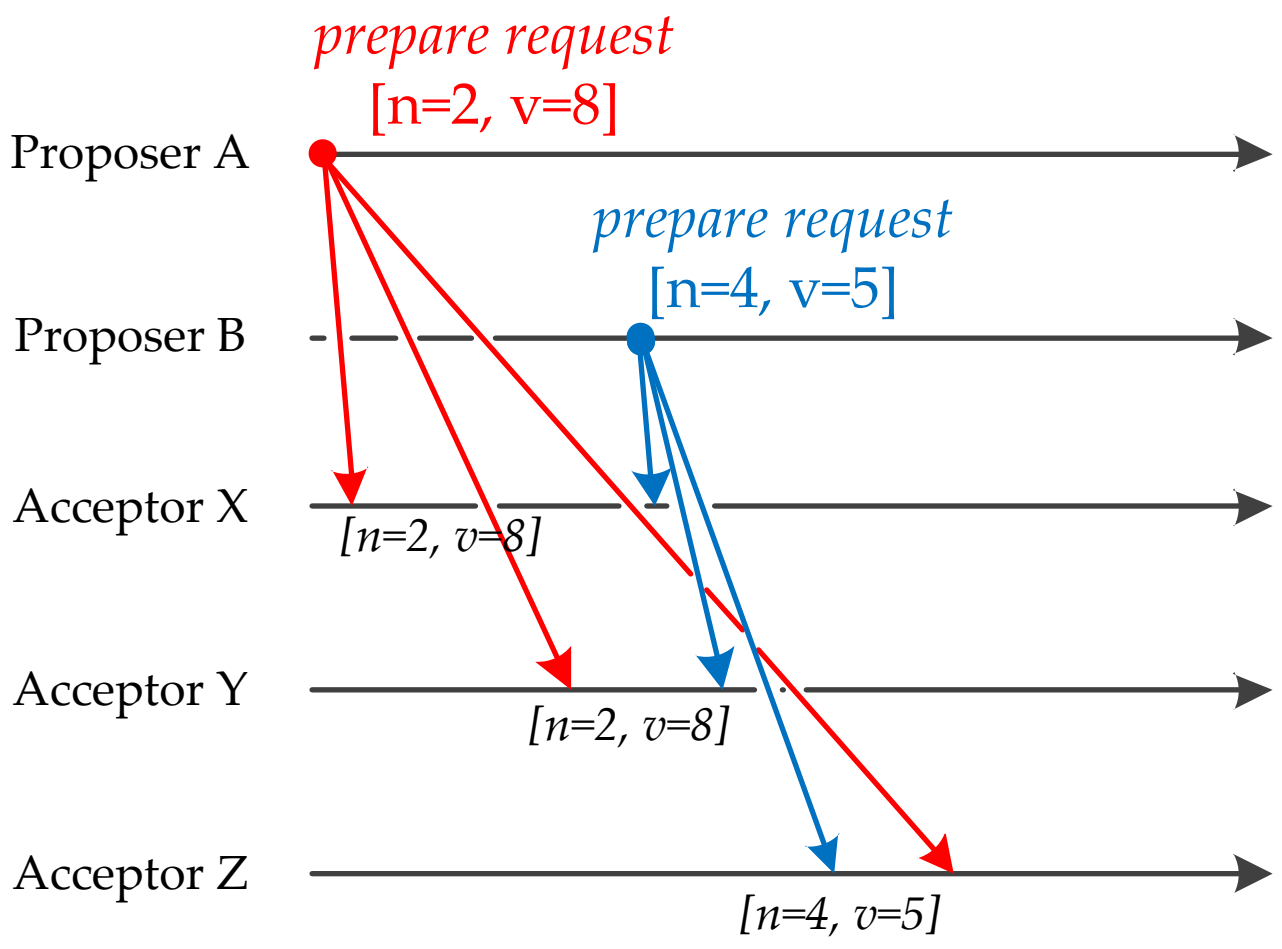

**Figure 5: Paxos. *Proposers A* and *B* each send prepare requests to every *acceptor*. In this example *proposer A*'s request reaches *acceptors X* and *Y* first, and *proposer B*'s request reaches *acceptor Z* first.**

If the acceptor receiving a *prepare request* has not seen another proposal, the acceptor responds with a *prepare response* which promises never to accept another proposal with a lower proposal number. This is illustrated in Figure 6 below, which shows the responses from each acceptor to the first *prepare request* they receive.

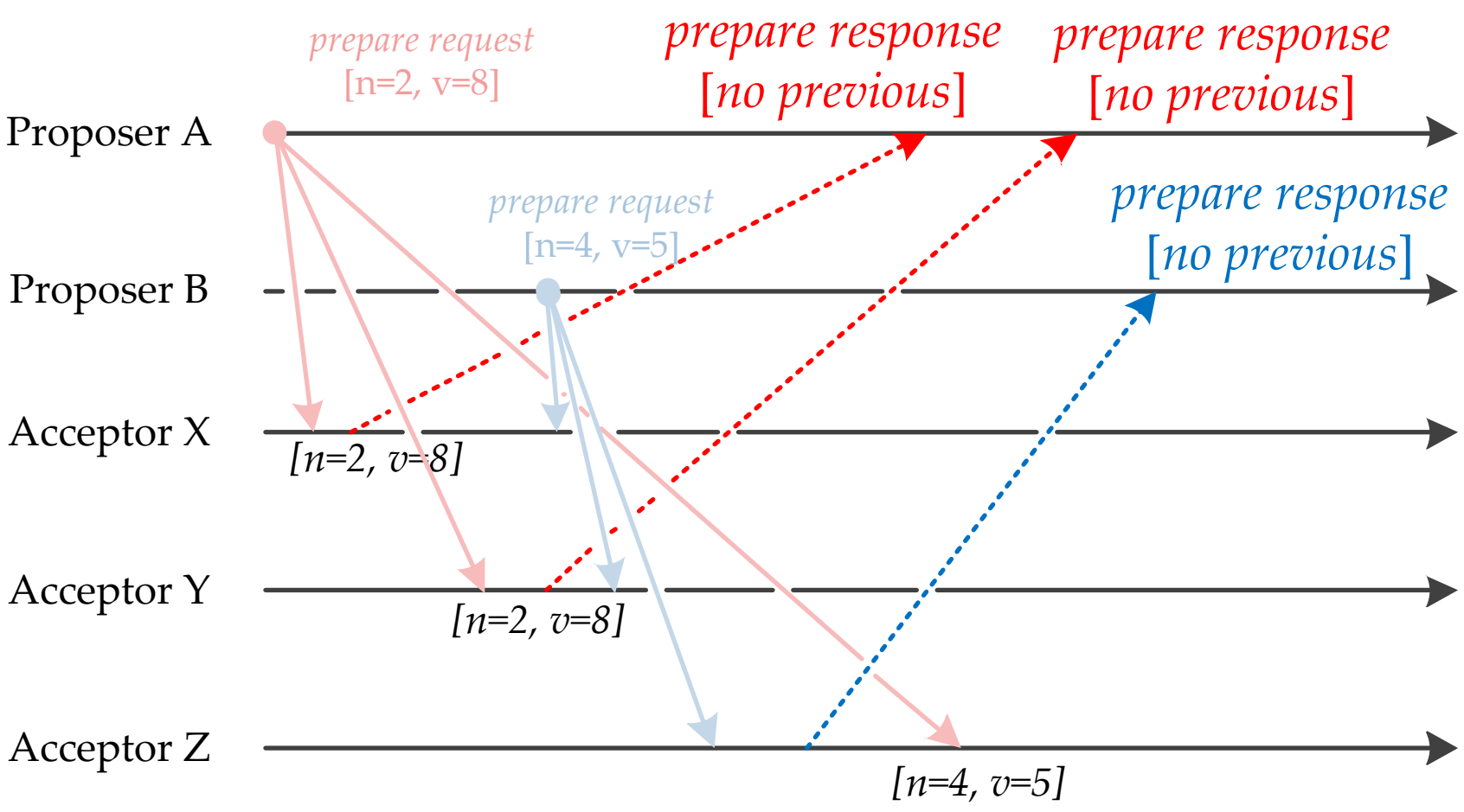

**Figure 6: Paxos. Each *acceptor* responds to the first *prepare request* message that it receives.**

Eventually, *acceptor Z* receives *proposer A's* request[10], and *acceptors X* and *Y* receive *proposer B's* request. If the acceptor has already seen a request with a higher proposal number, the *prepare* request is ignored, as is the case with *proposer A's* request to *acceptor Z*. If the acceptor has not seen a higher numbered request, it again promises to

[10] It may not, but the algorithm is resilient to this.

ignore any requests with lower proposal numbers, and sends back the previous highest proposal number that it has seen along with the value of that proposal. This is the case with *proposer B's* request to *acceptors X* and *Y*, as illustrated below:

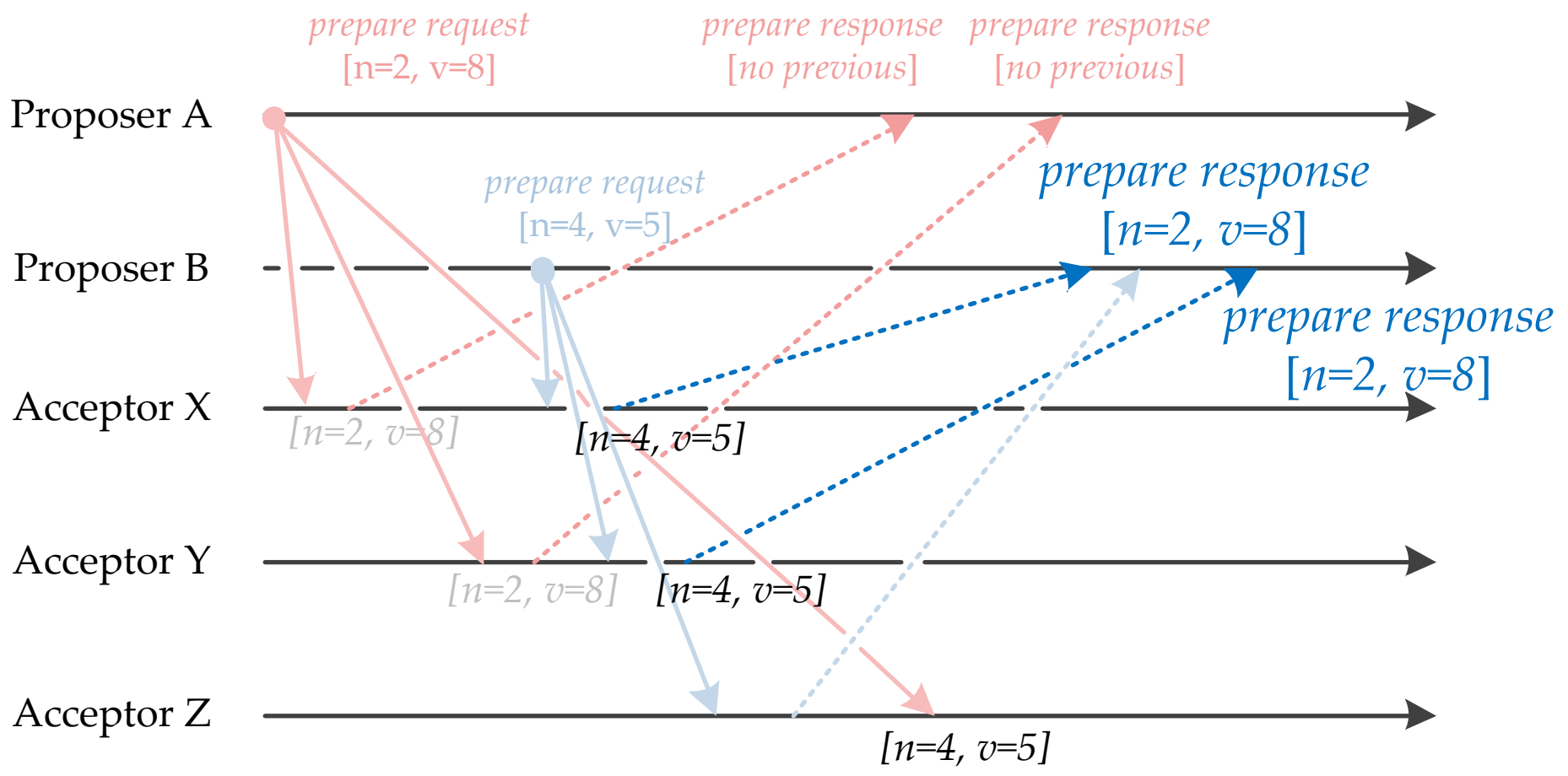

**Figure 7: Paxos. *Acceptor Z* ignores *proposer A's* request because it has already seen a higher numbered proposal (4 > 2). *Acceptors X* and *Y* respond to *proposer B's* request with the previous highest request that they acknowledged, and a promise to ignore any lower numbered proposals.**

Once a proposer has received *prepare responses* from a majority of acceptors it can issue an *accept request*. Since *proposer A* only received responses indicating that there were no previous proposals, it sends an *accept request* to every acceptor with the same proposal number and value as its initial proposal (*n*=2, *v*=8). However, these requests are ignored by every acceptor because they have all promised not to accept requests with a proposal number lower than 4 (in response to the *prepare request* from *proposer B*).

*Proposer B* sends an *accept request* to each acceptor containing the proposal number it previously used (*n*=4) and the value associated with the highest proposal number among the *prepare response* messages it received (*v*=8)[11]. Note that this is not the value that proposer *B* initially proposed, but the highest value from the *prepare response* messages it saw.

[11] Note that this is the highest proposal number that it ***received*** from *prepare response* messages. In this example, *proposer B* has a higher numbered proposal (*n*=4) than *proposer A* (*n*=2), but it has only ***received*** *proposer A's* proposal in response to its *prepare request*. If no previous proposals were returned by the *prepare response* messages, *proposer B* would use its own proposal (*n*=4).

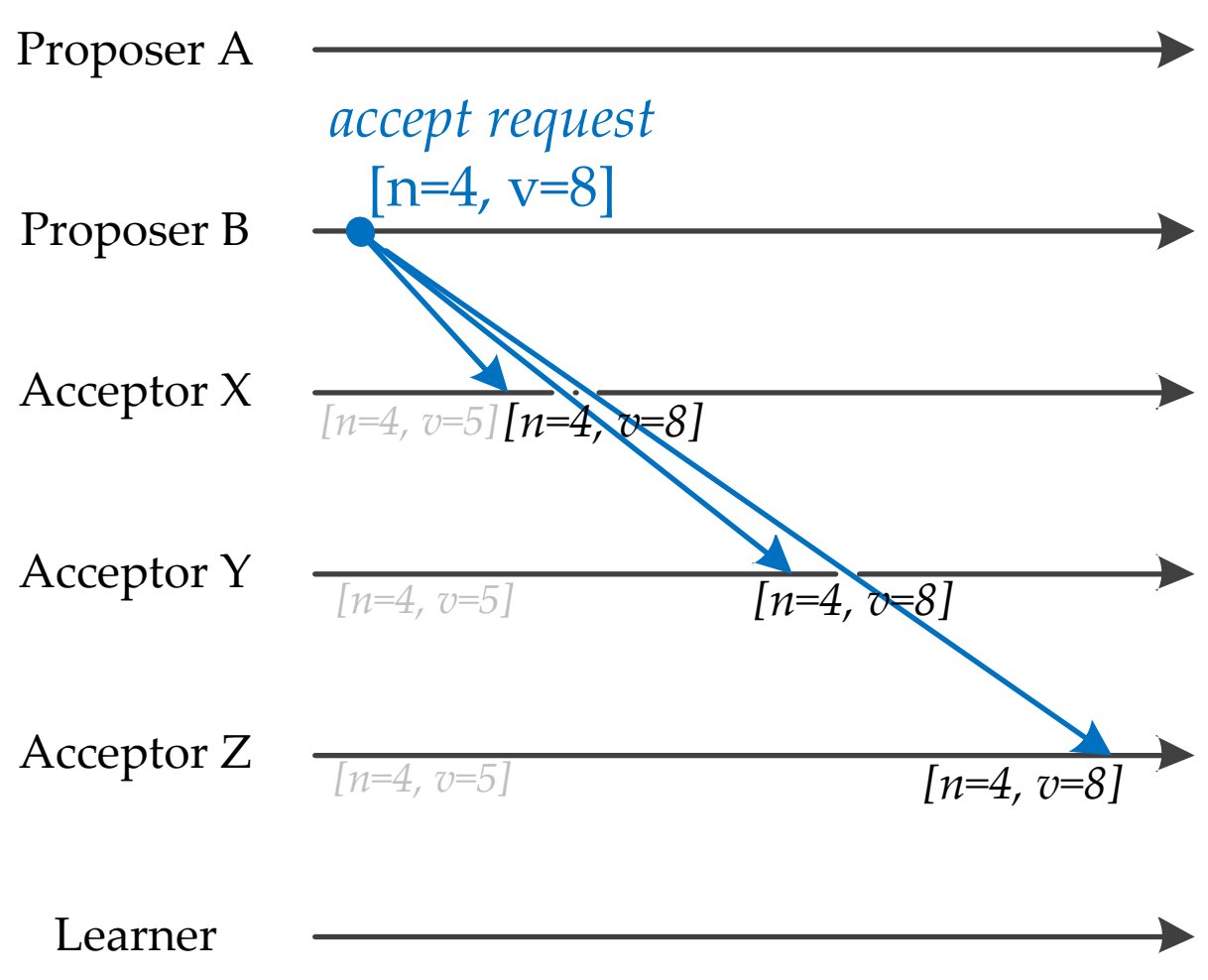

**Figure 8: Paxos. *Proposer B* sends an accept request to each *acceptor*, with its previous proposal number (4), and the value of the highest numbered proposal it has seen (*8*, from *[n=2, v=8]*).**

If an acceptor receives an *accept request* for a higher or equal proposal number than it has already seen, it accepts and sends a notification to every learner node. A value is chosen by the Paxos algorithm when a learner discovers that a majority of acceptors have accepted a value, as is illustrated below:

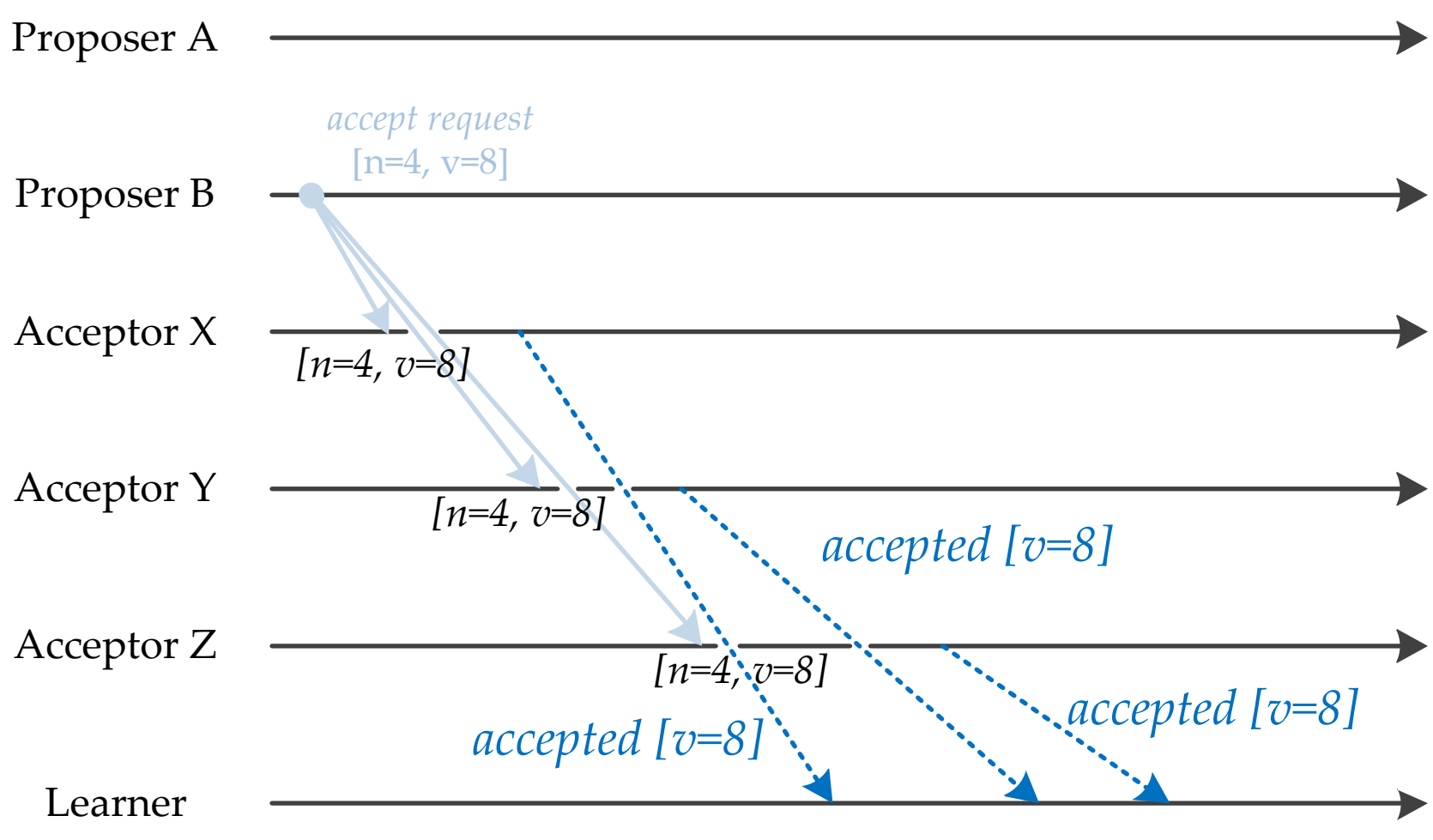

**Figure 9: Paxos. Each *acceptor* notifies every *learner* of the accepted value. It is possible to have multiple *learners*, though only one is used in this example.**

Once a value has been chosen by Paxos, further communication with other proposers cannot change this value. If another proposer, *proposer C*, sends a *prepare request* with a higher proposal number than has previously been seen, and a different value (for example, *n*=*6*, *v*=*7*), each acceptor responds with the previous highest proposal (*n*=*4*,

*v=8*). This requires *proposer C* to send an *accept request* containing `[n=6, v=8]`, which only confirms the value that has already been chosen. Furthermore, if some minority of acceptors have not yet chosen a value, this process ensures that they eventually reach consensus on the same value.

Various efficiency improvements to the standard *Paxos* algorithm are discussed in [16], [17]. For example, a *prepare* request is not necessary if the proposer knows that it is the first to suggest a value. The proposal for such a request is numbered *0*, so that it will be ignored if any higher numbered requests have been received.

*Paxos Atomic Commit*

The Paxos algorithm described above can be used to guarantee atomic commits with some modifications [19].

The transaction co-ordinator sends a message to every replica site asking whether they intend to *prepare* or *abort* a transaction. Each replica site acts as a *proposer* ***in its own instance of the Paxos algorithm,*** meaning they propose values to their own set of acceptors. A node sends an *accept* message to each *acceptor* with a proposal numbered *0* and a value of either *prepared* or *aborted*. Each *acceptor* then sends its response to the transaction co-ordinator, which acts as a *learner* for all instances of the Paxos algorithm, as illustrated below.

The transaction manager knows of a replica site's decision to prepare or abort when it receives a response from the majority of that site's *acceptors*.

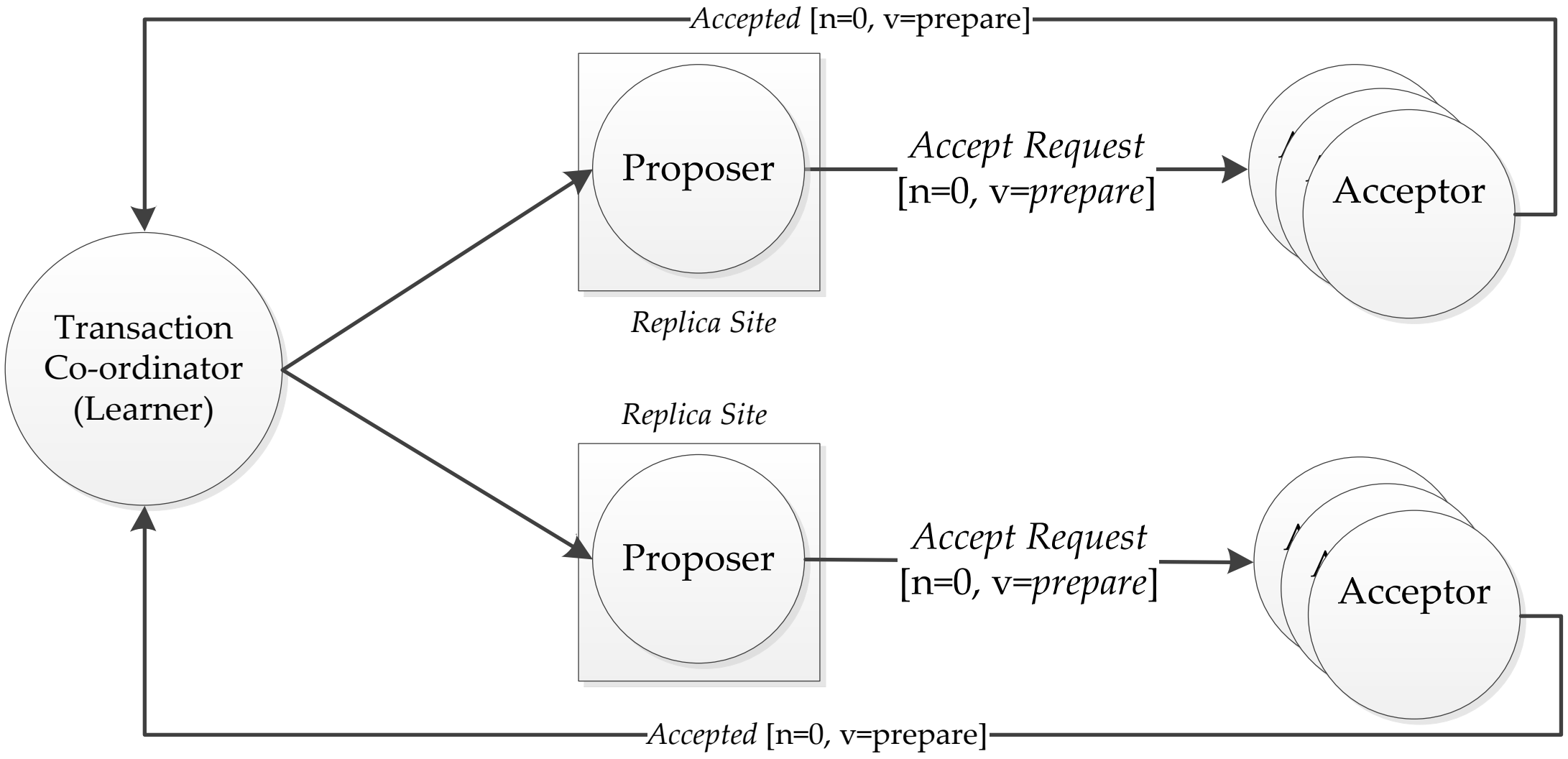

**Figure 10: The *Paxos* atomic commit algorithm in action. There are two replica sites, each running their own instance of Paxos, choosing a value of either prepared or aborted.**

*Two-phase commit* has been shown to be the degenerate case of the *Paxos* atomic commit, using only a single acceptor rather than a quorum [19]. The transaction co-ordinator in *two-phase commit* corresponds to the composite of an acceptor and a transaction co-ordinator in *Paxos*. *Paxos* is more resilient to failure in this case because the transaction co-ordinator can fail and be restarted elsewhere — it can recover lost state by contacting a majority of acceptors to discover whether a transaction was committed.

**Summary of Atomic Commit Protocols**

While these protocols are necessary for atomic transactions, they introduce an additional overhead in network communication compared to approaches which do not guarantee atomicity.

*Paxos* requires a minimum of $(2F + 3)N - 1$ messages, where $N$ is the number of participants and $F$ is the number of acceptors. In contrast, *two-phase commit* requires $3N - 1$ messages[12], and *three-phase commit* requires $5N - 3$ messages[13] [19].

[12] *3N-1* is a sum of: *1* message sent by a replica site to start the two-phase commit, *N-1* messages to send a prepare message to every other replica site, *N-1* prepare responses, and *N* commit messages [19].

Clustered DDBMSs typically use either *two-* or *three-phase commit* because it is rare for blocking to occur in practice [15], whereas *Paxos* is mostly used on cloud datastores where network partitions are considered more likely.

[13] *5N-3* is a sum of: the *3N-1* messages required by two-phase commit plus an additional two sets of *N-1* messages to send and respond to a pre-commit message.

### 2.3.3 Replication

Database replication is the process of making a copy of a table onto multiple instances. Homogeneous DDBMSs do this to improve their resilience both to failure, as it prevents the system from losing data when a single instance fails, and to scale, as it allows load to be spread over many machines. However, with replicas on many machines it is a challenge to ensure that data is consistent.

This section discusses the effect of using atomic commit protocols such as *two-phase commit* and *Paxos* to guarantee consistency, and the situations in which they should not be used.

#### *Approach to Updates*

The difficulty of maintaining consistency in distributed transactions is best highlighted by showing how a system responds to network partitions, so this section describes the way various approaches work in the event of a network partition. A system is considered inconsistent if two replicas of the same table do not contain the same data at some point in time.

Consider the network partition illustrated in Figure 11, where a group of two database instances, *B* and *C*, are partitioned and unable to contact the other database instance, *A*. Replicas of the table *X* are stored on every instance.

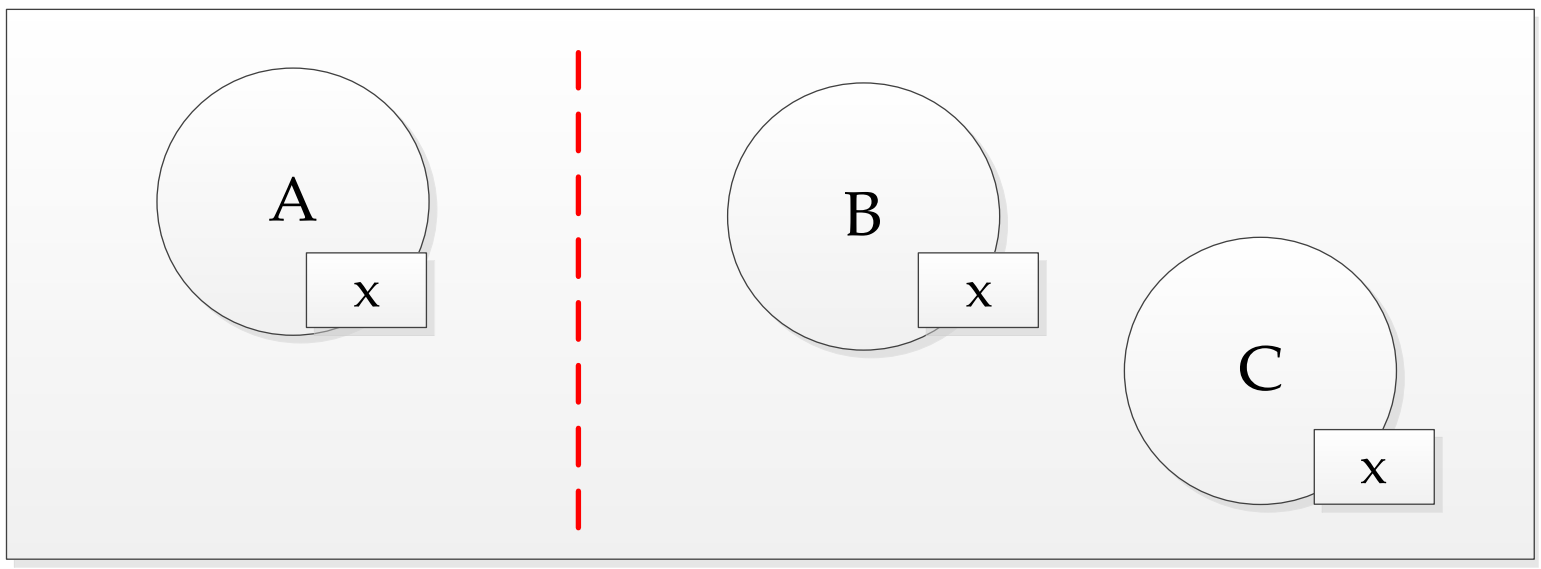

**Figure 11: An example of a network partition in a three-node system.**

There are three ways that databases typically handle updates in the presence of partitions such as this.

(1) Allow updates to be made to all copies of the table independently, meaning data is always available to any application that can contact an instance, but it may become inconsistent.
(2) Use a *central lock manager* to ensure consistency. Only nodes with access to the lock manager are able to read or update the data; nodes in the other partition cannot obtain a lock, so they cannot access any data.
(3) Use a *majority consensus protocol*. A majority of all nodes must be contacted to commit an update. In Figure 11, only the nodes in the right-sided partition (*B* and *C*) are able to do this.

With option two, nodes must be able to contact the lock manager for both reads and writes, whereas with option three, they must be able to obtain a majority consensus on writes, but reads may be allowed on the minority partition (though this allows reads of stale data, so not all systems permit it).

***CAP Theorem***

The *CAP Theorem* [20][21] formalizes these choices, stating that of three properties — *consistency*, *availability*, and *partition tolerance* — it is only possible to have at most two at a given point in time.

Consistency in this case refers to *atomic consistency*. This property exists provided that, to an application, each update operation appears as if it is completed in a single instant, even if multiple replicas are updated. The state of the database is inconsistent if there is a period where one replica has a different value to another.

Availability is the property that *"every request received by a non-failing node in the system … results in a response"* [21]. In this context availability is a global property, so the system is considered unavailable if one node in the system is unavailable as a result of a network partition.

Partition tolerance is provided if *"the network will be allowed to lose arbitrarily many messages sent from one node to another"* [21].

In the event of a network partition, a system can only be *available* (meaning all nodes are still able to answer requests) if *consistency* is sacrificed, because the *availability* property requires that every node is able to answer requests, which means that nodes on either side of a partition could commit conflicting updates. A system that maintains *consistency* in the event of a network partition must be *consistent* and *partition tolerant*, because some nodes are unavailable if they are on the wrong side of a partition. For example, if a majority consensus protocol such as *Paxos* is used, only the nodes that can reach a majority of other nodes are available.

Central lock managers, making use of protocols such as *two-phase commit*, are used to provide consistent and available systems. They do not provide partition tolerance because it is not possible for a node on one side of a partition to determine whether it needs to restart the lock manager, as it does not know if the lock manager has failed, or if it is active but partitioned. If it had failed the entire system would be unavailable until a new lock manager has been created, but a new lock manager cannot be created because the previous manager may still be running on the other side of a partition. There can only be one active lock manager, because having two makes it possible to alter data on each side of the partition, which means the system is not partition tolerant.

The three possible approaches are summarised and illustrated below:

- Available and Partition Tolerant (approach 1, possible inconsistencies)
- Consistent and Available (approach 2, central lock manager, or majority consensus protocol)
- Consistent and Partition Tolerant (approach 3, majority consensus protocol)

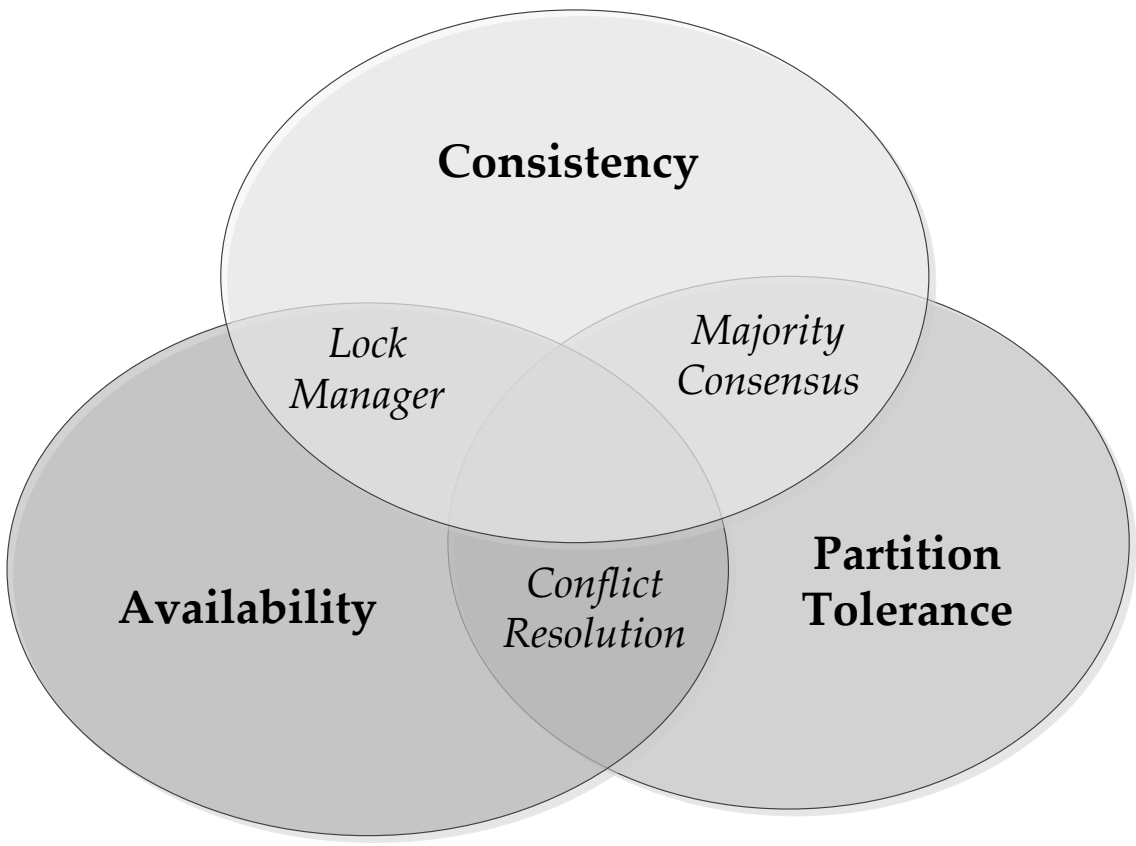

**Figure 12: A Venn diagram showing the options used by systems based on their approach to the CAP theorem.**

*PACELC*

Some researchers argue that CAP is too general because it conflates the way in which a system operates in both the presence and absence of partitions [22]. ***PACELC*** is a classification which separates the way in which a system operates in the presence and absence of network partitions: when there is a partition a system can be either *consistent* or *available*; but when there are no partitions a system can either be *consistent* or have *low latency* for queries[14]. This produces the acronym **P**artition: **A**vailable / **C**onsistent, **E**lse: **L**atency / **C**onsistent.

### Significance in Database Systems

For DDBMSs, consistency is often a rigid requirement, as it is needed to provide ACID compliant transactions[15], so these databases must sacrifice either *availability* or *partition tolerance* instead. Many clustered systems sacrifice *partition tolerance* because they can use methods such as redundant network interfaces to make the possibility of partitions very remote, whereas wide-area databases are more likely to see partitions and therefore often sacrifice *availability*.

For other databases, such as those run by large web companies, *availability* is a key requirement because even rare periods of downtime are not acceptable. These systems are also spread over many data centres, making partitions more likely. Consequently *consistency* is sacrificed, and typically replaced with the promise of *eventual consistency*, which is the best consistency guarantee possible for an *available*, *partition tolerant* solution.

---

[14] Latency is lower without consistency because the database does not need to block on locking or synchronization calls.
[15] Consistency in the context of CAP refers to *atomic consistency*, which is equivalent to the *atomicity* property in ACID [21].

CAP is used through the remainder of this thesis to motivate the design choices made by distributed DDBMSs.

***Method of Replication***

The choices made in relation to CAP determine how updates are managed — whether through a lock manager or a majority consensus protocol — and how replication is performed.

Consistent DDBMSs tend to use *synchronous* replication meaning an update must complete on every replica before its result can be returned to the application. Conversely, *asynchronous* replication, used by many eventually consistent databases, allows an update to be committed on one machine and execution started on a number of others before the result is returned[16]. Asynchronously replicated updates can be committed faster than synchronous updates because the database does not have to wait on all replicas before committing the transaction. However, if replicas are updated asynchronously, any replicas that are slow to be updated will contain out-dated data, lagging behind those replicas that were updated quickly.

Every replica in a synchronous system has the same content, so a query can access any replica to query the current version of a table. However, in an asynchronous system, each replica may have different content, because some may have completed an update while others are still processing it or are yet to receive it.

Despite this, there are two ways in which an asynchronous query system can guarantee queries enforce consistency. First, if queries are only directed at replicas that have completed the most recent update, then only current data will be accessed. This requires that the transaction co-ordinator (typically the instance that the application is connected to) knows which replicas have been updated. Alternatively, if the system requires that updates only commit once a majority of replicas have been updated, the node issuing the query can find the current version by querying every replica and using the result returned by the majority of replicas. Since a majority of replicas have the current version of the data (updates only commit once a majority has been updated), the response returned by the majority is guaranteed to be the current version.

---

[16] Some works call synchronous replication *eager replication*, and asynchronous replication *lazy replication* [95].

The time taken to execute an update is determined by the slowest replica being updated. Consequently, databases that require immediate consistency often require less than a majority of nodes for reads and updates, because this makes it possible to commit a transaction while the slowest replicas are still being updated [6].

*Replication Architectures*

Synchronous and asynchronous replication strategies require that an update is co-ordinated by a particular database instance. Some database architectures allow only a single instance to do this, in what is termed *master-slave* replication, while others allow numerous machines to co-ordinate updates, described as *multi-master* replication.

**Master-Slave Replication**

In master-slave replication one instance— the master — is responsible for handling all updates to the database. These updates are propagated to a number of other instances — slaves. If the master fails, one of the slaves can take its place as the new master.

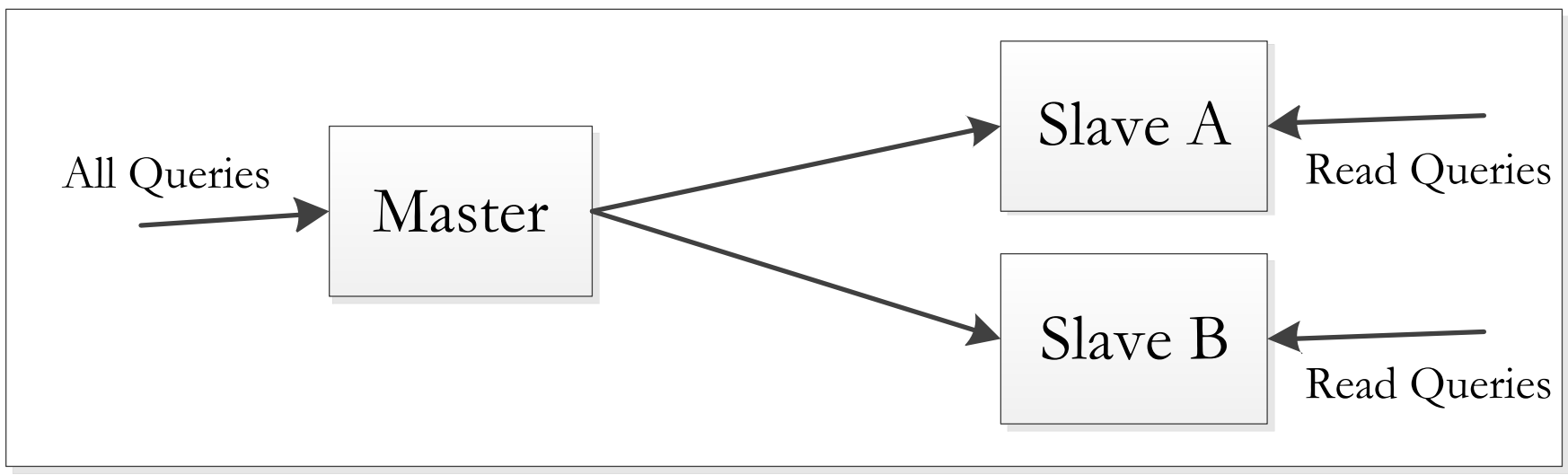

**Figure 13: A typical master-slave architecture**

Master-slave architectures scale for read-heavy traffic by allowing applications to read from any slave instance, but write to only the master (illustrated above in Figure 13). Spreading queries over slave instances reduces the load on the master instance, but it is not suitable for all applications — master-slave replication is often asynchronous, meaning under heavy load, read requests sent to slave instances may see stale data [1].

The only way of guaranteeing consistency with asynchronous replication is for reads to go through the master, which re-introduces the single-machine bottleneck into the system[17]. Synchronous replication can be used, though it is more common in multi-master architectures.

**Multi-Master Replication**

Multi-master replication allows updates to be sent to one of a number of masters, which propagates changes to the others. These masters must co-ordinate, often with a central lock manager, to ensure that conflicting updates cannot be committed. The example in Figure 14 illustrates a case where every instance is a master, though it is also possible that each of these masters have a number of slave instances.

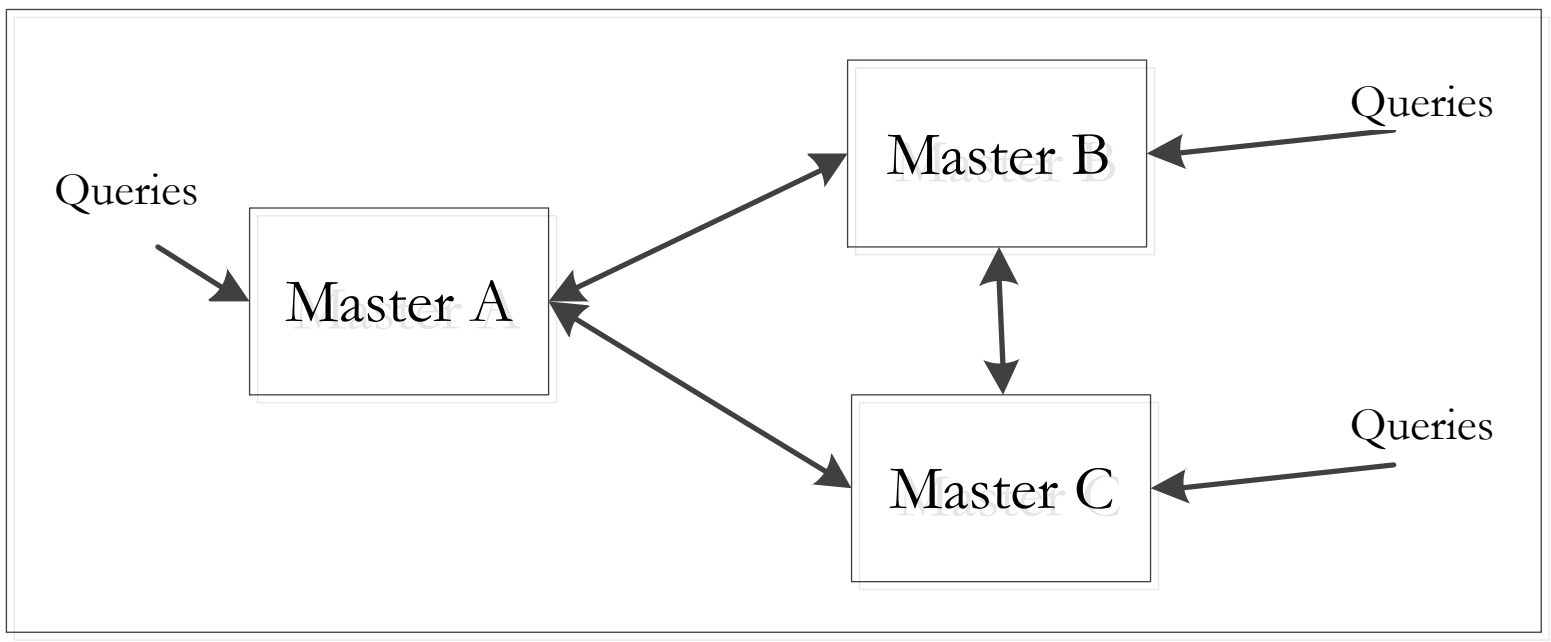

**Figure 14: A typical multi-master architecture.**

Multi-master systems typically use synchronous updates when they require that replicas are consistent, and asynchronous updates in cases where the speed of an update is more important.

[17] This assumes that there is only a single master, which is the case when the entire database is stored on every instance. An approach for using multiple masters, where each instance stores a subset of the database, is discussed in the next section.

In both approaches replicas can be updated in a number of ways. Asynchronous systems often write updates locally to a *log file* and then periodically send this file to slave instances where each update is replayed [1]. Some send *snapshots*, which represent the state of the database (or some part of it) at a point in time. Updates can also be sent either as unprocessed or compiled *SQL statements*, and re-executed on each instance.

***Granularity of Replication***

Multi-master and master-slave replication strategies describe how a *unit* of data is replicated, but not what this unit is. For some databases such as *MySQL* it is the entire database — all updates are sent to a single master instance and then propagated to one or more slaves (in the case of master-slave replication). For others, the database is split into multiple parts, with each part responsible for replicating its own data. Figure 15 illustrates the former approach (*full-database replication*) with a database instance storing three tables. The entire database is replicated to a single slave machine.

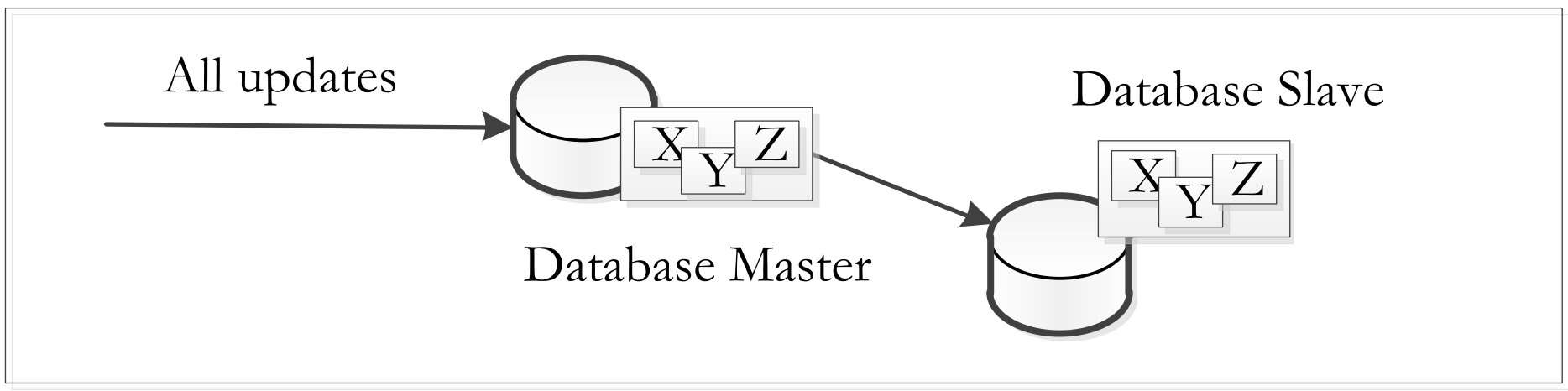

**Figure 15: An example of full-database replication in a master-slave configuration.**

The alternative to *full-database replication* is for each instance to store only a subset of the total database by dividing it into *segments*. Each segment contains a set of tables rather than the full database, so updates are sent to the master segment(s) responsible for a given table. There are now multiple masters, one for each segment, so segments can be spread over multiple machines, as Figure 16 illustrates.

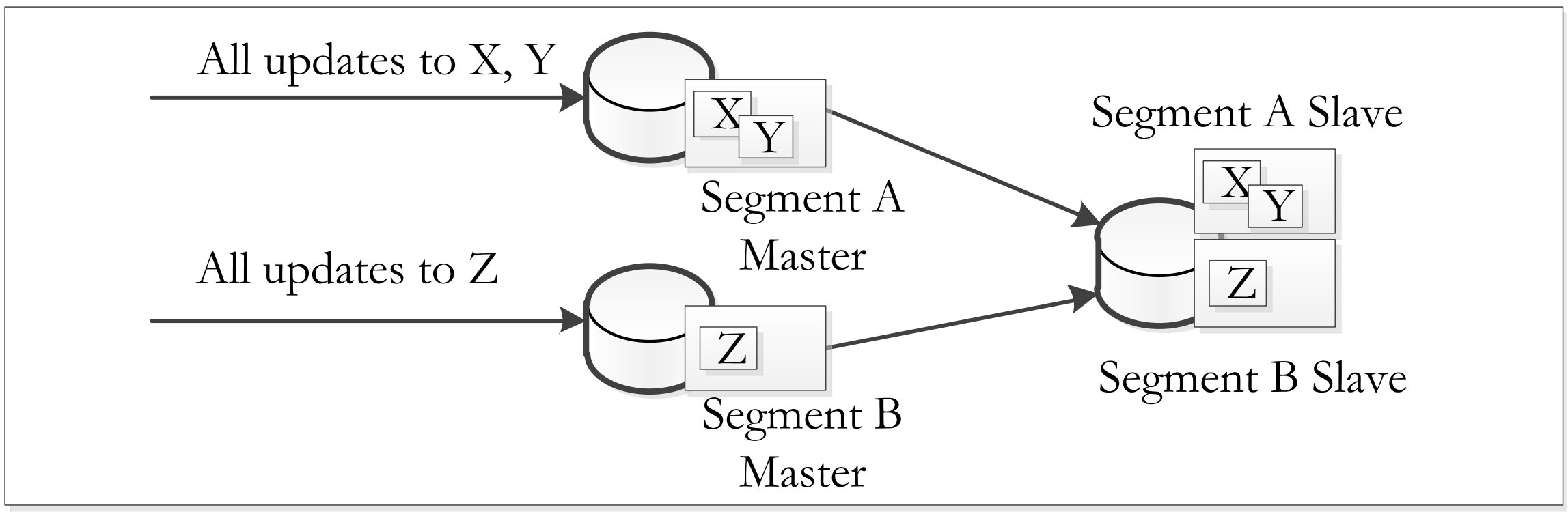

**Figure 16: An example of segment-level replication in a master-slave configuration.**

Another approach is *table-level replication*, the smallest unit of replication normally used. In this case one (or more) replica(s) of every table is designated as the master table, so updates are sent to the location of that master. Every table has its own master and its own slaves, as Figure 17 shows below.

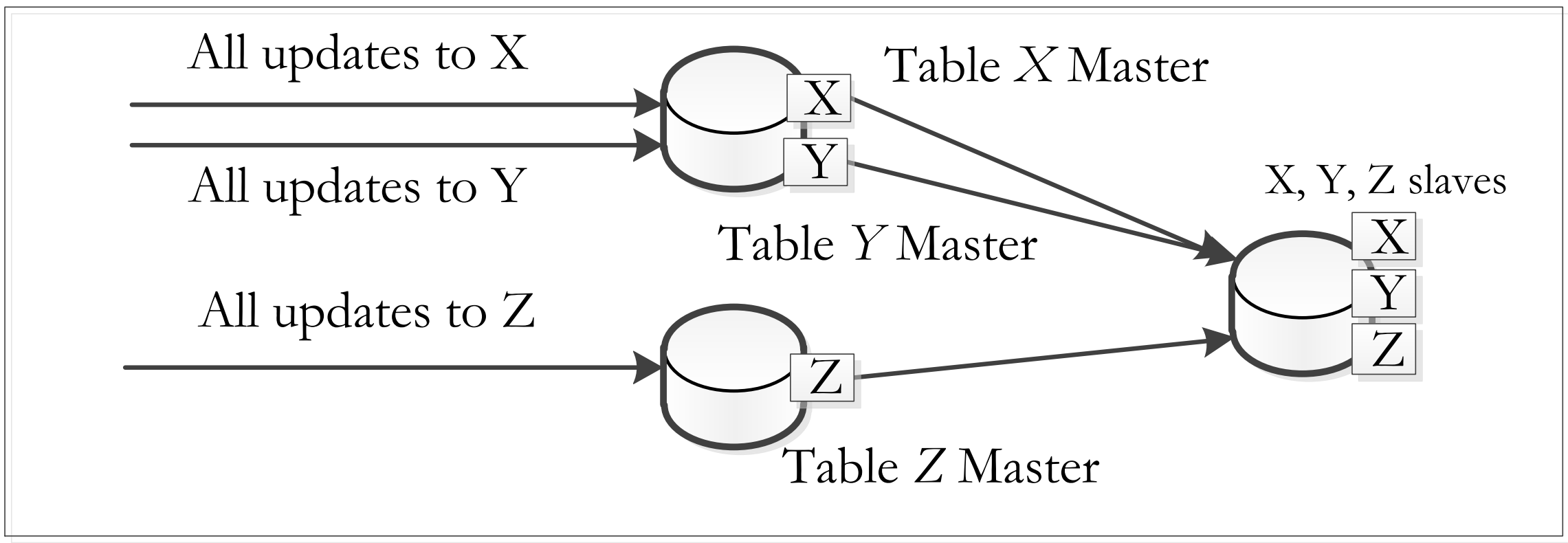

**Figure 17: An example of table-level replication in a master-slave configuration.**

For the remainder of this work these replication strategies are called *full-database*, *segment-level*, and *table-level* replication.

The smaller the unit of replication, the more control a database has over the placement of individual tables. *Full-database replication* gives almost no control, because every instance contains a copy of the entire database. *Segment-* and *table-level replication* schemes allow the database to be spread over many instances, which makes it easier to scale the database because each instance does not need to store every table. However, when tables are spread out, the database must implement a lookup mechanism to locate them for use in queries. The complexity of this lookup mechanism is dependent on the method used to place data, which is discussed in section 2.3.4.

**Sharding**

*Sharding* refers to intra-table partitioning of data, where *tables* are divided into partitions and stored on separate instances. This contrasts with the typical use of the term *partitioning* which describes how a *database* is divided by storing tables in separate locations. With a horizontal sharding scheme, shards are split based on a key, so a table containing user information may be split on surname, making separate machines responsible for surnames *A-F*, *G-M*, and *N-Z*.

Queries on a single table may span multiple shards, which can be beneficial for some queries (due to increased parallelism), but is often a disadvantage due to the increased network overhead of sending intermediate results between instances. As a result, databases such as *VoltDB* seek to avoid multi-instance transactions by co-locating related data where possible [23].

Shards are typically used with segment-level replication, as a shard or collection of shards can be used as a segment.

### 2.3.4 Placement

If a database uses *segment-* or *table-level replication* it must decide where to place data. Data can be clustered on a small number of co-located machines — giving low latency, but making it difficult to scale — or it can be spread out over many machines, producing the opposite effect. Accordingly, deciding where individual tables will be placed is a trade-off between scaling by spreading load and providing better query performance by co-locating related tables. This section describes two contrasting approaches to placement, *hashing* and *heuristics*.

***Hashing***

Hashing automates data placement by deterministically assigning a table (or set of tables) to a particular instance using a *hash function* [24].This is a function which takes an input value and produces an output — the hash value — which is the same every time that input is used (it is deterministic). The range of possible output values is called the keyspace.

In a typical system the keyspace will be large (for example, the popular SHA-256 hash function has a 256 bit keyspace), but for clarity the examples in this section explain hashing using a function with a keyspace of size 10, meaning any input to the example hash function will produce a value between 0 and 9.

The keyspace is typically illustrated as a ring, where the highest value in the keyspace 'wraps around' to the smallest value and an input is placed on this ring at the position of its hash value. In the example in Figure 18 below, four inputs (*A*, *B*, *C*, *D*) have been hashed and placed based on the hash value produced for each.

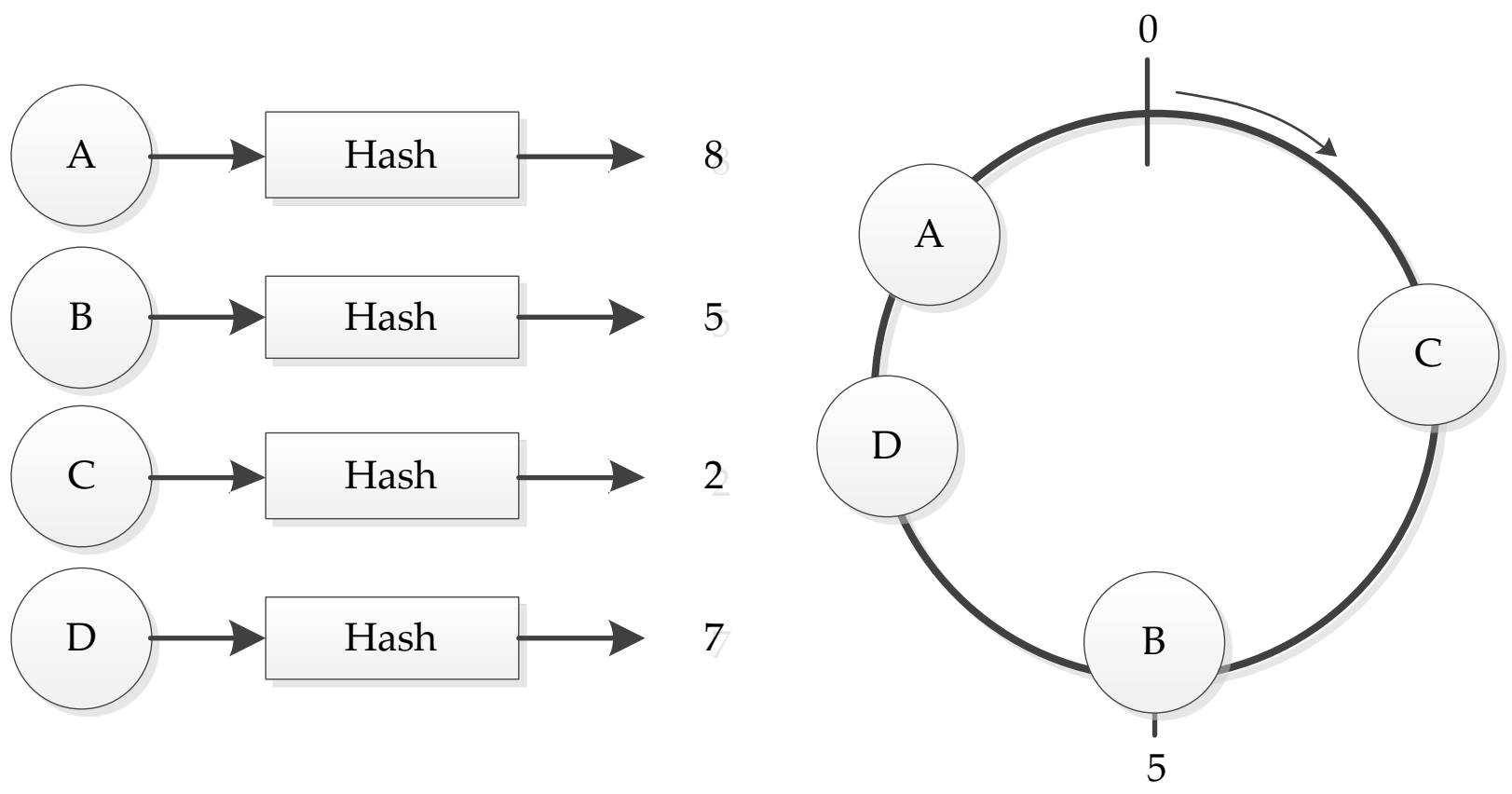

**Figure 18: An illustration of a number of objects and their position in the keyspace.**

This hashing mechanism can be used to assign tables to database instances. When each instance joins the system it is assigned a portion of the keyspace to manage. This is accomplished by hashing a value identifying the instance (typically its IP address and port) to place it in the keyspace. For example, in Figure 18, *A*, *B*, *C* and *D* could represent database instances. Each of these instances is responsible for the keyspace between itself and the previous instance in the ring, so *C* is responsible for all items with hash values between 9 and 2 (since the keyspace loops back at zero), *B* for 3-5, *D* for 6-7, and *A* for 8.

When a table is added it is also hashed on some identifier, such as its name. This is illustrated below, where, in an extension of the previous example, the tables *X* and *Y* have been added. *X* is placed on instance *A* because its hash value, 8, is in the range that *A* is responsible for. Similarly, *Y* is placed on instance *C*.

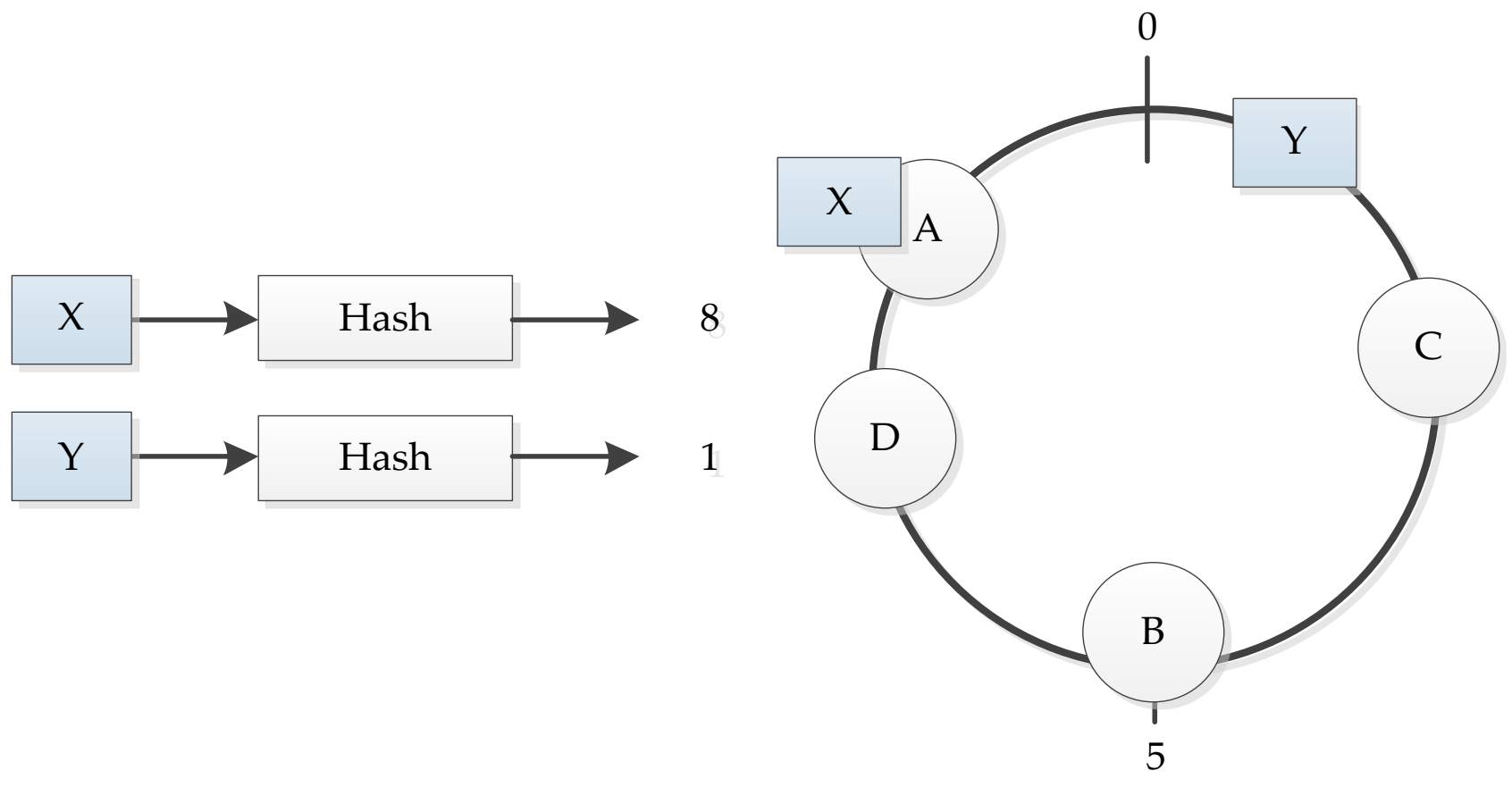

**Figure 19: Distributed hash tables. Two tables have been hashed (on their name) and given a hash value which places them in their respective positions in the keyspace.**

To perform a lookup for a table, the table's identifier is hashed again, producing the same hash value, and therefore identifying where the table can be located.

In production systems a hash function such as SHA-256 is used, giving a keyspace much larger than the above example, but using the same approach.

Because hashing automatically handles the placement of data, the lookup mechanism used to find tables is very simple; however this means that placement is entirely determined by hashing, rather than any information on table access patterns.

Hashing is more commonly used in cloud datastores such as *Amazon Dynamo*, but it is also used by various clustered databases including *Greenplum* [4]— both systems are discussed later in 3.1.

***Heuristics***

A database can use cost models to decide where to place data, rather than using a deterministic hash function. With this approach the database uses information such as query pattern data to co-locate commonly accessed tables, or the size of table data to decide where there is space to store a table. The use of cost models to place data is a *heuristic* approach, because placement decisions are imperfect (optimal data placement is an NP-hard problem [25]).

A heuristic approach is more complex than hashing because the database must manage indexes to map tables to instances and it must create a global *knowledgebase* on which to make placement decisions. A *knowledgebase* is the information the database has available to it from monitoring access patterns and other factors that relate to query performance. It is important that this information is accurate as it is used to optimize queries, but it is also expensive to collect, so there is a trade-off between the quality of the knowledgebase and the cost in maintaining it.

For global optimizations, information from each instance must be distributed to one or more instances storing the knowledgebase, which is costly in terms of the bandwidth required to send the information, and the costs of storing and analysing it. Local optimizations meanwhile incur no bandwidth costs but rely on less complete information because other instances may have information that is not being shared.

*Piazza* [26], a heterogeneous DDBMS, is an example of a system that uses a hybrid solution and shares knowledge amongst a subset of instances termed *spheres of co-operation*. This creates opportunity for optimization without broadcasting to all instances.

*Mariposa* [27], another heterogeneous distributed database, uses a microeconomic approach to knowledge sharing by pricing operations based on local knowledge and distributed information about available bandwidth. Each node keeps track of its own availability and reflects this knowledge in the price of data and the speed it advertises being able to process queries. Instances then pay each other in an in-system currency to perform operations such as replication. This is not a common approach, possibly because of the complexity involved in balancing such a free-market system.

### 2.3.5 Caching

Disk access may be a bottleneck for DDBMSs even with optimal data placement. To address this many systems use either integrated in-memory storage or external in-memory caching.

*Memcached* [28] is the most widely used of these external caching mechanisms. It is an in-memory key-value store that uses hashing to spread data out over multiple machines. It does not integrate into DDBMSs in any way, so applications using it must be programmed to first query *memcached* and then the database. Applications must also invalidate old entries in the cache, which adds additional complexity. *Memcached* is successful because in-memory access to data is much faster than disk access through a database. It is used by numerous large websites, including *Twitter* and *Flickr* [28], to manage read-heavy workloads.

Some DDBMSs such as *GenieDB* (discussed in 3.1.1) integrate an in-memory cache into the DDBMS making caching transparent to applications [5]. This means that some read queries do not require disk access, making them faster, but means that writes must be committed to two locations (the cache and the persistent store).

### 2.3.6 Concurrency Control

Databases use concurrency control mechanisms to restrict access to data when it is being modified, which is important with regards to the transactional guarantees they provide.

For instance, databases wishing to guarantee atomicity typically use a *shared-read, exclusive-write* approach, where many transactions can read table data at the same time, but a transaction must have exclusive access to make updates to a table. Shared-read, exclusive-write is a *pessimistic concurrency control* approach because locks are taken out ***before*** accessing data; the alternative, *optimistic concurrency control*, allows updates to be made in parallel, and checks ***after*** an update has been made whether it can be committed.

***Two-phase locking (2PL)*** is the most widely used *shared-read, exclusive write* approach. In this model, for a given transaction, all locking operations must occur *before* the first unlock operation[18]. By separating locking and unlocking into two phases it is guaranteed that no two transactions can interleave locking calls in a way that would break the *isolation* property of *ACID* [29].

There are numerous variations on two-phase locking. ***Conservative 2PL*** requires that all locks on tables are taken out at the beginning of the transaction, whereas ***Rigorous 2PL*** requires that all locks are held until after the transaction commits (or aborts) — the former collapses the expanding phase, while the latter collapses the shrinking phase. Each of these approaches guarantees *serializability*, meaning transactions are isolated, appearing as if they were executed in some serial ordering. Most of the clustered databases, including *MySQL,* discussed in 3.1.1 use two-phase locking.

***Timestamp-based concurrency control*** is a pessimistic approach that uses the ordering of unique transaction timestamps in place of conventional locks. When a transaction accesses an item, the system checks whether this transaction is older than the last one which accessed the same item. If it is older, the transaction proceeds; otherwise ordering is violated and the transaction is aborted. This can lead to cyclical restart of transactions and starvation, where the same transaction is aborted repeatedly. In this situation, one transaction is never able to update an item (even after many attempts), because each time it attempts the update another older transaction has already made a change, so the transaction must be aborted. Unless there is a mechanism to prevent this from happening it is possible that the same transaction will never commit. *Amazon Dynamo*, discussed later in 3.1, uses timestamp-based concurrency control.

***Multi-version concurrency control (MVCC)*** also incorporates timestamps by allowing several versions of an item to be stored. This allows the system to present a locally consistent, but potentially historic, version of the database, by reading only data older than a specified point. Fewer reads are rejected than with basic timestamp ordering, but applications must be designed to accept out-dated results. *Clustrix Sierra*, discussed later in 3.1, uses MVCC.

[18] The name *two-phase* refers to the two phases that result from this: the *expanding* phase where locks are acquired, and the *shrinking* phase where locks are released.

***Optimistic concurrency control (OCC)*** allows multiple transactions to read and update items without blocking. However, before a transaction is committed the database must check for conflicts – if any are found all but one of the conflicting transactions is rolled back. There is a cost involved in this rollback, making optimistic concurrency control a trade-off between the chance of conflicts occurring and the performance improvements gained by concurrent access.

There are therefore two factors that determine the choice of concurrency control mechanism: the chance that transactions will conflict, and the transactional guarantees required (summarised below).

| | **Approach** | **Conflict Prevention** | **Currency of Results** |
|---|---|---|---|
| **2PL** | Pessimistic | Locking | Single Version |
| **Timestamp-based** | Pessimistic | Locking through Synchronization | Single Version |
| **MVCC** | Pessimistic[19] | Reads on Historic Data | Historic Results Possible |
| **OCC** | Optimistic | Check on Commit | Single Version |

**Table 2: Summary of the approaches taken by various concurrency control mechanisms.**

To make a decision based on these factors, the designers of a DDBMS must have an idea how their database will be used. This is often possible because many DDBMSs are specialized, tailored for specific workloads.

[19] *MVCC* is similar to *timestamp-based concurrency control*, but it does not block on reads because it allows queries on historic data. If a write query is issued, it can only proceed if there are no earlier conflicting transactions, which makes it pessimistic. An optimistic version of this is possible if the database checks whether a transaction can commit after the write has been made.

### 2.3.7 Specialization

DDBMSs tend to focus on particular markets because this makes it possible to optimize for specific workloads. By eliminating unnecessary use cases it is often possible to improve the speed of a database for the target market, the most common of which are *On-Line Transaction Processing* and *Data Warehousing*.

***On-Line Transaction Processing*** (OLTP) is used to manage business tasks such as stock management and financial transactions, and is normally characterized by large numbers of very short transactions, mostly involving writes. Immediate consistency is important in OLTP because inconsistencies in the results of financial transactions and stock requests must be prevented.

***Data Warehousing*** is used to analyse large datasets, rather than record transactions. As a result, queries are generally more complex and longer running, and workloads are typically characterized by a high ratio of read queries to writes. Data warehousing databases often require consistency, but present a potentially historic version of the database to read queries, because the time it takes to execute these queries is more important than any guarantees that the data returned is current [30].

Specialization for these areas and others allows DDBMSs to optimize for a particular workload and ignore features that are not necessary. For example, because OLTP transactions are generally very short-lived, *VoltDB* — a database focused on such workloads [31] — is single-threaded, meaning it does not have to lock local tables and data structures. *C-Store* [30], a database focused on data warehousing, optimizes its storage format for reads rather than writes, because a read-mostly workload is expected.

Databases such as *MySQL* are described as *general purpose databases* because they are not optimized for a specific database workload. The standard version of *MySQL* can still be used for OLTP (and other workloads), but does not compete with the speed and scale that larger, more specialized, enterprise databases offer.

### 2.3.8 Benchmarking

Benchmarks are used to evaluate and compare the performance of databases in each of these target markets.

The most notable providers of these benchmarks, the *Transaction Processing Performance Council (TPC)* offer a set of benchmark specifications targeted at specific workloads [32], including the *TPC-C* benchmark which is targeted at OLTP DDBMSs.

### 2.3.9 Data Model

Most of the databases discussed in this work use an SQL interface and store data in a relational format. This section compares the relational data model to those used by *NoSQL* DDBMSs.

The data model of each type of database is illustrated with an example:

> *A database is needed to store information on football players and the teams they play for. Each player has a name, position, and team. Each team has a current league ranking.*

***Relational, SQL Databases***

Relational databases (such as *MySQL*) store data in collections of tables, whose structure and attributes must be specified before records can be added. In the football player example, a player has four attributes (*id*, *player_name*, *position*, *team_name*), while a separate *team* table records the *rank* of each team (illustrated below in Figure 20).

***player***

| id | player_name | position | team_name |
|---|---|---|---|
| 1 | Nelson | WR | Packers |
| 2 | Rodgers | QB | Packers |
| 3 | Kampman | DE | Jaguars |

***team***

| team_name | rank |
|---|---|
| Packers | 1 |
| Jaguars | 12 |

**Figure 20: An example of data stored in a relational database.**

There is an explicit relationship — a foreign key constraint — between the *team_name* attributes in *player* and in *team*, meaning a row in *player* must specify a *team_name* which exists in *team,* and a change to the team name must be made atomically in both tables. The database guarantees consistency (the *C* in ACID), which ensures that this relationship holds.

This approach encourages ***normalization***, which is the process of examining the relationships between attributes and making these explicit in the database schema [15]. This typically involves data being broken up into multiple tables, as in the *player-team* example in Figure 20. Rather than storing *team* information in the *player* table, a separate *team* table is used. This makes the structure of the data clearer (there is a many-to-one relationship between players and teams), and it eliminates duplication because team information is not repeated for every player that plays for the same team.

***NoSQL Databases***

Relational databases tend to support some subset of the SQL specification. Those that do not use SQL are often termed ***NoSQL*** databases — a broad term which covers a diverse range of typically non-relational data models. *Key-value*, *map*, and *document databases* are discussed here.

Most NoSQL databases are designed to store data in ***de-normalized form***, meaning that some redundancy is introduced to the schema to improve system read performance [15]. For example, if the player-team example was de-normalized, there would be a single player table storing all data, as with the example below from *CouchDB* [3], a document database which stores data in JSON[20]. Document databases store records as documents, rather than rows in a table, as illustrated below.

[20] CouchDB is discussed further in 3.1. JSON (JavaScript Object Notation) is a storage format designed to be human readable.

```
{  "_id":"1", "position: WR", "player_name: Nelson",
       "team_name: Packers”, "team_rank: 1" }

{  "_id":"2", "position: QB", "player_name: Rodgers",
       "team_name: Packers”, "team_rank: 1" }

{  "_id":"3", "position: DE", "player_name: Kampman",
       "team_name: Jaguars", "team_rank: 12" }
```

**Figure 21: An example of data stored in a document database.**

This introduces some redundancy, because team information is now duplicated for each player with the same team, but it means that queries accessing player and team information do not need to perform a join, which improves read performance. NoSQL databases such as *CouchDB* effectively preclude normalization by not supporting join operations and not providing integrity constraints or consistency guarantees across documents; they are not designed for the same highly structured datasets as relational databases.

These highly structured datasets and integrity constraints are one of the factors that make relational databases hard to scale. If a number of tables grow particularly large, a relational database may try to shard them over multiple machines, but it is difficult to co-locate the shards of different tables that are often queried together, because of the potential complexity of relationships between tables in the database. This results in more distributed transactions, which are significantly slower than single-machine transactions [33]. As a result, most relational databases require administrators to manually partition data to limit the number of distributed transactions, or they use full-database replication to ensure this is not a problem.

In contrast, NoSQL databases typically scale well for read queries because records have no explicit dependencies to other records. In the *player-team* example, it is trivial to execute a query to get the names of players and the rank of the team they play for, because all the necessary data is co-located. In a relational database this query requires a join between the *player* and *team* tables, and there are no guarantees that these tables are co-located.

Another significant difference between relational and NoSQL databases is their approach to the database schema. Many NoSQL databases are called *schema-free* meaning that they do not require the structure of a record to be specified in advance, unlike relational databases. Instead, an implicit schema is built up as records are added, and records can have different sets of attributes. For example, in the player-team example an extra attribute is needed to note any injuries a player has, but this is only needed for players that are injured. In a NoSQL database this can be added for those players and no others, unlike a relational database where either a new *injury* table needs to be created, or an injury attribute must be added for all players, even if they are not injured. Figure 21 shows an example of an extra injury attribute in *CouchDB*.

```
{  "_id":"1", "position: WR", "player_name: Nelson",
       "team_name: Packers”, "team_rank: 1" }

{  "_id":"2", "position: QB", "player_name: Rodgers",
       "team_name: Packers”, "team_rank: 1",
       "injury: Concussion" }

{  "_id":"3", "position: DE", "player_name: Kampman",
       "team_name: Jaguars", "team_rank: 12" }
```

**Figure 22: An example of the schema-free nature of document stores, where one document has an extra attribute for a player's injuries.**

Document databases are one example of NoSQL data models. *Cassandra* and *Bigtable* (discussed later in 3.1) are multi-dimensional map databases, which are effectively multi-dimensional key-value stores. They provide the same schema-free functionality used by *CouchDB*, but the difference in data model allows for different optimizations. *Bigtable*, for example, stores columns consecutively on disk, rather than rows. This optimization is discussed in 3.1 in relation to *C-Store*, another column-oriented database.

## 2.4 DISTRIBUTED SYSTEMS COMPONENTS

The previous section discussed the design dimensions of modern DDBMSs. This section describes other relevant work in distributed systems, not specifically related to databases.

### 2.4.1 Failure Detection

Distributed systems use failure detectors [34-35] to identify when processes have failed or are no longer accessible. This is important in systems with replicated data, because the undetected loss of a machine can result in data becoming unavailable, whereas if failure is detected, data can be re-replicated onto other machines to reduce the likelihood of unavailability.

Failure detectors are termed *unreliable*, because they make mistakes. A process may be active but partitioned or slow to respond, rather than dead[21], so, in the nomenclature of *Défago et al.* [36] a failure detector indicates whether a process is *trusted*[22] or *suspected* of failure.

***Method of Detection***

Failure detectors typically use heartbeat messages to determine whether a process is active. With this approach, the process being monitored — the client — sends a message to one or more observing processes to indicate that it is alive. The observing processes acknowledge the heartbeat to ensure that the client knows it is connected, which ensures that both the observer and the client know that the other is active and the network is not partitioned.

The observer suspects a client has failed when it has not received a message from the client for a specified period. The client assumes it is partitioned (and therefore considered failed by others) if it does not receive heartbeat acknowledgements from observer nodes after a specified period.

---

[21] In failure detector literature this problem is described in the context of asynchronous systems, where there is no bound on communications delays [96]. The same assumptions are made in this work.

[22] Indicating it is *thought* to be active.

This is the basic approach taken by most failure detectors, though a number of enhancements can be made. *Adaptive failure detectors* adjust the timeout period used by heartbeat detectors based on current network conditions [37]. These detectors monitor the latency and bandwidth between endpoints to determine the optimal timeout period, in contrast to standard failure detectors which do not adjust the timeout at runtime.

*Accrual failure detectors* such as the *φ accrual failure detector* [38] estimate failure on a continuous scale, rather than as a Boolean value (trusted/suspected). This value, the *suspicion level*, indicates the failure detector's degree of confidence that a process has crashed — the value accrues over time once the process has failed, tending towards infinity. Applications decide how to interpret the accrual value by choosing a threshold that, when reached, means that a machine is suspected of failure. A threshold that is too low will produce incorrect suspicions, whereas a threshold that is too high will take more time than necessary to detect failures.

***Uses of Failure Detection***

When a failure is detected, the machine suspected of failure is often removed from the system's *membership set*, meaning it is not considered part of the system and is not able to execute any application requests.

The node managing the membership set must decide whether to remove a process suspected of failure immediately, or to wait for further evidence that the process has failed and is not just running slowly [39]. If it acts too soon it may remove an active process, but if it waits too long it risks slowing the system by keeping dead processes in the active set.

### 2.4.2 Local Point of Presence

A *local point of presence* (*LPOP*) is a process that connects a local application to a remote distributed system. The application does not require knowledge of the distributed system's location because it only connects to the *LPOP*, which abstracts over the locality of the system. An *LPOP* process may have some or all of the functionality of the server instances it connects to, allowing for some client-side computation and storage.

It is simpler for an application to connect to a local process, because the location of this process does not change, in contrast to cluster machines which may fail or move. The *LPOP* can manage the changing locations of these machines instead of the application.

## 2.5 Resource Monitoring

Resource monitoring tools aim to give applications and users an overview of the available resources on a system. They are commonly used in shared computing environments such as grid computing where the resources of a machine are allocated amongst competing applications.

Physical machine resources such as the CPU, memory, disk, and network utilization are monitored. Monitoring tools typically report the current utilization of these resources, because performance information is generally only useful when it is fresh. However, they also aggregate and present summaries of the same information for longer-term trend analysis.

Figure 23 displays sample output from the *Ganglia* monitoring tool [40] that illustrates the types of resources that can be monitored.

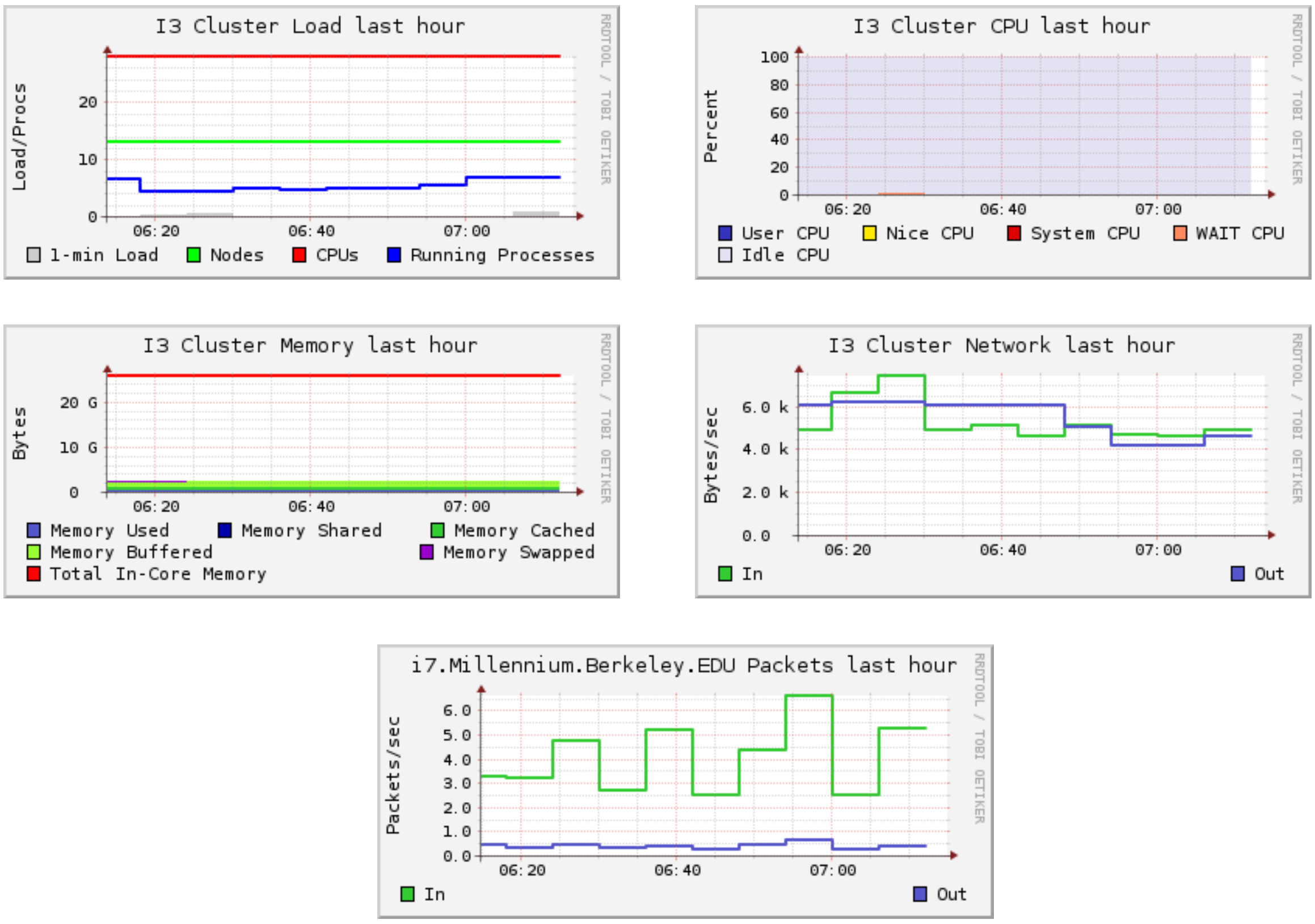

**Figure 23: Sample output from the *Ganglia* monitoring tool**

Monitoring information can, as in the Ganglia example above, be used to inform administrators of the load on their systems. Other tools focus on programmatic access to enable applications to adapt based on monitoring information. This section looks at how these architectures are typically structured.

### 2.5.1 Resource Monitoring Architectures

*Zanikolas et al.* [43] provide a comprehensive overview of monitoring architectures, which is summarised in this section. The components or processes that perform monitoring are called *sensors*. These sensors produce *events* which are either sent directly to the process requesting data, the *consumer*, or sent indirectly via a *producer*. Producers provide an interface over sensors to provide programmatic access to events, which means that multiple consumers can receive events (via producers), rather than only a single consumer receiving events when the sensor is connected to one directly.

The foundation of the taxonomy by *Zanikolas et al.* [43] is a categorization of monitoring architectures into four levels, described and illustrated below in Figure 24:

- Level 0 — *Basic, self-contained systems*. The sensor, which is obtaining monitoring results, communicates with the consumer of that data directly. There is no way to distribute data to multiple consumers.
- Level 1 — *Producers*. A producer component is added to the sensor, making it possible for multiple consumers to access the same data.
- Level 2 — *Re-publishers*. A re-publisher allows data to be consumed and then cached or manipulated, before being exposed as a producer again. These are compound producer-consumers.
- Level 3 — *Hierarchies*. Re-publishers are allowed to form hierarchies.

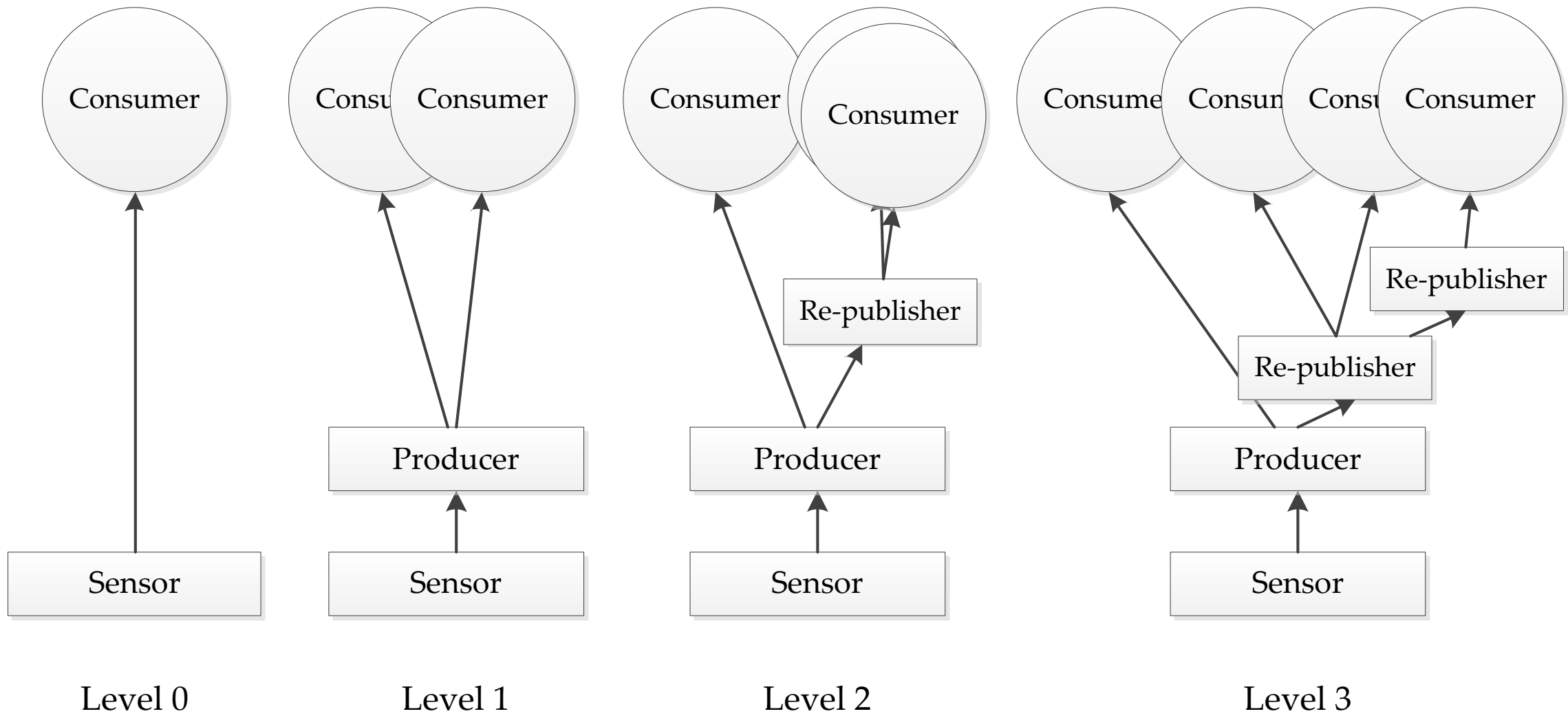

**Figure 24: The four levels of resource monitoring tools.**

In addition to these categories there are numerous multiplicities involving re-publishers and sensors. For instance, some systems allow only one re-publisher meaning the storage and access to data is centralized, whereas others allow for more, allowing data to be distributed further.

The majority of monitoring systems focus on monitoring the physical resources of a machine, such as CPU and memory. An alternative type of system called an ***application monitor***, observe the state of applications at runtime. These systems can be either *passive application monitors*, which send monitoring information to other tools and applications for processing and analysis, or *active application monitors* which do the same, but provide additional functionality to allow for the monitored application to be adapted at runtime, either through changes to configuration files or direct function calls to the application.

The difference between resource and application monitoring is illustrated in Figure 25, where a resource monitor (*left*) sends data through a series of producers to an analysis tool and a graphical display, and an active application monitor (*right*) sends data to an analysis tool, which then makes a call back to the monitored application.

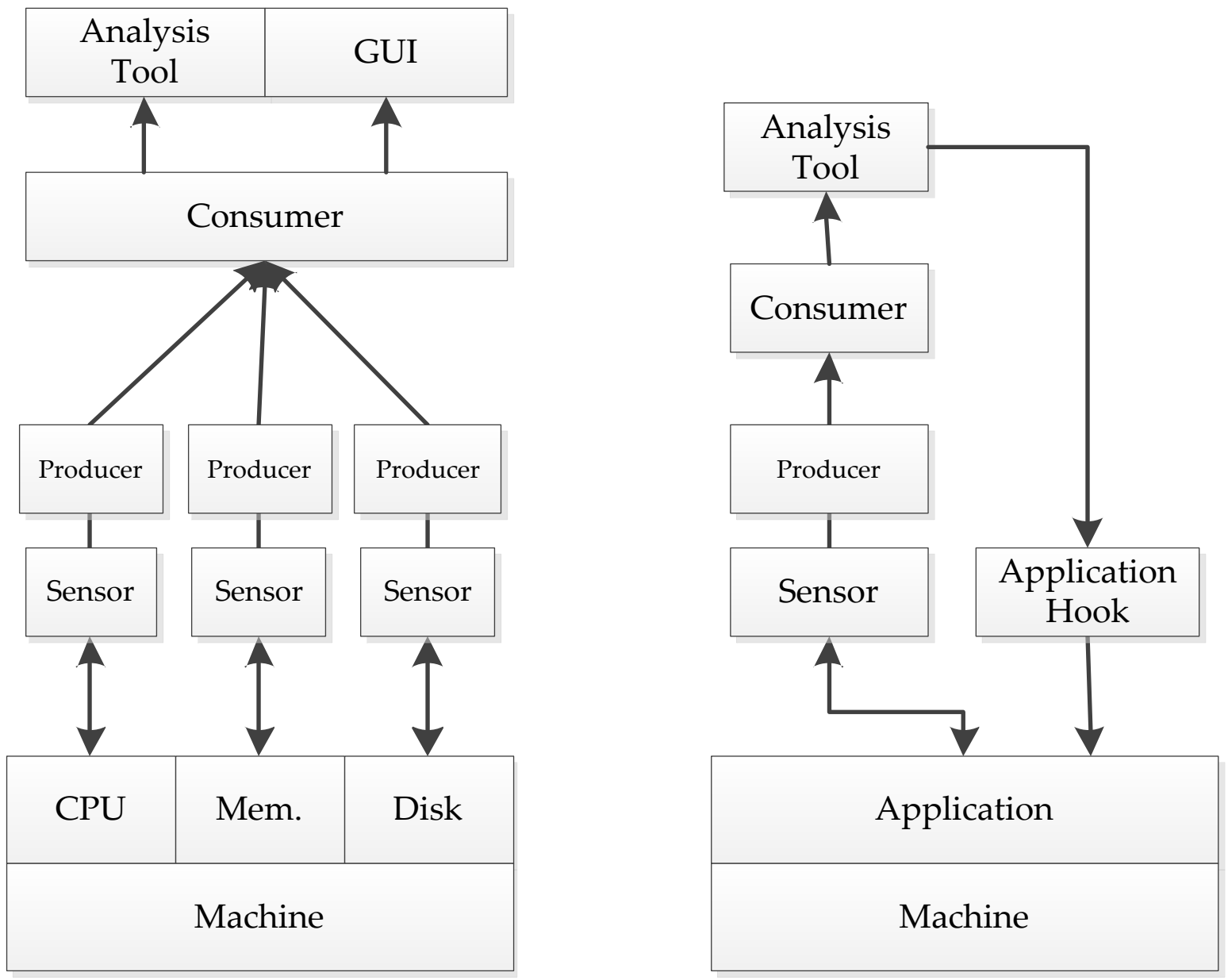

**Figure 25: (*left*) An example *resource* monitoring architecture. (*right*) An example *application* monitoring architecture.**

This section focuses on the architectures and characteristics of a number of solutions.

### *Publish-Subscribe*

*Publish-subscribe* is a model for event delivery [41], where the consumers of data (*subscribers*) are able to subscribe to events from the producers of data (*publishers*). In the Grid Monitoring Architecture [41], a monitoring system which uses this model, a *directory service* is used to advertise the location of active *publishers*, which enables discovery for *subscribers*.

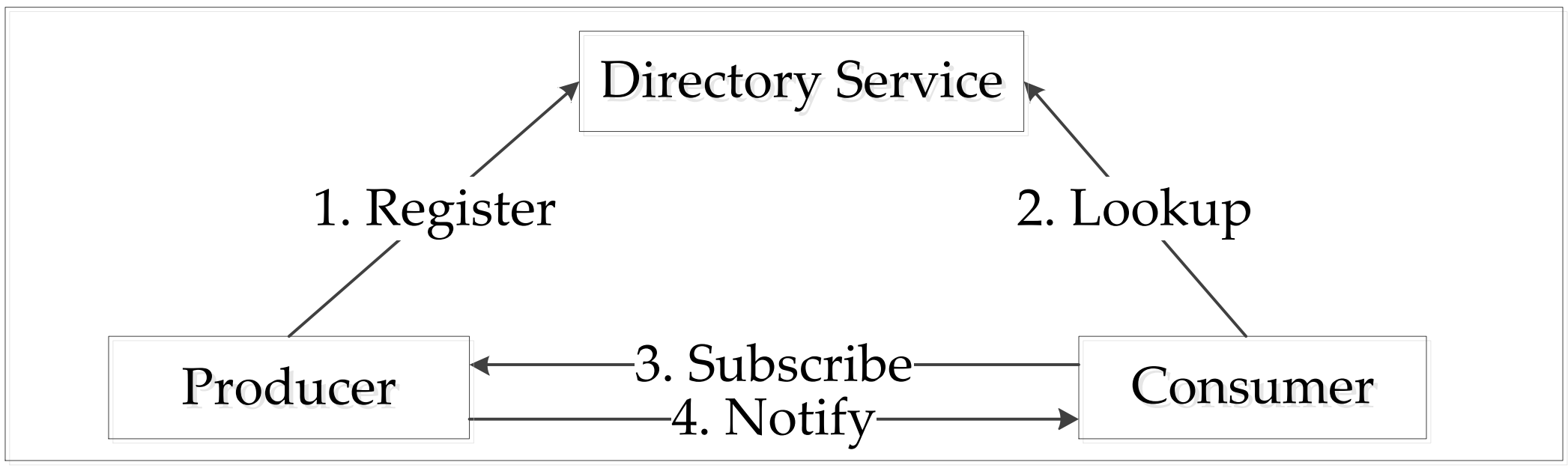

**Figure 26: Illustration of the publish-subscribe model.**

This design ensures that events are recent when they are received, because producers are able to *push* updates to interested consumers. In contrast, a system that *pulls* updates from producers, risks either receiving requests too infrequently, or overloading producers with too many requests.

*Publish-subscribe* is commonly used in monitoring systems because it enables these systems to naturally scale. Requests for data do not need to be concentrated on a single machine as data can be distributed across producers. Furthermore, to reduce contention on producers, many architectures also allow for hierarchies of re-publishers [43], which allows data to be dispersed amongst re-publishers which are then contacted by consumers.

***Replication***

Monitoring systems often replicate data to ensure that it is not lost after the failure of an individual machine. The simplest approach to doing this is to replicate data across every node in a system, either through multicast or use of a round-robin database[23]. This approach does not scale across multiple clusters because of communication and storage overheads, so a number of systems (such as *Ganglia* [40]) replicate only within clusters. In *Ganglia*, every cluster can then communicate with a re-publisher (called a *Ganglia meta-daemon* [42]), which may itself contact another re-publisher or connect with another cluster. This allows for hierarchies of re-publishers to form, connecting monitoring information across many clusters.

### 2.5.2 Querying Interfaces

Monitoring systems provide different interfaces to applications based on their intended use.

*Ganglia* and others provide a web-based GUI interface (as shown in Figure 23 previously) that allows system administrators to see an overview of a cluster's availability and utilization. *Ganglia* also sends data to consumers using in XML, which can be used by other applications. These applications may aggregate or relay monitoring information to users of the system, or to other components.

[23] These mechanisms are not discussed in this thesis.

Application monitoring systems not only allow applications to be monitored, but also provide mechanisms to allow the monitored applications to be adapted at runtime. This is typically achieved by allowing observing applications to make function calls to the monitored applications, or by allowing them access to configuration files for the monitored applications [43].

Some systems such as *R-GMA* [44] use a relational querying model which is designed to make querying resource information easier for those familiar with SQL.

### 2.5.3 Challenges

Monitoring systems necessarily affect the system that they are monitoring by consuming resources as they operate, but they aim to minimize this disruption as far as possible. This goal is noted by the authors of the *Autopilot* project [43] as they describe the trade-off between sending sensor events as raw data to remote machines, or processing them locally beforehand. Sending raw data requires less local computation but involves sending much more data, whereas pre-processing requires more computation but has a lower network overhead. In *Autopilot* both options are available, but processing locally is seen as less invasive.

## 2.6 AUTONOMIC COMPUTING

Autonomic computing is a vision of computer systems that manage themselves based on high level objectives set by administrators. This section summarises the work of *Kephart and Chess* [45].

The term *autonomic*, which originates in biology with the autonomic nervous system, captures the notion of a group of components that are managed automatically and adapted through a feedback loop rather than requiring input or configuration from users or applications.

Four aspects of self-management are discussed in relation to the components of a system [45]:

- Self-configuration, where high-level policies governing a system's installation and configuration are set.
- Self-optimization, where configurable parameters are set and altered at runtime to optimize performance.
- Self-healing, where problems in a component are detected and repaired.
- Self-protection, where attacks against the system or failures are detected and stopped.

*Autonomic elements* are components that provide this self-management in software systems. Each autonomic element has a feedback loop, which is structured in the manner shown in Figure 27.

Each autonomic element monitors and controls a *managed element*, which is a component within the system being managed. The process forms a cycle, where changes made by the *managed element* are monitored and used to inform future analysis and planning decisions.

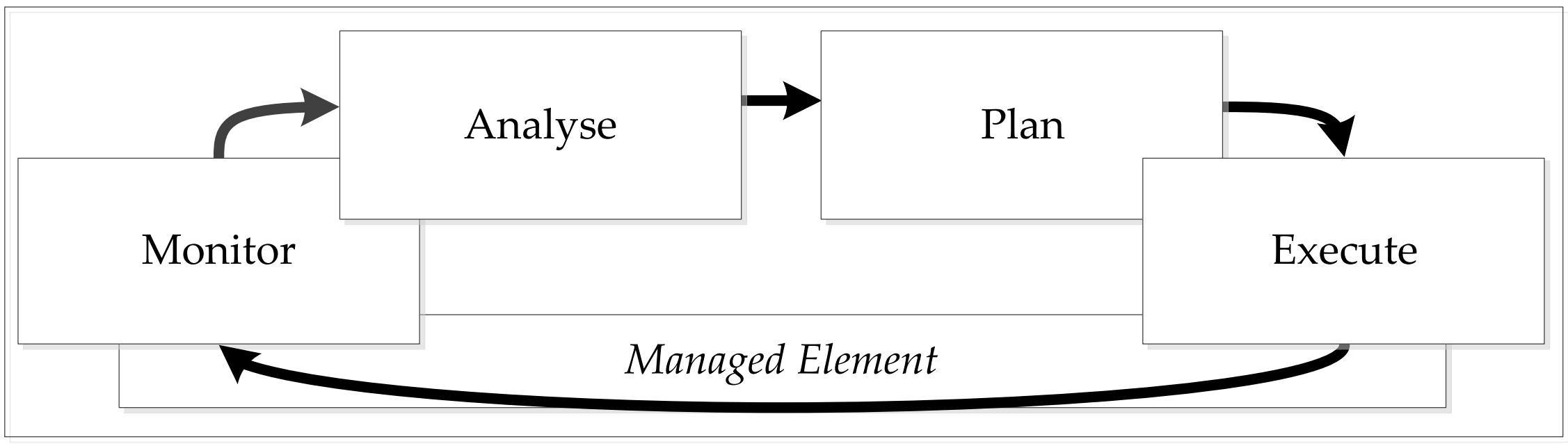

**Figure 27: An illustration of the autonomic cycle, as applied to individual managed elements.**

There are four stages to the autonomic cycle:

- *Monitor*. Sensors on the managed element produce monitoring information.
- *Analyse*. Monitoring information is analysed for patterns that indicate something must be changed.
- *Plan*. A plan is created, based on a perceived need to change the state or configuration of a component.
- *Execute*. The plan is executed by interacting with an *effector*, the part of a managed element that the autonomic element is able to communicate with, or the configuration options it is able to change.

The information produced by this process is stored in a *knowledgebase*, which represents an autonomic element's knowledge of the component it is managing.

## 2.7 WORKSTATION-BASED COMPUTING

Workstation-based computing is a term used in this thesis to describe software systems that make use of under-utilized resources on existing infrastructure in an organization. These systems run on machines such as workstations that are primarily used for user-centric tasks, instead of specially provisioned resources that are costly to purchase and maintain. Where the resources being used by these systems already exist, workstation-based computing can be more cost effective than alternative solutions.

Workstation machines are more likely to be restarted or become disconnected compared to servers [46], so most workstation-based systems are designed to perform highly parallelisable tasks that can be processed over multiple machines and re-computed in the event of failure. In this thesis this type of system is termed non-interactive.

***Non-Interactive Systems***

*Non-interactive* systems are batch-processing systems that take an input, perform a computation on the input, and then asynchronously return the result. They are particularly suited to workstation-based systems because each workstation only needs to be given the relevant input data and an application binary to perform computation. The only data stored on workstations is the transient data currently being processed as part of a computation and the state of the process performing the computation, which means that the failure of a workstation machine only results in the loss of intermediate results, which are reproducible, and not critical system data. The same work is often sent to multiple machines because there is no guarantee that a workstation will be available for long enough to complete a computation. By duplicating work, the system does not rely on any individual workstation, and is not significantly slowed down by the failure of a computation. This is called *redundant computing* [47].

To summarise, non-interactive systems exhibit the following characteristics: the focus of the system is on computation, the result of a job is returned asynchronously once the job has been completed, workstation machines store the intermediate results of computation, and the final result of computation are stored on external, non-workstation machines.

Non-interactive systems can also be distinguished by the locale of the workstation machines used in computation. *Opportunistic computing* is a class of non-interactive system where workstations and compute clusters within an organisation (or across similar organisations) are used to perform computations. *Condor*, a workstation-based scheduling system, is an example of opportunistic computing [48], and is discussed later in section 3.4.1.

*Volunteer computing* is another class of non-interactive system where the workstation machines being used to perform computations can be anywhere, provided they have an internet connection — users often volunteer the resources on their home machines, for example. Volunteer computing is often associated with applications such as *Folding@Home,* which uses volunteered resources to perform medical research which would otherwise require expensive super computers [49].

The primary distinction between these approaches is that the workstation machines in an *opportunistic computing* system are typically in the same local area network, whereas they are globally dispersed in a *volunteer computing* system. This has a number of implications. Firstly, *volunteer computing* systems tend to run shorter computations. This is because there are fewer guarantees that the workstations being used will be available for extended periods of time, as they are outwith the administrative control of the system's operators. Secondly, *volunteer computing* systems make more use of *redundant computing* than software designed for machine clusters, as many machines may not complete the computation they are given, and it is considered better to perform the same computation multiple times than it is to delay its execution by waiting on work that will never complete. Multiple results of the same computation can also be used to identify erroneous results, which are more likely in *volunteer computing* systems, where workstations are volunteered by anonymous users.

*Interactive Systems*

In contrast to non-interactive systems which perform asynchronous batch processing, *interactive workstation-based systems* provide services which give a synchronous response to requests.

We characterise an interactive system as a persistent, stateful program that applications are able to communicate with synchronously. Non-interactive systems tend to be transient, meaning only computation is performed on machines in the system, and the state that is stored can be removed and recomputed again elsewhere.

Running a workstation-based system with state is more difficult than running a compute job, where the critical state is not stored on workstations. An interactive system must ensure that its state is not lost when workstations become unavailable, and if data is replicated it may have to ensure that replicas are consistent.

### 2.7.1 Ad Hoc Cloud Computing

*Ad hoc clouds* are a type of interactive, workstation-based system where the underused capacity of workstation machines is used to run distributed applications that would otherwise run in local machine clusters [46].

Ad hoc clouds must automatically adapt to cope with the highly variable resource availability exhibited by workstation-based systems. To achieve this, the ad hoc clouds proposal in [46] suggests a supporting infrastructure for distributed applications, including a monitoring component which monitors local resources, and brokering component which manages the applications running in the cloud.

The remainder of the ad hoc cloud architecture focuses on *cloudlets*, the services running on the cloud. Each cloudlet is made up of multiple *cloud elements*, which are the instances of the cloudlet running on individual machines in the cloud. Using the nomenclature of this thesis, a database system is a cloudlet and a database instance is a cloud element. Cloud elements are able to interact with other cloud elements and with the cloud infrastructure components on each machine.

## 2.8 SUMMARY

### 2.8.1 Database Design

Distributed databases are distinguished by a number of factors in their design.

The ***heterogeneity*** of a database defines whether it is a clustered set of *instances* or a disparate group of *databases*.

The ***hardware*** on which a database is run determines whether the system will scale out or scale up, while the ***locale*** of this hardware determines the bandwidth and latency between database instances. Locale itself affects the expected ***churn*** of database nodes, and the ***transactional properties*** the database provides.

The method of ***replication*** a database uses is decided on the basis of these transactional properties and expected churn. Replication architectures are determined by this and described in terms of the ***unit*** of data that is replicated and the ***synchronicity*** of replication itself.

When data is created or replicated, the database makes a ***placement*** decision, which determines where data should be placed, typically using either hashing or heuristics.

Because there are so many design dimensions that each affect database performance, there is a trend towards ***specialization***, where a database is targeted at specific workloads.

### 2.8.2 Distributed Systems

Distributed systems often use ***failure detectors*** to determine whether a process has failed. These are imperfect, because it is not possible to distinguish between failed processes and slow processes in an asynchronous system such as the internet.

When applications are notified of possible failures they may choose to adjust their ***membership set***, the set of instances that are considered active within the system.

Some systems use a ***local point of presence***, a process running on the same machine as an application using the system, to make it easier for applications to connect to remote instances, and to allow the system to perform some tasks client-side, rather than at the system's remote instances.

### 2.8.3 Resource Monitoring

***Resource monitoring*** applications monitor the resources available on a set of machines, or the activity of an application running over these machines. Monitoring data can be used to provide an overview of availability to administrators, to perform automated analysis of trends in availability, or to allow applications to automatically adapt to changes in availability at runtime.

### 2.8.4 Autonomic Computing

***Autonomic computing*** is a vision of computing where applications are self-managing. This is achieved through the autonomic cycle: application components are monitored, monitoring information is analysed and used to create a plan of action, which is then executed. The results of this execution are monitored, and the cycle continues.

The next chapter discusses the state-of-the-art in each of the areas discussed above.

# Chapter 3: Related Work

# 3 Related Work

The work discussed in this chapter is considered relevant to the design of an interactive workstation-based database system. It is divided by function, including sections on database architectures, more general distributed systems architectures, and resource monitoring tools.

## 3.1 Distributed Databases

This section presents the architectures behind various DDBMSs, with specific sub-sections on *clustered*, *cloud-based*, and *NoSQL* databases.

### 3.1.1 Clustered Databases (SQL)

***PostgreSQL/PGCluster***

***PostgreSQL*** [2] is a popular open-source DDBMS that uses full-database, asynchronous, master-slave replication.

Various middleware systems offer other forms of replication, and one such, ***PGCluster***, is discussed in detail here, as it is used in the evaluations in *chapter 7*.

*PGCluster* uses synchronous multi-master replication to replicate to a cluster of *PostgreSQL* instances [50]. These instances, called *data nodes*, receive requests through a *load balancer*, which distributes load on read requests, and a *replication server* which manages locking and the propagation of updates. This is illustrated with an example below.

When an update is sent to the *load balancer*, it is sent onward to one of the *data instances* (the arrow marked 1 in Figure 28). The *data instance* sends the update to the *replication server* (2) which obtains a lock for the relevant table(s). When a lock is obtained, the *replication server* sends the update to every *data instance* to be executed (3), each of these *data instances* sends a commit message to the *replication server* once they have completed the update (4). Finally, the result of the update is returned to the *load balancer (5).*

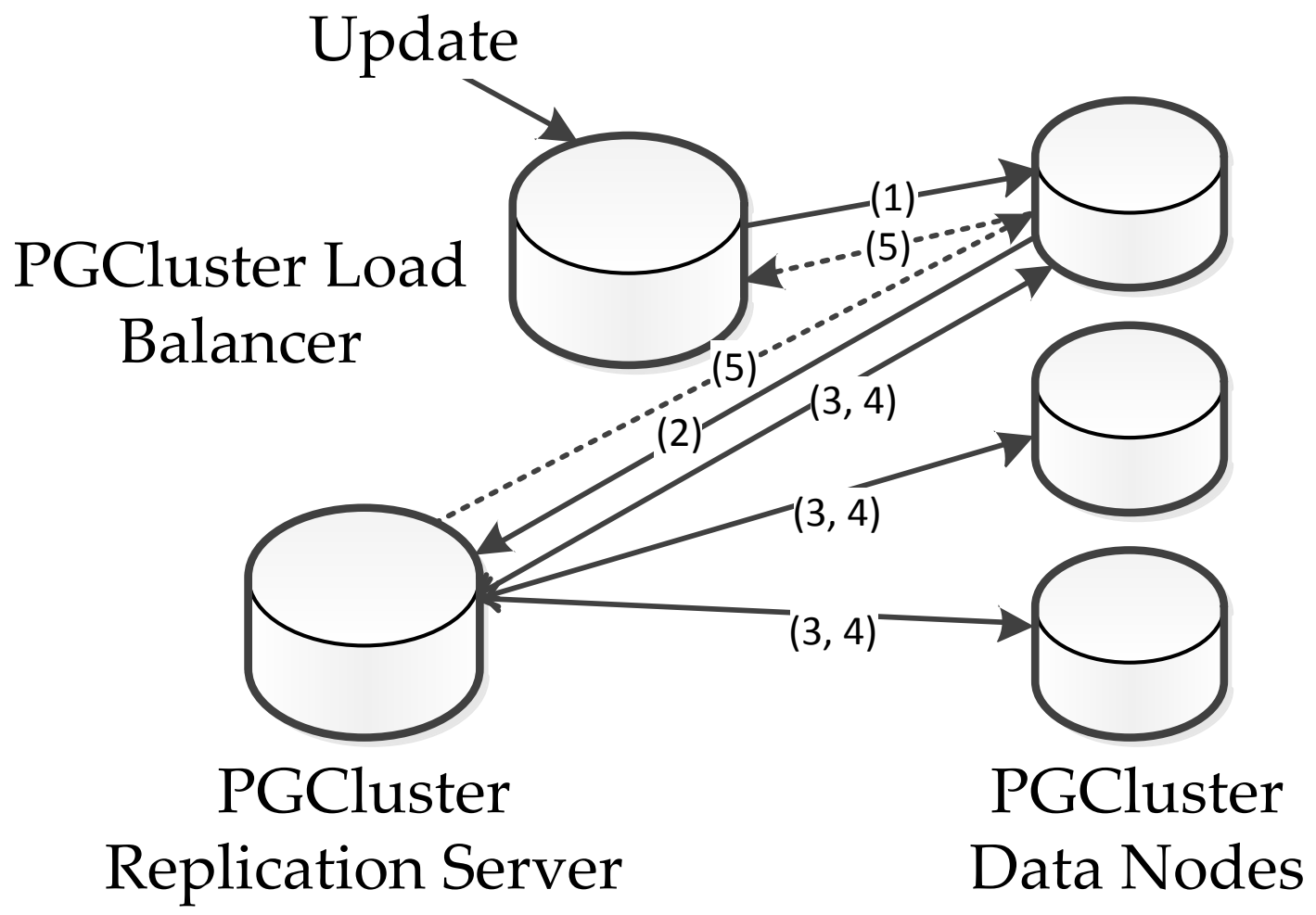

**Figure 28: The architecture of PGCluster, illustrated with an example query flow.**

While this is the typical approach taken, *PGCluster* is a multi-master database, meaning queries can be sent directly to either of the *data nodes*, in addition to the load balancer. Each of these nodes uses the *replication server* for locking.

A cluster can be set up with multiple load balancers to ensure that there is no single point of failure [51], though applications must be aware of the location of each of these load balancers to continue operating when one fails.

***MySQL***

***MySQL*** [1] is another popular open-source DDBMS that is commonly used as the backend database for web applications. It is notable for supporting various pluggable storage engines, which decide how and where data is stored, and manage related functionality including transaction management and indexing.

The standard storage engines provided by *MySQL* are non-distributed, meaning they only store data on the machine running the *MySQL* instance. However, *MySQL* also supports distributed storage engines, which act as a bridge to storage on multiple remote machines. Figure 29 illustrates this distinction below.

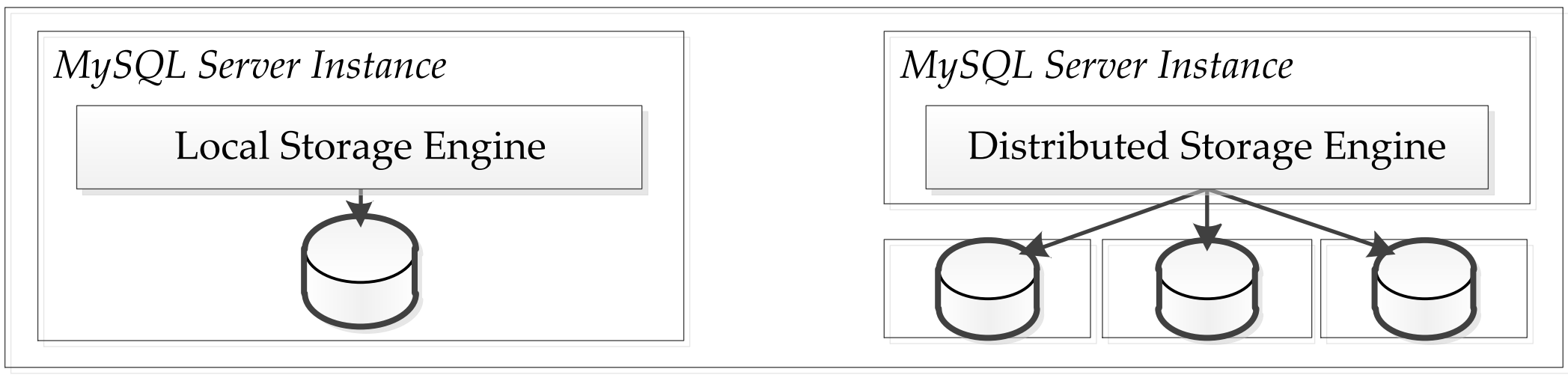

**Figure 29: MySQL. A comparison between local storage engines (left) and distributed storage engines (right).**

*MySQL* databases can be replicated by using a master-slave replication mechanism between separate *MySQL* instances, or by using a distributed storage engine that supports replication.

*MySQL's* master-slave replication mechanism supports only asynchronous, full-database replication. If the master instance fails, manual intervention is required to replace it with a slave, and all remaining slaves must be wiped and reloaded with a copy of the new master's data (to keep them in sync), which results in downtime [4].

Non-distributed storage engines are limited because the master-slave replication mechanism supports only full-database replication, so each instance has to be able to store the whole database. In contrast, the storage engines described in the next two sections, *MySQL Cluster* and *GenieDB*, use segment-level replication, enabling a database to scale out.

### *MySQL Cluster*

***MySQL Cluster*** is a distributed, in-memory storage engine for *MySQL* [52]. Data can be asynchronously written to disk, but it is stored entirely in-memory at runtime.

There are three kinds of nodes in a *MySQL Cluster* system, illustrated in Figure 30 below:

1. *MySQL server instances* manage access to the cluster. These are standard MySQL instances (as described above) which use *MySQL Cluster* as their storage engine.
2. *Data instances* store tables, which are replicated across multiple instances.
3. *Management server instances* handle system configuration, including cluster membership.

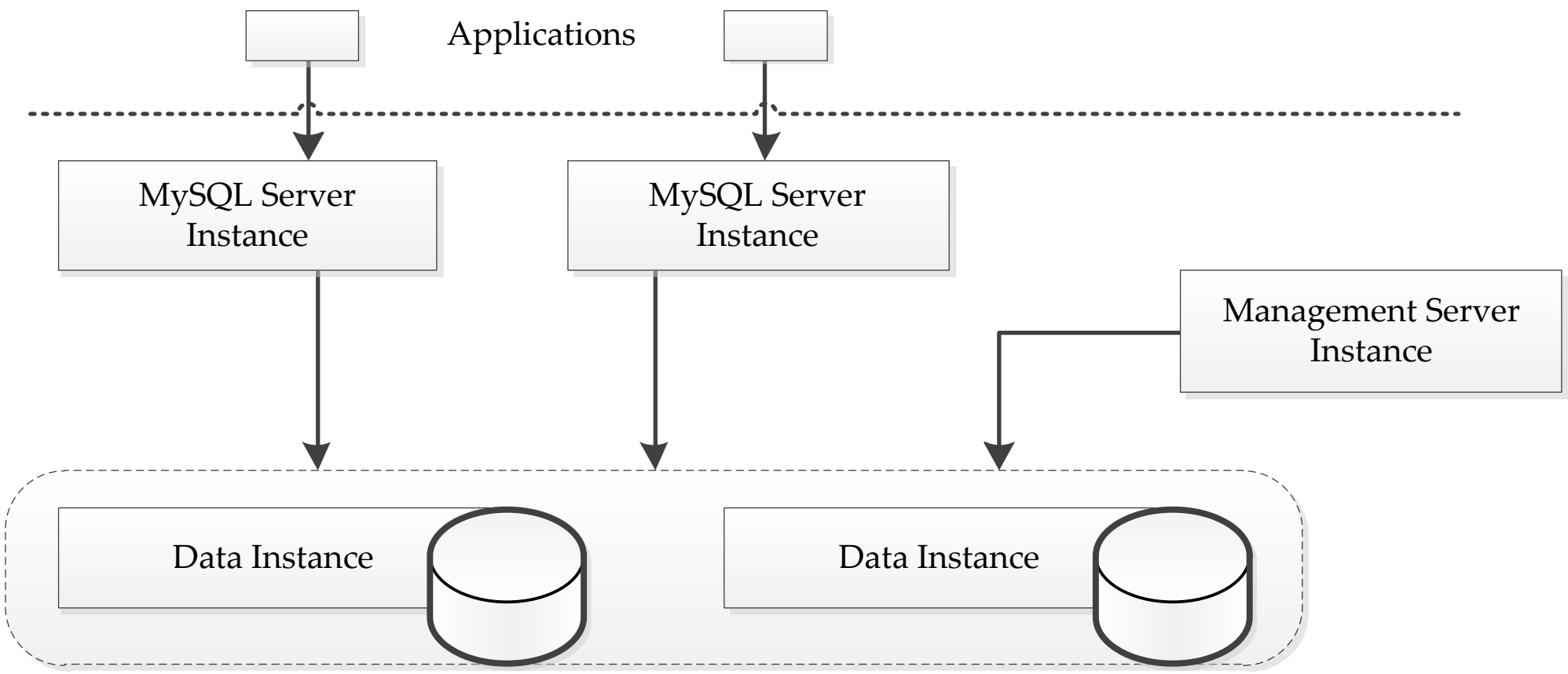

**Figure 30: The architecture of *MySQL Cluster*.**

*MySQL Cluster* uses segment-level replication to allow data to be partitioned over multiple data nodes, which makes it possible to incrementally scale the database by adding nodes as demand for space increases.

A multi-master replication scheme is used, where the set of data instances that hold replicas of a given segment are said to be part of the same *instance group*. Updates are synchronously replicated to all nodes in an *instance group*.

*MySQL Cluster* is consistent and partition tolerant in the context of the CAP theorem, because in the event of a partition (or node failure) data instances must be able to communicate with a majority of data instances in their instance group to continue to operate [53]. If there is no majority, but half of the instances in an instance group are available, an arbitrator is given a vote to allow one partition to continue to operate. The arbitrator is typically a specially designated management server [53].

If a single data instance fails, other data instances in the same *instance group* are used to execute the remainder of a transaction. *MySQL Cluster* is able to guarantee immediate consistency in the event of single node failure because it uses synchronous replication.

Management server instances can fail without making the system unavailable, because they are only used at start-up and during system re-configuration. New instances cannot be added without management servers, but existing instances continue to function. At start-up they are used by instances to get information on the system’s configuration [53].

New nodes can be added incrementally to scale out the system, but existing nodes must be restarted for these changes to take effect. Existing data must be manually repartitioned to take advantage of the extra space.

***GenieDB***

***GenieDB*** [5] is a distributed storage engine for *MySQL* that uses both an in-memory cache and a persistent, disk-based storage.

*GenieDB* uses full-database, synchronous, multi-master replication, meaning every instance is able to replicate changes to every other instance.

It uses optimistic concurrency control, meaning transactions check before they commit (but after the update has been written) whether a conflicting update has been made. If there is a conflict the transaction rolls back; otherwise it commits. Timestamps are also used to recover from failure — when a machine restarts it asks all the other servers for updates that occurred after the timestamp of the last update it recorded.

Updates are made to an in-memory cache and to a replicated, disk-based, datastore. This allows *GenieDB* to provide either immediate or eventual consistency: immediate consistency is guaranteed by requiring that an update commits only once it has been persisted to the disk-based datastore, whereas only eventual consistency is guaranteed if updates commit once they are written to the in-memory cache and replication has begun on the datastore — in the latter case, if the in-memory cache fails, an update committed on the in-memory cache may still be taking place on the persisted datastore, making the result of an interleaving read query inconsistent. Applications are able to configure which option is used on a per-query/update basis. This allows applications to provide immediate consistency where it is needed, and to lower the latency of queries and updates when it is not needed. *GenieDB* is *consistent & available* with immediate consistency, and *available and partition tolerant* with eventual consistency [54].

*GenieDB* manages the consistency of the in-memory cache, so applications do not have to perform the types of cache invalidation procedures associated with *memcached*.

Updates are queued at each instance, rather than on the instance that initiated the update, to prevent a bottleneck on the initiating instance. If the size of this update queue increases to a certain point on any instance, the instance stops accepting updates for a while, to allow existing updates to complete.

*GenieDB* extends the SQL specification to allow administrators to optimise how data is stored. A technique called *record coalescing* gives two views of the database: a logical view as seen by the application, and a physical view as stored on the database. Multiple tables in the logical view can be assigned a so-called *table group*, indicating that they should be seen and stored as a single *super-table* in the physical view [5]. This is called *horizontal coalescing*, and is used by the database to make optimized placement decisions. Tables that are part of the same *table group* will often be read or written together, so when a query touches one of these tables, the record returned is a combination of multiple tables in the logical view, but a single table in the physical view. This is advantageous, because it allows tables to be stored contiguously if they are commonly accessed together.

***Clustrix Sierra***

***Clustrix Sierra*** is a clustered distributed DDBMS designed for On-line Transaction Processing (OLTP) workloads [55].

*Sierra* uses segment-level replication to spread data over multiple instances. Tables are sharded and these shards are hashed (on the values of a specified set of attributes) to place data on a particular segment. When nodes join or leave, shards are automatically repartitioned to balance load.

*Sierra* instances store data on disk, and use an in-memory cache for read requests. Each table is stored in this in-memory cache on the instance that holds the table's primary copy. Consequently, updates are made synchronously to every replica, but reads are sent to the primary instance because it has the cached copy of the table. This could become a bottleneck, so the location of the primary copy can be changed dynamically to balance load.

*Sierra* uses a majority consensus protocol which means that more than half of the nodes in a cluster are active before any queries can be processed [56]. This makes it consistent and partition tolerant in the CAP theorem.

Data can also be replicated between different *Sierra* clusters asynchronously, which allows off-site back-ups to be made without slowing local queries and updates. This means that in the event of a catastrophic cluster failure there will be backups of data available, though these backups may not contain recent updates.

***H-Store / VoltDB***

***VoltDB*** is a clustered distributed database targeted at OLTP. Originally an academic project named ***H-Store*** [57][58], it has now been commercialized as *VoltDB* [31].

*VoltDB* uses segment-level replication to divide tables amongst instances. This partitioning must be done manually, because the system's design optimizes for single-segment transactions [23] and automating data placement is difficult [25].

*VoltDB* is a radical example of specialization, removing many features common to DDBMSs that are deemed unnecessary for OLTP.

Data is stored entirely in-memory, because most OLTP databases are relatively small in size and disk access is relatively slow. Rather than using local logging to provide durability (which would require disk writes), data is synchronously replicated onto other instances in the cluster. Each database instance is also single-threaded, because a typical OLTP transaction is generally very short-lived.

The disk buffer pool, needed to buffer pages to disk, is not required as the database runs entirely in-memory. Similarly, disk-based logging is not needed because synchronous replication is used to provide durability. Latching — the locking of index structures — is unnecessary because each instance is single threaded, so it is only ever possible for one thread to be accessing a data structure at any point. Finally, table locking is not needed when transactions only touch a single segment, for the same reasons as latching[24]. In an experiment on *Shore* [59], an open source DDBMS, *H-Store* researchers found that these features accounted for 88% of the instructions executed per-transaction [60].

*VoltDB* provides only a stored procedure SQL interface, meaning all queries and updates must be pre-compiled before being run. This is possible because in an OLTP environment it is expected that all queries are already known in advance. It is advantageous because the intermediate results of transactions can be processed at the database rather than on the client, which can reduce the number of round-trips required between application and database [61].

[24] It is the responsibility of the database administrator to minimize multi-segment transactions by selecting an appropriate data placement.

The designers of *VoltDB* state that the system shouldn't be designed to be partition tolerant at the expense of other functionality, because partitions are extremely rare [23]. It is consistent and available instead.

Applications connect to *VoltDB* through a *local point of presence* running local to the application. The point of presence maintains connections to a subset of nodes in the *VoltDB* cluster and issues round-robin requests to them, moving to the next instance after each request [62].

*VoltDB* is an extreme example of specialisation in DDBMSs. It shows that by focussing on a particular type of workload (OLTP) it is possible to eliminate many of the features that are considered necessary for general-purpose DDBMSs. VoltDB is 100 times faster than MySQL in single node performance when running a variant of the TPC-C benchmark [23][25].

***C-Store / Vertica***

***C-Store*** is a column-oriented database optimized for data warehousing [30]. It has been commercialized as ***Vertica*** [63].

The name *C-Store* refers to the representation of tables on disk: traditional row-oriented databases store records contiguously on disk, whereas column-oriented databases store attributes contiguously. Row stores offer comparatively efficient inserts as all the attributes from a newly added record are stored together on disk, whereas column stores are read-optimized as queries tend to require that whole columns of attributes are read sequentially, allowing attributes not used in a query to be ignored. The latter read-optimized approach is suited to data warehousing where writes will typically be infrequent compared to queries.

Compression can be used effectively in DDBMSs to reduce the storage size of a database without significantly impacting query performance [64]. This is particularly true of column stores, because compression techniques such as run-length encoding work more effectively over sorted data, and consecutive entries in sorted columns are more likely to be similar than consecutive entries in rows [65]. In addition, it is sometimes possible to query directly over compressed data, thus avoiding the cost of decompression [66]. *C-Store* does this where possible [67].

[25] This result favours VoltDB because it is run without replication, and without multi-query transactions, both of which would reduce its transaction throughput.

*C-Store* replicates data by storing collections of columns called *projections* over multiple instances[26]. A projection may contain some or all of the columns of a table or set of tables, and is sorted on one or more of these columns. *C-Store* is able to store each projection in a different sort order and with different collections of columns, to optimize for different types of queries in addition to improving the availability of data. This allows queries to be directed towards the replicas with the most appropriate sort order, which improves query performance [30].

***Greenplum***

***Greenplum Database 4.0*** is a distributed database targeted at data warehousing [4]. It has a similar architecture to *MySQL Cluster*: a group of *master instances* provide an application interface, and a number of *data instances* use segment-level partitioning to store data.

*Greenplum* is consistent and available, because it uses a central lock manager. It is designed to run on bespoke hardware which uses redundant network interfaces to make network partitions unlikely to occur.

Data is sharded and placed on segments using hashing. The key used to hash records is based on the values from a specified set of columns called a "hash distribution key". For each record inserted, a hash of the values for this key is used to determine where to place the record.

Tables can also be partitioned based on ranges of values. For example, a table can be partitioned by month (where month is an attribute) so that queries involving a particular month only need to be sent to a single segment.

*Greenplum* uses synchronous segment-level replication, which means that when an instance fails, a replica on another machine can take over with no downtime.

Like *GenieDB*, *Greenplum* allows administrators to optimise the storage format it uses. Data can be stored in a traditional row-oriented table optimized for writes or reads, or in a column-oriented table optimized for reads. These settings can be applied on a per-table basis.

[26] Projections are effectively shards, but of columns, not rows, as is typically the case.

*Greenplum* allows different forms of disk-based storage to be used. For example, commonly accessed tables can be stored on SSD media with other tables stored on standard disks. This is an area of emphasis for systems such as *Greenplum* that are purchased with high performance bespoke hardware, but it is increasingly applicable to databases designed for off-the-shelf hardware.

***Analysis of Clustered Databases***

Clustered DDBMSs tend to be consistent and available, since their designers assume that network partitions are rare in local area networks. If network partitions do occur, these DDBMSs may lose consistency.

Segment-level replication is popular because it allows data to be scaled out over many machines, unlike full-database replication which requires that every machine is capable of storing the entire database.

Hashing is commonly used for data placement because it allows databases to quickly place and retrieve data, and to easily repartition on the addition of new hardware. The locale of placement is less important in clustered databases because every machine is co-located. In contrast, locale is more important in wide-area databases, because placing data on a remote instance may greatly increase the latency of queries. The focus of placement optimization on clustered systems is on co-locating related data on the same machine, which can greatly reduce the number of queries that touch data on multiple machines [23]. This is why databases such as *Greenplum* allow database administrators to specify what columns to use when hashing records.

Optimal data placement is an NP-Hard problem [25], so clustered DDBMSs use mechanisms such as record coalescing to allow administrators to specify how data should be segmented, rather than trying to make automatic inferences.

*C-Store*, *VoltDB*, and *GenieDB* look to optimize the speed of clustered DDBMSs through specialization — *C-Store* and *VoltDB* do this by focusing on very specific target workloads, and *GenieDB* allows applications to switch to eventual consistency when possible. This reflects a trend of providing applications with the most minimal service needed, to optimize the speed at which this service can be provided.

### 3.1.2 Cloud-based Databases

***Amazon Relational Database Service***

***Amazon Relational Database Service (RDS)*** is a version of *MySQL* that is modified to run on virtual machines in Amazon's *EC2* cloud. It automates parts of the set-up and maintenance of instances, and creates off-line backups to a cloud datastore.

*RDS* uses standard *MySQL* functionality, which is the same as discussed in section 3.1. It does not automate repartitioning, so a system administrator must manually re-configure instances after the failure of the master.

***Xeround***

***Xeround*** is a distributed *MySQL* storage engine designed to be run on *Amazon EC2* [10]. It automatically repartitions data across instances, which means it is able to scale out without administrator intervention when new *EC2* instances are added.

*Xeround* instances use a similar structure to *MySQL Cluster*: standard *MySQL Server* instances provide an application interface, and *Xeround data instances* replicate data using segment-level replication.

To provide consistency and partition tolerance, *Xeround* uses a majority consensus protocol for locking. An update commits once it has completed on a majority of replica locations, and continues to execute asynchronously on the remaining replicas.

In the event of hardware failure, *Xeround* automatically re-replicates to maintain a system-wide replication factor, and, to protect against the catastrophic failure of all active replicas, it periodically writes back-ups to external storage.

***Yahoo Scalable Data Platform***

***Yahoo Scalable Data Platform (SDP)*** is a database clustering middleware which provides synchronous, full-database replication for *MySQL* instances in a data centre [68]. It is intended for web applications that are small enough to store data on a single machine and also require synchronous multi-master replication, which is not supported by standard *MySQL*.

Updates between instances are executed as ACID transactions. This is achieved by using synchronous replication and two-phase commit to provide atomic updates across instances.

*Yahoo SDP* is notable because it is a middleware solution, rather than a complete DDBMS. Database middleware systems are typically used to create heterogeneous distributed databases where disparate schemas are connected to enable cross-schema queries. Instead, Yahoo's goal is to provide a system that supports replication between individual *MySQL* instances. This task is simplified because *MySQL* uses full-database replication, so *SDP* does not need to make any decisions regarding data placement — updates are simply sent to every instance.

***Analysis of Cloud-based Databases***

Cloud-based databases are similar to clustered DDBMSs, with a few exceptions. Cloud-based databases place more focus on automating basic configuration tasks, which reflects the *software-as-a-service* philosophy of cloud providers[27].

Both *Amazon RDS* and *Xeround* automate the ability to create off-site backups, which indicates a greater focus on the possible failure of cloud-based instances than is seen with clustered instances. This is necessary because failed machines are outwith the administrative control of the organizations using them, so a systems administrator is not able to access or repair a failed machine; they can only restart the instance on another machine in the cloud.

*Xeround* is partition tolerant — it uses a majority consensus protocol for concurrency control — which is not an approach normally taken in clustered DDBMSs, where administrators can ensure that partitions are unlikely to occur.

Cloud-based databases are useful for applications with variable demand because they are easy to scale out to add greater capacity. New machines can be added on demand, unlike clustered databases which require machines to be provisioned and installed locally when needed.

[27] Software-as-a-service refers to a branch of cloud computing where a specific service is offered as a utility, rather than lower-level components such as virtual machines.

The main limitation of cloud-based databases is their distance from the applications sending queries and updates. Unless the applications are also hosted in the cloud, there is a significant latency and lower bandwidth between each system, compared to the case where both are hosted within an organization.

### 3.1.3 NoSQL Databases

***Key-Value Databases***

***Dynamo*** is a key-value datastore developed by *Amazon* to run in its data centres and support applications such as customer shopping carts [6]. It is designed to be "always-writable", because availability is Amazon's primary goal for customer-facing services. As a result, *Dynamo* is *available* and *partition-tolerant*, in terms of CAP. It uses a quorum consensus protocol to guarantee eventual consistency, and places data using hashing.

The hash function *Dynamo* uses to partition data does not guarantee that load will be spread uniformly over instances, which means that it is possible for some instances to be heavily loaded while others are unused. To compensate for this, *Dynamo* can assign each instance multiple segments of the keyspace using *logical instances*. Every logical instance is assigned an individual segment of the keyspace, and multiple logical instances can be stored on each database instance. Database instances responsible for storing more logical instances will take on more data, balancing load more effectively.

For concurrency control, *Dynamo* uses a quorum consensus protocol which allows applications to specify different consistency guarantees based on their requirements. Three configurable values ($R$, $W$, and $N$) can be tuned to achieve this. $R$ is the minimum number of nodes required for a successful read operation, $W$ is the minimum number of nodes required for a successful write operation, and $N$ is the number of nodes maintaining replicas of a given item. This approach produces a consistent system when both R and $W$ are greater than $\frac{N}{2}$, but Dynamo can also be run without any consistency guarantees ($R + W < N$) to reduce latency.

*Map-Based Databases*

***Bigtable*** is a distributed storage system used by Google for projects that require petabytes of storage [69]. It stores data in a multi-dimensional map, where each value is indexed by the following keys:

- Row name — the key for a particular row.
- Column name — column which is used within a row.
- Timestamp — version of the data.

The *timestamp* allows Bigtable to store multiple versions of the same data.

Replication in *Bigtable* is handled by the underlying file system – the ***Google File System*** (*GFS*) [70] — unlike *Dynamo*, which controls data placement directly.

*GFS* can be used as an append-only file system, which makes applications such as *Bigtable* write-optimized.

Data is stored in segments called *tablets*, which are shards of *Bigtable* tables. To locate data, *meta-data tablets* record where tablets are located, and the rows for which each tablet is responsible. Finally, a special meta-data tablet, the *root tablet*, stores the location of all meta-data tablets. Tablets are stored on *tablet servers* [71], as illustrated below.

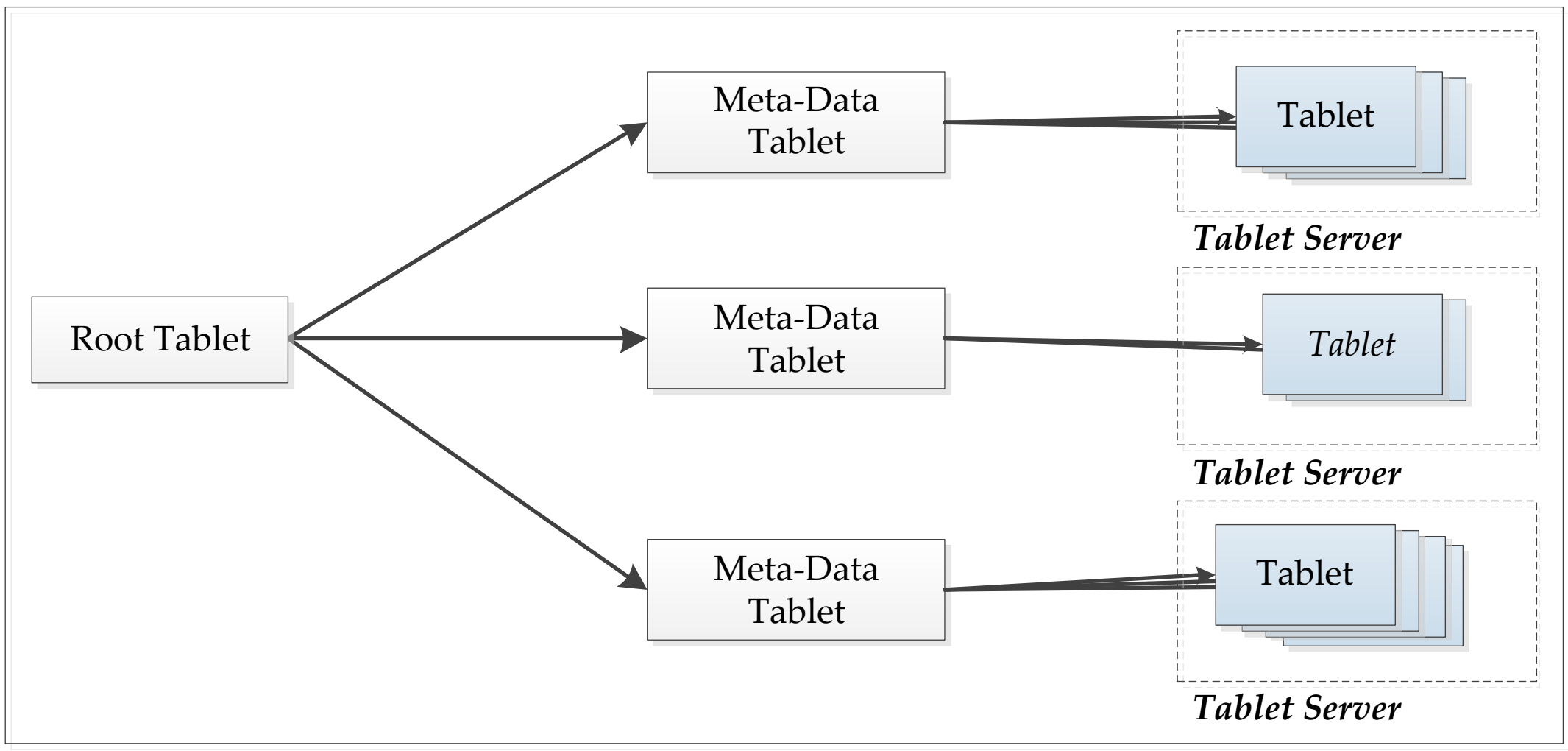

**Figure 31: Illustration of Bigtable's storage architecture, where data (stored in tablets) is located via special meta-data tablets.**

This architecture enables *Bigtable* to locate data by querying the root and meta-data tablets. It requires that there is only one root tablet, and that this tablet must maintain a consistent view of tablet locations, which would not be possible if two root tablets were created on either side of a network partition. This is not a problem with hashing approaches, which use a hashing function to locate data, rather than requiring the location of data to be stored in a known location.

To ensure there is only a single root tablet, *Bigtable* uses an external mechanism, the *Chubby* lock manager [72]. *Chubby* is a persistent lock service which is implemented as a file system: applications atomically create and read from directories and files, which are used as locks.

The *Bigtable* root tablet is stored in *Chubby* along with the database schema and the location of tablet servers. One Chubby instance is elected to be a *master*, which makes it responsible for the root tablet. In this role it assigns tablets to tablet servers, updating the state of the root and meta-data tablets in the process (illustrated below).

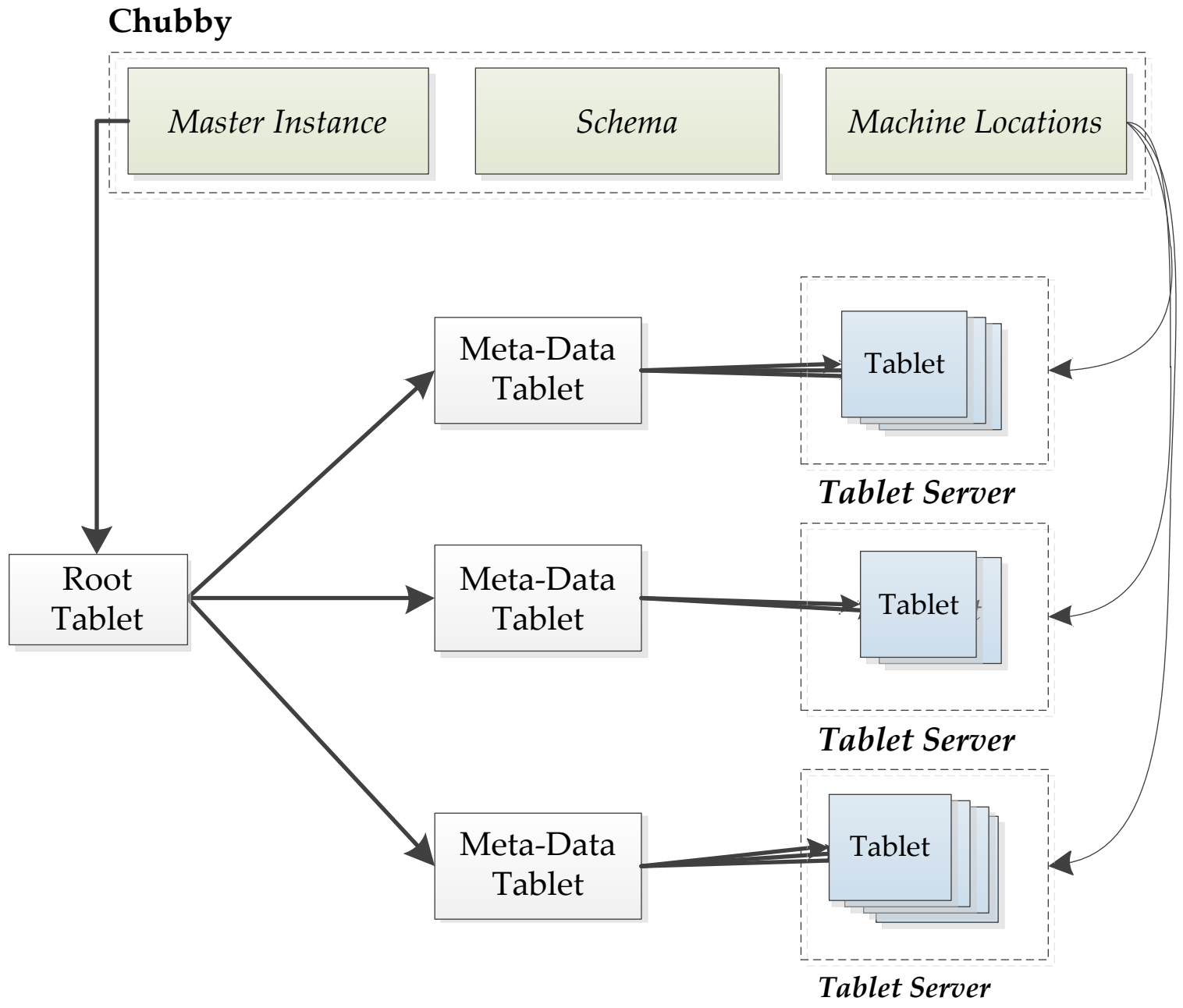

**Figure 32: An overview of Bigtable's architecture. Data items (tablets) are stored on *tablet servers*, which are referenced by special *meta-data tables*. *Chubby* stores references to the *root meta-data tablet*, and the location of every instance.**

*Chubby* replicates state, including the root tablet, over many machines. It uses *Paxos* to ensure that these replicas are consistent, and to elect a single master instance, which ensures that *Bigtable* consistently assigns tablets to tablet servers. A majority of instances must be available for *Chubby*, and therefore *Bigtable*, to be active.

In *Bigtable*, a tablet server must be able to connect to the *Chubby* instance to be active. This achieves two purposes: it allows the *Chubby* master to detect and react to failure, and it ensures that *Bigtable* is consistent and available. When a tablet server starts up, a uniquely named file is created in a special *Chubby* directory which lists all active servers. This file is used as a lock, and is held until the server's connection to *Chubby* is terminated. If the lock is lost, as would happen in the case of a network partition, the server shuts down. This is illustrated below, where Server *A* is partitioned.

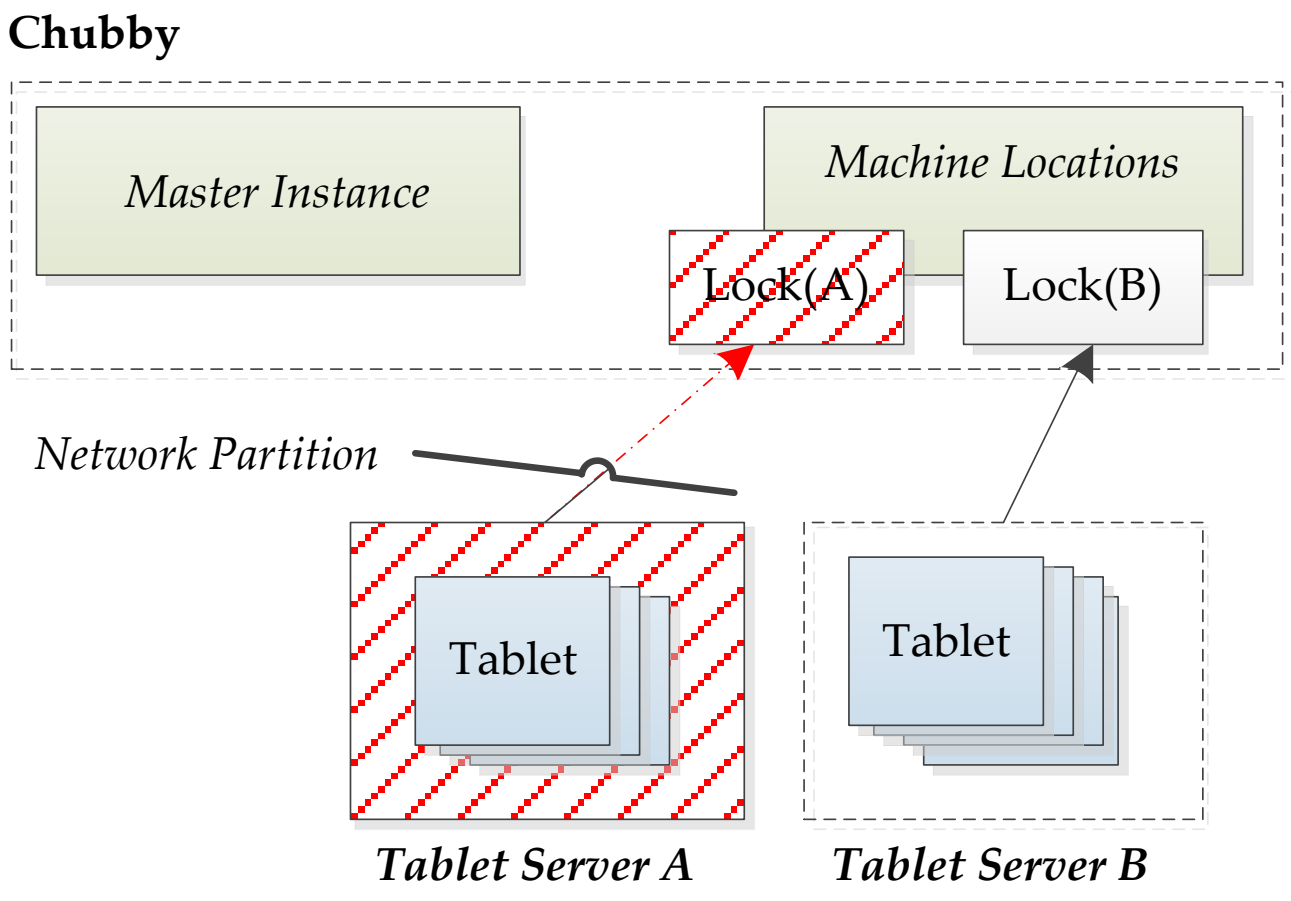

**Figure 33: An example of a network partition in Bigtable. Tablet server A shuts down when it cannot contact the *Chubby* instance.**

The master periodically contacts tablet servers to check whether they still hold a lock on *Chubby*. If they respond indicating they no longer hold a lock, or fail to respond after a number of attempts, the master attempts to obtain the lock itself. If it manages to obtain the lock, the tablet server has lost its connection, so the master removes the lock and re-partitions tablets onto other servers. This is the case with the example in Figure 33.

If the master cannot communicate with *Chubby* to obtain the lock, it is assumed that the majority of *Chubby* instances have failed, so the master shuts itself down. With this approach, changes to system membership are made quickly (when contact to *Chubby* is lost), unlike *Dynamo* which assumes that failure is transient and relies on manual intervention to remove instances.

A new master must be started when the current master fails. Once *Chubby* has agreed on the new master, the master communicates with all running servers (found via the *Chubby* directory listing) to obtain the current assignment of tablets to tablet servers. When no master instance is active the database can still be queried, but no new tablets can be assigned.

***Megastore***

*Megastore* is a storage system built over *Bigtable* that allows for SQL-like ACID transactions over data that is synchronously replicated between data centres [17]. It is discussed in this section due to its relation to *Bigtable*, though it is not technically a NoSQL system.

*Megastore* uses synchronous segment-level replication, where segments may contain shards of an extremely large table. For instance, email data may be sharded based on user accounts, or mapping data may be sharded by region.

ACID transactions are enabled by default on single-segment queries, but disabled on multi-segment queries due to the latency overhead involved in maintaining consistency. If ACID transactions are enabled, two-phase commit is used for cross-partition transactions, but an alternative system using asynchronous messaging is recommended because it reduces the latency of queries. *Paxos* is used to replicate the data in each segment.

***Cassandra***

***Cassandra*** is a multi-dimensional map database developed by Facebook [73]. In contrast to *Bigtable*, *Cassandra* does not store versioned data, and it uses hashing to place items on instances arranged in a conceptual ring.

*Cassandra* balances load across nodes by moving lightly loaded instances to different positions in the keyspace, to alleviate heavily loaded instances. This approach contrasts with *Dynamo*, which balances load by allowing instances to be responsible for multiple segments of the keyspace.

On start-up, nodes use a configuration file which lists a number of known nodes within the cluster. The starting node connects to one of these nodes, and joins the cluster by selecting a random position in the ring that identifies the portion of the keyspace for which it is responsible. This ring position is sent to other active nodes using a gossip protocol, and is persisted to *Zookeeper*, a service used to maintain configuration information and to perform leader election [74]. This process is illustrated below.

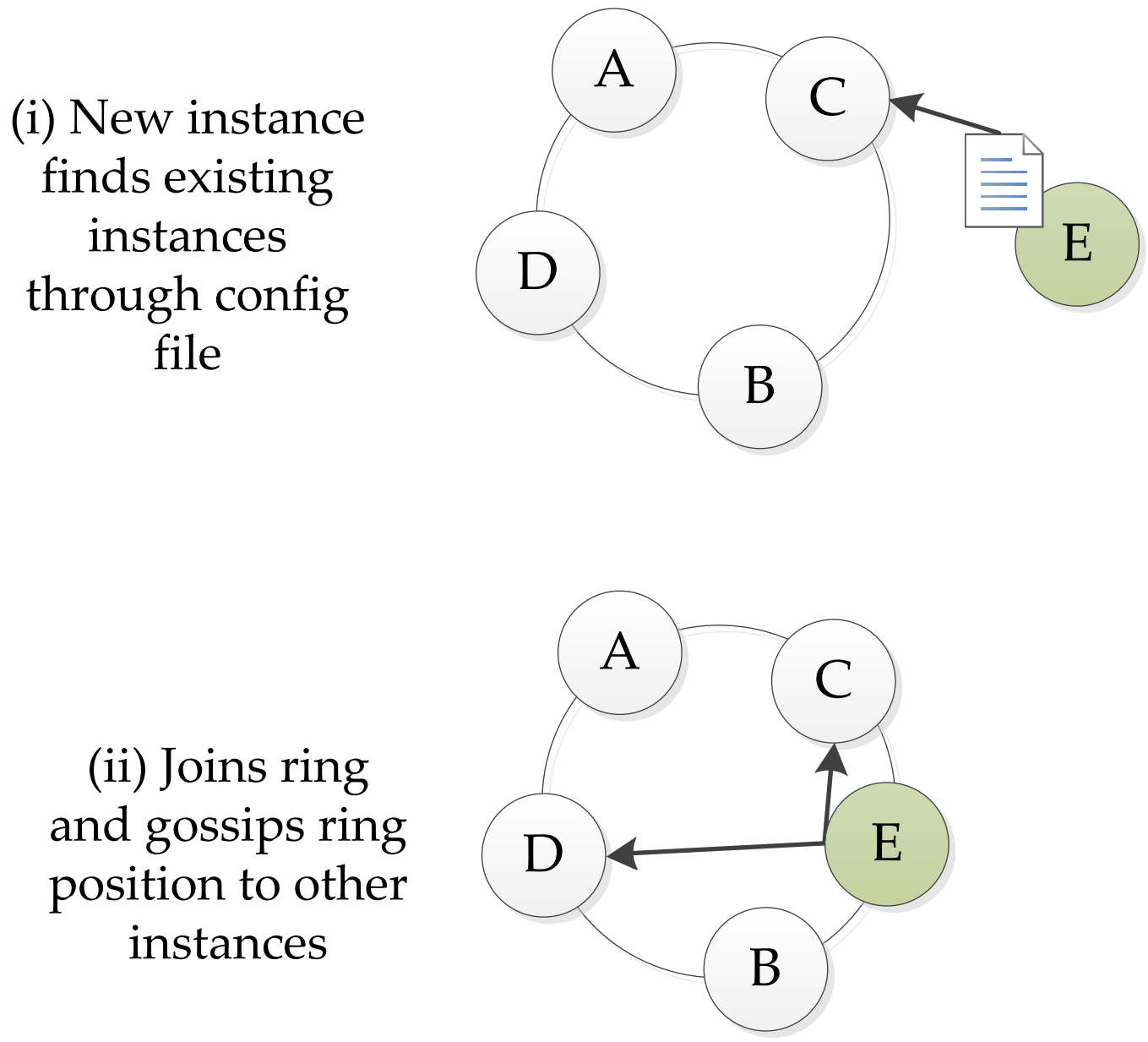

**Figure 34: The process of bootstrapping a new node in Cassandra.**

The configuration file used on start-up lists the locations of a number of *Cassandra* instances, so the starting node is not reliant on any one instance to join the system. *Zookeeper* can be used to maintain this configuration file, or it can be stored externally. These options reflect the likelihood that nodes within *Cassandra* will fail, so it is not possible to rely on a single set of nodes being available while a new node bootstraps.

***Document-Oriented Databases***

***CouchDB*** [3] is a document-oriented database that uses full-database replication [75]. Data is stored and returned to applications in JSON objects.

*CouchDB* replication is asynchronous, and can be initiated either by an update or by an application call (when the application wants to replicate). When replication is initiated, a *CouchDB* instance compares the contents of its local database to that of the replication target, and then transfers any updates that it has not yet committed locally. Every instance maintains a sequence number that is incremented on update, allowing other instances to identify when a new, unseen update has been made.

*CouchDB* supports ACID transactions on local instances, but is designed to allow instances to continue to operate even when disconnected from other instances in the DDBMS. Conflicting updates can be made to each disconnected database, and these conflicts can be merged later, using a conflict resolution process when the instances next connect to each other. Conflict resolution is managed automatically by *CouchDB*: each document has a revision history and a unique document ID, and the document with the largest history is chosen. If documents have revision histories of the same size, the document with the highest document ID is chosen. This mechanism is arbitrary, so it is recommended that applications override this process to preserve the semantic integrity of documents [75]. To do this an application must commit a new document which represents the merge of the two conflicting documents.

***CouchDB Lounge*** is a clustering solution for *CouchDB*. It allows data to be partitioned over multiple *CouchDB* instances by hashing the document ID, which is unique for every *CouchDB* document. *CouchDB* does not support automatic partitioning of data, but it is possible to repartition manually. A partition is moved using replication, by replicating an instance's state to an empty database [75].

***Analysis of NoSQL Databases***

NoSQL databases tend to be targeted at applications which value scale and availability more than consistency. This allows the replication scheme used by these databases to avoid costly distributed commit protocols and to execute updates asynchronously, which both reduce the latency of queries.

*Megastore* is designed to scale like a NoSQL database, but with the relational data model of an SQL database. This approach is limited because *Megastore* is designed to support ACID transactions for single partition queries only. While cross-segment ACID transactions are possible, they are not recommended because of the cost involved in doing so at a large scale. This approach is espoused by *Pat Helland* [76], who states in a position paper that atomic transactions cannot span multiple entities for a system to be *"almost-infinitely scalable"* (meaning it can linearly scale with any load). Helland's approach is to use an asynchronous, idempotent messaging system between entities, and to only guarantee serializability within entities.

Similarly, *VoltDB* is designed for workloads that rarely require multi-segment transactions. When transactions are particularly short-lived (most in-memory OLTP transactions take less than a millisecond [57]) the cost of any network communication greatly outweighs the other costs of executing the transaction [33]. The next section discusses *VoltDB* and a number of other systems in terms of their method of failure detection.

## 3.2 FAILURE DETECTION

### 3.2.1 VoltDB

Every instance in a *VoltDB* cluster (which is typically less than 100 instances in size) transactionally agrees on the set of failed instances — these instances store system membership and use a broadcast mechanism to reach system-wide agreement on failures [77]. Each *VoltDB* instance records system membership and the location of segments and segment replicas.

Instance failure is suspected if, after a specified period of time, a heartbeat message is not acknowledged by an instance. Every instance periodically broadcasts the set of instances suspected of failure, to update this information on other active instances. This broadcasting mechanism ensures that out-dated failure information is ignored by requiring that each instance compares the set of failed instances it receives to the set of failed instances recorded locally. If there are new failed instances, the instance receiving the updated set starts the failure notification process again by rebroadcasting the now larger set of failed instances. Instances drop failure messages if they do not contain a common subset of failed nodes, because this indicates that the sending instance has out-dated information. It is assumed that the sending instance will eventually catch up and rebroadcast the correct set, so each instance will eventually reach agreement on the set of failed nodes.

### 3.2.2 Boxwood

***Boxwood*** is a system developed at Microsoft Research that provides abstractions for storage infrastructures [78]. It provides a B-tree interface, allowing for basic lookup, insert and delete operations, and a chunk store interface that allows for variable-sized items to be written. Its mechanism of failure detection is discussed in this section.

*Boxwood* nodes send periodic heartbeat messages to each other, which are used as means of failure detection. Each node has a set of observer nodes which are the other nodes in the system responsible for exchanging heartbeat messages using the method described in section 2.4.1. *Boxwood* uses multiple observer processes rather than one, so a majority of observer processes must have received heartbeat messages for a monitored process to be considered alive. This means that a Boxwood node is not considered active if the majority of observer nodes are on the other side of a network partition.

### 3.2.3 Dynamo

*Dynamo* uses a distributed, gossip-based mechanism for membership and failure detection [6].

Changes to system membership must be made manually, because in a *Dynamo* data centre, node outage is often transient. The add/remove request can be made to any *Dynamo* instance, which writes membership changes to a local persistent store. This information is then propagated using a gossip-based protocol, which gives instances an eventually consistent view of system membership. Each instance communicates with a random peer every two seconds to reconcile their membership list.

Every *Dynamo* instance uses a failure detector to detect the failure of those nodes that it may have to communicate with during query execution. When a failure is detected, the instance uses other instances holding replicas of the required data to execute queries, and periodically checks whether the failed instance has recovered. Failure information is not shared with other instances and is not used to update system membership. This eventually consistent view of system membership is considered sufficient.

Dynamo's approach to failure detection and system membership is unusual among the DDBMSs discussed here. Administrators manually update system membership, which is acceptable because the typical replication factor of a data item is sufficient to tolerate the failure of multiple machines, and such failure is uncommon in the data centres in which *Dynamo* runs.

### 3.2.4 Chord

***Chord*** [79] is a peer-to-peer lookup protocol that is used to place items on nodes through hashing (as described in section 2.3.4). On joining a *Chord* system, each node creates a hash of their IP address, which is used to place them on the network, a logical ring ordered by the hash. Each node knows of its successor and predecessor — the nodes which come immediately before and after it in the ring, as illustrated below.

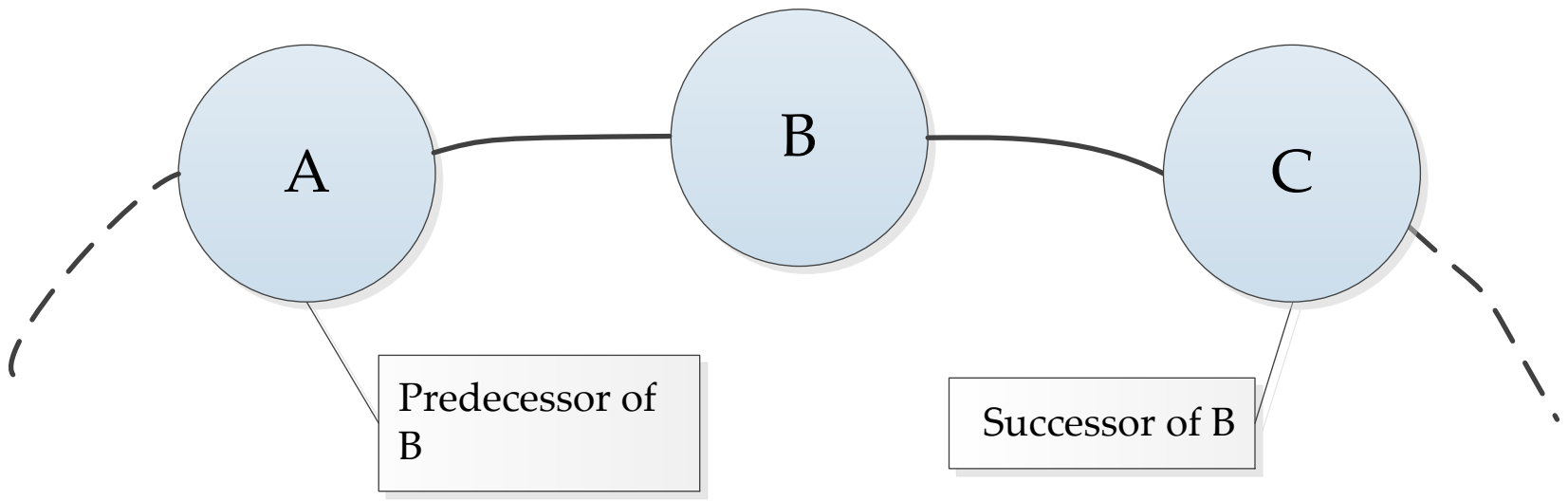

**Figure 35: An illustration of three *Chord* nodes that make up part of a *Chord* ring. *A* is the predecessor of *B*, and *C* is the successor of *B*.**

To preserve the integrity of this ring, Chord uses a stabilization mechanism which is periodically executed. In the above example, node *B* sends periodic messages to its successor, node *C*, to check that *C* still considers *B* its predecessor. There are two situations in which this is no longer the case: if *C* has failed, in which case it does not respond to the message, or if a new node has joined the ring in-between *B* and *C*, in which case *C* responds to *B*'s message with the address of the node it believes to be its predecessor. In both cases the stabilization mechanism is used to find *B*'s successor.

As a side-effect of the stabilization process, a Chord node issues a local event when its successor or predecessor changes. This can be used to notify an application of possible failure, assuming each instance of the application is linked to an individual Chord node. For example, in Figure 35, if there is a database instance running alongside each Chord node, the failure of *C* eventually results in a successor change event on *B.* The database instance running on *B* receives a notification of this change and is able to react accordingly. It may remove the database instance on *C* from system membership, or create replicas elsewhere of the data that was stored on *C*.

The advantage of this approach is that no central co-ordination is required. However, even if the ring eventually stabilizes, it is possible that some node failures may be missed in a situation (such as a power failure) where multiple machines fail at once. This will occur if three consecutive instances fail at the same time, as illustrated below in Figure 36, where, *B*, *C* and *D* fail.

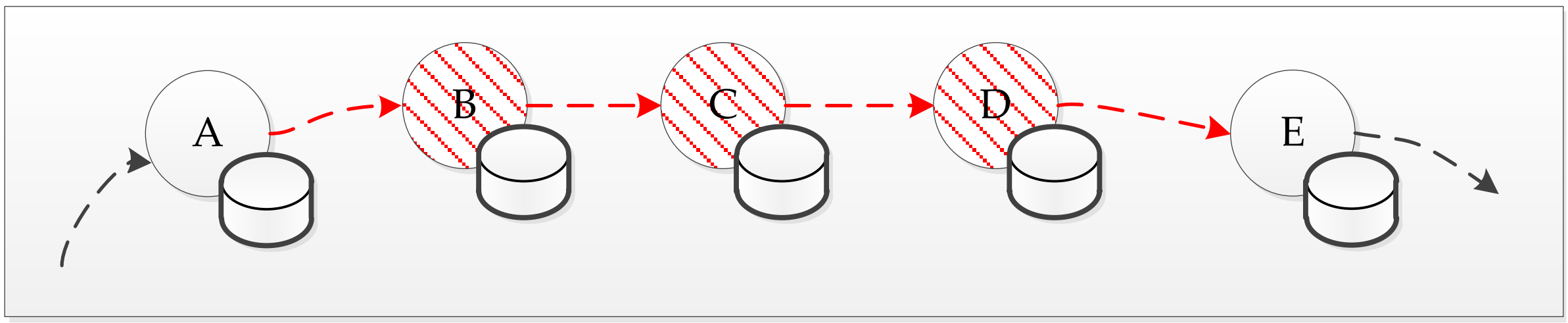

**Figure 36: An illustration of part of a *Chord* ring, showing three nodes (*B*, *C*, and *D*) that have failed.**

In this scenario, where instances know only of their immediate predecessor and successor, Chord will not detect the failure of *C* using the stabilization mechanism. Node *E* will notice a predecessor change, and node *A* will notice the failure of *B* and generate a successor change event, but no remaining active nodes will notice the failure of *C*, because the change events announcing the possible failure of *C* would be generated on *B* and *D*, two machines that have also failed.

In practice, *Chord* maintains a *successor list* containing an instance's immediate successor and *n* further successors, which means that this problem does not typically occur with the failure of only three instances. However, if the number of machines failing exceeds *n+1* instances, failures may still be missed. Those that are not detected by *Chord* will eventually be detected when the application tries to access a particular node. Consequently, *Chord* is a useful lightweight failure detector, but is not suitable for applications that rely on the failure detector to detect all failures.

### 3.2.5 Analysis of Failure Detection

The importance of failure detection to a system is based on a number of factors, including the likelihood of failure occurring, the impact of failure, and the need for quick detection of failure.

***The likelihood of failure occurring***. Systems such as *Dynamo* assume failure is rare, so failure detection is not handled automatically within the system.

***The impact of failure***. Clustered databases such as *VoltDB* rely on a small number of machines for replication, so the failure of a single machine may have a significant impact on system performance. Others, including non-interactive workstation-based systems, have an expectation of failure, so failure detection is not needed because they are designed to cope with unresponsive machines.

***The time to detect failure***. Failure detectors may incorrectly assume that a process has failed if it is slow to respond or it becomes partitioned. Some systems need to detect failure quickly despite the risk of having a false positive (*VoltDB*, *Boxwood*), while others are able to wait because the failure of an individual machine does not immediately impact system performance (*Dynamo*).

## 3.3 RESOURCE MONITORING

Resource monitoring is necessary for workstation-based systems, as it enables applications to determine where computation and storage is available, and provides feedback when a process consumes too many resources. This section discusses a number of existing approaches to gain an overview into typical architectures. It does not attempt to comprehensively review the literature in this area.

***Ganglia***

*Ganglia* [40] is a hierarchical monitoring system for compute clusters. It collects monitoring data on each machine and broadcasts to other nodes in the same cluster. Any node in the cluster can be queried for resource information, as can the re-publishers of cluster data. Ganglia re-publishers aggregate data and act as producers for higher-level re-publishers, allowing for a cluster hierarchy to be developed.

Monitoring information can be queried programmatically or viewed through a web-based interface, illustrated previously in Figure 23. *Ganglia* is intended as a tool to provide system administrators with an overview of resource availability over clusters.

*Ganglia* does not have a registry for node discovery, because it is targeted at relatively static clusters. Instead, an external mechanism must be used to make nodes aware of system membership.

***Hawkeye***

*Hawkeye* is a monitoring system that is used in conjunction with the workstation-based task scheduling system *Condor* (discussed later in 3.4.1). Every node in a *Condor* cluster runs a local monitoring agent, which is the producer of events. These events are sent to a central manager, which indexes the current state of nodes to allow applications to query availability through a command-line API or a web-based front-end. The central manager is a centralized re-publisher [43].

*Hawkeye* can also monitor specific events on each machine. This allows the local monitor to issue an event when, for example, available disk space goes below a threshold. This *threshold detection* is useful in a workstation-based system such as *Condor*, because it provides a mechanism for alerting the application when resources are becoming too scarce on the local machine. In *Condor* it may trigger the suspension or migration of a process.

***Remos***

*Remos* [80] is an example of a *network resource monitoring system*. It provides a query-based interface that allows applications to query information on network links between applications and on the structure of the network topology itself. The latter provides information on the connections between nodes in terms of link capacity, available bandwidth, and latency.

### 3.3.1 Application Monitoring

***OCM-G***

*OCM-G* [81] is an active application monitoring system. It provides services to interact with running applications, and to execute actions based on specific events.

*OCM-G* sensors are attached to applications through hooks in code or explicit function calls. The application-specific events generated by these are sent to a producer and onto a hierarchy of consumers, which expose an interface to allow client applications to retrieve monitoring data.

***Autopilot***

*Autopilot* [43] is an active application monitoring tool. To change application state, an *Autopilot* component called an *actuator* is integrated into the application, allowing *Autopilot* to modify application state and make function calls. Requests to the actuator are initiated by remote clients, which are the consumers of application monitoring data.

### 3.3.2 Analysis of Monitoring Systems

The monitoring systems covered in this section address a variety of monitoring tasks, from resource monitoring to application monitoring and adaption. Many of these are targeted at grid computing, but others such as *Hawkeye* are aimed at monitoring resources on workstation machines.

Grid-based monitoring tools provide methods for analysing monitoring information programmatically and viewing trends graphically. In contrast, *Hawkeye* is used primarily to inform applications when resource availability reaches a specified threshold. This is needed to ensure that workstation-based applications do not interfere with local processes.

*Hawkeye* is similar to monitoring tools such as *Autopilot* that allow applications to be modified. However, where *Hawkeye* informs an application of a specific event, *Autopilot* clients directly alter the state of the application.

## 3.4 WORKSTATION-BASED SYSTEMS

### 3.4.1 Condor

*Condor* [48] is a non-interactive workstation-based scheduling system which makes use of unused workstation resources by allowing long running computations to be run remotely. Provided that processes are self-contained and linked to *Condor* libraries at compilation, they can be *checkpointed*, meaning they can be paused for a short period, or moved between machines. Data produced as a result of computations is stored on the machine which submitted the compute job and not on the machine running the job, to prevent the job from monopolising resources on the remote machine.

*Condor* uses the *Hawkeye* monitoring system (discussed in section 3.3) to monitor local resources. *Hawkeye's* threshold detection is used to suspend processes when insufficient resources are available on the local machine to continue computation.

### 3.4.2 BOINC (Folding @Home)

***Folding@Home*** [49] is a volunteer computing project that distributes computation for research into protein folding, which is linked to Alzheimer's and Parkinson's disease.

Volunteer members of the public are encouraged to install a client application on their home machines to perform this computation when their computer is idle.

*Folding@Home* is one of a number of projects which uses the ***BOINC*** framework, which provides mechanisms for making use of idle resources on client machines [47]. *BOINC* projects have a *master URL* that is used by clients to sign up for a project. The master URL resolves to a directory service of *scheduling servers*, which give clients computation to perform during idle periods. Finally, when a computation is completed, the client uploads the result to a specified *data server*.

The *BOINC* client can be run on user machines as a screensaver, a background service, or a standard application. When the machine is idle, the client requests a *workunit*, which contains the data required for computation, along with the compute, memory and storage requirements needed to perform the computation.

*BOINC* makes extensive use of *redundant computing* to reduce reliance on individual machines completing a computation, and to identify erroneous results. Each workunit may be sent to thousands of machines.

### 3.4.3 Analysis of Workstation-based systems

Both *BOINC* and *Condor* are non-interactive systems with a focus on computation. Neither stores any data on the workstation machines being used, instead making use of external storage for the final results of computations.

*Condor* focuses on larger, longer-running computations that are expected to eventually complete, in contrast to *BOINC's* smaller workunits which run over machines that are likely to be available for shorter periods. As a result, *BOINC* makes much more use of redundant computing, to ensure that workunits complete somewhere.

There are no current examples of interactive, workstation-based systems.

## 3.5 Resource-Aware Systems

Resource-aware systems are those that take into account the availability of resources in their execution.

### 3.5.1 Resource-Aware Query Planning

*Lang and Patel* [82] propose that energy consumption be considered a first-class performance goal when planning and processing queries in a DDBMS. They consider two optimizations to help reduce energy consumption: allowing databases to explicitly order the processor to operate at a lower voltage when it is not needed, and queuing queries where possible, so that the number of repeat queries to the database can be reduced by aggregating requests.

An evaluation of these approaches showed a 49% reduction in energy consumption against only a 3% increase in response time for certain workloads [82].

### 3.5.2 Resource-Aware Data Placement

No existing work has considered introducing resource availability as a factor in database data placement decisions. However, the *Lang et al.* [83] discuss how replication schemes can be used to re-balance load when machines are shut down. The premise of this research is that machines should be shut down when they are underutilized to improve the energy efficiency of the system, but that this is only possible if the system is able to evenly re-balance load across the remaining machines.

The authors use a replication technique called *chained declustering* [84] to balance load evenly across instances. Instances are ordered in a ring topology, as with the hashing approaches described earlier. Each instance is assigned an equal number of segments (portions of a keyspace) to manage, and replicas of each segment are stored on the preceding instance in the ring. Requests for data go to the primary instance responsible for a segment ($N_x$), but if this segment is shut down requests are routed to the replica on the previous instance, $N_{x-1}$. To prevent instance $N_{x-1}$ from becoming overloaded, requests for the data items for which instance $N_{x-1}$ is primarily responsible, are redirected to $N_{x-2}$. This evenly balances when it is carried out on all remaining instances.

*Chained declustering* requires that nodes are shut down in a specific order, to prevent data from becoming unavailable. When an instance shuts down, the ring is broken and there are two *end instances* which only point to an active instance in one direction. If either of these nodes fails, some data segments will be lost [83]. Consequently *chained declustering* is not suitable for a workstation-based database, where node availability is not determined by the system, but by the users of the workstation machines or by power failure.

## 3.6 REVIEW

The previous sections have discussed concepts in database design and provided a summary of existing work. This section presents a critical overview of the systems presented so far.

### 3.6.1 Interactive, Workstation-Based Systems

There has been limited work towards interactive, workstation-based systems.

Existing work on workstation-based computing is limited to computationally intensive, non-interactive tasks. In these systems the storage capacity of workstation machines is only used to store the intermediate results of computation, rather than system-critical state. This makes the failure or unavailability of workstations easier to manage, because work can be transferred elsewhere without loss of data (other than intermediate results, which can be re-computed), but it limits the range of uses of workstations and requires that external servers are provisioned to store the results of computations.

*As a result, there is a need for research into the viability of interactive, workstation-based systems and the architectures needed to support them.*

### 3.6.2 Database Design

Existing DDBMSs are not well placed to make use of workstation-based environments for the reasons identified below.

***Static System Membership***

Databases supporting ACID transactions are designed to operate over static, rarely changing, sets of machines.

Clustered DDBMSs run over small numbers of static machines, so when one of these machines fails, it is not replaced by another existing machine. Instead, the system continues to operate at slightly reduced capacity (lower replication factor), until an administrator repairs or replaces the machine. This contrasts with machines in workstation-based systems, which are more likely to be unavailable for extended periods. A database running in this type of system must be able to automatically react to the failure or unavailability of machines, instead of waiting on system administrators.

The workstation-based system must make use of machines whenever they are available, even if the set of available machines is continually changing, whereas the clustered system relies on the same set of machines being available for long periods. The ACID compliant DDBMSs described in this chapter are not designed to handle frequent changes in system membership. Even *Dynamo*, an eventually consistent data store, requires manual intervention to repartition data.

*Consequently, there is a need to investigate the viability of running ACID compliant DDBMSs over highly transient sets of machines.*

***Highly Heterogeneous Architectures***

Clustered DDBMSs are typically not designed for highly heterogeneous architectures, where the resources on one machine may differ significantly from the resources on another. It is unlikely, for example, that one machine will have 20 GB in available storage, whereas another machine will have 100 GB. Similarly, each machine typically runs the same operating system.

In contrast, workstations within an organization often have vastly different hardware resources and run different operating systems. It is more challenging to design a database to run over this type of system because the database must account for differences in the capacity of machines to make full use of them.

Of the systems described in this section, some are designed to handle these problems. *Dynamo* and *Cassandra* adapt their hash partitioning scheme so that each machine can be made responsible for different volumes of data depending on available space. In *Dynamo*, instances can be placed into multiple points in the keyspace, whereas in *Cassandra* lightly loaded instances can be moved in the keyspace to take on more data.

Databases using full-database replication (*MySQL* and *GenieDB*) are not designed to run over machines with vastly different storage capacity, because each machine must be able to store the full database. Other systems are purchased as combined software-hardware solutions (*Greenplum, Clustrix Sierra*), and are therefore designed for very specific hardware configurations.

In-memory databases such as *MySQL Cluster* and *VoltDB* rely on uninterruptible power sources to make the failure of the majority of instances at a single point in time unlikely. These systems could not reliably run on workstation machines, which do not have uninterruptible power sources and will fail during power cuts.

*There is a need to analyse the database architectures that are appropriate for workstation-based deployment.*

***Data Placement Decisions***

DDBMSs are not designed to run over machines that have changing resource availability based on the activity of other processes on each machine.

When placing data, the databases discussed in this chapter look to co-locate related tables and spread load evenly over machines. None of these systems take into account the reliability of the machines storing data, because they assume that machines within clusters or data centres have an equal (low) chance of failure. They also ignore the utilization of resources on each machine, so a machine with fluctuating resource availability is treated the same as a machine that is often idle.

*There is a need to investigate a new data placement algorithm which makes use of information on the availability and utilization of resources.*

***Static Instance Placement***

The ACID compliant DDBMSs discussed in this chapter are not designed to continue operation when machines change IP address, because they assume a relatively static collection of machines.

*CouchDB* is the only one of these DDBMSs designed to handle transient connections and changing IP addresses. However, with this approach it only supports eventually consistent transactions, and it requires application intervention for conflict resolution. Any application using *CouchDB* must also specify where an instance has moved, because it provides no mechanism for automatically handling address changes.

*There is a need to consider the challenges and requirements for an ACID compliant DDBMS that runs over machines (such as laptops) which often change network addresses.*

***Administration Cost***

Setting up and maintaining a replicated database system is a difficult task which requires significant manual effort. New machines may need to be provisioned to support this system, and when a machine fails manual intervention is often required.

Many of the systems described above (*Greenplum*, *GenieDB*, *Oracle RAC*, *Clustrix Sierra*) are commercial solutions which require bespoke hardware supported by trained system administrators and dedicated support staff.

Of the systems often used by smaller organizations, *MySQL* and *CouchDB* make it relatively simple to set up databases on commodity hardware. However, to create a replicated *MySQL* or *CouchDB* installation requires more effort, and replicated instances of these systems must be actively maintained because the failure of a single instance can result in the system becoming unavailable, or backups not being kept. As a result, to run and maintain these systems requires significant active user involvement which is likely to deter many organizations from creating replicated databases.

*There is a need for a system which removes the deficiencies which make it challenging to maintain existing replicated databases, and a need to create new replicated databases.*

### 3.6.3 Conclusion

This chapter has identified two key deficiencies in existing research. Namely, there are no current interactive workstation-based systems, and existing applications that could run over such systems are not currently designed to do so.

***There are no interactive workstation-based systems.*** There are no systems that make use of the full capacity of workstation machines. No systems provide synchronous services that run over these machines. Instead, existing work focuses on non-interactive workstation-based systems, which allow applications to run asynchronous computations over workstation machines.

***Existing applications are not designed for workstation-based environments.*** Clustered distributed systems such as databases are not designed to run workstation-based systems in their current form. As this chapter has discussed, existing systems:

- Run over relatively static sets of machines
- Run over highly homogeneous machines
- Do not take resource availability into account when placing data
- Do not automatically handle machines which change IP address at runtime

Despite these issues, database systems are well positioned to make use of the resources available on workstation machines, given their need for storage capacity and computation. The highly structured nature of database data and the need for fast responses to queries provides a suitable test of the viability of interactive workstation-based systems.

In addition, it can be costly and difficult to set up and manage a replicated instance of a database system. This motivates the need for a database that is able to automate replication while making use of the existing infrastructure of an organization.

***This thesis investigates the viability of interactive workstation-based systems by designing a workstation-based database system.***

The next chapter looks at the requirements of such a system.

# Chapter 4:
# Requirements of a Workstation-based Database System

# 4 Requirements of a Workstation-based Database System

The primary objective of this work is to evaluate the viability of an interactive workstation-based DDBMS. This chapter presents the requirements for such a system, split into a number of categories.

*Database requirements* specify what the DDBMS must provide to be considered comparable to existing clustered solutions. For the workstation-based system to be viable it should have comparable performance to these existing solutions.

*Architectural requirements* specify what functionality is needed to run over workstations. These requirements are categorized by the specific need which they address — this thesis only evaluates a specific subset of these categories.

## 4.1 Database Requirements

The aim of this work is to create a DDBMS comparable to clustered databases such as *MySQL*. As a result, the requirements in this section reflect the need to provide standard functionality such as ACID transactions and an SQL interface.

**D1 Interface to Applications**

*The database provides a JDBC SQL interface.*

Clustered systems such as *MySQL* provide SQL interfaces that can be queried using various types of connectors. The database should support at least one of these connectors, JDBC.

**D2 Transactional Requirements**

*Transactions are ACID compliant.*

Clustered systems such as *MySQL* support ACID compliant transactions. For the database to be evaluated against these systems it must also provide this support.

**D3 Deployment**

*The database can be started in a replicated set-up without extensive manual configuration.*

It is relatively simple to install a non-replicated database instance such as *MySQL*. The database should be as simple to install as a non-replicated system, because a single non-replicated instance running over workstation machines is of little value given the reliability of these machines.

### D4 Performance

*The database has comparable performance to existing clustered DDBMSs.*

To be considered a viable alternative to existing clustered systems, the database should provide comparable performance.

## 4.2 ARCHITECTURAL REQUIREMENTS

The architectural requirements of the proposed database are guided by the characteristics of typical workstation-based systems, which are summarised in comparison to clustered systems in Table 3 below.

The typical deployment of this system, as discussed in Chapter 1, is over a number of workstation machines within an enterprise, replacing DDBMSs currently run within a server cluster in the same enterprise.

| Cluster-based System | Workstation-based System |
|---|---|
| Large storage capacity, specifically provisioned | Small storage capacity (on each machine), large overall storage capacity |
| Software can make full use of resources available on machine | Software must yield resources to users when needed |
| Machines are generally highly available, and the system administrator is in full control of maintenance | Machines may become unavailable due to user activity and can be restarted by users |
| Small number of machines available | Potentially large number of available machines |
| Static IP addresses | IP address may change at runtime, particularly for laptops |

**Table 3: A comparison of workstation-based and clustered systems.**

The database's architecture must be designed to accommodate these traits. Table 4, below, extracts some informal requirements based on the characteristics of workstation-based systems in Table 3:

| **Characteristics of Workstation-based Systems** | **Implied Requirements** |
|---|---|
| Small storage capacity (on each machine), large overall storage capacity | Allow the database to be partitioned over many machines rather than storing the entire database on a single machine |
| Software must yield resources to a user when needed | Resource monitoring to ensure the database is able to identify when it can use local resources |
| Machines may become unavailable due to user activity and can be restarted by users | Replicated state and no single point-of-failure, to cope with machine failure |
| Potentially large number of available machines | Autonomic system able to identify and make use of machines as they become available |
| Possibly dynamic IP address, particularly for laptops | Database identity and table locations handled independently of IP address |

**Table 4: The implied requirements of a DDBMS running over workstations.**

This section categorises requirements into the following groups:

- General. Architectural properties required for any workstation-based system.
- Fault Tolerance. Functionality required for the system to withstand the failure and periodic unavailability of workstations, whilst still maintaining regular operation.
- Resource-Awareness. Functionality required for the system to manage its resource utilization so the current operations of a workstation are not disrupted.
- Autonomics. Functionality required for the system to manage itself, without manual intervention from a system administrator.

The *general* and *fault tolerant* requirements are necessary for a workstation-based system to operate itself. The requirements of *resource-awareness* and *autonomics* are necessary for the system to be able to operate without disrupting other applications, and without requiring extensive manual administration, but they are not necessary for the operation of workstation-based system itself.

The remainder of this section extracts a formal set of requirements from this analysis.

*General*

### A1 Self-Contained

*The system is able to run in its entirety on a set of unreliable machines.*

Non-interactive, workstation-based systems typically rely on external storage to store the results of computations. An interactive system should be able to run using only the resources of unreliable machines such as workstation machines. This ensures that new machines do not have to be provisioned to run the system.

### A2 Capacity

*A database can grow bigger than the capacity of a single machine.*

Large DDBMSs typically allow data to be spread over multiple machines because the database may not fit on a single machine. This is particularly important in a workstation-based system where hard disks have not necessarily been provisioned to store large datasets and may only have a relatively small available capacity.

### A3 Heterogeneity in Platforms

*A database instance can run on multiple operating systems.*

Workstations within an office are often heterogeneous with respect to the operating system they run. To make use of the capacity of all of these machines, a workstation-based system should be able to run on each of the most common operating systems (Windows, OSX, Linux).

*Fault Tolerance*

### A4 Resilience to Failure

*The database is able to withstand the failure of individual machines running database instances.*

The database should not require manual intervention to recover from machine failure, as it is expected that a workstation-based database will have lower availability than a typical clustered DDBMS, and instances may fail too often to rely on manual recovery. It should be *k*-safe, indicating that it is able to continue operating up to the failure of *k* machines, where those machines store replicated data relating to a specific process. The value of $k$ — effectively the number of replicas to be created — should be configurable, but is limited by the number of machines that are active in the database system at a point in time.

### A5 Mobility

*The database is able to handle database instances changing IP addresses.*

It should be possible for database instances to restart with a different IP address and still be able to operate correctly. Clustered DDBMSs are typically not designed to support this requirement.

It is more likely that a database running over workstation machines (which may include laptops) will have to cope with this problem because machines may use dynamic IP addresses, or they may be moved and restarted. A laptop, for example, may be taken from work and connected to a home internet connection.

***Resource-Awareness***

### A6 Local Resource Monitoring

*The database is able to monitor the availability of resources on every node in the system.*

To utilize unused resources effectively on existing machines the system must be able to identify what resources are available. Local monitoring is needed to establish the availability and utilization of resources such as CPU, memory, and disk.

### A7 Network Resource Monitoring

*The database is able to monitor the bandwidth and latency of connections between instances.*

Locality is important in DDBMSs to prevent network latency becoming a bottleneck in query execution. If a database is too large to be stored on a single machine it is desirable (in terms of network latency) that tables are located on nearby machines to ensure that join queries touching multiple instances are not prohibitively slow.

### A8 Local Resource Analysis

*The database is able to analyse local monitoring information to determine when insufficient resources are available to complete a query, and when data or processes must be moved to other machines.*

When resources on a machine are constrained by other user processes or activity, the database should be able to react by moving tables and processes (such as lock managers) to other machines.

The system should be able to identify patterns in availability that can later be used to predict when movement of data and processes may be necessary. For example, a machine may be shut down or restarted every evening, or a user may make heavy use of local resources every day during work hours and no use of them in the evening.

### A9 Global Resource Analysis

*The database is able to make use of monitoring information from a number of database instances to make decisions on data placement.*

Given a potentially large number of machines in an enterprise it is important that the database is able to place tables on an appropriate machine(s) based on their availability and locality.

Global analysis must also be used in placement decisions to manage the conflicting goals of maintaining locality amongst related tables to improve query performance, and balancing load, both to prevent overloading a single machine, and to ensure data is resilient to failure. It should be able to identify what machines are reliable and what machines are not, to avoid placing replicas on a set of unreliable machines.

***Autonomics***

### A10 Replication

*The database is able to automatically replicate data with no user involvement.*

It should be possible for a user with little experience in creating highly-available systems to use this database to make their data accessible to users on remote systems, resilient to the failure of individual machines, and scalable to increased demand and larger datasets.

They should not have to manually partition data, replicate state, adjust replication factors, or move data or restart processes on failure.

### A11 Opportunistic Utilization

*The system is able to incorporate new resources into the database at runtime.*

The DDBMS should be able to make use of new resources as they become available. It is likely that the set of available machines will constantly change over time, meaning it is not possible to rely on a small set of machines.

### A12 Using Resource Awareness

*The system is able to use resource monitoring data to autonomically manage the placement of data and processes.*

The DDBMS must be able to autonomically manage the placement of data and processes, and determine when data and processes should be moved as resource availability changes.

## 4.3 CONCLUSION

This chapter has identified the requirements for a workstation-based database system.

The design of such a system, named D2O, is discussed in the following chapter.

# Chapter 5: The Design of D2O

# 5 The Design of D2O

D2O is an interactive, workstation-based distributed database system designed to meet the requirements presented in the previous chapter. It is resource aware, which enables it to run over a dynamically changing set of machines, and it is fault tolerant, enabling it to recover from the failure or unavailability of individual machines.

## 5.1 Overview

There are two independent components in the system's architecture, a database component and a bootstrapping component.

The database component provides the database functionality required in *chapter 4.* Tables can be replicated across D2O instances, and are accessed through an exclusive write, shared read locking mechanism. In this respect D2O is similar in design to clustered databases such as *MySQL* and *PostgreSQL*.

D2O differs from these DDBMSs in that it is intended to be run over a less reliable set of machines. As a result, it has various features designed to improve its resilience to failure that are more commonly seen in larger clustered databases and in cloud-based data stores. For example, updates are fully synchronous, meaning that a table will still be available and consistent despite the failure of a single machine. This approach is similar to *VoltDB* and other multi-master databases.

The second component in the system, a bootstrapping component called a *locator server*, also reflects the fact that the database component must be run over an unreliable set of machines. A new node joining the system must be able to find and join the set of extant instances, even though this set can completely change over time. This bootstrapping problem is managed by the *locator server*, which takes an approach similar to that of *Paxos*, and is discussed in 5.7.

The design of D2O is a combination of features from clustered database architectures (to support ACID compliant transactions), and cloud datastores (to cope with the frequent failure of machines). This chapter discusses its architecture and the justification for the design decisions that were made.

## 5.2 THE ARCHITECTURE OF D2O

A D2O instance is designed to run on a collection of machines in an organization and on the machine of any database user, where it acts as a local point of presence. Tables can be created and stored on any one of these instances, as they are all equal peers. Figure 37 illustrates this concept from the perspective of a single user's machine: this machine runs a local D2O instance, and this instance is connected to the other identical instances which are running on other users' machines with spare compute capacity.

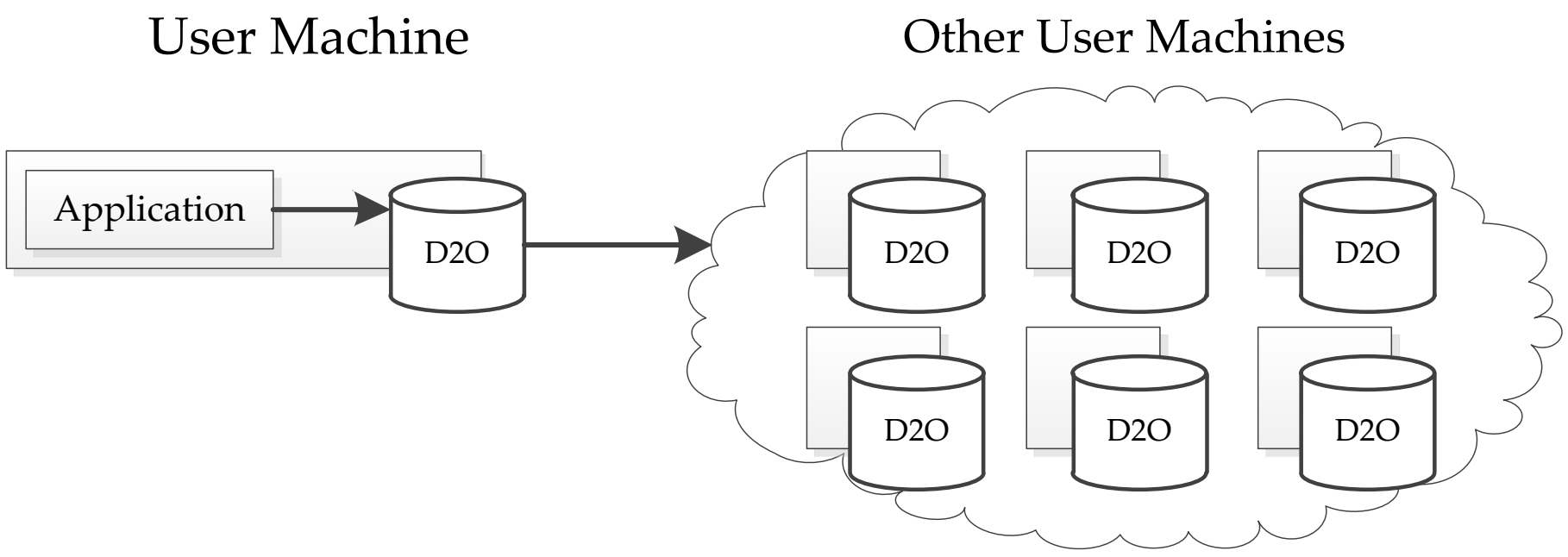

**Figure 37: A vision of D2O's use within an enterprise.**

The lock management of tables stored on these instances is distributed, meaning there is no single, centralized lock manager. Instead, each table is managed by a ***Table Manager***, a process that is responsible for lock management of that table. The *Table Manager* also records the location of every replica of its assigned table, so it can find what replicas must be updated and what replicas can be queried given an application request. Circular wait is prevented by requiring that locks are requested in a fixed order. Two-phase commit is used to commit updates.

To enable discovery of existing tables and to mediate the creation of new tables, the system maintains a ***System Table*** which holds references to all extant *Table Managers*.

The System Table is needed to find extant Table Managers, though references to these Table Managers may also be cached locally by database instances. There is only ever one active System Table in the system, and one active Table Manager for each database table, though the state of these components is synchronously replicated, which allows them to fail without rendering the database system permanently unavailable[28] — they can be recovered elsewhere using this replicated state.

### 5.2.1 Use Case

The above architecture is illustrated in this section with two use cases. In each case, users submit queries to the database instance on their machine and the query is executed at the most appropriate replica(s). This architecture is illustrated in Figure 38.

*Machine B* is currently responsible for maintaining the System Table, which maintains references to the Table Managers for tables *X* and *Y*. The Table Manager for *X* maintains references to replicas of the table on machines *C* and *D*, while the Table Manager for *Y* maintains references to the table on *machine C*. A user making queries from *machine A* has no knowledge of the location of the System Table, the Table Managers, or the data. They only need to know how to connect to the local point of presence, the D2O instance on *A*. This ensures that an application can connect to D2O as if it were a standard local database system, without needing to know about the topology of instances within the system or the distribution of data.

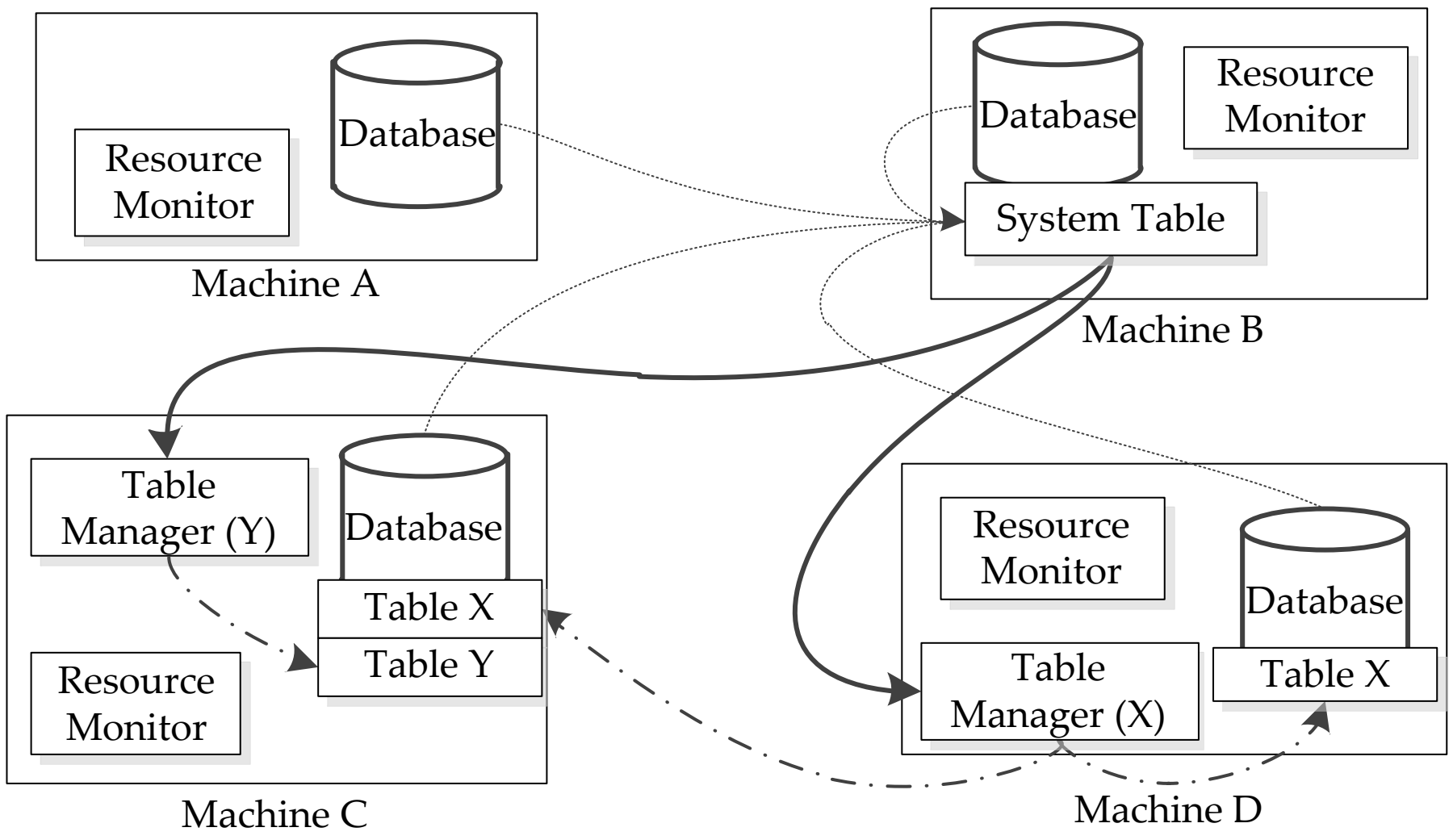

**Figure 38: An overview of D2O's Architecture, including Table Managers and the System Table. There is also a resource monitor on each machine.**

[28] Table Manager locking information is not persisted because in the event of failure any running transactions are rolled back.

**Example Use Case (Query)**

To illustrate the architecture of the system consider how a basic join query is executed by the database system.

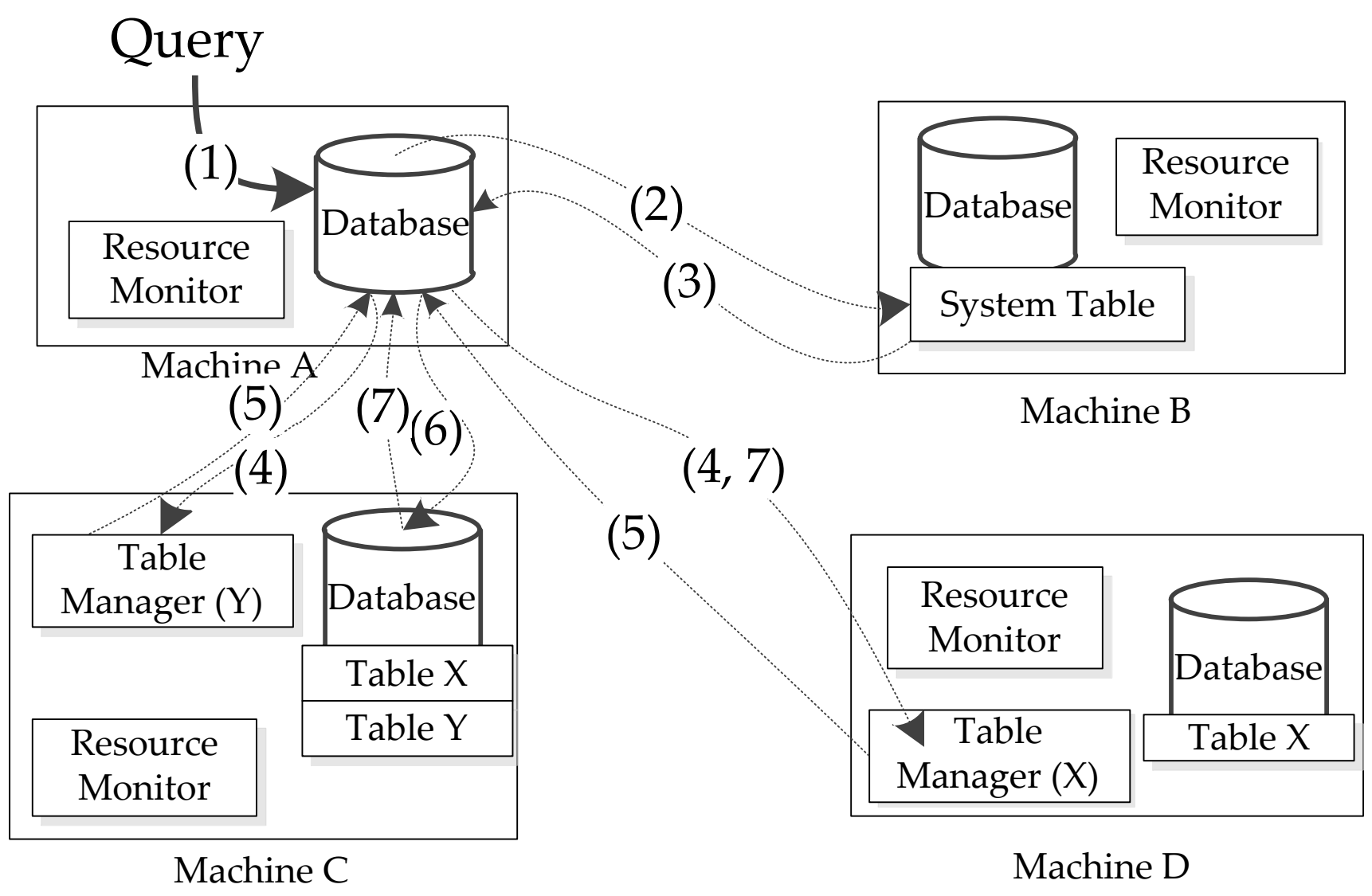

**Figure 39: An illustration of a join query being performed in D2O.**

1. A user submits a query via a database interface on their machine, *A*.

```
SELECT * FROM X, Y WHERE X.a_id = Y.a_id;
```

2. Their local database instance (on machine A) parses the query and sends a request to the System Table to discover the location of the *X* and *Y* Table Managers[29].
3. The System Table returns the location of these Table Managers on machines *C* and *D*.
4. The user's local database instance (which now has references to both Table Managers) requests read locks on both tables from their managers.
5. The Table Managers return locks and meta-data describing where replicas can be found.
6. The query is sent to machine *C*, which holds both tables, and is then executed. The decision about which machine executes the query is based on monitoring information relating to computational availability on machines and on other resource monitoring[30].

[29] The mechanism for discovering the System Table is discussed in section 5.7.
[30] While this is true of the design of H2O it is not currently true of the implementation. Monitoring information is used to determine replica placement, but not in the selection of replicas for read queries.

7. Once the query has been executed, the co-ordinating instance (the instance which initiated the query) commits the transaction and releases the locks for both tables on the Table Managers. The result of the query is then returned to the user.

**Example Use Case (Update)**

This example illustrates how updates are handled.

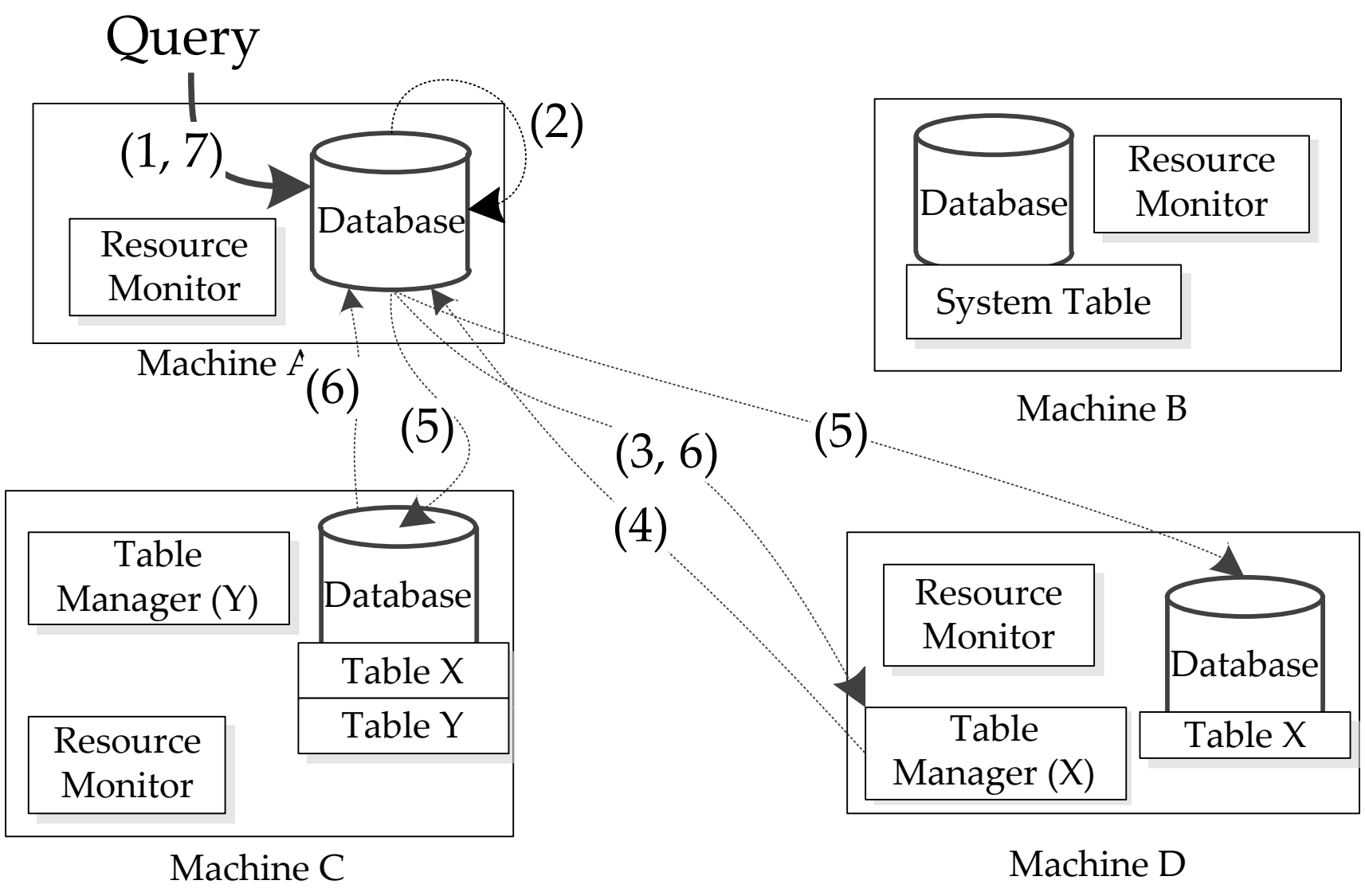

**Figure 40: An illustration of an update being executed in D2O.**

1. A user submits an update via a database interface on their machine, *A*.

```
INSERT INTO X VALUES (21, 'Charles Woodson');
```

2. Their local database instance (on machine *A*) parses the query and obtains a locally cached reference to the Table Manager for *X* (which exists because it was accessed in the previous query).
3. The user's local database instance requests a write lock for *X* from its Table Manager.
4. The Table Manager returns locks and meta-data describing where the table data can be found.
5. The update is sent to each replica and executed.
6. Once all replicas have sent responses to *A* indicating that the update has been successful (the *PREPARE* message of three-phase commit), *A* issues a *PRE-COMMIT* message then a *COMMIT* message to all replica sites and to the Table Manager, completing the update.
7. The result of the update (an integer representing the number of rows changed or added) is then returned to the application.

These examples illustrate a number of the features of D2O, including:

**Interface**. Applications connect to an D2O database through a JDBC interface provided by the local point of presence, which is itself a regular D2O instance.

**Locking**. A pessimistic two-phase locking mechanism is used to guarantee isolation of transactions. Each table has its own lock manager, the *Table Manager*, and another lock manager, the *System Table,* is used to manage the global schema.

**Replication**. D2O uses table-level replication to give control over the placement of data onto database instances. Each *Table Manager* is responsible for handling the replication of its own table's data.

**Resource monitoring**. A resource monitor runs on every machine running a D2O instance. The monitor reports monitoring information by storing it in the local D2O instance.

The remainder of chapter 5 describes this architecture.

## 5.3 SYSTEM TABLE ARCHITECTURE

The System Table maintains the global schema of a D2O database. It is responsible for handling requests to create new tables and requests to find Table Managers for existing tables. When a *CREATE TABLE* request is made to the System Table it must check whether a table with this name already exists.

There can only ever be one active System Table — an invariant that must be upheld to ensure consistency. If there were more than one System Table it would be possible to create two tables with the same name, which would violate consistency. *This invariant is guaranteed by the locator server architecture described in section 5.7.*

If a Table Manager fails, it is the System Table's responsibility to restart a replacement. This means that the System Table must be aware of the current location of all active Table Managers and the location of all replicas of Table Manager state. *The mechanisms used by D2O to detect failure are discussed in more detail in section 5.10.*

The System Table also aggregates monitoring information summarising the availability of all machines in the instance of the database system. This allows it to rank machines based on their resource availability, and is used when deciding where to place replicas. *Resource monitoring is discussed in more detail in section 5.11.*

To summarise, the System Table stores the following information:

- Location of all extant Table Managers
- Location of all replicas of Table Manager state (Table Manager replicas)
- Location of all D2O instances
- Summaries of monitoring information, showing machine availability

The System Table can be dynamically migrated between database instances at runtime. This can be done either programmatically or through a *MIGRATE SYSTEM TABLE* call from an application.

## 5.4 Table Manager Architecture

A Table Manager is created for a table whenever a *CREATE TABLE* statement is committed[31] — its state is synchronously replicated to other instances, but only a single active Table Manager is created. The system chooses where to store replicas using resource availability information, as discussed in 5.11. When a user issues a query or update to a table, the Table Manager must then be contacted to obtain a lock.

Lock requests specify what type of access is required (shared or exclusive). The Table Manager returns the type of lock that it has granted and the location of all replicas for a table, allowing the requesting instance to find the tables it needs to update or query. *The mechanics of this locking approach are discussed in section 5.6.*

Like the System Table, Table Managers can be migrated at runtime onto other machines. This can be initiated programmatically by the database instance itself, or via a *MIGRATE TABLE MANAGER* call from an application.

[31] As soon as a CREATE TABLE operation is completed, the database commits the current transaction, even if it is configured not to commit until a *COMMIT* command is executed (i.e. *auto-commit* is turned off). This is the approach taken by most databases, including Oracle [97] and MySQL [98].

## 5.5 REPLICATION

D2O synchronously replicates table data to *n* database instances, where *n* is a configurable value. This ensures that, if *n* replicas are successfully created, queries will proceed and access current data despite the failure of up to *n-1* replica sites.

### 5.5.1 Granularity of Replication

Table-level replication is used to replicate data, meaning each Table Manager is responsible for replication of the table data it manages. This gives the database more flexibility in data placement than full-database-level replication, because the set of replicas stored on one instance does not have to be the same set of replicas that are stored on another instance. In a workstation-based system this is important because storage capacity may vary significantly between machines, and load balancing is more difficult if the full database must be moved from machine to machine.

Table-level replication introduces some complexity, because a Table Manager is required for each table, in contrast to a per-segment or full-database manager. However, it is more flexible than segment-level replication, because the set of tables that are stored on a machine can be varied dynamically as the load on the machine changes. The contents of a segment are typically not altered dynamically, and where they are (Dynamo [6], Cassandra [73]), hashing is used to determine placement. This approach does not provide the level of control needed for a workstation-based system, where a placement decision is made based on resource availability and not just capacity.

### 5.5.2 Method of Replication

Replication is performed *per-update*, meaning when an update is executed it is sent to each replica location synchronously, and executed. This is illustrated in Figure 41, below. In this example a database instance, *A*, is updating a table, *X*. By obtaining a lock from the Table Manager, *A* also obtains the list of locations where *X* replicas are held. It then sends the query to each replica and executes a three-phase commit to ensure they remain consistent.

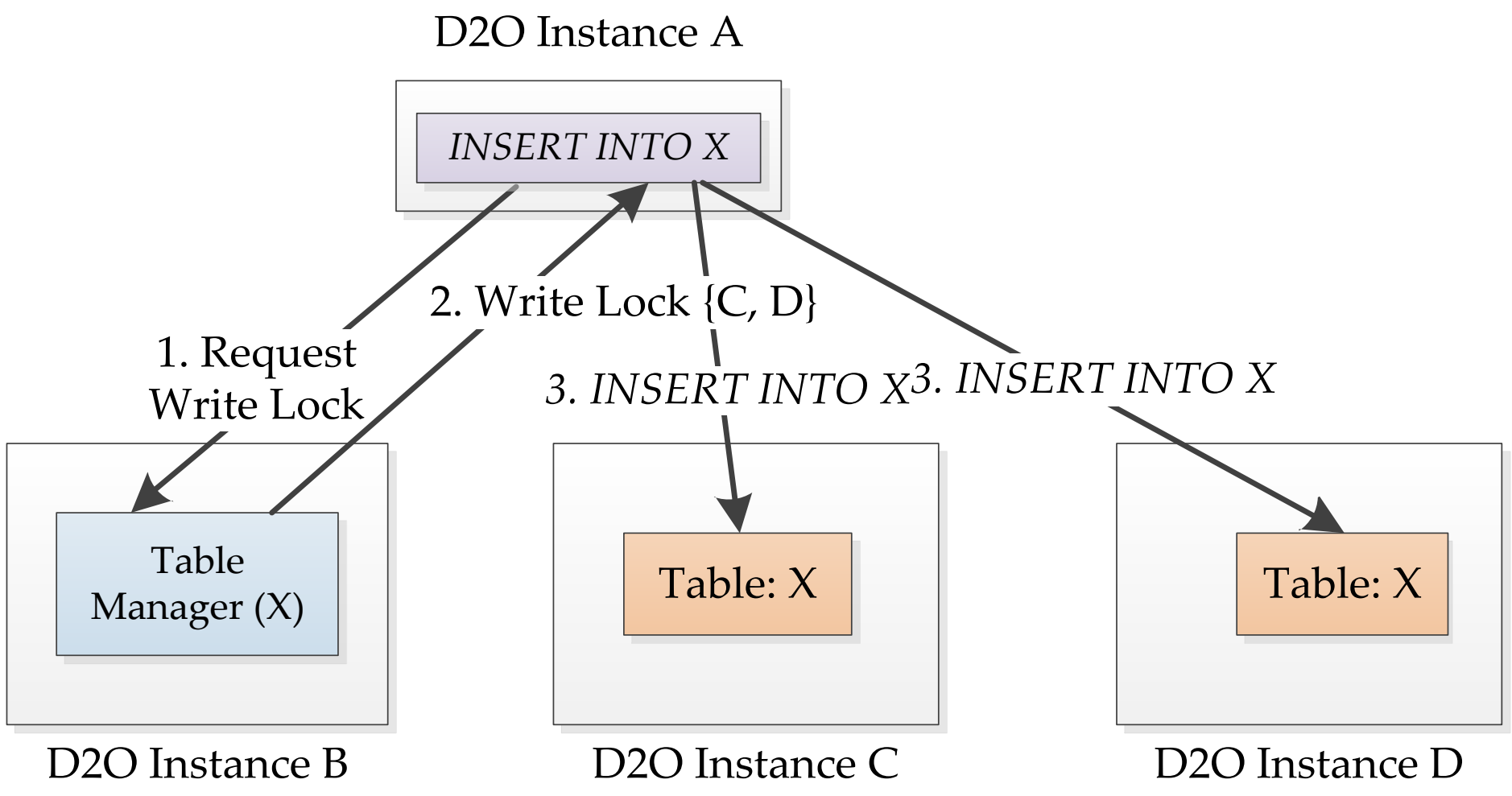

**Figure 41: An illustration of D2O's method of lock acquisition through Table Managers, and update propagation to replica sites.**

By default, replication is synchronous, but if a replica site has failed, an update can still commit by completing on the remaining replica sites. D2O also supports partially synchronous replication, where only a configurable number of replica sites, *r*, must complete an update for it to be committed — so if *r* replicas are available after a failure, or if some remaining replicas are slow to complete, an update is able to commit. The remaining replicas are eventually updated if they are active, but they cannot be used for read queries until this occurs.

Table Managers can trivially manage replication because they are also used as lock managers for individual tables, meaning they are able to manage when and how updates take place. Locking is discussed in more detail in the next section.

## 5.6 ACID Transactions

To recap *Chapter 2*, an ACID compliant database must ensure that transactions are atomic, consistent, isolated, and durable. D2O is designed to provide ACID transactions.

Durability is provided by logging updates to disk, which ensures that they are recoverable after machine failure if the hard disk of the failed machine is not corrupted. In addition, hard disk failure is not fatal, because updates are synchronously replicated across multiple machines.

D2O uses *pessimistic two-phase locking* to provide isolation. To perform an update an instance contacts the Table Managers of each table involved in the update and obtains an exclusive lock. Queries request a shared lock from each Table Manager.

Isolation is guaranteed by the two-phase locking invariant that all locks must be taken out before any locks are released. This prevents interleaving updates accessing shared state.

*Three-phase commit* is used to commit or rollback transactions. The co-ordinator of the three-phase commit operation is the instance which initiated the transaction, while the participants are the replica sites and the System Table or Table Managers involved in the transaction.

Three-phase commit is used due to the increased likelihood of co-ordinator failure, which would result in updates being blocked if two-phase commit was used.

The System Table commits database-level operations[32], and the Table Manager commits all table-specific operations[33]. The next section discusses the role of the Table Manager in locking.

### 5.6.1 Locking

D2O creates a Table Manager for every table in the instance of the database system. This means that a query involving three tables must request locks on all three table managers. However, when multiple Table Managers are located on the same instance, the Table Managers can be combined into a single composite manager, which allows a single lock request to be made to multiple tables.

Table-level locking is a good fit with table-level replication because the lock manager (the Table Manager) can trivially support replication by informing instances requesting write locks what sites need to be updated. It is used in D2O for two further reasons. First, Table Managers can be moved closer to the site of frequent lock requests to improve response time. In a typical database cluster this is not an issue, because the latency between instances is extremely low, but this is not necessarily true of workstation-based systems. Second, it

[32] Any query involving the creation or deletion of tables and schema, including: *CREATE TABLE*, *DROP TABLE*, *CREATE SCHEMA*, and *DROP SCHEMA*.

[33] Any query specifically involving an operation on a table, including: *INSERT*, *UPDATE*, and *DELETE*.

ensures that no single machine is responsible for lock requests, which removes a point of contention in a system where machines are likely to have limited resources available.

A majority consensus protocol cannot be used, because such an approach requires the system to know how many nodes are in the system (which is not possible with dynamic membership and the potential for network partitions), so that an instance knows when a majority has been reached.

## 5.7 Creating and Joining a Database System

The term *database system* captures the notion of multiple *database instances* connected to each other with a single global schema. When a D2O instance starts, it needs to be told which database system it is part of, and must be able to find other active instances that are also part of this system. This is complicated by the fact that D2O must be able to operate over machines exhibiting high churn, making it possible that an instance that is stopped and later restarted may join a database system with an entirely new set of instances, none of which it had known about previously (this is termed the *observation problem*[34]). It must not be possible under any circumstances for multiple diverging global schemas to emerge, as the system would no longer be ACID compliant.

D2O provides a way of definitively identifying which instances (if any) are currently active in the database system, and more importantly which instances hold a current copy of the global schema — the System Table — preventing divergent schemas from being created. This section describes the architecture used to achieve this.

### 5.7.1 Locator Servers and Descriptor Files

A set of external servers — called *locator servers* —store the location of the active System Table and of replicas of System Table state. They are similar in design to *Paxos*, in that they allow the database system to reach consensus on a single active System Table, but they also store the aforementioned locations of System Table state and enable instances to obtain a lock while creating the System Table.

Locator servers are used by D2O instances which are starting to find the System Table, which allows them to join the database system, and used by active D2O instances to find System Table replicas, which allows them to recreate the System Table when it has failed. Locator servers present the following interface to these instances:

```
void setSystemTableReplicaLocations(DatabaseURL[] replicaLocations);
void setActiveSystemTable(DatabaseURL systemTableLocation);
DatabaseURL getActiveSystemTableLocation();
```

[34] The observation problem in the context of this thesis relates to the knowledge of an individual observer. A database instance knows only about what it observes during its runtime. If, when the database is no longer active, all of these observations (such as the location of System Table replicas) become invalid, it does not know anything of use when it restarts.

```
DatabaseURL[] getSystemTableReplicaLocations();
boolean obtainLockToCreateSystemTable(DatabaseURL requester);
void commitSystemTableCreation(DatabaseURL locationOfSystemTable);
```

**Figure 42: Interface for Locator Servers.**

Multiple locator servers are used to ensure that they do not become a single point of failure. A D2O instance must be able to contact the majority of these servers when requesting or updating the location of System Table state, to ensure that this architecture is partition tolerant. To prevent deadlock they must be accessed in a fixed order.

The URI of each locator server is stored in a *database descriptor file*, which is given to D2O instances on start-up. Locator servers are expected to rarely change location (they are not designed to run on workstation machines), so descriptor files need to be changed infrequently. When there is a change, all D2O instances must be restarted and given the new descriptor file — there is no mechanism for doing this automatically in D2O as the descriptor file is maintained outwith the system.

This architecture is illustrated below, where a descriptor file (*top*) contains the address of each locator server (*middle*). These locator servers maintain references to the D2O instance running the System Table, and to the D2O instances with replicas of system table state (*bottom*).

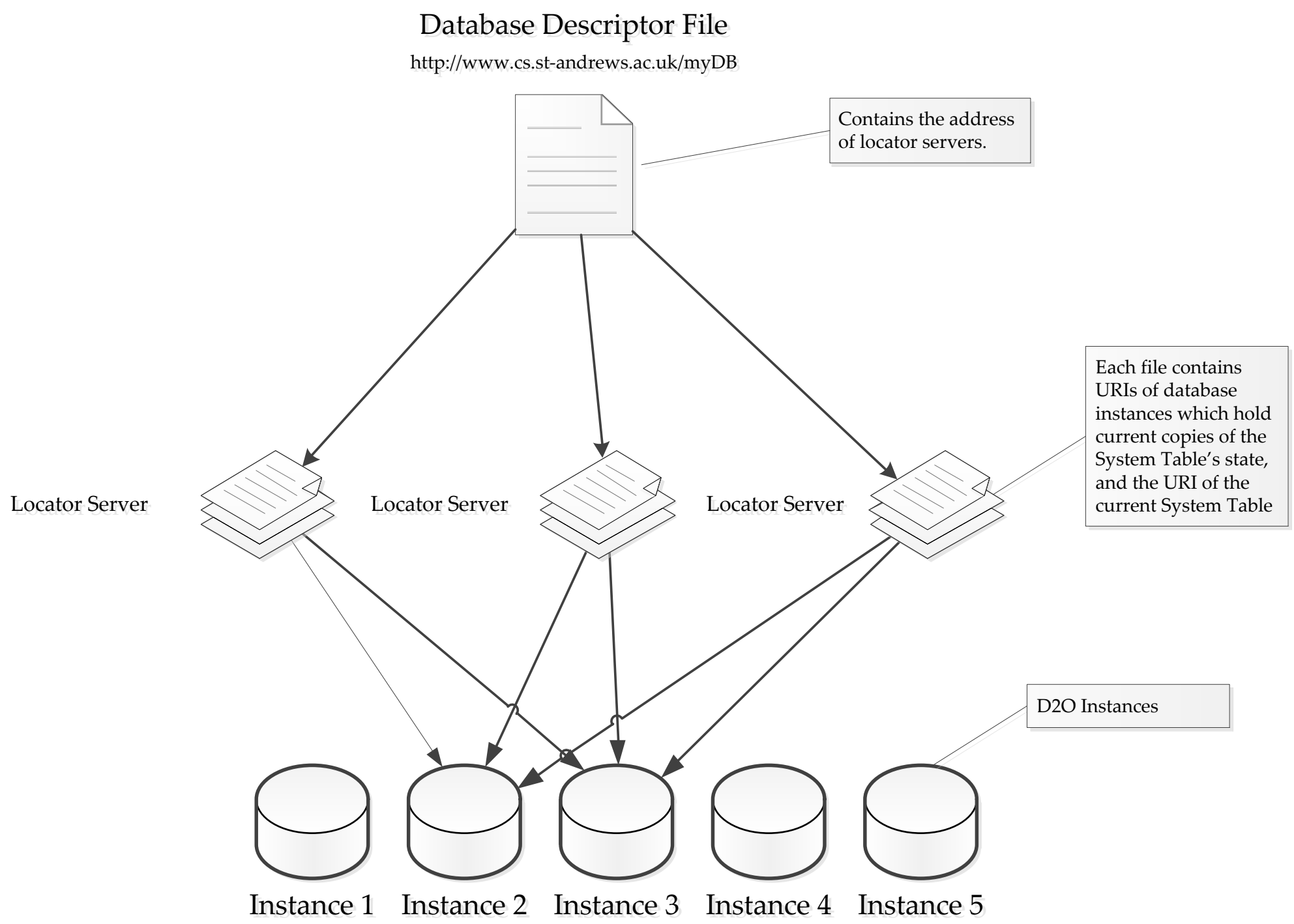

**Figure 43: An overview of the descriptor-locator architecture used by D2O.**

The locator-descriptor design ensures that D2O is not vulnerable to the *observation problem* because an instance can restart and see an entirely different set of running D2O instances, and still connect to the database system by finding running instances through the locator servers.

This design does not meet the *self-containment* requirement specified in *Chapter 4*, because a majority consensus protocol does not work in a system with dynamic membership. An instance cannot know the total number of instances in a system with dynamic membership, because of the potential for network partitions. Without this knowledge the instance cannot know whether a majority has been achieved, so the system cannot reach agreement on which instance is the System Table. Since D2O is designed to support dynamic membership the instances implementing this majority consensus protocol must therefore be run on a static set of machines.

### 5.7.2 Locator Servers as used by D2O Instances which are starting

When a D2O instance starts up it attempts to connect to the locator servers to find the current System Table. There are four possible scenarios that can occur:

1. The majority of locator servers know of an existing System Table and the System Table is active.

2. The majority of locator servers have a reference to an existing System Table, but the System Table is not active.
3. A minority of locator servers are active.
4. The majority of locator servers are active, but do not know of any existing System Table or System Table replicas.

Each of these points is discussed below.

*The locator servers know of an existing System Table and the System Table is active*. This is the typical case. On start-up, an instance discovers the System Table location from the majority of locator servers and initiates a connection to the System Table, which results in it joining the database system.

*The locator servers have a reference to an existing System Table, but the System Table is not active*. If a D2O instance cannot contact the previously active System Table it contacts one of the instances holding replicas of System Table state, which are also located by the locator servers. The instance sends a request to one of these replica-holding instances to restart the System Table. If none of the replica holding instances can be contacted, the database instance cannot join the database system.

*A minority of locator servers are active*. If an instance can only contact a minority of locator servers it cannot be certain where the System Table is currently located — because a network partition might result in a majority being available elsewhere — so it cannot join the database system.

*The locator servers are active, but do not know of any existing System Table or System Table replicas*. When a new database system is started, locator servers have no state. An instance connecting to these locator servers can obtain a lock giving it exclusive rights to create the System Table (for a period up to a timeout). The instance commits the creation of the System Table by adding its location to each locator server.

An instance must contact the locator servers to check that an active System Table does not currently exist, before it is able to request a lock to create a new System Table. To ensure that this does not create a race condition — between the initial check for the current System

Table's location and the lock request to create a new System Table — each locator server maintains an *update number*, which is incremented on every update to its state. Locator servers respond to all requests with this update number and instances send the last update number they receive along with new requests. Requests based on old information[35] are rejected.

The full extent of these interactions is illustrated in a Mealy machine diagram in Figure 44, below. This diagram show the process by which an individual D2O instance starts up, beginning at the *Start D2O Instance* state.

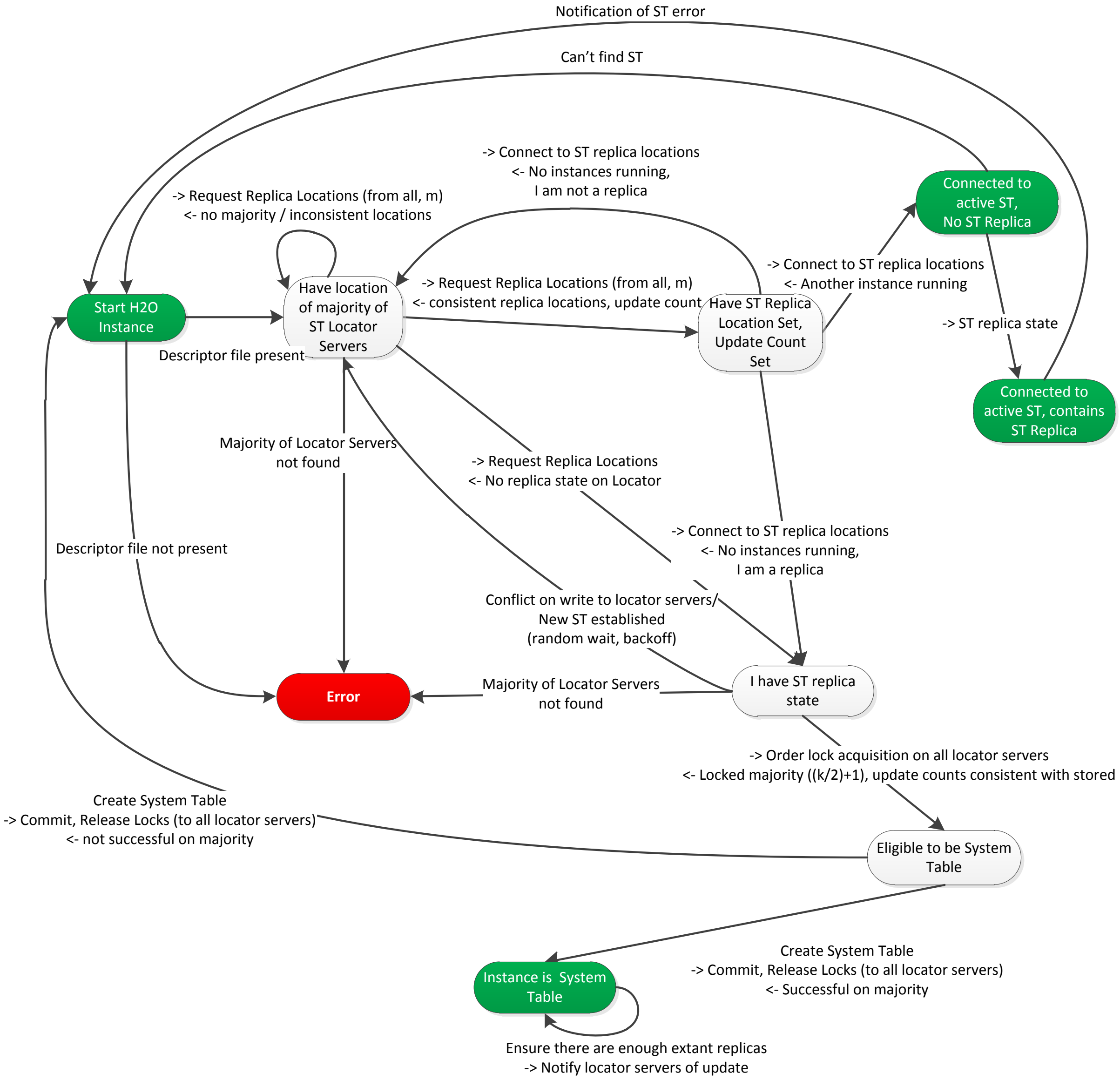

[35] Where the current update number is higher than the requesting instances last known update number.

**Figure 44: A Mealy machine diagram showing a D2O instance's interaction with locator servers when it attempts to find or start a System Table.**

### 5.7.3 Locator Servers as used by running D2O Instances

To provide partition tolerance, the majority of locator servers must be contacted by an active D2O instance before an update to the System Table can be committed. This prevents a situation where, during a network partition, a D2O instance believes that the System Table has failed even though it is still active on the other side of the partition. If the instance then attempts to recreate the System Table, two divergent System Tables can emerge, each operating on one side of the partition.

In D2O divergent System Tables cannot be created, because the System Table must contact a majority of locator servers on every commit of meta-data and table data state. The locator servers authoritatively state where the current System Table is located — they effectively decide which side of the partition can continue to operate — so only one System Table is able to commit updates at a given time.

Table Managers do not need to contact locator servers when they commit an update, provided all instances holding replicated Table Manager state are able to execute the update. If one of these replica sites cannot be contacted, the Table Manager must contact the majority of locator servers to ensure that it is not partitioned. This is necessary, because the replica site that cannot be contacted may be on the other side of a network partition, where a System Table has been recreated. The re-created System Table could then create a new Table Manager using the state on the partitioned replica site, creating divergent Table Managers. By contacting the majority of locator servers, the Table Manager ensures that it is on the same side of the partition as the System Table, which prevents divergent Table Managers because the System Table can contact the current Table Manager (meaning it will not try to recreate it).

This design ensures that the system is *consistent* and *partition tolerant*, but introduces another network request to every update. An option is provided to disable this feature, which removes the additional network request and makes the system *consistent* and *available*. All of the clustered systems and some of the cloud systems described in *Chapter 3* assume that network partitions are unlikely, so they are *consistent* and *available*.

To maintain a consistent and partition tolerant system without adding an extra network request, the locator server can give a D2O instance a lease, which guarantees that this instance is the sole owner of the System Table for the duration of that lease. This means that the System Table only has to contact the locator servers periodically to renew the lease, rather than every time a transaction needs to be committed.

The downside of this approach is that when a System Table fails, there is a period where another System Table cannot be created — the system must wait for the lease to expire.

## 5.8 IDENTIFYING DATABASE INSTANCES

Each D2O instance is identified by a URI which is an extension of the standard JDBC URL used by other database systems. An example URI is shown below:

`jdbc:d2o:tcp://archive.cs.st-andrews.ac.uk/d2o:9090/db_files/24e9bj81ff3`

This URI format has a number of component parts, as explained below (changeable parts emphasised):

jdbc:d2o:tcp://***<hostname>***:***<jdbc_port>***/***<location_on_disk>***/***<instance_name>***

| Element of URI | Use |
|---|---|
| `jdbc:d2o:tcp:` | Standard prefix for a D2O database. It is also possible to run D2O as an embedded in-memory database, in which case *tcp* is replaced with *mem*, and no hostname or port needs to be specified. |
| Hostname | The IP address or hostname of the machine running the D2O instance |
| JDBC Port | The port on which the instance's JDBC server is running (this accepts requests from user applications) |
| Location on Disk | The location of the database on disk, relative to the current working directory of the D2O instance, or relative to the user's home directory (using the ~ symbol)[36]. |
| Instance Name | The name of this D2O instance, which must be unique in the database system |

**Table 5: An explanation of the use of each element of the D2O JDBC URI.**

There are two critical parts to this URI: the address of an instance's JDBC server (*hostname* and *JDBC port*), which is used by applications to connect to the database system, and the name of the database instance (*instance name*), which uniquely identifies the database instance. Because each instance name is unique, D2O instances are able to operate correctly when the local machine's IP address changes.

[36] This mechanism does not work if the current working directory of the database files change during the lifetime of the database system.

Many machines, particularly laptops, change IP addresses routinely, either because they are taken to a different location (e.g. from work to home) or because they are assigned a dynamic IP address each time they start. In both cases a change of address makes the D2O instance's JDBC hostname and port invalid, meaning the URI cannot be used to locate a database instance. To address this, locator servers store the full URI of each instance, including the unique instance name. The port and IP address of the database are used by other instances to initiate a connection, but the unique instance name is used to identify an instance. This allows other instances to locate data stored on an instance, even if it was previously run at another address.

When a D2O instance changes IP address it updates its location on the locator servers and the System Table, so other D2O instances will be able to locate it. The System Table uses this information to ensure that replica locations are maintained, which allows other instances to find Table Managers or replicas at the new URI when they perform a System Table lookup operation.

Applications connected to D2O do not need to be aware of address changes because they connect to a local point of presence, a D2O instance running locally. Because the point of presence is on the same machine as the application, *localhost* can be used as a hostname, meaning it is unaffected by address changes.

This approach is similar to that taken by *VoltDB*, which use local daemons to create an abstraction over the machine in the cluster to which queries are made. In contrast, D2O's local point of presence is a fully functional database instance, so it is able to store replicas and run Table Managers. The intention of this approach is to reduce the cost of sending lock requests by enabling Table Managers to be moved closer to the requesting database instance where possible.

## 5.9 FAULT TOLERANCE

The circumstances in which D2O can recover from failure depend on the replication factor of the data and meta-data involved. The following section assumes a system where table state (of a given table) is replicated *n* times, all Table Manager state is replicated *t* times, and System Table state is replicated *s* times. In D2O each of these values is configurable.

If the sites storing all *n* replicas of a table fail, then the table cannot be queried or updated until one of those replicas recovers. A configurable setting determines how many replicas must be active to commit an update, though the default is one.

If the site of the active *Table Manager* or *System Table* fails, then a transaction involving these processes must wait until they are re-instantiated on another machine.

If the sites of all *t* Table Manager replicas (including the active Table Manager) fail, a table cannot be queried until one of these sites recovers.

If the sites of all *s* System Table replicas (including the active System Table) fail, then new tables cannot be created, existing tables cannot be dropped, and instances cannot execute queries that involve the discovery of Table Manager references until the System Table is recreated.

If at least one replica of the Table Manager or System Table does *not* fail and the majority of locator servers can be contacted, a new Table Manager or System Table can always be recreated. Only a single replica is needed to recreate these processes, but the majority of locator servers must be contacted to ensure that a partition does not exist, to prevent the case where divergent Table Managers or System Tables are created on either side of a partition.

This means that for D2O to be considered partition tolerant, a majority of locator servers must be running for Table Managers and the System Table to commit any updates. This consistent, partition tolerant design is similar to *Bigtable*, which requires that Tablet Servers maintain a connection to *Chubby* to be able to continue to serve tablets. Other systems such as *Xeround*, which use consensus protocols, typically require a majority of instances to be available to complete an update. Since these systems use a consensus protocol in place of traditional locking, they do not have to recreate a central lock manager (such as the Table

Manager and System Table) if it fails. A majority of instances must be available for a majority consensus protocol to work, so some failure is tolerated. However, this approach is not suited to a workstation-based system because it requires that each instance is aware of the full set of instances in the system, as this allows it to determine when a majority is reached. When the system contains a dynamically changing set of instances this is not possible: if the set of instances is updated, an instance cannot determine whether this new set contains all active instances, or just those on one side of a partition; if it is not updated then it will eventually become impossible to reach a majority if half of the instances in the system become unavailable.

## 5.10 Failure Detection

The failure of a D2O instance can be detected either by another instance issuing a query involving an instance which fails to respond, or by a primitive failure detector implemented using *Chord*.

When a potential failure is detected during a query's execution, the database detecting the failure informs the System Table. If the System Table cannot be contacted, the detecting instance initiates the System Table recovery mechanisms described in 5.7. If it can be contacted, the System Table checks whether the instance suspected of failure has actually failed (it is considered failed if the System Table is unable to communicate with it). If it has failed the System Table is responsible for re-instantiating any Table Managers that were on the failed instance onto other instances.

This approach removes an instance from the membership set if it is suspected of failure by the System Table. It is possible that the suspected machine is merely slow to respond or experiencing a transient failure, so this approach could remove an active machine or one that is likely to return quickly. However, D2O is designed under the principle that it is better to over-react to these cases than it is to wait and risk losing data[37].

If a failure is detected by a machine executing a transaction, the transaction will take longer to execute as recovery procedures are executed. To make this situation less likely, D2O uses another mechanism for failure detection — the stabilization mechanism of *Chord* [79].

Every D2O instance creates a Chord node, so if there are $n$ active instances there are also $n$ active Chord nodes, all connected in a single ring per database system. This means that each D2O instance has a predecessor instance and a successor instance in the *Chord* ring.

When a *Chord* node joins or leaves the database system, successor and predecessor change events are generated by *Chord* on the predecessor and successor of the joining/leaving node. This is used to detect failure, though a change event can occur for reasons other than failure. For instance, a new node joining a ring or a network partition splitting a ring will cause change events to be generated, even though neither is a node failure.

[37] Waiting to remove instances makes failure more likely, because further failures after the initial failure continue to lower the number of available replicas.

### 5.10.1 On Predecessor Change

When an instance receives a predecessor change event it compares the identity of the new predecessor (its hash value) to that of the previous predecessor. It suspects failure if the new predecessor has a hash value less than the previous predecessor (which indicates that the previous predecessor is no longer in the ring), otherwise it is assumed that a new node has joined the ring between the instance and its previous predecessor.

When an instance suspects failure, it contacts the System Table to inform it of the suspected instance. The System Table then attempts to contact this instance. If it is not accessible to the System Table it is considered dead; if it responds to the System Table it is considered active and no action is taken, even though the other instance, which reported the suspected failure, is not able to contact it. In the case that it is not accessible by the System Table, any Table Managers that were active on this instance are restarted at other locations. This process is initiated by the System Table.

### 5.10.2 Shutdown

When a database instance shuts down it initiates a hand-off process to pass extant Table Managers, and possibly the System Table, onto other database instances. It also contacts the Table Managers of any tables that have replicas stored locally to allow them to create another replica elsewhere.

## 5.11 AUTONOMIC MANAGEMENT

D2O is designed to use information on the availability of resources on workstation machines to make data placement decisions. This is intended to improve the D2O's query response time — a typical goal of DDBMSs — and to improve resource utilization — a requirement of workstation-based systems.

This section discusses how this is achieved in the context of autonomic management.

### 5.11.1 Autonomic Elements in D2O

Autonomic management is used in four areas of D2O:

A. The *placement of objects*, including replicas of table data, Table Managers, and the System Table.
B. *Replica choice* during query planning, determining which replicas are used to answer read requests.
C. *Threshold analysis* of local resource utilization, using resource monitoring to stop the database from overloading any one machine.
D. *Replication factor monitoring*, checking periodically whether a sufficient number of replicas exist for each table.

Figure 45 illustrates how some of these elements fit into the autonomic cycle.

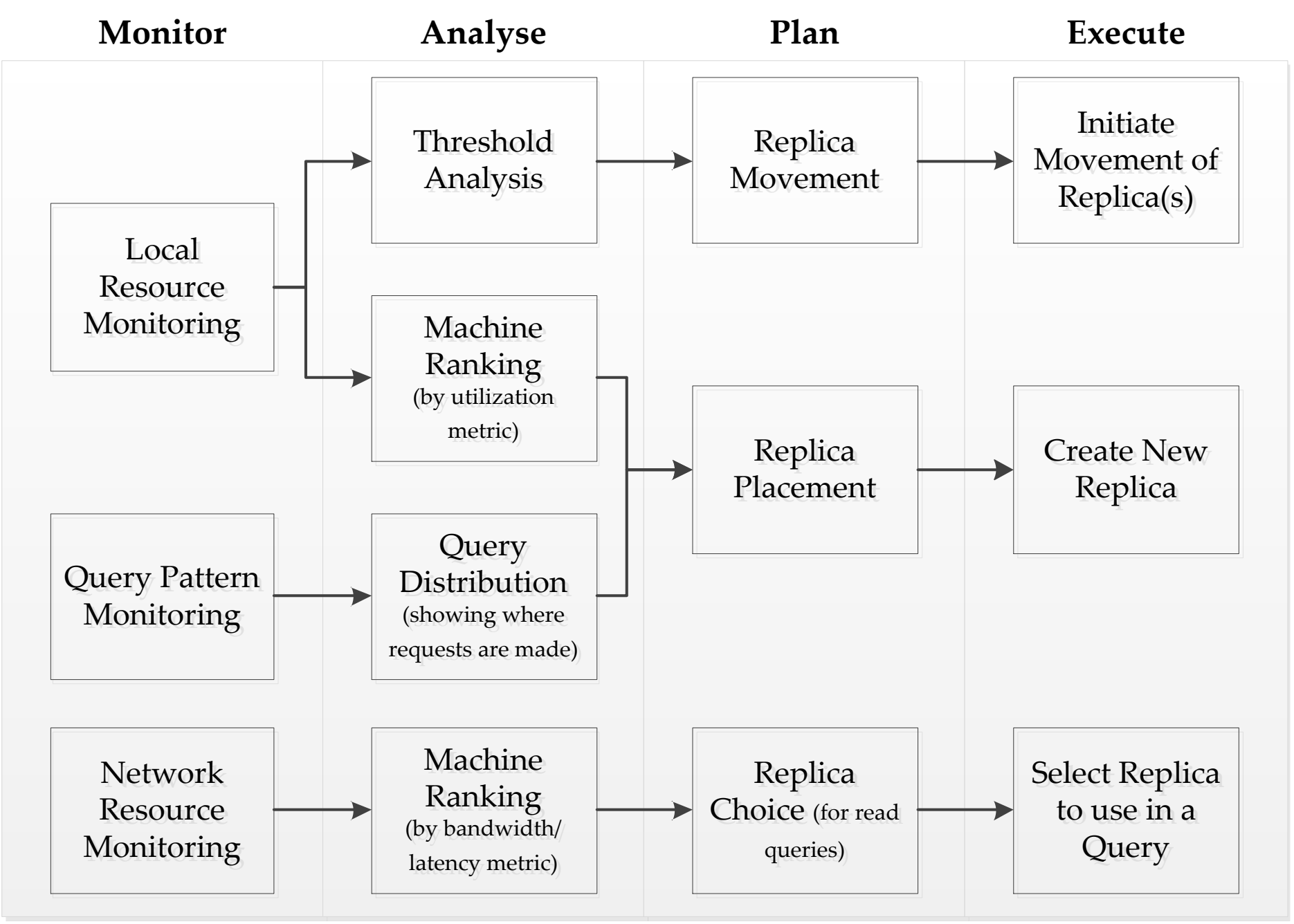

**Figure 45: An overview of the elements in D2O's autonomic architecture, and the ways that they are used.**

The remainder of this section is split into a discussion of the *monitor*, *analyse*, *plan*, and *execute* phases for each of these elements. The diagram to the right of each section header is a condensed version of Figure 27, which illustrates the place of each component in the autonomic cycle.

### 5.11.2 Autonomic Cycle: Monitor

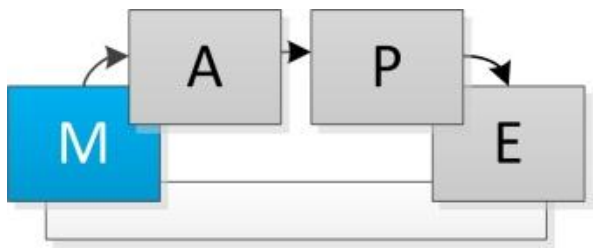

D2O performs monitoring of local machine resources and of database components.

*Local resource monitoring* is used to identify when resources such as disk space and CPU capacity become scarce — a process termed *threshold analysis* — and to categorise the current availability of resources on each machine so that they can be compared. Machine ranking based on this comparison is used to determine where replicas should be placed.

*Network resource monitoring information* establishes the latency and bandwidth of connections between D2O instances, and is used to determine which replicas are used to answer queries.

D2O also performs *query pattern monitoring* to record query patterns such as the read-write ratio of requests, and the locations most queries are made from. This information is used to decide where to place replicas, in addition to the machine ranking produced from resource monitoring.

*Monitoring Architecture*

Machine and network resource monitoring is performed by a resource monitoring tool named ***NUMONIC*** [85]. A *NUMONIC* instance runs on every machine alongside a D2O instance, collecting data periodically and storing it on the local D2O instance.

For database monitoring, D2O records the instance making query and update requests on each Table Manager. This architecture is illustrated below in Figure 46, with monitoring components highlighted in blue. Note that all monitoring information is stored locally on the machine performing the monitoring, or at each Table Manager for query monitoring.

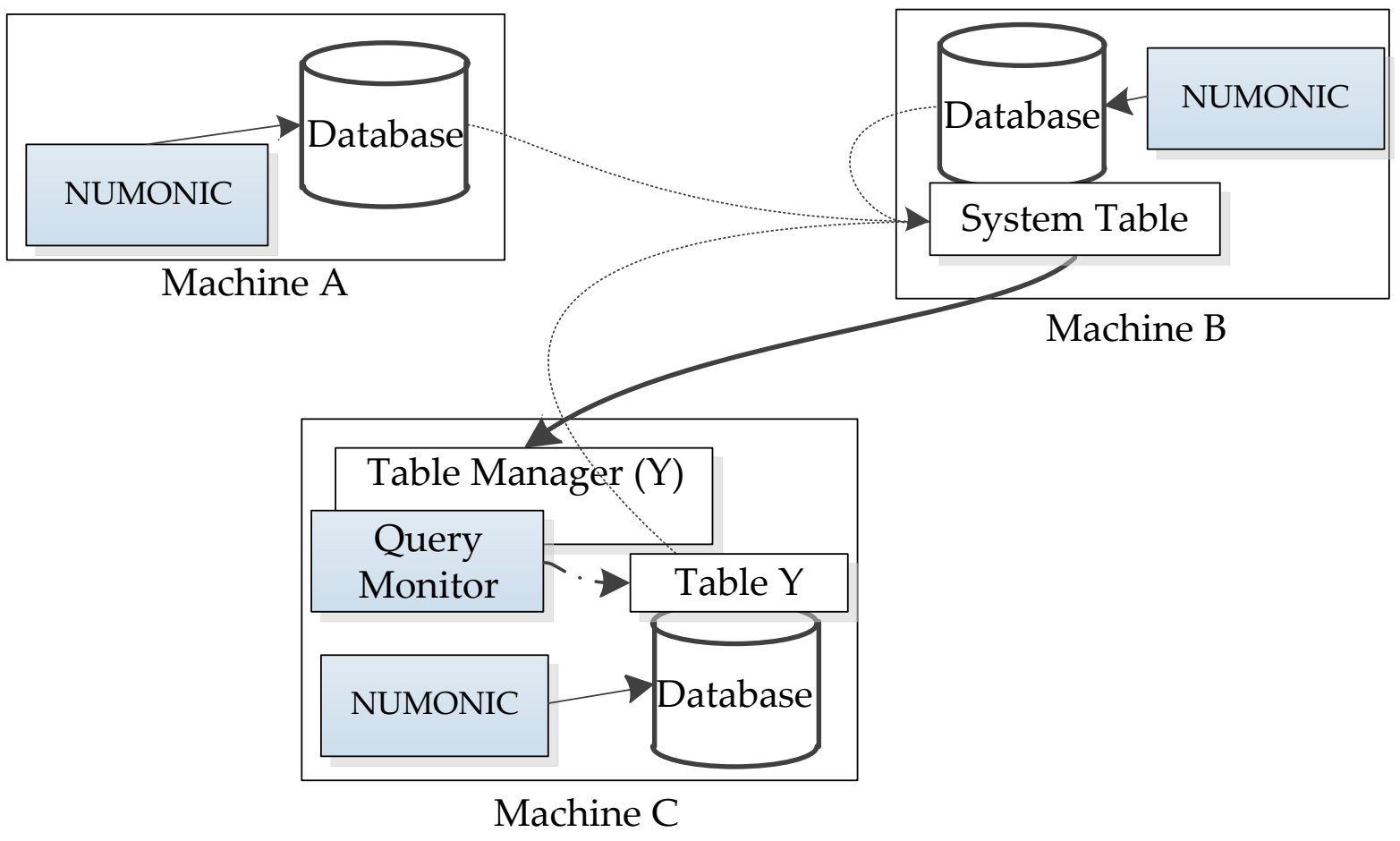

**Figure 46: An illustration of the monitoring components in D2O.**

*Monitored Resources*

*NUMONIC* monitors both static and dynamic resource information. Static information, such as CPU capacity, never changes during the runtime of a *NUMONIC* instance, and is unlikely to change at all. It is recorded when the system starts up. Dynamic information, such as CPU utilization, changes continuously and requires constant monitoring to obtain a current overview of the system.

The specific static and dynamic information monitored by *NUMONIC* is listed below in Table 6 and Table 7 respectively.

| Resource | Unit |
| --- | --- |
| CPU Capacity | MHz |
| CPU – Number of Cores | *N/A* |
| Memory Capacity | Mbytes |
| Disk Capacity | Mbytes |
| Operating System | Name + Version |

**Table 6: The static machine information obtained by NUMONIC.**

| Resource Monitored | Measurement Taken | Unit |
|---|---|---|
| CPU | Utilization | % |
| Memory | Utilization | % |
| Disk | Utilization | % |
| | Writes | Kbytes/s |
| | Reads | Kbytes/s |
| Process Activity | Start-up/Shutdown Events | *Unix Time of Event* |
| D2O Process | CPU Utilization | % |
| Network[38] | Bandwidth | Kbytes/s |
| | Latency | ms |
| | IP Address | *IP Address* |

**Table 7: The dynamic monitoring data stored by each D2O instance.**

Dynamic monitoring is used to analyse two different aspects of a machine's utilization. Some information, such as free disk space, is useful as an immediate indicator of system state. Other information, such as CPU utilization, can also be used for longer term pattern analysis. Finally, event data, including start-up and shutdown events, can be used to develop a view of each machine's availability over time.

All dynamic monitoring data is summarised before being stored in the local D2O instance, with the exception of event data which records the timing of events. Summarisation is discussed in more detail below.

***Summarising Monitoring Data***

NUMONIC summarises monitoring data and stores the result on the local D2O instance. A typical monitoring setup involves taking sensor readings every five seconds, and then summarising and storing these measurements every minute, though the timing of this is configurable. The database stores the minimum, maximum, mean, and median utilization of a resource for each summary.

[38] Connectivity to other known instances.

Summaries are periodically trimmed to prevent the database from being filled with monitoring results. New data is therefore given a higher weighting, as it is assumed that recent results are better indicators of future trends than older results.

On failure, summarised information is not lost because it is stored in the local D2O instance in regular database tables.

### 5.11.3 Autonomic Cycle: Analyse

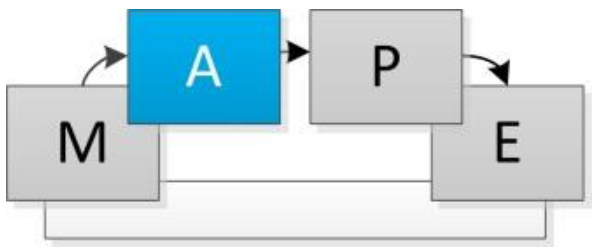

D2O analyses monitoring data in four ways:

A. ***Global Monitoring Analysis*** is used to produce a global ranking of machines based on their availability.
B. ***Threshold Analysis*** is used to monitor resource utilization on each instance, and to produce an event if utilization exceeds a specified threshold.
C. ***Network Resource Analysis*** is used to produce a local per-instance ranking of machines based on the bandwidth and latency of their connections to other instances.
D. ***Query Pattern Monitoring*** is used to produce a distribution of the instances querying particular tables.

The relationship between monitoring and analysis is summarised in Table 8 below.

| Type of Monitoring | Location of Monitors | Location of Analysis |
|---|---|---|
| Global Monitoring Analysis | Database Instance | System Table |
| Threshold Analysis | Database Instance | Database Instance |
| Network Resource Analysis | Database Instance | Database Instance |
| Query Pattern Monitoring | Table Manager | Table Manager |

**Table 8: The relationship between monitoring and analysis components in D2O, based on their locale.**

There are broadly two types of analysis described above. *Active analysis*, such as threshold analysis, involves interpreting monitoring information and initiating the planning phase of the autonomic cycle when one of the managed elements is no longer functioning as expected. *Passive analysis*, such as *global monitoring analysis*, involves interpreting monitoring information that will be used later in the course of the system's normal operation. Both involve using monitoring information to improve the operation of D2O, but active analysis initiates changes whereas passive analysis influences decision making.

The remainder of this section describes each analysis component in more detail.

***Global Monitoring Analysis***

Locally collected monitoring information is summarised and then collated on a single machine — at the System Table — to allow instances to be ranked based on availability. Each D2O instance periodically sends a summary of its own machine's resource utilization to the System Table, which then ranks machines based on a metric. This process is illustrated in Figure 47, and the metric is discussed below.

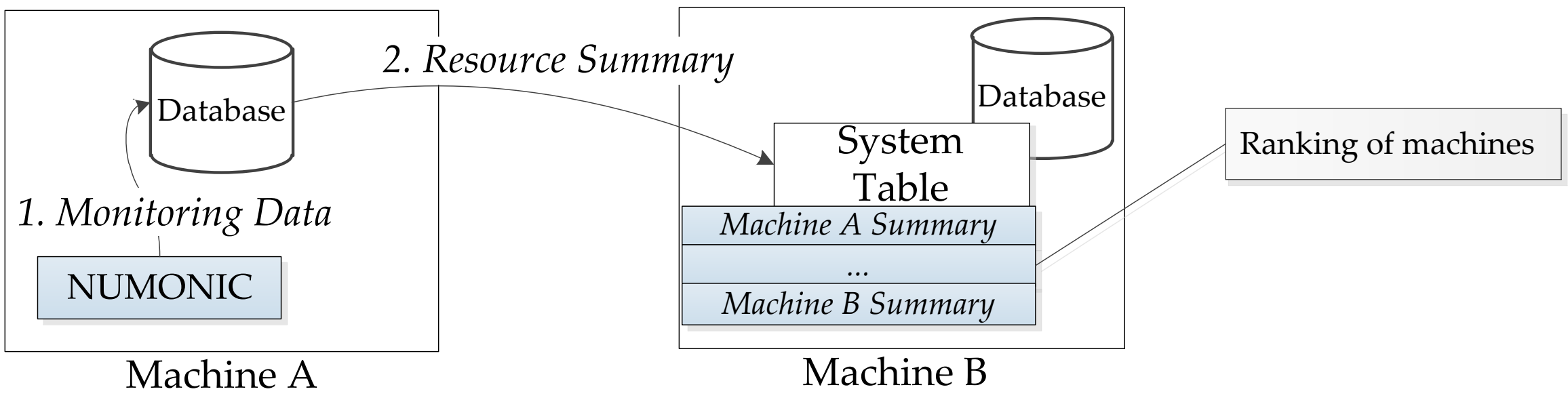

**Figure 47: An illustration of the process through which local resource monitoring information is sent to and summarised on the System Table.**

The data sent to the System Table is a summary of each machine's availability over a period, typically around five minutes.

With the exception of start-up and shutdown events, which show when a D2O instance starts and stops, each resource is summarised by producing the following values:

- Mean
- Median
- Minimum
- Maximum

The sampling rate of the summary data (the number of measurements per minute) is also sent, as is the period over which sampling took place.

*Ranking Metric*

The System Table uses monitoring information to rank machines, using a formula (shown below in Equation 1) that assigns a weighting to each resource indicating its importance relative to other resources, and produces a single value representing each machine's availability. The weighting given to each resource is specified in a metric which is passed to the System Table when a request to rank machines is made.

$$machine_ranking(metric) = calc(cpu, metric) + calc(mem, metric) + calc(disk, metric)$$

$$calc(res, metric) = \frac{res_{capacity}}{res_{max}} \times (1 - r_{utilization}) \times metric_{weighting}$$

**Equation 1: The metric used to rank database instances.**

The overall ranking for a machine is determined by computing a value for each resource (CPU, memory, and disk utilization) based on a provided metric and adding them together.

The *calc* function, which computes the per-resource value, takes the capacity of a resource (e.g. 4 GB memory) and normalizes this value on a scale of 0 to 1, where the machine with the largest capacity is given the value 1. This is multiplied by the percentage of this capacity that is free and by the metric weighting assigned to that resource. The metric weighting is a number between 0 to 1 for each resource that is used to weight resources by their relative importance. For example, if a particular task is CPU intensive, a metric may weight *CPU* as 1, and *memory* and *disk* as 0.5, meaning the availability of CPU resources is given more importance in the ranking of machines.

***Threshold Analysis***

D2O uses monitoring information locally on each machine to perform *threshold analysis*, which detects when particular resources are becoming scarce — when utilization exceeds a threshold. When this occurs an event is produced which triggers the creation of an autonomic plan.

The goal of threshold analysis is to identify when the D2O instance is running out of resources, or when it is using too many resources at the expense of other applications.

All of the thresholds being monitored are derived from the monitoring information provided by *NUMONIC*, listed in Table 7 (page 137). The following table gives some examples of the types of thresholds that can be monitored and their intent:

| Threshold | What this may mean |
|---|---|
| CPU Utilization higher than 70% | A machine is heavily utilized, and activity from the D2O process may start affecting other users of the machine[39]. |
| Memory utilization higher than 80% | Same as above. |
| Disk Activity higher than 4000 writes/sec | The hard disk is heavily utilized so writes of new updates on this machine may be slower than expected. |
| Available disk space lower than 4 Gb | There is so little hard disk space that adding more table data onto this instance may start slowing down the system as a whole. |

**Table 9: A number of examples of the types of thresholds that D2O monitors.**

As with the examples above, a *threshold value* — the value that the system aims not to exceed — is specified for each resource. Threshold values can be specified on a per-instance basis because there are various types of machines that run D2O instances. For example, an instance running on a dedicated server may set thresholds that will never be exceeded, because the instance is allowed to make full use of the machine's resources. In contrast, an instance running on a workstation will have lower thresholds, because the machine is intended to be used primarily for other tasks. With a lower threshold D2O will only make use of machines that are not being heavily used by other applications.

It is possible to fully customize threshold settings, but three configurations are included by default because they cover common scenarios:

| Machine Type | Typical Characteristics |
|---|---|
| **Dedicated Server** | No limits on resources that can be used. This machine is set up primarily |

[39] If the D2O process is causing the high CPU utilization, this stops it from dominating the available resources on the machine. If other processes are causing the high CPU utilization, this stops H2O from interfering with these processes.

| | |
|---|---|
| | to run a D2O instance. |
| **Shared Server** | High thresholds, because there is no need to allow interactive user sessions. The machine is set up to be used by a number of applications, so D2O does not have complete control of all resources. |
| **Workstation** | Low thresholds, because the running of the database should not interfere with a user's use of the machine. In addition to user interaction, the machine may also be running numerous other applications. |

**Table 10: Descriptions of the types of machines that are supported by pre-configured threshold settings.**

When a D2O instance is set up on a machine, the type of threshold configuration used is expected to be specified by the system administrator, but it could also be set at runtime by a user, as the use of the system changes.

The process of using these configurations to analyse monitoring data is illustrated below in Figure 48.

Monitoring data is sent to a local threshold analysis component as it is received by *NUMONIC* (i) and compared against the thresholds specified in the configuration file being used by the local D2O instance (ii). For each threshold that is exceeded an event is generated and sent to the autonomic planner (iii), which is discussed in more detail in section 5.11.4.

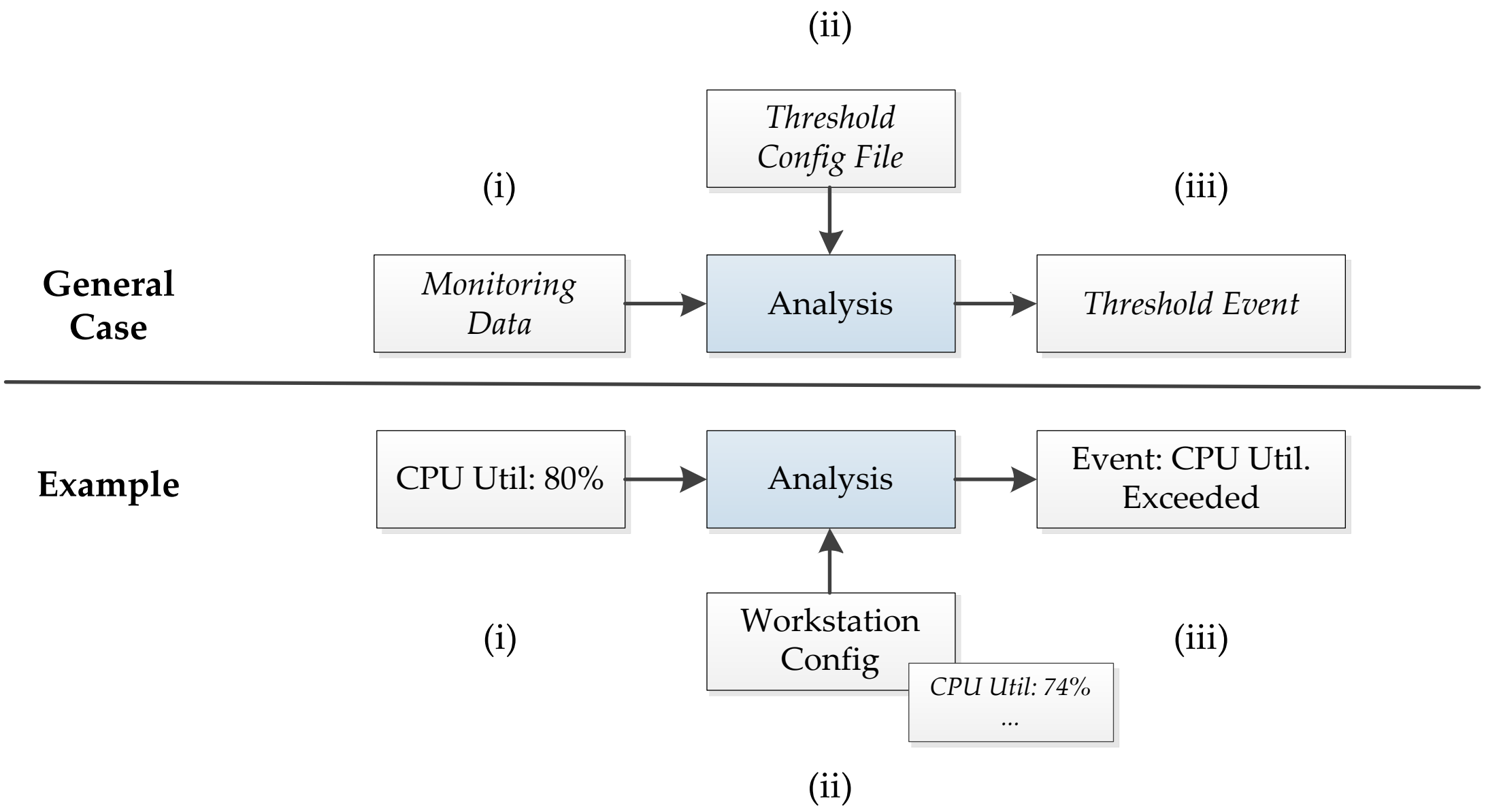

**Figure 48: An example of the process used to monitor thresholds and create events when they have been exceeded.**

*Network Resource Monitoring*

The bandwidth and latency of point-to-point links is monitored from each instance in the database to every other instance. This data is then analysed to produce a ranking of machines based on the quality of their connection to the local instance.

*Query Pattern Monitoring*

Every lock request made to a Table Manager is monitored by the Table Manager itself, allowing the location of requests to be tracked. This enables machines to be ranked based on the volume of requests they have made involving a given table, and is used to determine where new replicas should be placed. It is seen as an advantage to place replicas close to the location of requests.

Query pattern information is periodically trimmed to remove dated entries.

*Summary*

The components of the analysis stage of the autonomic cycle described above are illustrated in the following diagram in grey. The only monitoring information that is distributed as part of the analysis stage is the machine summary data sent to the System Table.

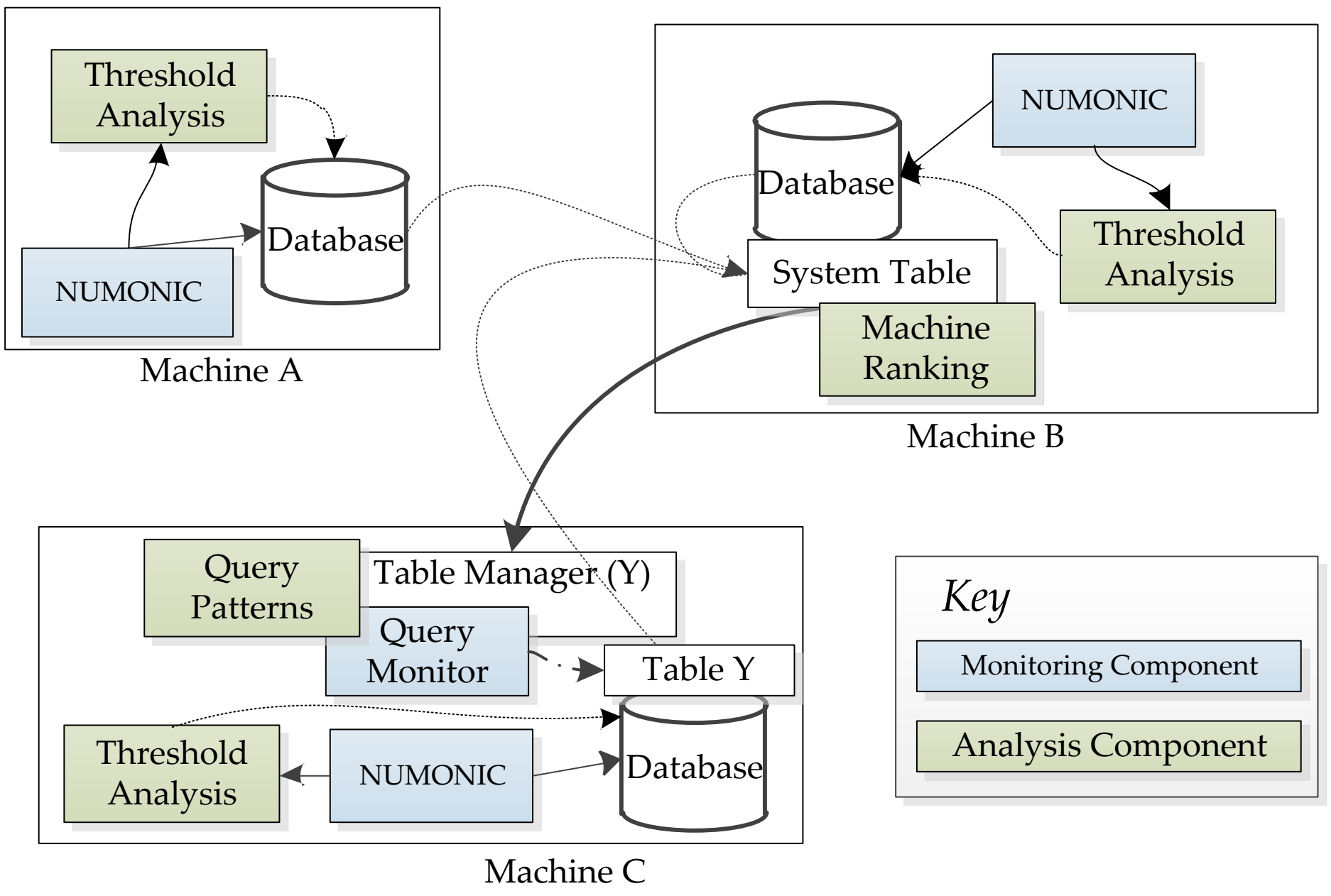

**Figure 49: An illustration of the autonomic architecture of D2O, showing monitoring and analysis components.**

### 5.11.4 Autonomic Cycle: Plan

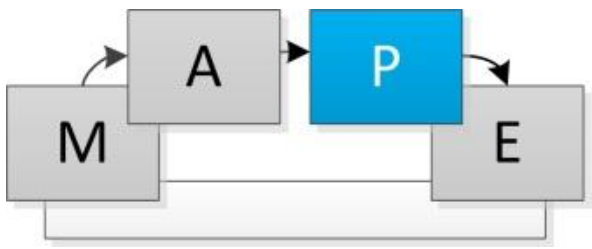

The planning phase of the autonomic cycle begins once monitoring data has been analysed and interpreted. *Active analysis* triggers the planning phase by creating an event when a resource such as CPU utilization reaches a specified threshold. *Passive analysis* initiates planning on a database event such as replica creation. Regardless, both cases make use of the analysis performed in the previous phase.

D2O is designed to perform planning of:

A. *Replica placement*. Machine utilization information (global resource analysis, as per section 5.11.3) and query monitoring information is used to decide where replicas should be placed. A replica placement plan is created in two cases: when a new replica is needed and when an existing replica is to be moved.
B. *Query planning*. Network resource analysis is used to determine which replica should be used in answering a read query.
C. *Replica movement*. Threshold analysis, based on local resource monitoring, initiates planning on the movement of replicas.

This section goes into more detail about the planning involved for each of these decisions.

***Replica Placement***

D2O uses autonomic planning to select the location of new replicas, and to decide where existing replicas should be moved. It does not decide where the initial replica of a table will be created — the initial replica is always created on the machine making the *CREATE TABLE* request.

When a request is made to create a new replica (from the user, or an autonomic planner), the Table Manager contacts the System Table to obtain the machine ranking information discussed in the previous section. It also queries the query monitoring information that is held locally. These elements are used to determine the most appropriate location for the replica. Other factors such as the latency and bandwidth between instances are not taken into account, to reduce the quantity of information that must be shared globally[40].

The Table Manager makes the final decision on replica placement, based on the aforementioned machine ranking and query information. The System Table could potentially influence this decision by altering the ranking of machines for particular requests. This involves placing heavily loaded instances lower in the ranking[41], or omitting them completely. The System Table currently only returns the ranking produced by the availability metric printed in *Equation 1*.

The Table Manager uses the ranking information from the System Table in combination with its own monitoring information, which records the location from which queries to the table are made. If, for example, the majority of requests come from a single node, then this instance is given a higher ranking than other instances.

***Query Planning***

When an application sends a read query to a D2O instance, the instance contacts the Table Manager(s) for the table(s) involved in the query. Each Table Manager returns a lock for the request along with a list of the D2O instances that store replicas of the table.

There are two areas where planning is designed to be used in this process. First, the Table Manager removes the locations of replicas that it deems are overloaded. Next, the requesting D2O instance compares the remaining replica locations with network monitoring data to decide on the most appropriate replica to query.

For reasons discussed in *Chapter 6 (Implementation)*, autonomic plans are not currently used in query planning, though they are intended to be used in the manner described above.

---

[40] Network information is $O(n^2)$ in relation to storage, for $n$ machines, and also requires more information to be sent between instances. It may improve the placement decisions made by H2O, but it is left as future work.

[41] In this context 'heavily loaded' refers to the number of Table Managers that are active on an instance. A machine which is heavily loaded in terms of its resource utilization will already have a low ranking.

***Replica and Process Movement***

In contrast with replica *placement* which determines where a new replica is placed, replica *movement* determines when existing replicas must be moved. This decision is initiated by threshold analysis.

If the planning phase of replica movement is started, D2O has already identified the need to take some action. The planning phase involves deciding what action to take.

The table below summarises the reasons why replica movement may be necessary, and the actions that will be taken:

| **Threshold Reached** | **Action** | **Intention** |
|---|---|---|
| Low Disk Space Available | Move replica(s) | Reduce D2O's use of local disk space |
| High CPU Utilization | Move Table Manager(s), System Table | Reduce the number of D2O components that could contribute to high utilization |
| High number of disk writes/reads | Move replica(s) | Reduce contention with other applications by making disk writes/reads on this instance less likely |

**Table 11: A summary of the actions taken by D2O on specific thresholds being reached.**

It is important that a D2O instance does not make too many changes — for example, moving all replicas off an instance, when only moving a single instance was required — so the planning of replica movement is designed to take into account how much space is needed and the size of each replica (each instance is aware of the size of replicas that are stored locally). In addition the cost of movement must be taken into account.

Similarly, in the movement of Table Managers and System Tables, D2O is not designed to move all the active processes on an instance. Instead, it queries the Table Managers and System Table (if active on this instance) and establishes which is the most active (in terms of lock requests made). The number of processes that are moved onto other machines is determined by the level of CPU utilization and the threshold specified by the database.

This approach is not without problems. Clustering Table Managers on a single instance may improve query performance for queries accessing each of these tables, because only a single network request is needed for all lock requests. This means that moving individual Table Managers off machines may degrade query performance by increasing the number of network requests that need to be made.

D2O does not co-ordinate placement decisions globally, so numerous machines may decide at the same time to overload a single instance. To prevent this a negotiation phase is used to limit replica movement requests. Without negotiation, an instance, *X*, issues a non-negotiable command that a replica should be created or moved to a remote instance, *Y*. D2O adds a negotiation phase to these requests, so that *X* sends a request to create or move a replica to *Y* along with a value indicating its level of need — i.e. how important it is that *Y* accepts its request. *Y* compares this value to one of its own, indicating how willing it is to accept new replicas. If X's *level-of-need* value is greater than Y's *1-willingness_to_help* value, then the replica is stored on Y; if it is less than Y's *1-willingness_to_help* value, then X makes a request to another instance, gradually increasing its *level-of-need* value. It makes this request to every appropriate remote instance until it finds somewhere with a sufficient *willingness-to-help*, or until it reaches a maximum *level-of-need*, at which time no remote instance can reject its request.

This approach is a simpler bargaining mechanism than that used by the micro-economic-based DDBMS, *Mariposa* [86], because D2O is a homogeneous DDBMS, unlike Mariposa. This means that data is always placed somewhere — because an instance will eventually be forced to accept a replica, if necessary — but instances have a chance to use locally available information on their availability to inform data placement decisions.

***Summary***

D2O's planning architecture is summarised below, with planning components illustrated in orange. *Replica movement* is initiated through threshold analysis, but uses the more general *replica placement* planner to decide where replicas should be moved, while *query planning* makes use of local network monitoring information to decide which replica to query.

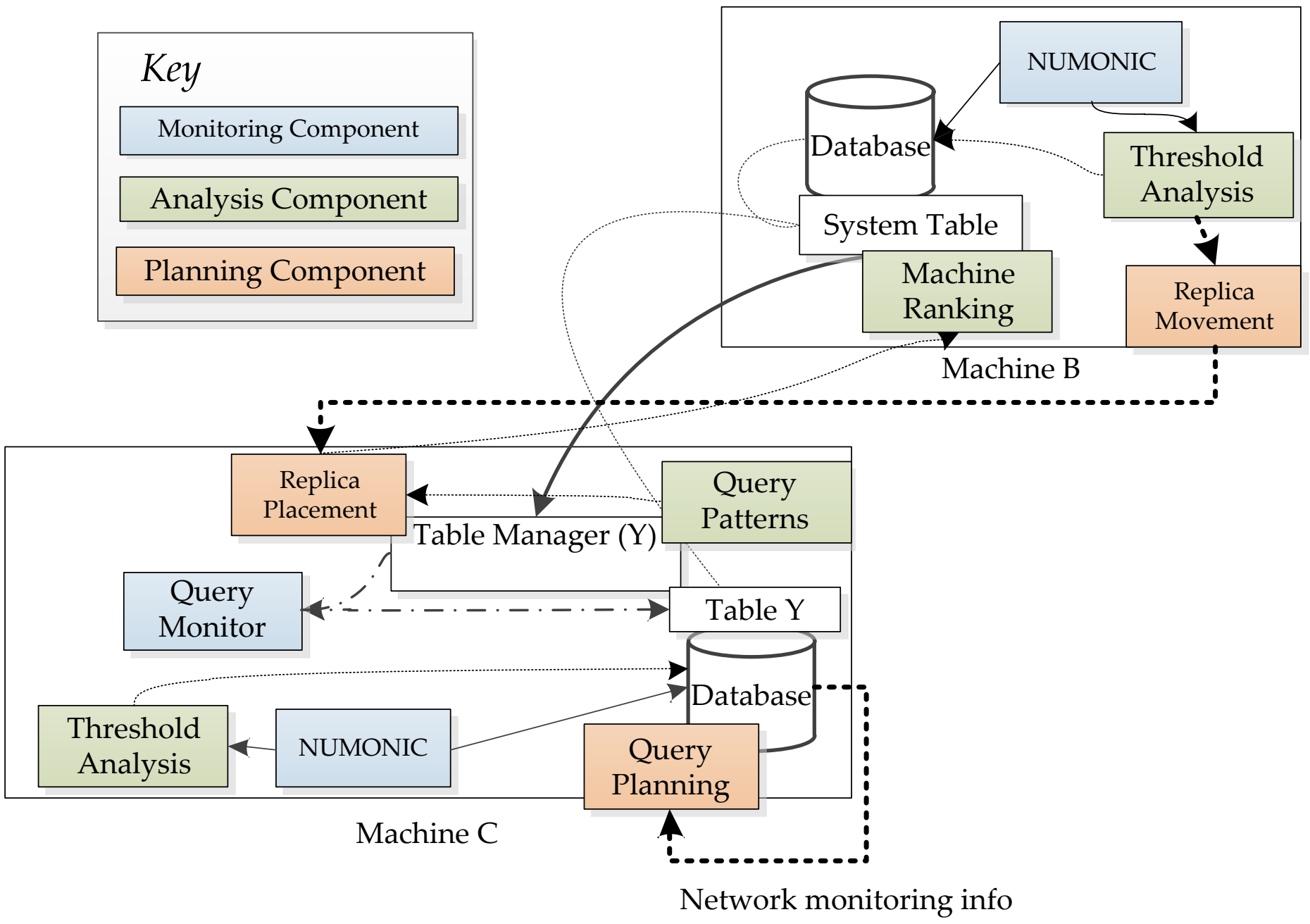

**Figure 50: An illustration of the D2O's architecture with respect to monitoring, analysis, and planning components.**

Once a plan has been created, it is executed. This stage of the autonomic cycle is described below.

### 5.11.5 Autonomic Cycle: Execute

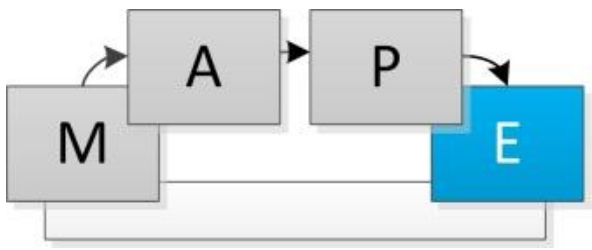

The execution phase of the autonomic cycle simply involves executing the plans described in the previous section, listed below:

- Deciding where to place table data, Table Managers, and the System Table.
- Determining which replica to use during query planning.
- Moving data or processes as a result of threshold analysis.
- Creating new replicas if an insufficient number exist for a particular table.

## 5.12 SUMMARY OF D2O

This chapter introduced the architecture of an autonomic, resource-aware distributed database system named D2O.

D2O meets all of the requirements of Chapter 4 (Requirements), but for *self-containment*, which is not possible because D2O makes use of locator servers which are designed to run on machines more reliable than workstations.

D2O uses pessimistic locking to provide atomicity, and data is synchronously replicated to ensure that it is durable even in the event of permanent machine failure. D2O is also able to recover from the failure of its transaction managers — the System Table and Table Managers — though extant transactions involving the failed managers must be restarted.

D2O uses resource monitoring to establish when data or processes should be moved off a machine, to prevent it from becoming overloaded. This functionality is designed to run without administrator intervention, using an autonomic framework.

The next chapter discusses a partial implementation of D2O, which is used to evaluate the performance and fault tolerance of its design.

# Chapter 6: $H_2O$: The Implementation of $D_2O$

# 6 H2O: The Implementation of D2O

This chapter discusses H2O, an implementation of D2O which meets the database and fault tolerance requirements outlined in *Chapter 4 (Requirements).* The implementation of these requirements is prioritised over the autonomic, resource-aware functionality described in 5.11, as this functionality is the focus of this thesis. They are more important in establishing the viability of a workstation-based database system, for reasons outline in *Chapter 4 (Requirements)*.

Future work on autonomic, resource aware functionality is discussed in *Chapter 8 (Conclusion)*.

The H2O implementation is faithful to the design of D2O with two exceptions: the aforementioned lack of autonomic, resource-aware functionality, and the use of two-phase commit rather than three-phase commit. The latter change is discussed later in 6.3.3.

## 6.1 H2

H2O is based on ***H2***, a non-distributed DBMS written in Java [87][42]. H2 provides the following notable features of a local DBMS:

- Implementation of JDBC API.
- Support for the SQL specification
- Atomic, Consistent, Isolated Transactions
- Table-level locking
- Write-ahead logging

The support of the SQL specification and a JDBC interface satisfies *requirement D1* for a workstation-based database system.

*Durability*

The support of atomic, consistent, isolated transactions partially fulfils *requirement D2*, but transactions are not durable by default — if a machine loses power or suffers a hardware failure it is possible that some recently committed transactions will be lost. H2 uses a *write delay*, which means that transactions do not immediately flush to disk and instead wait a specified amount of time (500 milliseconds by default). This is done because of the cost of calling `fsync` (which is intended to force data to be flushed to disk), and because many systems do not guarantee that `fsync` will flush data to disk [88]. For instance, many hard drives do not obey `fsync` when a request is made to flush data to disk, and in Mac OS X `fsync` does not attempt to flush hard drive buffers. By setting a write delay of 500ms, H2 is able to execute multiple transactions (assuming they take less than 500ms to execute) without the cost of flushing to disk on each commit.

---

[42] It is based on H2 version 1.1, released in February 2009.

H2O sets the write delay to zero and calls `fsync` on every commit, which means it provides durability in systems that correctly implement `fsync`. The effect of this change on query performance is discussed in *Chapter 7 (Evaluation)*.

In *VoltDB*, updates are not immediately persisted to disk — instead, durability is provided by synchronously replicating data. This is possible because *VoltDB* is designed to run on servers with uninterruptible power supplies, in contrast to H2O. A power cut to a set of *VoltDB* servers would not result in data loss, because the servers could persist to disk after the event, but a power cut to H2O instances would result in immediate failure and — if data was not immediately flushed to disk — the loss of data.

*Replication*
H2 supports full-database replication, but this feature is not used in H2O. Instead it is heavily modified to implement the design discussed in *Chapter 5*.

*Locking*
H2 uses read-committed locking by default, meaning write locks are held until the end of a transaction, but read locks are released as soon as a query has completed. These locks are local to each instance, so H2O implements its own additional locking mechanism using Table Managers. This locking provides serializable isolation, which guarantees the *isolation* property of ACID by holding all locks are held until the end of a transaction.

### 6.1.1 H2O's Impact on H2's Architecture

In H2O, functionality related to logging, indexing and storing data is unchanged from H2. Some components related to query parsing and execution have been modified to add support for replication.

H2O implements entirely new components to support global locking and replication. The locator server design is also implemented in its entirety.

The extent of these modifications are illustrated below, where the darkness of the background colour indicates the degree to which the code was written specially for H2O. H2 is designed as a non-distributed database, so only the H2O-specific functionality shown in this diagram is designed specifically to manage distribution; the columns showing H2 functionality represent separate instances.

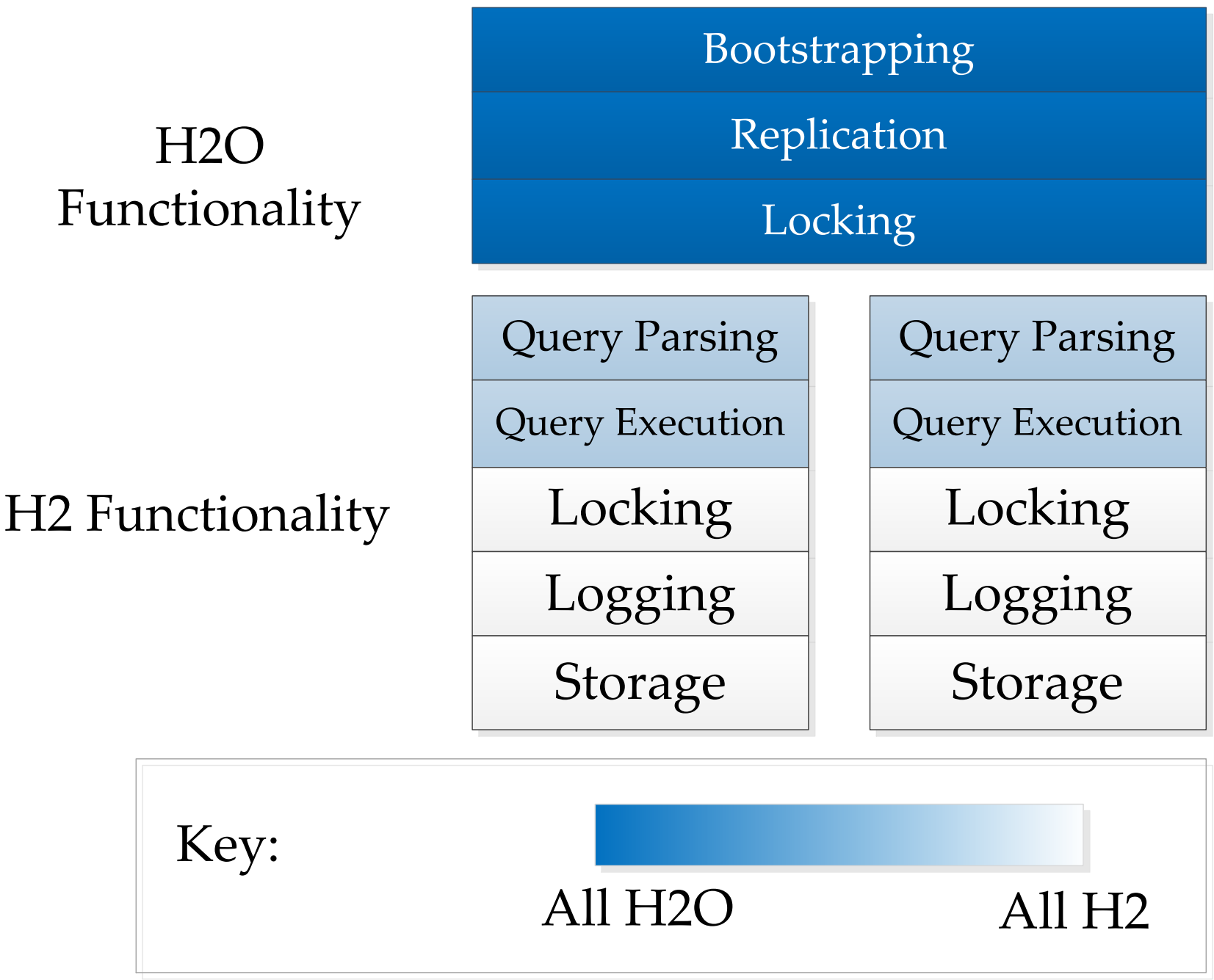

**Figure 51: An illustration of the relationship between H2 and H2O.**

The query parsing and execution components are modified to the extent that updates are now sent to multiple machines synchronously (rather than executed locally), and a new set of queries are supported (for the creation of replicas and the migration of the System Table and Table Manager).

The re-use of these H2 components introduces some inefficiencies in this implementation. For example, in H2O, queries are first parsed to determine where they need to be sent, and then re-parsed again at the location where they are to be executed. If a prepared statement is used, it is parsed and converted to a regular query before being sent to all replica sites and re-parsed. These limitations are the result of time constraints on development, rather than any technical limitations.

The remainder of this chapter discusses the implementation of H2O-specific functionality.

## 6.2 INTER-PROCESS COMMUNICATION

Earlier versions of H2O used Java RMI for communication between database instances, but RMI fails when a machine's IP address changes at runtime[43], so it was replaced in later builds with a JSON RPC library [89].

## 6.3 TABLE MANAGERS AND THE SYSTEM TABLE

Table Managers and the System Table are run as separate Java threads on the database instance that they are running on. Their state is stored in a number of in-memory data structures to provide fast lookup, and synchronously persisted to disk to provide durability. Information on who holds a lock is held in-memory at each Table Manager, but not written to disk. Therefore, on Table Manager failure, a transaction must wait for a new Table Manager to be created to re-acquire a lock.

The Table Manager and System Table state that is replicated is stored in the database itself, to provide the same consistency guarantees as regular tables. However, replication of meta-data tables is treated differently to regular tables, as is discussed below.

### 6.3.1 Replication

While table data is replicated using *table-level* replication, meta-data is replicated using *segment-level* replication. Table-level replication is used for *table data* because it gives the database the flexibility to run over machines with substantially different storage capacity. This is important when tables can grow to gigabytes in size, but it is not a problem with meta-data replication because meta-data table replicas are small (typically around 10KB), so they can be replicated in groups.

This reduces the complexity of meta-data replication because each Table Manager does not determine data placement of its own state; the H2O instance manages this decision.

Meta-data replication is managed by a single *replication manager* per node, each of which manages all System Table and Table Manager meta-data local to that node. This means that the replication manager of an instance running two Table Managers, *A* and *B*, is responsible for deciding where to replicate the state of *A* and *B*. The state of both Table Managers is replicated together to the same set of instances.

[43] If a machine's IP address changes, the local RMI registry returns remote proxy objects that throw exceptions when called.

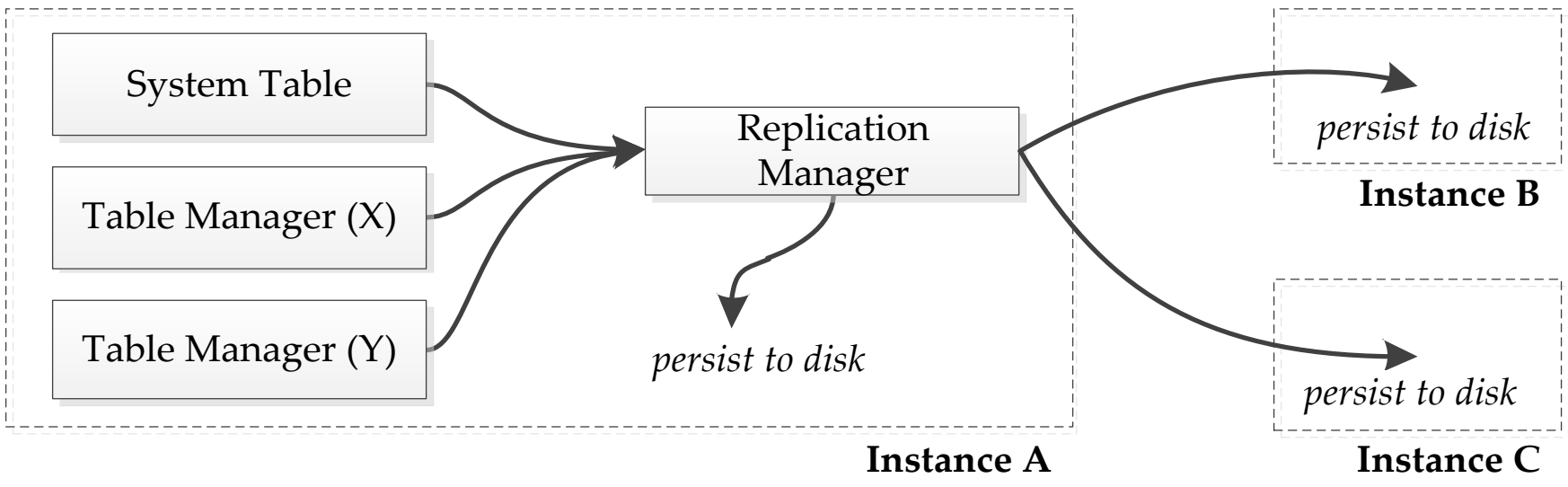

**Figure 52: Local Replication Manager Approach**

A distinction is made between *meta-data* replication (described here), and the *table data* replication (described in 5.5), to prevent a design where Table Managers need their own Table Managers to manage the state they persist. *Table data* replication and placement is determined by the appropriate Table Manager, whereas *meta-data* replication and placement is determined by the database instance local to the Table Manager or System Table.

### 6.3.2 Locking

H2O uses two-phase locking as described in the D2O design. It is possible that other locking mechanisms such as an optimistic locking scheme would be more appropriate in some cases, but two-phase locking is used in H2O as it is appropriate for the OLTP use case which is used to evaluate its performance in *Chapter 7*.

H2O does not implement the composite Table Manager functionality described in 5.6.1 (where Table Managers can be combined to combine lock requests). This functionality was not considered necessary for the evaluation of H2O's performance, which tests the degenerate case where multiple distinct lock requests must be made. However, in a production system this feature would likely be desirable, as it provides the benefits of segment-level locking (fewer lock requests) with the flexibility of the current table-level approach.

### 6.3.3 Updates

H2O uses two-phase commit rather than three-phase commit to execute updates. This decision was made due to H2's existing implementation of two-phase commit (the only design decision made during implementation which deferred to H2).

### 6.3.4 Deciding Where to Place Replicas

If an H2O instance is running an active System Table or Table Manager process, the instance determines where to place replicas of the meta-data for these processes by querying the System Table for a sorted list of other available H2O instances. This list is sorted using a metric produced from the compute and storage capacity of each instance (the *global monitoring analysis* discussed in 5.11.3), allowing the instance to place replicas on the most appropriate available machine.

H2O does not implement the bargaining mechanism discussed in 5.11.4.

### 6.3.5 Recovery from Failure

When a System Table or Table Manager fails it can be recreated from the persisted state on another machine. To achieve this, a new System Table or Table Manager object is created and its state is read from the meta-data tables (via SQL queries) stored on an active instance. The location of every Table Manager replica is stored on the System Table, and the location of every System Table replica is stored on the locator servers.

## 6.4 REPLICATION FACTOR

H2O allows the database user to set their desired replication factor for regular tables and meta-data tables. A Table Manager or H2O instance will attempt to reach that replication factor immediately, but if not enough machines are available it will periodically retry until the replication factor is reached.

## 6.5 LOCATOR SERVER

The locator server architecture is implemented as it is described in Chapter 5. This implementation is similar to *Paxos*, in that a group of nodes are used to reach agreement on a single value, the location of the current System Table.

They differ in that the locator servers give the first instance to obtain a lock on the majority of instances the chance to create the System Table, whereas *Paxos* does not guarantee that the first proposal will be accepted, only that a single value will be chosen. The purpose of the *Paxos* proposal number is similar to that of the locator server update count. Both approaches guarantee *safety*, but not *progress*.

The maintenance of System Table replica locations is similar to *Chubby*, the lock manager used to record active tablets in *Bigtable*. *Chubby* uses *Paxos* to update the state of the servers, and the majority of *Chubby* instances must be active for the database system to continue to create tables. None of the clustered systems described in *chapter 3* use this approach, because they run over relatively static sets of machines that can be manually configured and started.

*Cassandra* uses a similar approach to the descriptor files described here, storing a configuration file which lists a number of known nodes within the cluster. This file can be stored and maintained outside the system, or by *Zookeeper* [74]. The descriptor file is always stored outside D2O, because it is expected that the references it contains to locator servers rarely change, in contrast to the references maintained by the locator servers which change often. This contrasts with *Cassandra* where configuration files store the locations of Cassandra instances directly.

## 6.6 AUTONOMIC MANAGEMENT

To recap section 5.11, H2O is designed to perform autonomic management of replica placement, replica choice (for read queries), threshold analysis, and replication factor. This section discusses the current implementation of each of these features in H2O.

H2O uses a ranked list of database instances on the System Table to determine *replica placement*, however monitoring information is not used in determining *replica choice* for read queries due to the previously discussed focus on the implementation of fault tolerance over autonomic, resource-awareness. Currently each instance caches a ranked list of instances based on their resource availability, but this information is not used to choose replicas. Instead the replica at the site of the Table Manager is used, or another replica is randomly chosen if no replica exists on the same site as the Table Manager. The original design of H2O uses monitoring information along with information on the bandwidth and latency between machines, and the relationship between tables.

*Threshold analysis* is performed using the *NUMONIC* resource monitoring tool, which is discussed in the next section. The planning and execution phase of this threshold analysis is not currently implemented in H2O, though threshold monitoring and analysis is performed.

The *replication factor* of H2O tables and meta-data tables is continually monitored and new replicas are created when machines become available.

## 6.7 NUMONIC

*NUMONIC* is the resource monitoring tool used by H2O to monitor the resources on each machine running an H2O instance.

It uses the *SIGAR* [90] monitoring library to take measurements of CPU, memory, disk, and network utilization. These measurements are aggregated to produce average utilization over a specified period of time, and sent to a reporter class which has registered interest in receiving results.

In H2O, each instance registers a reporter class with a local *NUMONIC* instance. The use of the monitoring data received through this reporter class is discussed in section 5.11.2.

## 6.8 Evaluation of Requirements

The requirements introduced in *Chapter 4* specify what is required of a workstation-based database system. This section recaps these requirements and evaluates how they are met in H2O. The *resource-awareness* and *autonomics* requirements are only partially implemented in H2O, as the implementation focuses on the *database* and *fault tolerance* functionality which is evaluated in *Chapter 7*.

### 6.8.1 Database Requirements

***D1 Interface to applications****: Provide a JDBC SQL interface.*

- ✓ H2O provides a JDBC SQL interface.

***D2 Transactional Requirements****: Transactions are ACID compliant.*

- ✓ Transactions are atomic, consistent, isolated, and durable.

***D3 Deployment****: The database can be started without extensive manual configuration.*

- ✓ If a user has a descriptor file, they can start an instance that automatically joins a database system and starts replicating its state and that of others.

***D3 Deployment****: The database has comparable performance to existing clustered DDBMSs.*

- o This requirement is evaluated in the first experiment in chapter 7.

### 6.8.2 Architectural Requirements

***General***

***A1 Self-Contained****: The system is able to run in its entirety on workstation machines.*

- × This requirement is not met because locator servers are designed to be run on machines more reliable than workstations. If locator servers were not used knowledge of System Table locations would have to be stored within the workstation-based system, which means that bootstrapping instances must be aware of the location of a currently active instance.
  In H2O, locators are used to make it easier to start new instances, which is specified in *requirement D3*.

***A2 Capacity****: The database can grow bigger than the capacity of a single machine.*

✓ Table-level replication gives the database the flexibility to partition a single database over multiple machines.

***A3 Heterogeneity in Platforms:*** *A database instance can run on multiple operating systems.*

✓ H2O is written in Java and is able to run on most common operating systems. It has been tested on Windows, OSX and a number of Linux distribution (CentOS and Ubuntu).

***Fault Tolerance***

***A4 Resilience to Failure***: *The database is able to withstand the failure of individual machines.*

✓ Synchronous replication is used to ensure that the database can continue to operate unless the instances holding all the replicas of a table fail.

***A5 Mobility***: *The database is able to handle database instances changing IP addresses.*

✓ Databases are identified by a unique ID rather than their IP address. The current location of an instance is stored by the System Table and by the locator server (if it stores System Table state), so an instance can be found after an address change.

***Resource-Awareness***

***A6 Local Resource Monitoring***: *The database is able to monitor the availability of local resources.*

✓ *NUMONIC* is used to monitor resources on every machine running an H2O instance.

***A7 Network Resource Monitoring***: *The database is able to monitor bandwidth and latency.*

✗ Network resource monitoring is not currently implemented.

***A8 Local Resource Analysis***: *The database is able to determine when resource utilization changes.*

✗ Monitoring data is not currently used to move processes or replicas of data.

***A9 Global Resource Analysis***: *The database is able to rank instances by availability.*

✓ Monitoring data is collated at the System Table and used to rank machines. This is used to decide where to place replicas.

***Autonomics***

***A10 Replication***: *The database is able to automatically replicate data with no user involvement.*

- ✓ Each Table Manager periodically checks whether enough replicas exist. New replicas are created if there are not enough active replicas.

***A11 Opportunistic Utilization***: *The system is able to incorporate new resources into the database.*

- ✓ When an H2O instance joins a database system it starts sending monitoring data to the System Table, so that it can be ranked against other instances and used to store replicas of data on other instances.

***A12 Using Resource Awareness***: *The system is able to autonomically use resource monitoring.*

- × Resource monitoring data is not currently used to implement any of the autonomic functionality described in *Chapter 5*.

### 6.8.3 Conclusion

The implementation of H2O that is discussed in this chapter meets most of the requirements of an interactive workstation-based database system.

The lack of network resource monitoring (*requirement A7*) potentially impacts the performance of read queries, but it is not critical to the experiments in the next chapter.

The next chapter describes two experiments which are designed to test other aspects of H2O's effectiveness as a workstation-based system.

# Chapter 7: Evaluation

# 7 EVALUATION

This chapter describes two experiments which evaluate the viability of interactive workstation-based systems in terms of fault tolerance and performance.

In the first, H2O is compared against existing clustered DDBMSs to establish whether a workstation-based database can be an effective replacement for a clustered system in terms of transaction throughput.

In the second, H2O is evaluated under various failure scenarios to establish the effect of failure on transaction execution time and the replication factor of data.

These experiments were chosen as they evaluate two key aspects which determine the viability of a workstation-based system: the performance of a workstation-based architecture in comparison to comparable clustered systems, and the ability of the workstation-based architecture to respond to failure. The aspects not evaluated — the functionality described in the *autonomic* and *resource-aware* requirements in *Chapter 4* — are discussed as future work in *Chapter 8*.

## 7.1 EXPERIMENTAL FRAMEWORK

All of the experiments described in this chapter are run on a cluster of machines, rather than on workstations, to ensure that the reliability and availability of machines is controlled by the experimental framework rather than external forces. Each machine in this cluster has the following specification:

- **CPU:** Intel(R) Xeon(TM), 2.40GHz Processor, 2 CPUs, 512KB Cache
- **Memory:** 2GB RAM
- **Hard Disk:** 20 GB capacity, IDE, 5400 RPM
- **Network:** 1Gbit Ethernet
- **Operating System:** Centos 5.5

## 7.2 Experiment 1: Database Benchmarking

The aim of *experiment 1* is to evaluate the performance of *H2O* against existing clustered DDBMSs. For *H2O* to be a viable alternative to these systems, its performance must offer comparable performance in answering queries.

This experiment evaluates the effectiveness of each of these databases with different replication factors ranging from 1 to 11.

### 7.2.1 Workload

A variant of the *TPC-C* benchmark, *BenchmarkSQL* version 2.3.3 [91], is used to evaluate the performance of each database. The *BenchmarkSQL* program has been modified to remove a number of vendor-specific calls and to allow for non-interactive use[44].

*TPC-C* is an OLTP benchmark which measures performance in terms of transactions per minute. It models a wholesale supplier managing orders through five types of transactions, including the creation of new orders, and updates to customer balances, delivery information and stock level data [92]. *BenchmarkSQL* can be run for a specified number of minutes, during which time it repeatedly executes transactions against these tables.

OLTP is a suitable target workload because it requires ACID transactions, and consequently makes use of the features that *H2O* is designed to support.

### 7.2.2 DDBMSs to Evaluate

In this experiment H2O is evaluated against a number of other DDBMSs. These systems are listed below along with details of their replication scheme.

| Database | Type of Replication | Granularity of Replication | Storage |
|---|---|---|---|
| **MySQL** | Master-Slave | Full-Database | Disk |
| **PGCluster** | Multi-Master | Full-Database | Disk |
| **H2O** | Multi-Master | Table-Level | Disk |

**Table 12: The databases being evaluated in the benchmarking experiment.**

*MySQL* and *PGCluster* are comparable to H2O in that they are used for the type of applications running in small clusters that *H2O* is designed to serve running on workstation machines. While *MySQL* does not use a comparable replication scheme, it serves as a useful baseline on which to base *H2O's* benchmarking results.

[44] The source code for this program can be found at: http://archive.cs.st-andrews.ac.uk/h2o

The systems discussed in *Chapter 3* that use segment-level replication — *Greenplum* and *MySQL Cluster* — are either hardware-based solutions that cannot be run on off-the-shelf machines, or in-memory databases that are not directly comparable to *H2O*. As a result, no segment-level systems are included in this experiment.

### 7.2.3 Hypothesis and Criteria for Success

It is expected that *MySQL* will perform much better than *H2O* because it uses asynchronous replication. This should mean that *MySQL's* performance does not degrade as the replication factor is increased, whereas with *H2O* and *PGCluster* — both synchronous replication systems — performance will degrade as the database must perform more work as more instances become involved in the execution of a transaction. However, since updates are performed concurrently over multiple machines, performance should not degrade linearly.

For *H2O* to be considered as a viable alternative to clustered systems it must have comparable performance to *PGCluster*, which is the synchronous replication system being compared in this experiment.

### 7.2.4 Experimental Setup

The experiment is run in a specially created evaluation framework, which automatically creates clustered instances of each database and executes the *BenchmarkSQL* benchmark against each of them using the same experiment parameters[45].

For each evaluation run, the *BenchmarkSQL* application connects to a single database instance to issue queries and updates. Each database has a different architecture, so their precise setup is explained below.

***MySQL Setup***

For each experiment, a single *MySQL* master server and *n-1* slave servers are run, where *n* is the replication factor of each table. For a replication factor of three, there is a single master and two slaves. The benchmark is executed against the master instance, as illustrated in

[45] The source for the script which automates these experiments can be found at: http://archive.cs.st-andrews.ac.uk/h2o

Figure 53, and a binary log of updates is asynchronously sent to slave instances. Each MySQL instance has a full copy of the database.

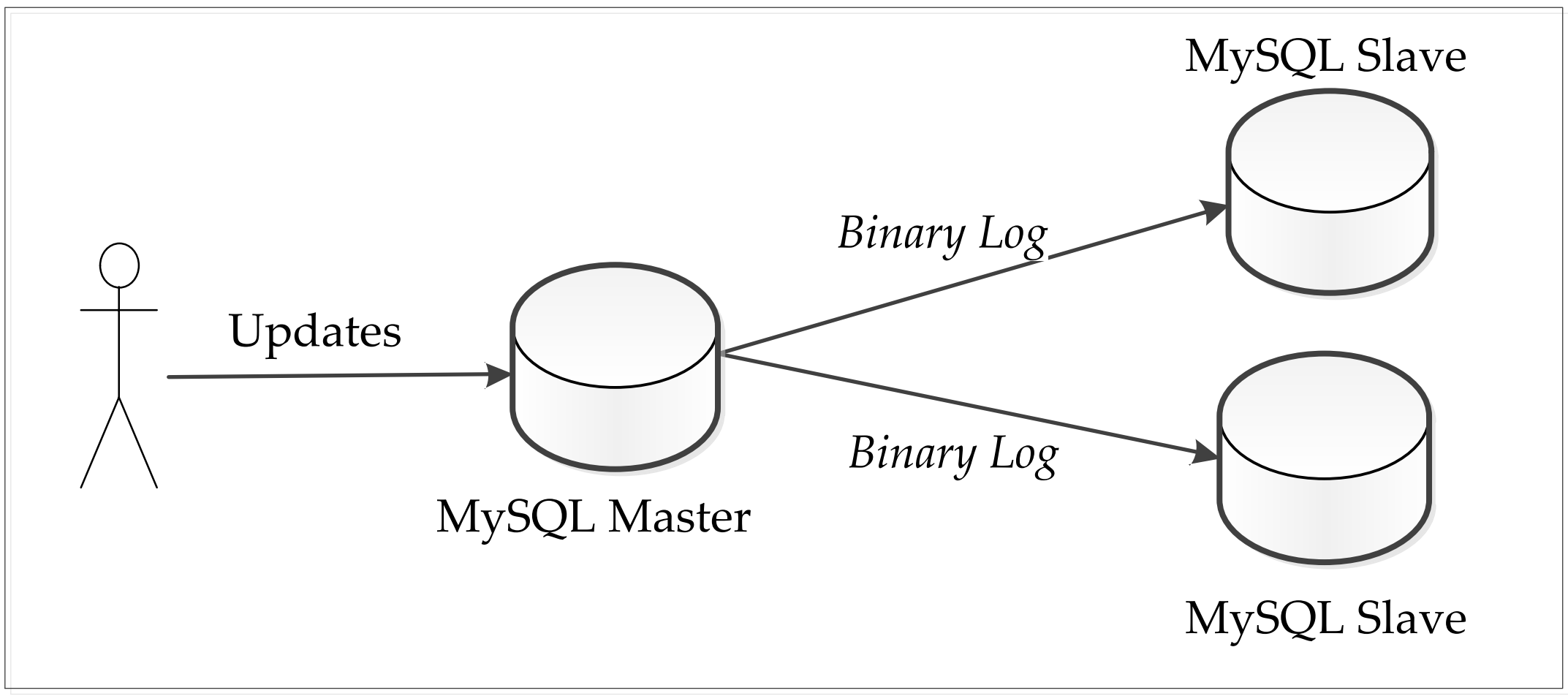

**Figure 53: An illustration of the *MySQL* architecture that the *BenchmarkSQL* benchmark is executed against. In this example the system has a replication factor of three.**

The system uses the default *MySQL* configuration file (*my.cnf*), with some additional settings related to replication. This configuration file is listed in *appendix 1.1*.

The default configuration file is used for each database for two reasons. First, it ensures that there are no special optimizations made to one system that are not made to the others. Second, it reflects a goal in the design of *H2O* that it should be as simple as possible to create a replicated database system. Consequently, each system is set up with a minimum level of configuration.

***H2O Setup***

For each experiment, *n* instances of *H2O* are run, where *n* is the replication factor of each table. The benchmark is executed against a single *H2O* instance, which manages the System Table and all Table Managers for the tables involved in the benchmark. This architecture is illustrated in Figure 54. Updates are sent to the H2O instance running the Table Managers and the System Table, and these updates are then sent synchronously to each replica site to be executed. Each H2O instance has a full copy of the database.

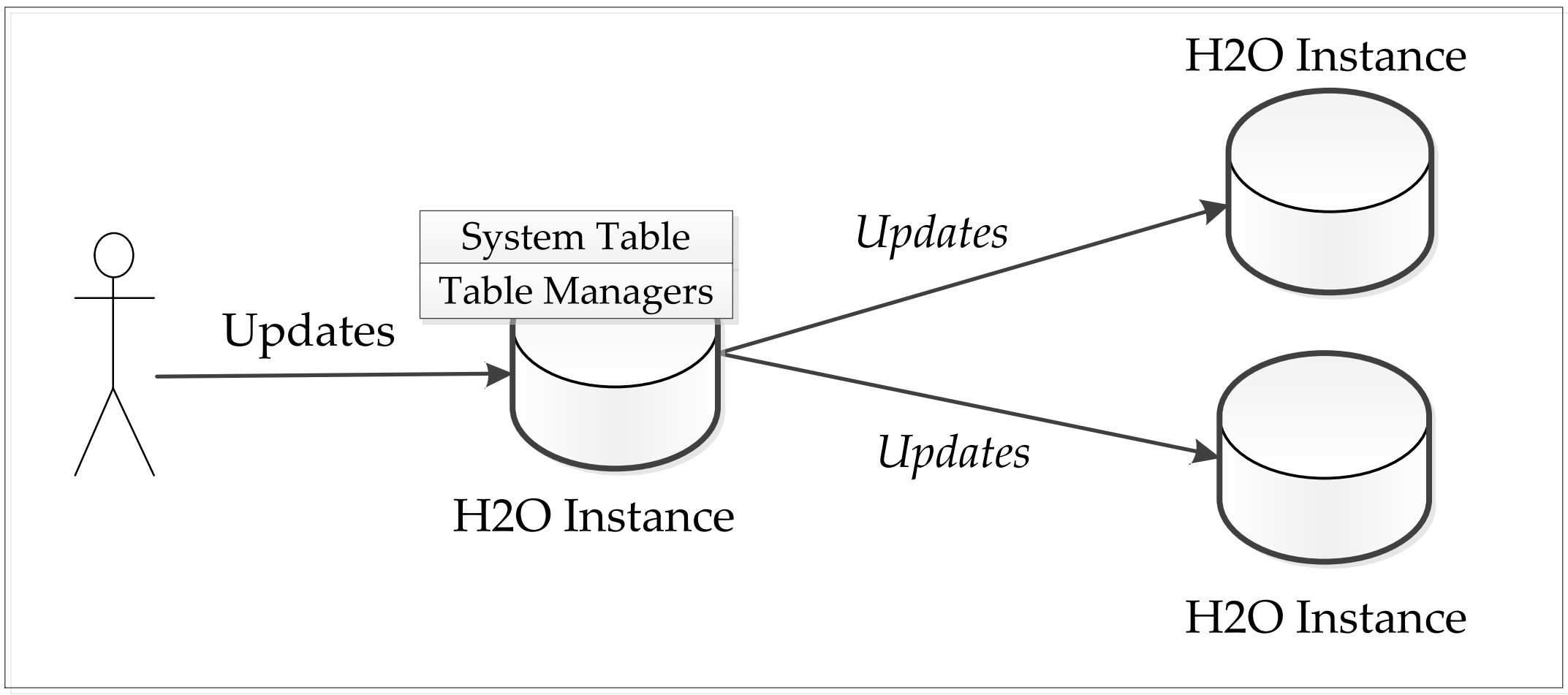

**Figure 54: An illustration of the *H2O* architecture that the *BenchmarkSQL* benchmark is executed against. In this example the system has a replication factor of three.**

In this experiment, *H2O* is run as a consistent and available database, meaning the System Table does not contact the locator server after every update. This should have no effect on the benchmark's performance, because all tables are created before the benchmark starts (so there are no System Table updates).

### *PGCluster Setup*

In *PGCluster*, data instances are used to store data, so there are $n$ data instances, where $n$ is the replication factor. In addition, a replication server and load balancer are also running, so there are $n$+2 instances in the system.

*PGCluster* can be used with or without a load balancer. The load balancer is designed to balance load across data instances, which may be useful when the system is under heavy load. However, it introduces additional indirection in answering queries, which may not always be necessary. The alternative solution is to issue queries directly against a data instance — this still replicates data, but read queries are only issued to a single data instance. The architecture of the second approach, not using the load balancer, is illustrated in Figure 55.

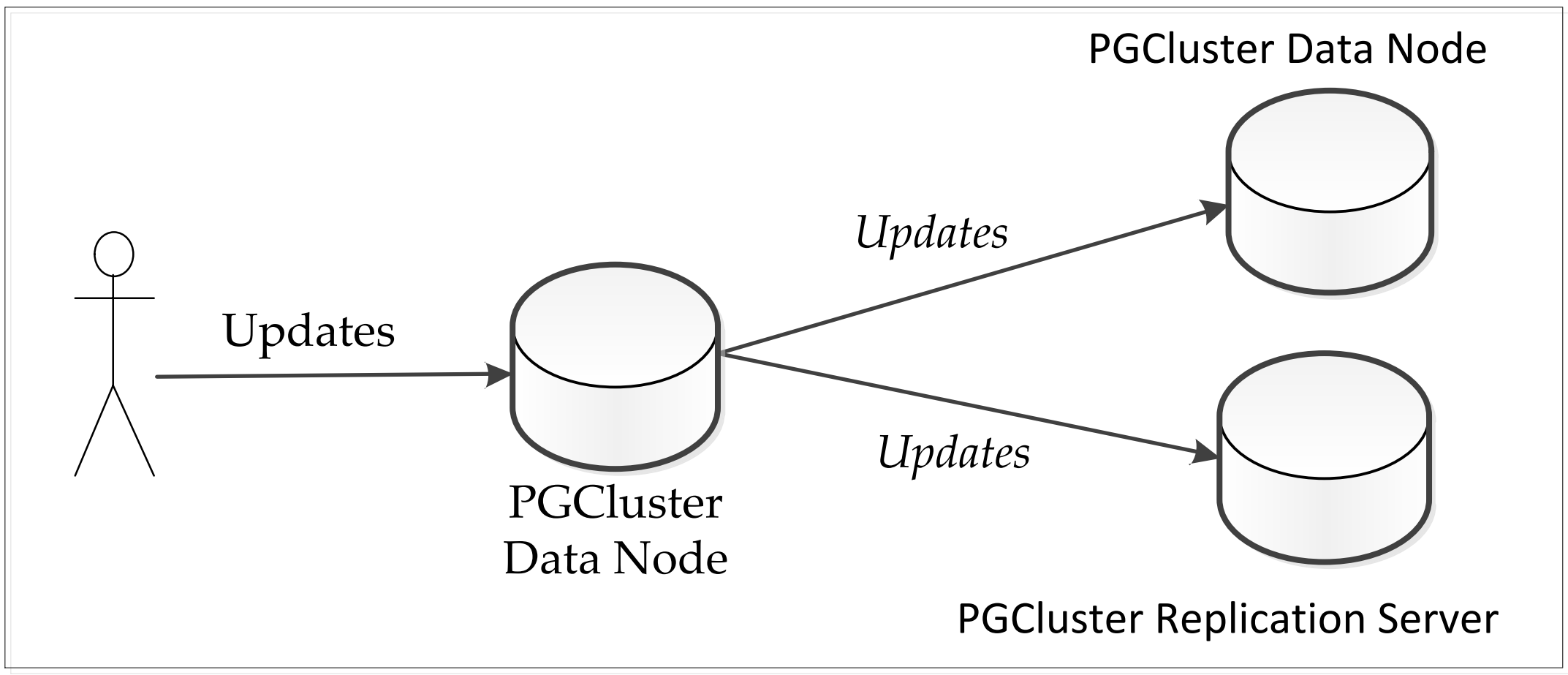

**Figure 55: An illustration of the *PGCluster* architecture that the *BenchmarkSQL* benchmark is executed against. In this example the system has a replication factor of two.**

To determine the most effective architecture for *PGCluster*, the *BenchmarkSQL* benchmark was run for eight minutes and repeated 30 times for both configurations (queries against the load balancer, queries against a data node)[46]. This produced the following results:

| Configuration | Samples | Avg. Transactions Executed | Throughput (trans/min) | Standard Error |
|---|---|---|---|---|
| Query Data Node | 30 | 1719.9 | 214.98 | 0.59 |
| Query Load Balancer | 30 | 1669 | 208.62 | 0.75 |

**Table 13: Experiment 1. The results of executing BenchmarkSQL against different configurations of PGCluster.**

The data node configuration produces higher transaction throughput, so this is the architecture used in this experiment. Both sets of experiments and those described below use the default *PGCluster* configuration file (*postgresql.conf*), which is listed in *appendix 1.2*.

### 7.2.5 Experimental Parameters

***Establishing Benchmark Duration***

An initial run of the benchmark was carried out to establish the run duration of the benchmark. To achieve this, the benchmark was executed against each database for

[46] The rationale behind the run time and number of repetitions is explained below in 7.2.5.

durations ranging from one minute up to thirty minutes and plotted against the transaction throughput, as shown in the graph below:

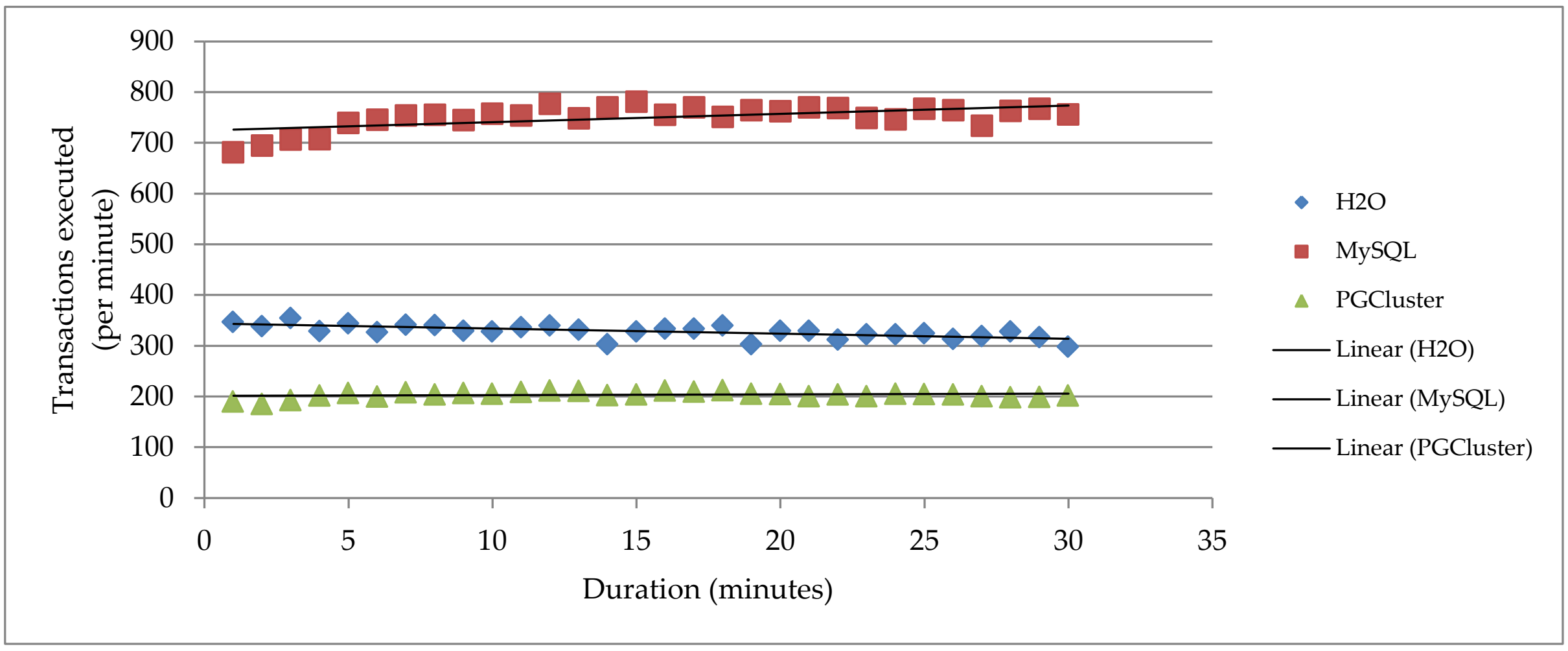

**Figure 56: Experiment 1. Transaction throughput measured against benchmark duration for each database system.**

This shows that the transaction throughput remains relatively constant after eight minutes, so the remainder of experiments in this section sample results from benchmarks that run for *eight* minutes.

### *Standard Error*

Another set of experiments were undertaken to establish the standard error of these results. The benchmark was repeated *30* times, with each benchmark running for *eight* minutes. These results show that the standard error of transaction throughput (transactions per minute) does not decrease sharply after more than 20 samples are used. The graph below plots how the standard error changes for each database as the number of samples taken in the experiment is increased.

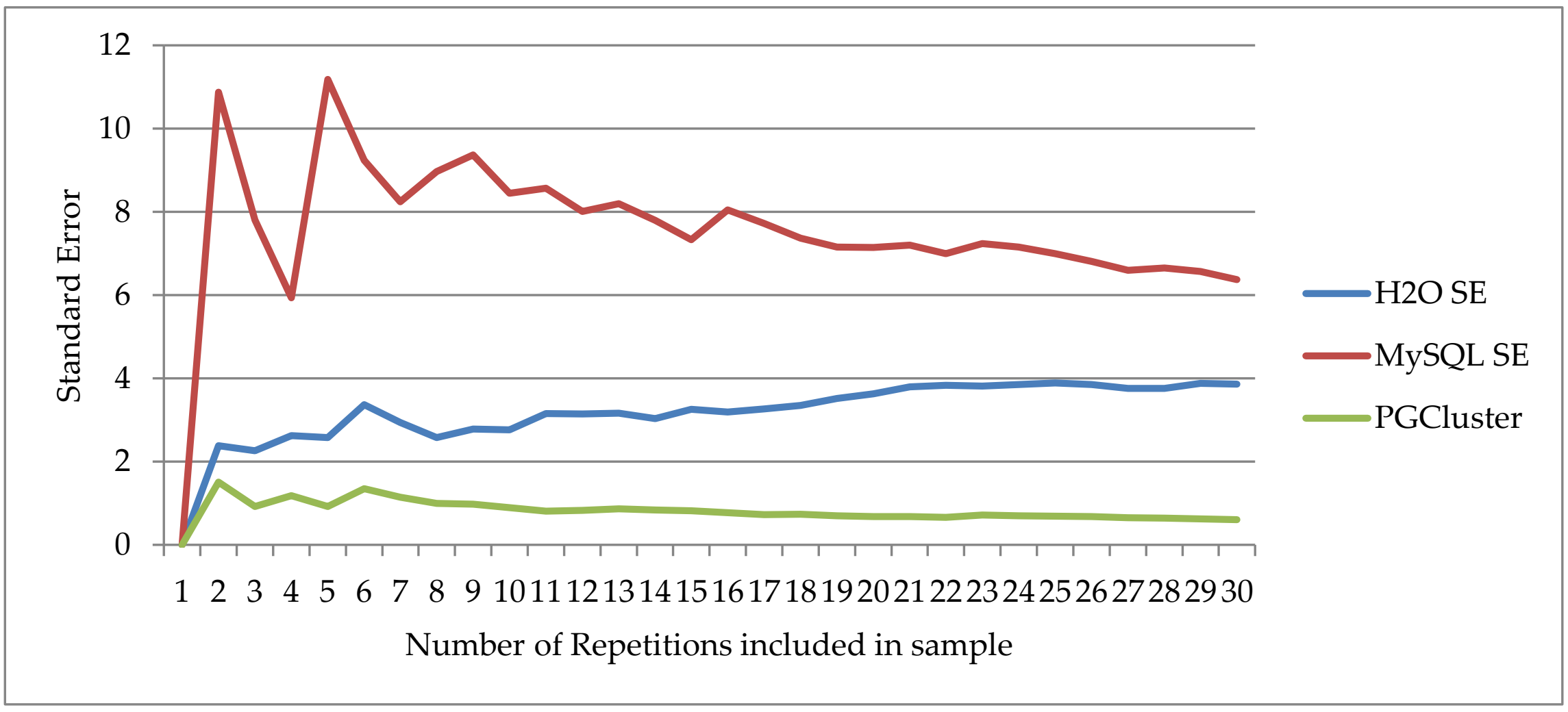

**Figure 57: Experiment 1. The standard error of samples, as the number of samples is increased for each database (eight minute samples).**

In *experiment 1* the benchmark is run against database configurations with replication factors of 1, 2, 3, 5, 8, and 11, with 20 samples taken for each replication factor.

### 7.2.6 Results

The results of *experiment 1* are summarised below in Table 14 and in Figure 58, both of which show how transaction throughput changes as the number of replicas is increased.

| | **Average Throughput per Minute (20 samples)** | | |
|---|---|---|---|
| Number of Replicas | MySQL | PGCluster | H2O |
| 1 | 754±2 | 256±1 | 375±9 |
| 2 | 735±7 | 225±1 | 250±2 |
| 3 | 716±11 | 198±1 | 224±2 |
| 5 | 723±6 | 166±1 | 176±2 |
| 8 | 711±7 | 140±0 | 141±2 |
| 11 | 668±9 | 129±1 | 122±2 |

**Table 14: Experiment 1 results. The effect of increasing replication factor when running the BenchmarkSQL benchmark against each database.**

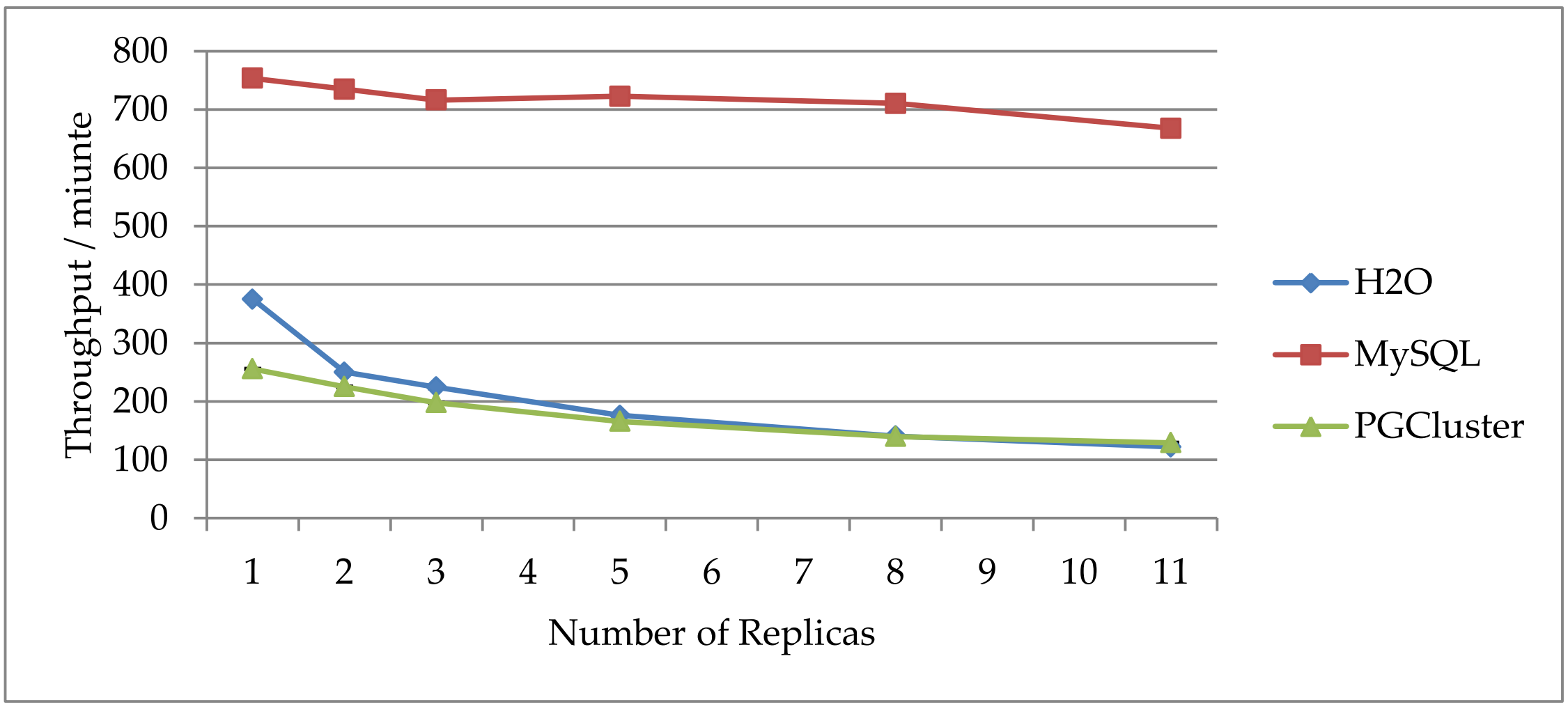

**Figure 58: Experiment 1 results. The transaction throughput achieved by each database when executing a benchmark over varying numbers of replicas[47].**

*MySQL* performs well in comparison to *H2O* and *PGCluster*, largely because it uses asynchronous replication. This explains why transaction throughput on *MySQL* is largely unaffected by the degree of replication.

*H2O* and *PGCluster* perform significantly better when only one replica exists (compared to multi-replica evaluation runs) because transactions are not distributed. As the number of replicas increases their performance degrades, because they must wait for every instance to execute the transaction before it can be committed.

The synchronicity of replication does not explain why *H2O* is significantly slower than *MySQL* with only a single replica, though there are a number of other factors which may explain this:

In *H2O*, lock requests to local Table Managers uses the same JSON RPC mechanism that is used when Table Managers are remote, which adds an extra overhead to these requests. There is no single lock manager, so transactions querying multiple tables must contact each Table Manager separately using JSON. In addition, *H2O's* implementation may simply be inherently less efficient than MySQL, which has had significantly many more man hours spent on development and optimization.

[47] The standard error is shown for each data point, but it is so small that it is not visible in a number of cases.

*H2O* and *MySQL* call `fsync` on every commit, to provide durability in the event of the simultaneous failure of multiple machines (an issue discussed in chapter 6) Systems such as *VoltDB* are able to provide fault tolerance without writing to disk if each database instance is run on machines with uninterruptible power sources, so the simultaneous failure of multiple machines is extremely unlikely. Figure 59 (below) shows the effect of the call to `fsync` on transaction throughput in H2O. When H2O is run with no write delay (`fsync` is called immediately on every commit) it runs roughly 100 transactions/second slower than when it is run with a write delay of 500 milliseconds (`fsync` calls are made every 500 milliseconds). The reasoning behind write delay is discussed in 6.1.

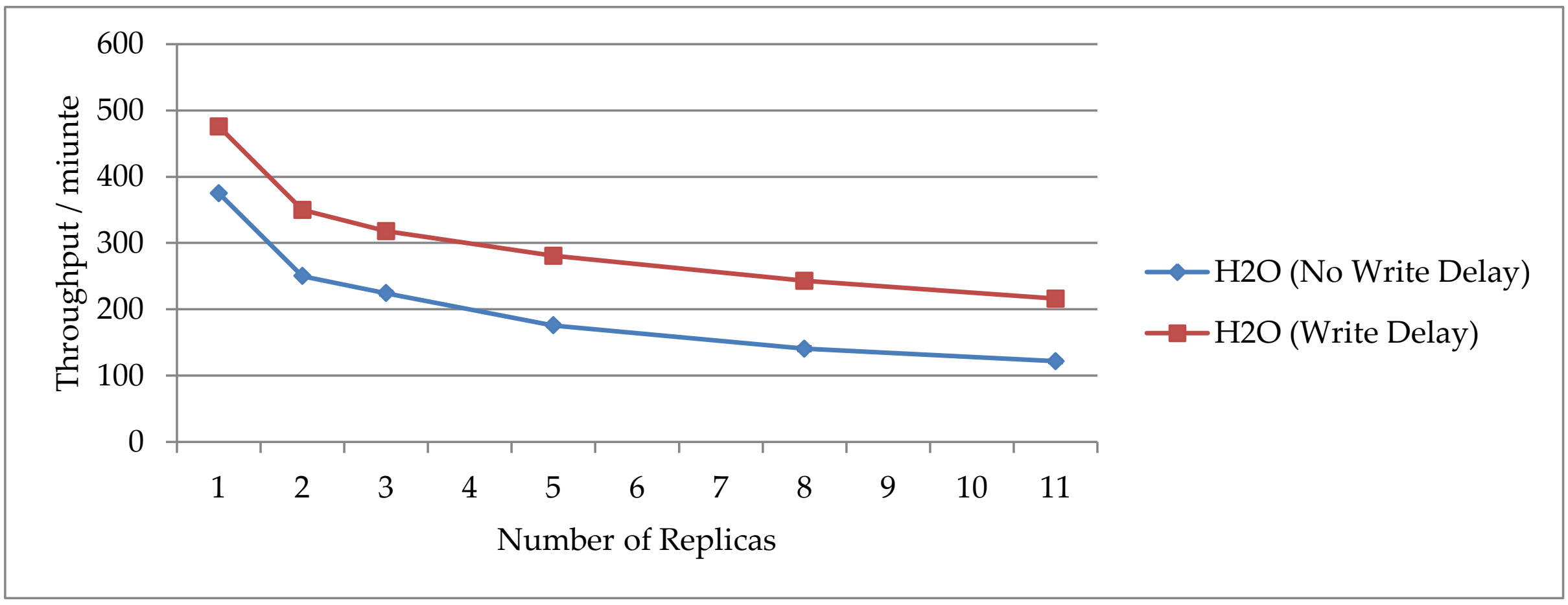

**Figure 59: Experiment 1. H2O's benchmark performance with and without write delay enabled.**

### 7.2.7 Conclusion

***Costs of Workstation-based Approach***

The results of this experiment highlight synchronous replication and lack of write delay as the likely primary costs of running in a workstation-based environment.

Synchronous replication adds overhead because the database must wait for every replica to be updated before it commits. The effect of this is shown in Figure 58, where *MySQL* — which asynchronously replicates data — performs consistently better than both *H2O* and *PGCluster*, both of which synchronously replicate data. *MySQL's* performance does not significantly degrade as the number of replicas is increased, in contrast to *H2O* and *PGCluster*. The effect of this is shown below, where the throughput at each replication factor is shown relative to the throughput with two replicas.

| | Database | | |
|---|---|---|---|
| Number of Replicas | MySQL | PGCluster | H2O |
| 1 | 103% | 114% | 150% |
| 2 | ***100%*** | ***100%*** | ***100%*** |
| 3 | 97% | 88% | 90% |
| 5 | 98% | 74% | 70% |
| 8 | 97% | 62% | 56% |
| 11 | 91% | 57% | 49% |

**Table 15: Experiment 1. A comparison of the throughput achieved at different replication factors, compared against the throughput achieved with a replication factor of two.**

The effect of write delay on transaction throughput is shown in Figure 59. As with synchronous replication, the database is slowed by forcing an `fsync` call on every commit (no write delay).

***Limitations of Experiment***

This experiment focuses on comparing the transaction throughput of *H2O* against two existing DDBMSs. It does not evaluate the effect of running each system on a heterogeneous system with varied latency and bandwidth between machines, and it does not show the effect of partitioning data over multiple machines.

It is likely that varying the available bandwidth and latency between machines would reduce the transaction throughput of *H2O* and *PGCluster*, leaving *MySQL* relatively unaffected. However, the consistency of *MySQL* data may be affected, as queries to slave instances may see stale data and the failure of the master may result in data loss (this experiment evaluates neither of these factors).

*MySQL* and *PGCluster* do not provide the ability to partition data over database instances, and *H2O* does not do so in this experiment to ensure that comparable evaluation setups are used to obtain benchmark results. The database created by *BenchmarkSQL* in this experiment grows as large as 300MB, which is relatively small in comparison to many production databases. In this situation it is reasonable to assume that a single H2O instance can store a copy of the entire database, but for larger datasets, H2O's transaction throughput should be evaluated with partitioning enabled, to provide results that reflect an actual use case in a workstation-based environment.

***Summary***

This experiment shows that in a reliable, low latency network, *H2O* has comparable throughput to an existing synchronous replication database (*PGCluster*), but significantly slower throughput than an asynchronous replication database (*MySQL*).

The closeness of *H2O's* performance to *PGCluster* shows that *H2O's* architecture has comparable performance to an existing synchronous replication database. However, the costs of synchronous replication are also shown, as the performance of both *H2O* and *PGCluster* degrades when more replicas are added.

To maintain durability in a workstation-based system, *H2O* flushes transactions to disk when they commit. This introduces an additional overhead that reduces throughput by roughly one hundred transactions per minute, which gives comparable performance to *PGCluster*.

The evaluations in this section do not test H2O running on workstation machines, but these results are relevant because they show that H2O's architecture (which is designed to meet the challenges of running over workstation machines) is viable and has comparable performance to similar synchronous replication systems.

## 7.3 EXPERIMENT 2: AVAILABILITY DURING MACHINE FAILURE

This experiment uses an evaluation framework to show how H2O responds to failure. To evaluate the effectiveness of H2O, a workload is run on the database, and the system's response to various failure scenarios is measured in terms of its transaction throughput and the time it takes to recover from each failure.

These evaluations are designed to test H2O's response to failure in a number of isolated failure scenarios (they test specific types of failure individually, rather than the response to various failures over a long period of time). Failures are simulated on a specially provisioned cluster of machines rather than workstations, to limit uncontrollable factors and to improve the reproducibility of the evaluations.

*MySQL* and *PGCluster* are not evaluated in this experiment because neither is able to automatically recover from the failure of a database instance.

Failure scenarios are specified in *co-ordination scripts*, which contain a sequence of actions that involve database instances starting and failing at specified intervals. Each co-ordination script is repeated 10 times to illustrate the typical recovery time of the database.

### 7.3.1 Experimental Setup

This experiment uses a specially created evaluation framework, which is comprised of two components: *workers*, which are started on each machine in the compute cluster (as described in *7.1*), and a *co-ordinator*, which is used to run *co-ordination scripts* and communicate instructions to *workers*. *Workers* are started using a deployment tool named *MADFACE*[48].

The *co-ordinator node* issues commands from a *co-ordination script* to each *worker* to start, stop or terminate an H2O instance. It is also able to send a *workload* containing a set of queries to be executed by that *worker*. When a *workload* has been executed, its result — the query throughput for each second of its execution — is sent back to the *co-ordinator*, which collates results. This architecture is illustrated below in Figure 60, where multiple machines are started, but a workload is only executed on one of these machines.

---

[48] The source for *MADFACE* can be found at: http://archive.cs.st-andrews.ac.uk/h2o

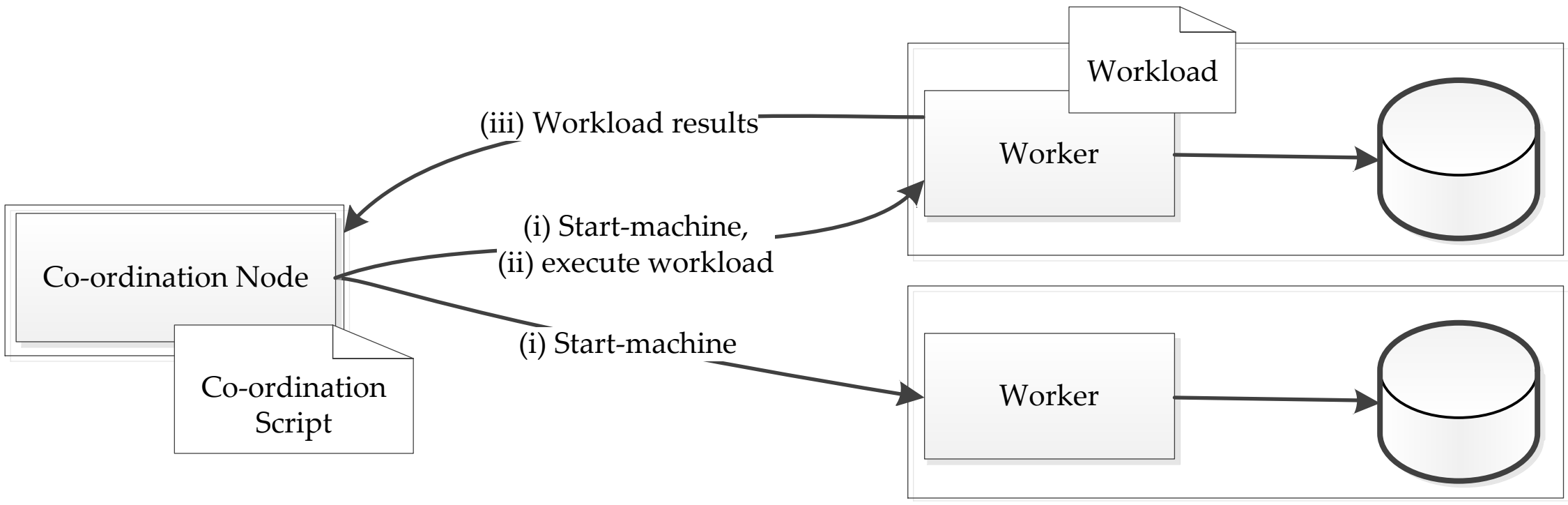

**Figure 60: Architecture of the Co-ordination Framework used in Experiment 2.**

If a failure occurs when a transaction is executing, H2O attempts to recover and re-execute the transaction, so the only transactions that should fail are those that H2O is not able to recover from (for example, when there are no available replicas of a table).

### 7.3.2 Workload

A workload is run by a single worker against its local H2O instance. Two types of workload are used in this experiment.

The first, designed to put load on the Table Manager, repeatedly runs a transaction which executes an *INSERT* query and a *SELECT* query against a single table. These queries test the System Table by repeatedly requesting locks, and they test the availability of replicas by updating and requesting data.

The second workload is designed to put load on the System Table, so it performs repeated *CREATE* and *DROP TABLE* operations. These are the only operations that always require contact with the System Table, and by being repeatedly executed they test the System Table's availability and its capacity to execute queries.

Both workloads are shown in *appendix 2.*

In this experiment these workloads are run in separate tests. The queries from each workload are sent to a single H2O instance that acts as the local point of presence, and no data is replicated onto this instance, so all the transactions it executes must be sent to other instances. This makes it possible to terminate machines which store data and Table Managers without terminating the machine that is receiving queries.

### 7.3.3 Co-ordination Scripts

Co-ordination scripts are used through this chapter to describe specific failure cases, so their syntax is described below, with excerpts from an example script. This script starts two H2O instances, creates a table and begins executing a workload against it, then terminates one of these machines before restarting it later. It is described in fragments through the rest of this section, but is shown in full below.

```
{start_machine id="0"}
{start_machine id="1"}
{create_table id="1" name="workloadTable"
schema=" id int, str_a varchar(40)" prepopulate_with="300"}
{sleep="20000"}
{1} MIGRATE SYSTEMTABLE
{0} {execute_workload="readWorkload.workload"
duration="60000"}
{sleep="20000"}
{terminate_machine id="1"}
{sleep="20000"}
{start_machine id="1" block-workloads="true"}
{check_meta_repl_factor expected="3"}
{check_repl_factor name="workloadTable" expected="2"}
```

**Figure 61: Example Co-ordination Script.**

The `start_machine` command starts an H2O instance on any available worker. An `ID` is used to identify each instance, which enables later commands to be executed against specific instances. The co-ordination script blocks until the instance has started.

The first machine started is always the site of the System Table (unless it is migrated later in H2O), and no data is replicated on this machine. Thus, when the following fragment is executed, two instances are started on two separate machines.

```
{start_machine id="0"}
{start_machine id="1"}
```

The `create_table` command specifies the machine on which the table should be created (*id*), the name of the table (*name*), its attributes (*schema*), and how many rows should be created in this table (*prepopulate_with*). When the following fragment is executed, a table named *workloadTable* is created on the machine with *id=1*, and this table is pre-populated with 300 rows of randomly generated data.

```
{create_table id="1" name="workloadTable"
    schema=" id int, str_a varchar(40)" prepopulate_with="300"}
```

SQL commands can be executed against individual instances by specifying the ID of the machine in brackets, followed by the command to be executed. In the following fragment *machine 1* executes the *MIGRATE SYSTEMTABLE* command, which moves the System Table to *machine 1*.

```
{1} MIGRATE SYSTEMTABLE
```

The `sleep` command stops the script from continuing for a specified number of milliseconds.

```
    {sleep="20000"}
```

The `execute_workload` command starts the execution of a workload on the instance identified by the number in brackets, and continues to execute for the `duration` specified. Workloads run asynchronously, so the co-ordination script continues to execute commands while they execute. In the following fragment the workload contained in the file *readWorkload.workload* is executed for 60 seconds.

```
{0} {execute_workload="readWorkload.workload"
    duration="60000"}
```

The `terminate_machine` command specifies that the H2O instance with the given ID should be terminated (by terminating the database instance's process). This is typically preceded by a `sleep` command, which in the following example means that the workload has been executing for 20 seconds before the first machine is terminated.

```
{sleep="20000"}
{terminate_machine id="1"}
```

Machines can be restarted with the `start_machine` command. An additional `block-workloads` parameter can be included to suspend all workloads while the machine is started. Without `block-workloads`, the `start_machine` command blocks the co-ordination script while the machine is being started, but any active workloads will continue to execute. The `block_workloads` command ensures that all workloads are suspended while the machine starts, so the machine always becomes available at the specified time in the co-ordination script, rather than at a point determined by its start-up time.

In the following fragment *machine 1* is restarted after another 20 seconds and workloads are suspended until it has started up.

```
{sleep="20000"}
{start_machine id="1" block-workloads="true"}
```

Finally, the `check_repl_factor` commands are used to check that the replication factor of a table or meta-data table is as expected at a given point in the script. These commands are used to check that the system has started and is operating as expected during the execution of a script. The following fragment checks that System Table state is replicated three times and that there are two replicas of *workloadTable*.

```
{check_meta_repl_factor expected="3"}
{check_repl_factor name="workloadTable" expected="2"}
```

If these assertions fail, the script executor terminates with an error code, informing the test framework that the test run was unsuccessful.

### 7.3.4 Experiment Parameters

This section describes the parameters of the script files used to test H2O's effectiveness at handling failure.

These tests evaluate how long it takes H2O to recover from four failure scenarios. In these scenarios the failed machine(s) may hold:

1. A single replica of table data only
2. An active Table Manager and table data[49]
3. The active System Table only
4. An active Table Manager, table data, and the System Table

With regards to replication, there are five scenarios that can occur once an instance has failed:

A. There are no other machines available to re-replicate data (increasing the replication factor up to its previous level)
B. There are a number of other machines available onto which data can be re-replicated

[49] It is assumed that an instance holding an active Table Manager also has a copy of that table's data.

C. There are no machines available immediately, but a new machine starts up
D. There are no machines available immediately, but the failed machine restarts
E. There are no other active replicas of table/meta-data state

This experiment tests how H2O reacts to the failure of machines in each of these cases. The numbered points above (*1-4*, *A-E*) are used to concisely describe each test.

The number of machines involved in each of these tests is kept to a minimum, to reduce the number of variables in each test. For tests in this experiment, the stated number of machines does not include the machine being used to execute the test workload (the local point of presence). This machine does not store any table data or meta-data used in the test.

### 7.3.5 Scenarios Being Tested

This section categorizes the states that H2O could go into as a result of these tests. The icons displayed next to each of these states are used later to explain the expected outcome of each test.

The failure of a machine(s) may have one of four effects on the database system:

T1.

[50]

The failure does not affect the ability of the system to execute transactions successfully.

T2. 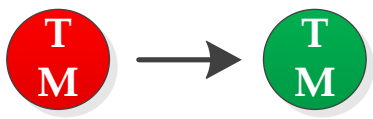

The failure temporarily stops the database from being able to execute transactions, due to the failure of the Table Manager or System Table. However, there are replicas available which are used to re-instantiate these processes.

T3.

The failure of this machine stops the database from being able to execute transactions, because the Table Manager/System Table has failed. There are no additional replicas, so these processes cannot be restarted.

[50] In the icons in this section, *TM* stands for *Table Manager*. When a failure affects the System Table these icons use the abbreviation *ST*, which stands for *System Table*.

T4.

The failure of this machine stops the database from being able to execute transactions, because no replicas of data/meta-data are available to execute the query. There are no additional replicas, so the database cannot fix this problem.

The failure of a machine(s) also has an impact on replication factor, of which there are three possibilities:

R1.

The replication factor of the table decreases.

R2.

The replication factor of the table temporarily decreases, then increases to its previous level almost immediately.

R3. 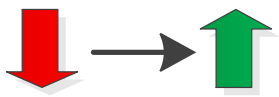

The replication factor of the table decreases, then increases to its previous level, but only when another H2O instance is started.

The following table lists the complete set of tests carried out for this experiment, and uses the possible scenarios listed above to describe the expected outcome of each test. The *no. of machines* column lists the number of machines running *at the start* of the script's execution, *not including* the machine that is used by the script executor to run the workload.

| Categorization | No. of Machines | Repl. Factor | Expected Outcome |
|---|---|---|---|
| 1A | 2 | 3 | 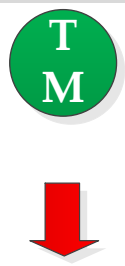  |
| 1B | 3 | 2 | 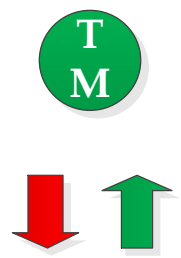  |
| 1C | 2 | 3 | 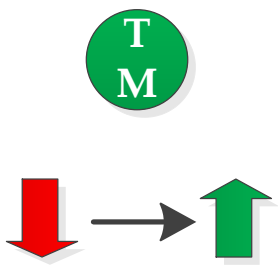  |
| 1D | 2 | 3 |   |
| | | | 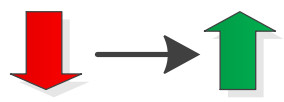 |
| 1E | 2 | 3 | 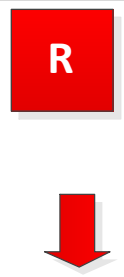 [51] |
| 2A | 2 | 3 | 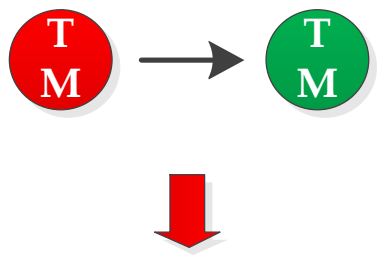 |
| 2B | 3 | 2 | 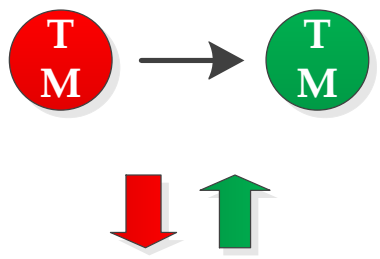 |
| 2C | 2 | 3 | 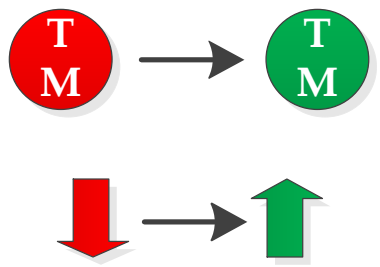 |
| 2D | 2 | 3 | 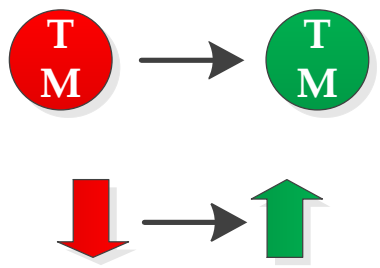 |
| 2E | 1 | 3 | 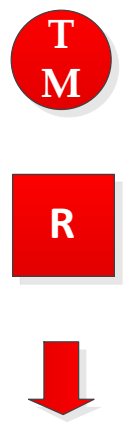 |
| 3A | 2 | 3 | 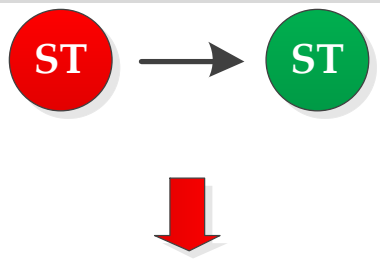 |
| 3B | 3 | 2 | 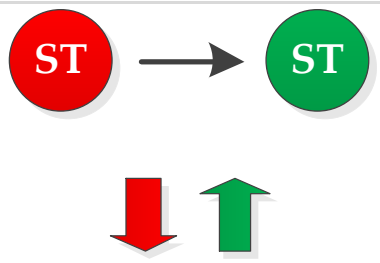 |
| 3C | 2 | 3 | 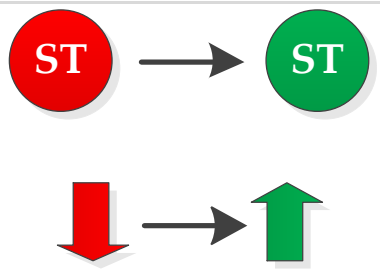 |
| 3D | 2 | 3 | 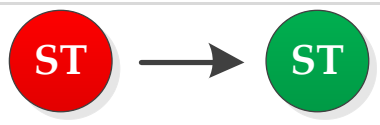 |

[51] The machine that fails in this test only contains a replica of table data and not the Table Manager, despite there only being a single machine listed in the *no. of machines* column. This is because the Table Manager is migrated to the instance which is executing the workload, which is not counted.

| | | | |
|---|---|---|---|
| | | | 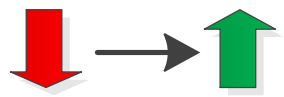 |
| 3E | 1 | 3 | 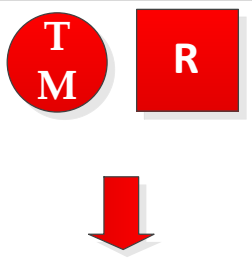 |
| 4A | 2 | 3 | 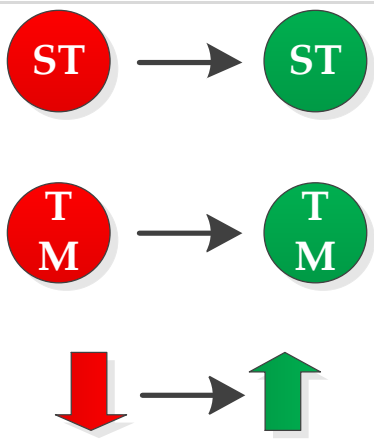 |
| 4B | 3 | 2 | 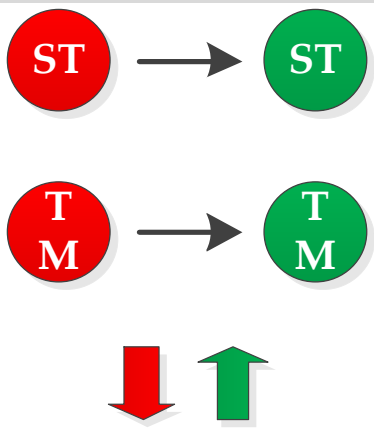 |
| 4C | 2 | 3 | 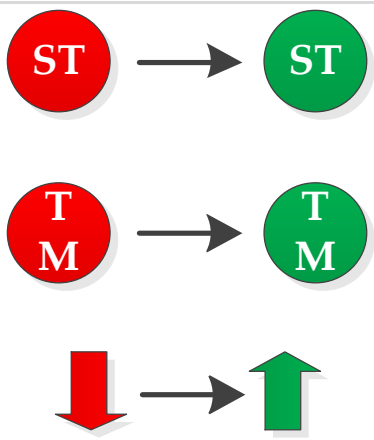 |
| 4D | 2 | 3 | 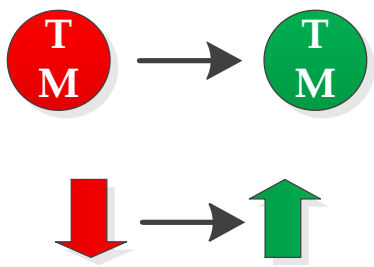 |
| 4E | 1 | 3 | 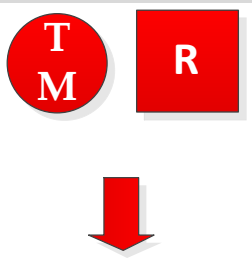 |

**Table 16: Experiment 2. A complete list of the tests being carried out, and a description of the expected outcome of each test.**

The co-ordination scripts for each of these scenarios are listed in full in the appendix on page 233.

### 7.3.6 Results

The results of each test in experiment 2 are illustrated in line graphs showing the throughput per second for the runtime of the test. Each of the lines in these graphs represents a single run of the test script, and there are 10 runs of the test script in each test.

The intention of each of these tests is summarised before each set of results, with both an illustration of the co-ordination scripts execution and the script itself. The results of each test are analysed to establish how long it typically takes the database to resume committing transactions after a failure has occurred — this is termed the *recovery time* of the database.

***<u>Test 1A</u>***

This test evaluates the ability of the system to tolerate the failure of a single replica holding instance when no other machines are available on which to re-replicate data. This is shown below in a drawing which illustrates the script's execution, and the state held on each machine over time.

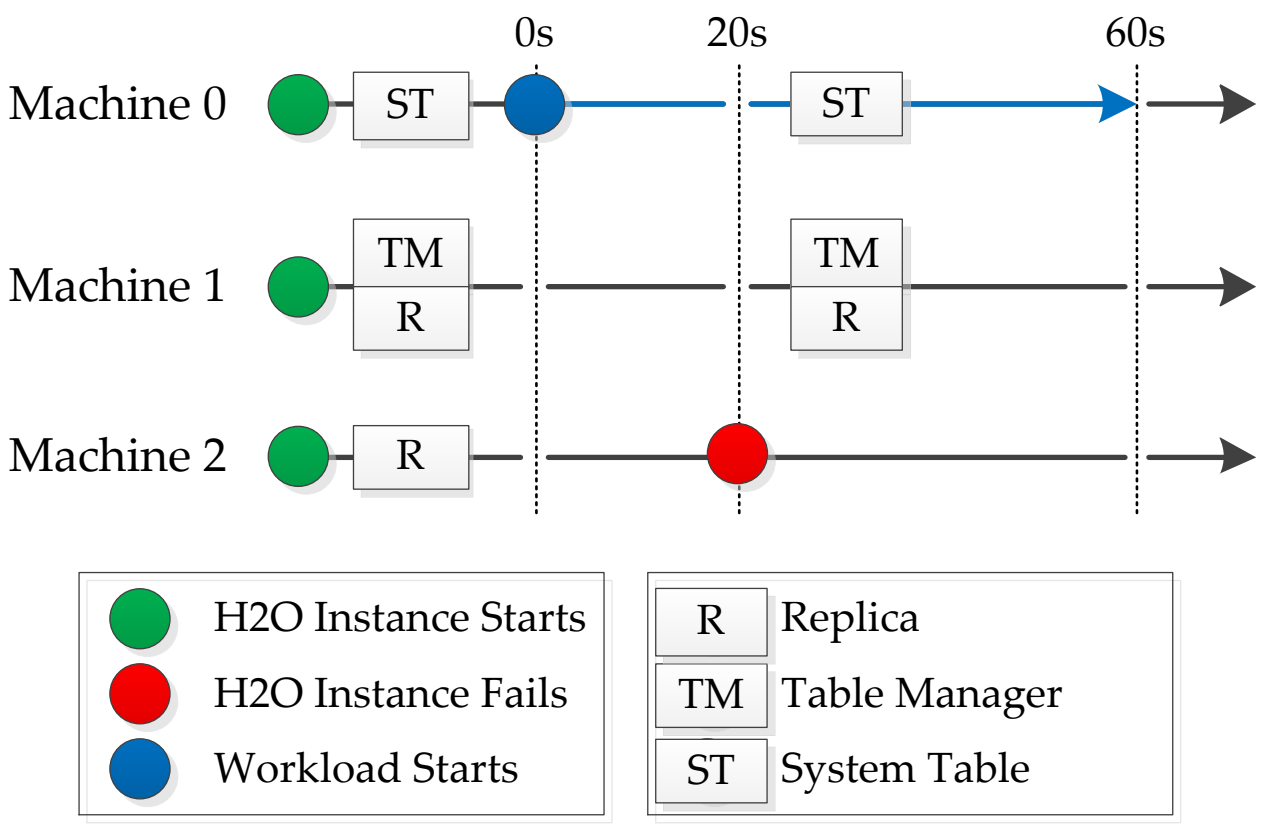

**Figure 62: Execution over 1A over time.**

***Script***

```
[Global Parameters: System-wide replication factor = 3]
{start_machine id="0"}
{start_machine id="1"}
{start_machine id="2"}
{create_table id="1" name="workloadTable" prepopulate_with="300"}
{0} {execute_workload="short.workload" duration="60000"}
{sleep="20000"}
{terminate_machine id="2"}
```

***Summary***

Replicas of the *workloadTable* are initially held on *machine's 1* and 2 (*machine 0* never holds any replicas). After twenty seconds this script terminates *machine 2*, which holds a replica of *workloadTable*.

***Results***

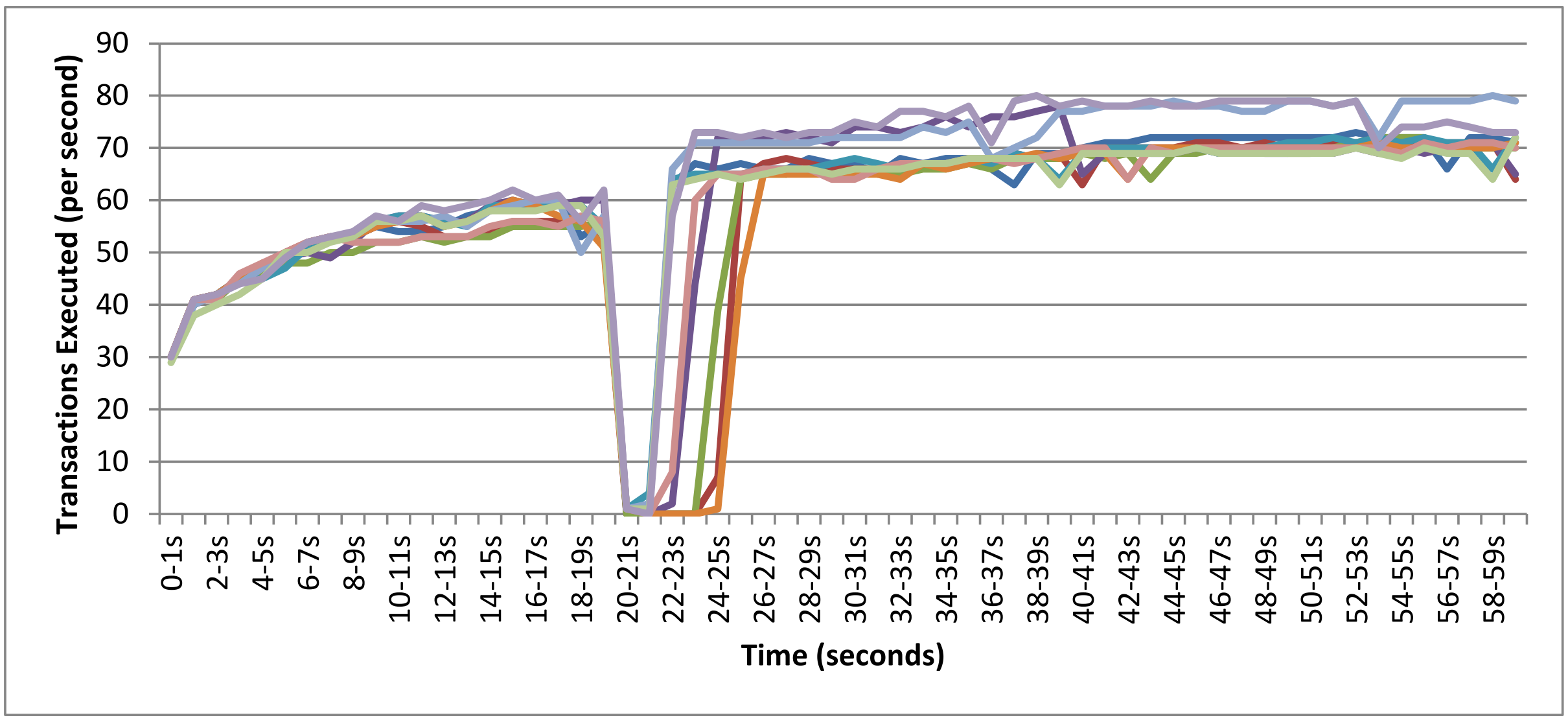

**Figure 63: Results of experiment 2, test 1A.**

When *machine 2* is killed off 20 seconds into the execution of the script, the number of transactions the system executes per second decreases to zero for somewhere between 1 and 5 seconds (an average of 2.3 seconds). This happens despite another replica being available, because when an update is performed, the database attempts to contact each replica synchronously. The delay in restarting transactions is the time it takes for the database to recognise that a replica cannot be contacted and abort the transaction.

This illustrates the cost of performing fully synchronous updates, where a response must be received from every participating replica for a transaction to proceed.

After the system recovers from this failure the transaction throughput increases. On average 56.5 transactions were executed per second between 10 and 20 seconds, compared to an average of 68.1 transactions executed per second between 25 and 35 seconds. This increase occurs because the replication factor of *workloadTable* is now lower, so fewer replicas need to be contacted when an update is made.

***Test 1B***

This test evaluates the ability of the system to immediately re-replicate data to a new instance when a single replica holding instance fails.

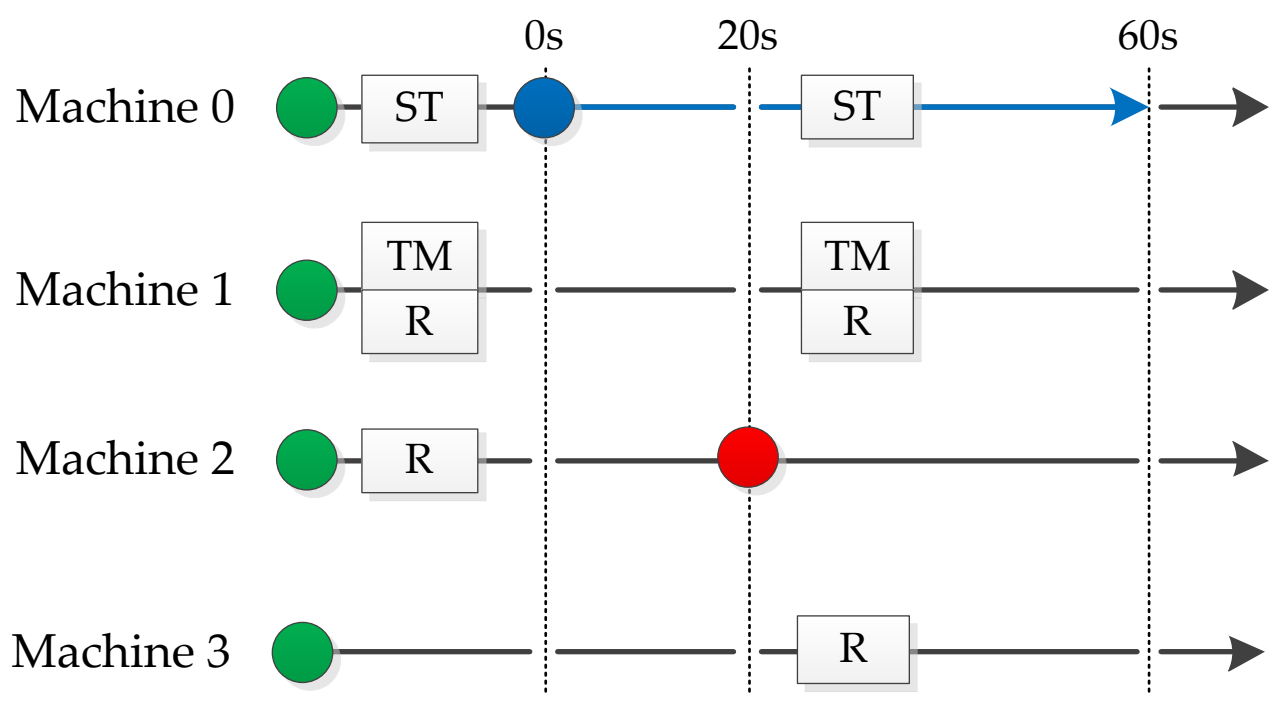

**Figure 64: Execution over 1B over time.**

### *Script*

```
[Global Parameters: System-wide replication factor = 2]
{start_machine id="0"}
{start_machine id="1"}
{start_machine id="2"}
{start_machine id="3"}
{create_table id="1" name="workloadTable" prepopulate_with="300"}
{0} {execute_workload="short.workload" duration="60000"}
{sleep="20000"}
{terminate_machine id="2"}
```

### *Summary*

After twenty seconds this script terminates *machine 2*, which holds a replica of *workloadTable*. The only replica of *workloadTable* is on *machine 1*, so the system replicates data and meta-data to *machine 3* to increase replication factor back to two.

### *Results*

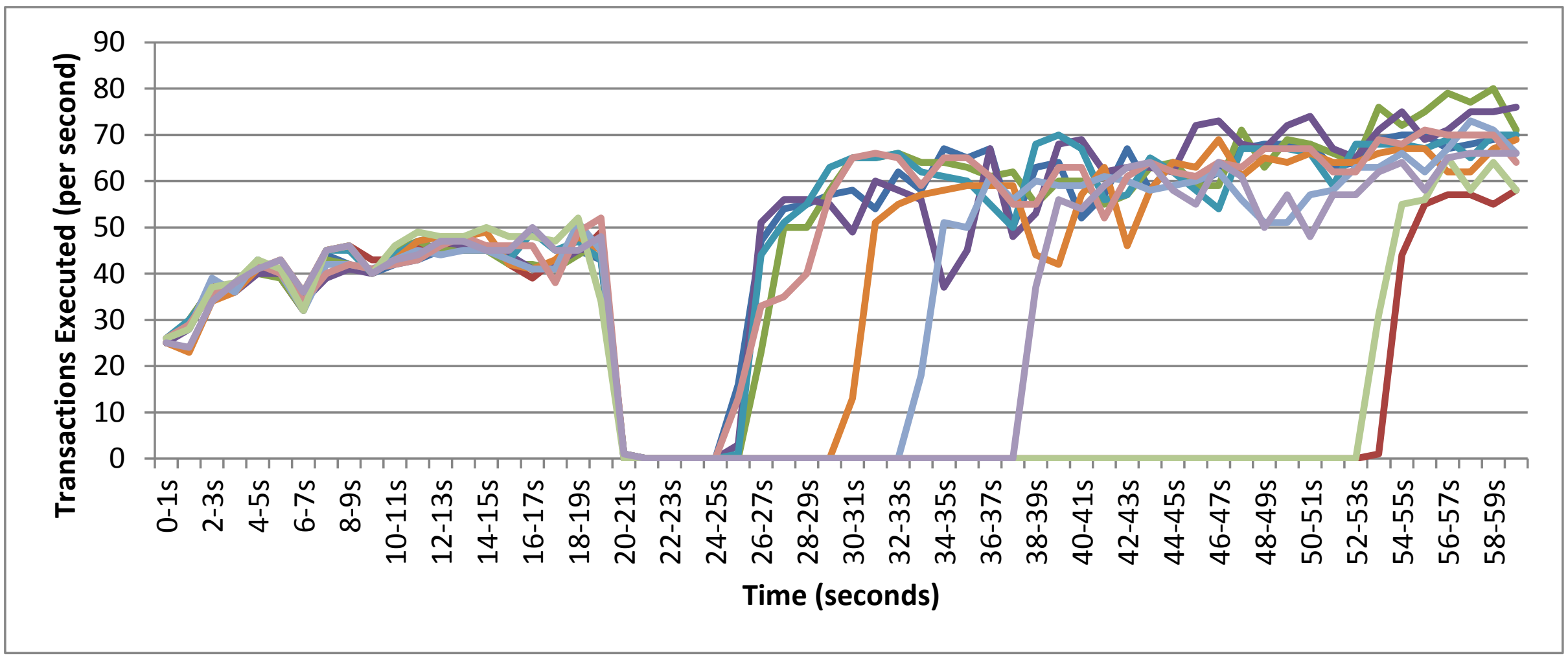

**Figure 65: Results of experiment 2, test 1B.**

The results of *test 1B* show an increase in the post-failure recovery time compared to *1A*, and more variability between test runs. On average H2O takes 13.3 seconds to recover from the failure of *machine 2*, with a median of 8 seconds.

The increase in recovery time is due to the *CREATE REPLICA* operation performed to create a new replica on *machine 3*. This operation requires an exclusive write lock to ensure that a consistent replica is created, which means that no workload transactions can execute during this period.

It is not clear why there is an increase in the variability of these results.

***Test 1C***

This test evaluates the ability of the system to re-replicate data after a machine failure, to a new machine which starts 20 seconds later.

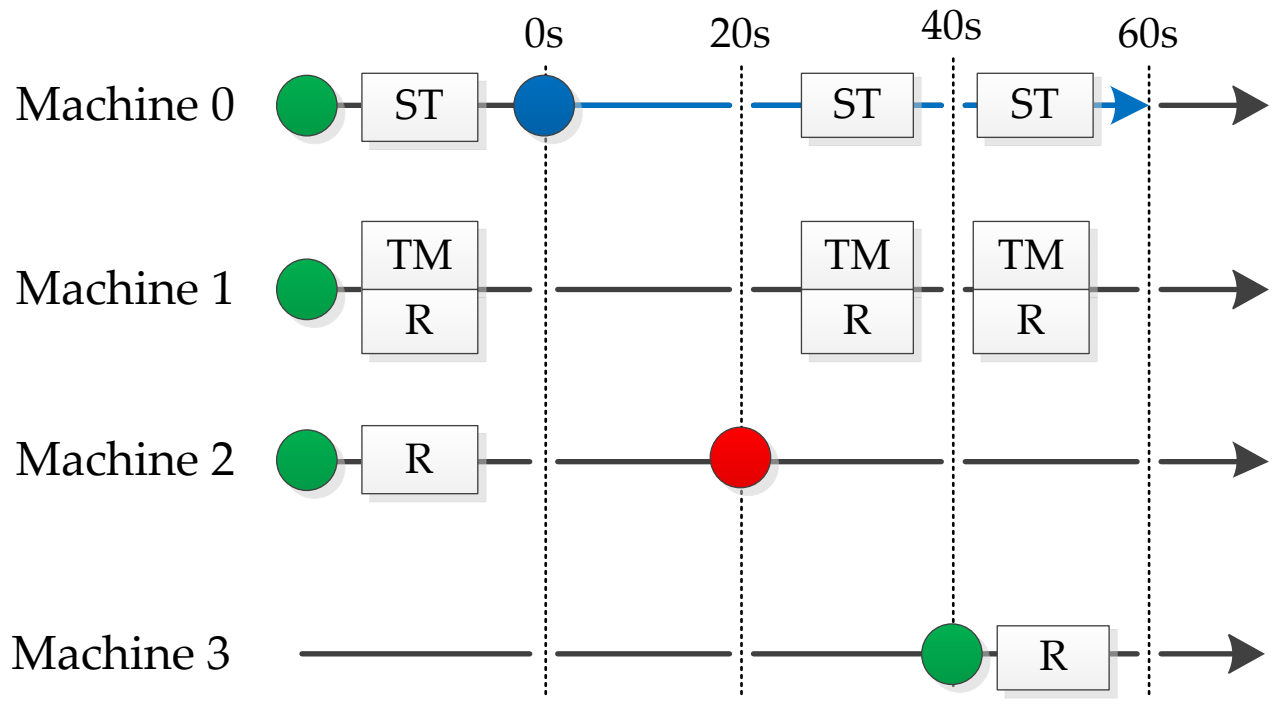

**Figure 66: Execution of 1C over time.**

***Script***

```
[Global Parameters: System-wide replication factor = 3]
{start_machine id="0"}
{start_machine id="1"}
{start_machine id="2"}
{create_table id="1" name="workloadTable" prepopulate_with="300"}
{0} {execute_workload="short.workload" duration="60000"}
{sleep="20000"}
{terminate_machine id="2"}
{sleep="20000"}
{start_machine id="3" block-workloads="true"}
```

***Summary***

After twenty seconds this script terminates *machine 2*, which holds a replica of *workloadTable*. After another twenty seconds a new machine is started and the system replicates to this machine.

***Results***

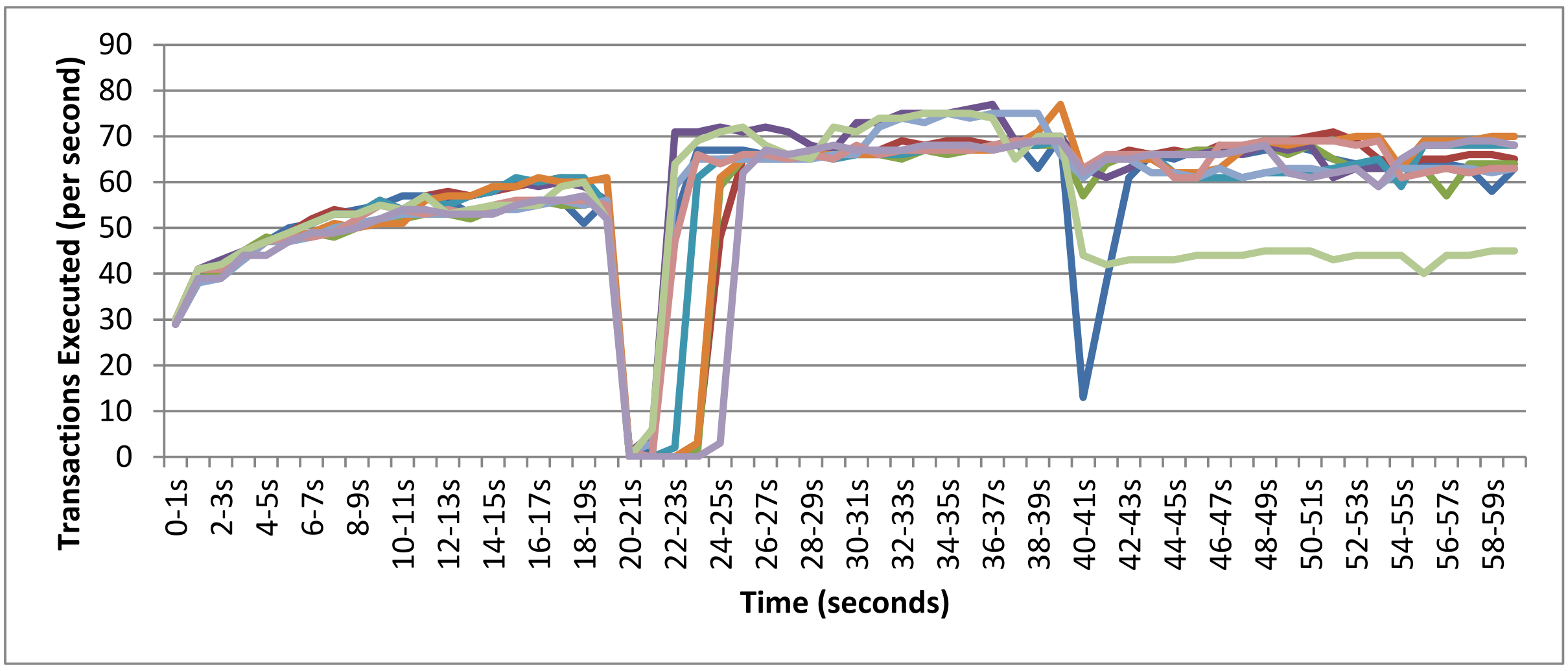

**Figure 67: Results of experiment 2, test 1C.**

The first 40 seconds of this experiment are the same as in *test 1A*, with a single machine being killed off, and the replication factor of the table decreasing to one. In this test a new machine is started after 40 seconds, and the addition of this machine results in a new replica of *workloadTable* being created.

When machine 3 starts, the average throughput per second drops from 69.2 (between 30-40 seconds) to 62.1 (between 40-50 seconds), a result of the table's replication factor increasing.

Two test runs suffer a larger drop-off in performance when the new machine starts. One of these runs (shown in blue in Figure 67) recovers after an initial drop-off, possibly due to a delay in the time it took to replicate *workloadTable*. The other run (shown in a light green) is consistently slower once the new machine starts (averaging 43.7 transactions per second) — it is unclear why this occurs.

***<u>Test 1D</u>***

This test evaluates the ability of the system recover from the failure of a machine, and to re-replicate to this machine when it is restarted.

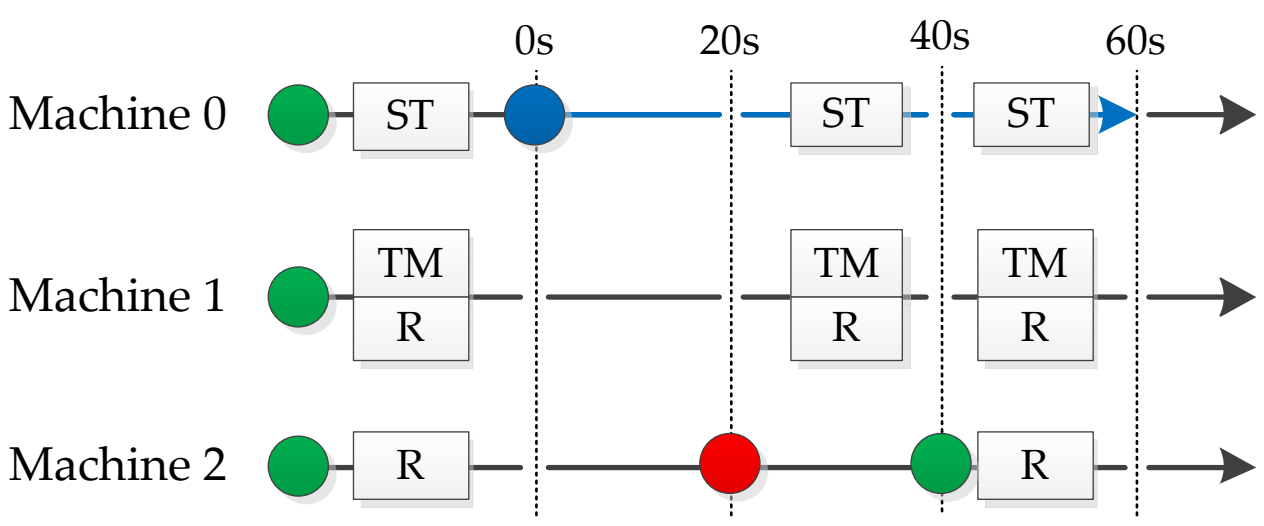

**Figure 68: Execution of 1D over time.**

### *Script*

```
[Global Parameters: System-wide replication factor = 3]
{start_machine id="0"}
{start_machine id="1"}
{start_machine id="2"}
{create_table id="1" name="workloadTable" prepopulate_with="300"}
{0} {execute_workload="short.workload" duration="60000"}
{sleep="20000"}
{terminate_machine id="2"}
{sleep="20000"}
{start_machine id="2" block-workloads="true"}
```

### *Summary*

After twenty seconds this script terminates *machine 2*, which holds a replica of *workloadTable*, but after another twenty seconds *machine 2* restarts and a replica is recreated there.

### *Results*

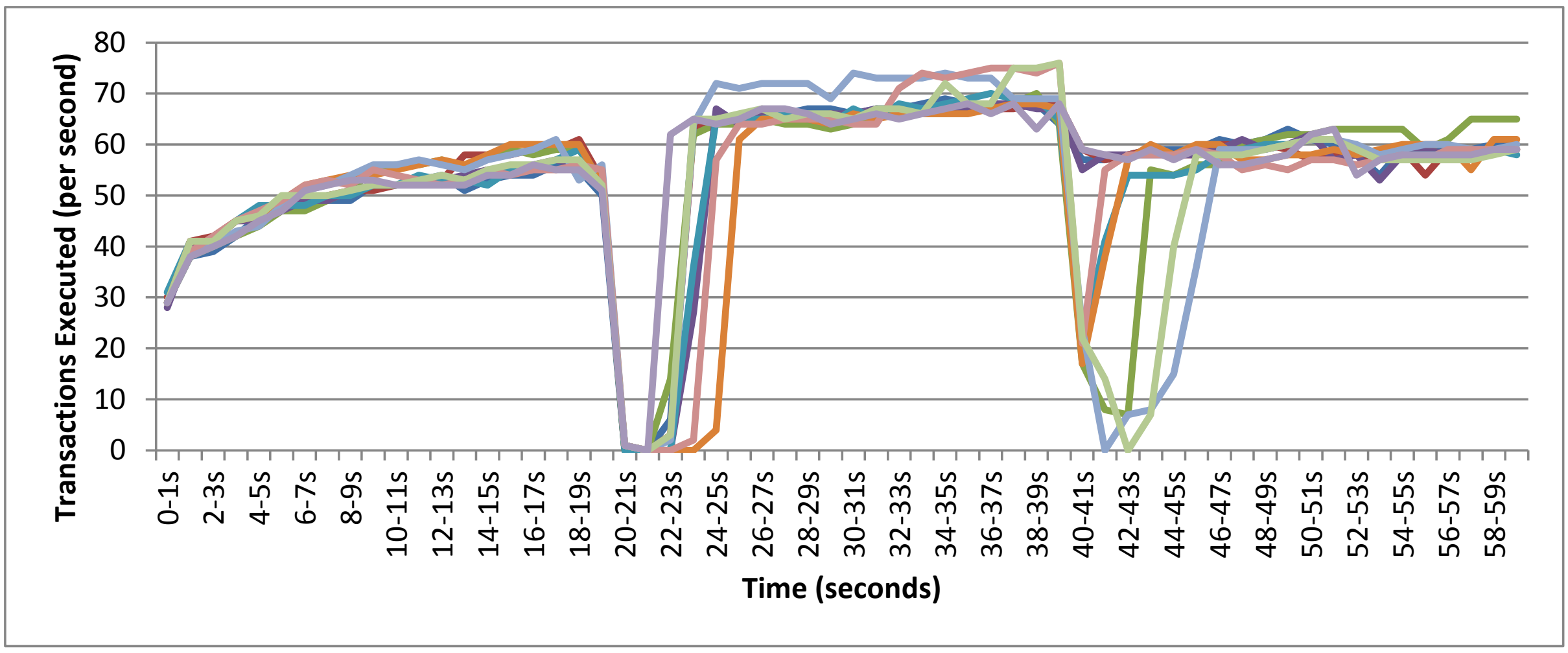

**Figure 69: Result of experiment 2, test 1D.**

This test is the same as 1C, but the machine that is started at 40 seconds is the same machine that is killed off at 20 seconds.

In 1D, throughput drops dramatically for a period of 0 to 4 seconds (average of 1.5) when the failed machine restarts, unlike 1C. This drop-off is a facet of the current implementation of H2O, which first checks whether the newly started machine has a replica of the test table (it does) and then completely drops this replica if it is outdated. A new, up-to-date replica is subsequently created.

A more efficient implementation would not completely drop the outdated replica, and would replay only the updates that occurred after the replica was last updated. This would remove the need to completely drop the replica, and would reduce the volume of data sent between machines when creating the new replica, as the size of all updates to the table is less than the total size of the table in this case.

### Test 1E

This script evaluates how the system responds to the failure of a machine which holds the only replica of a table.

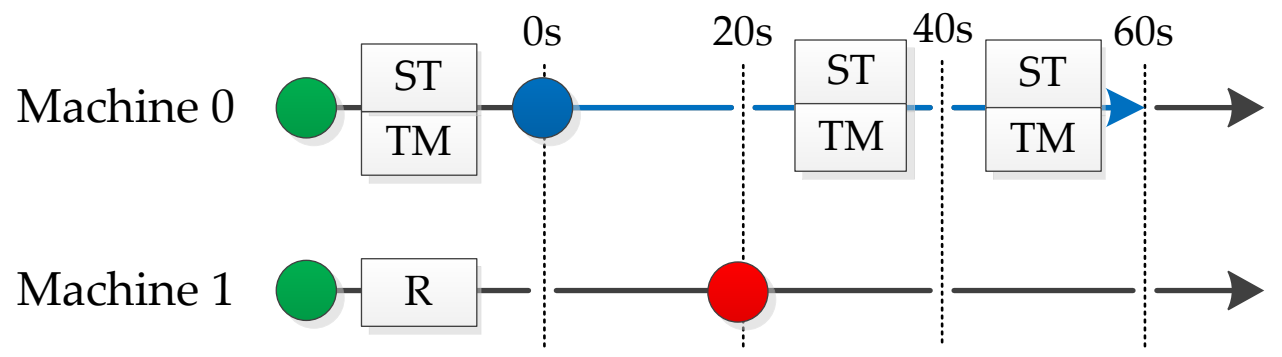

**Figure 70: Execution of 1E over time.**

### *Script*

```
[Global Parameters: System-wide replication factor = 3]
{start_machine id="0"}
{start_machine id="1"}
{create_table id="1" name="workloadTable" prepopulate_with="300"}
{0} MIGRATE TABLEMANAGER workloadTable;
{0} {execute_workload="short.workload" duration="60000"}
{sleep="20000"}
{terminate_machine id="1"}
```

### *Summary*

After twenty seconds this script terminates *machine 1*, which holds a replica of *workloadTable* (the Table Manager has first been migrated to *machine 0*), at which point there are no active copies of *workloadTable*.

### *Results*

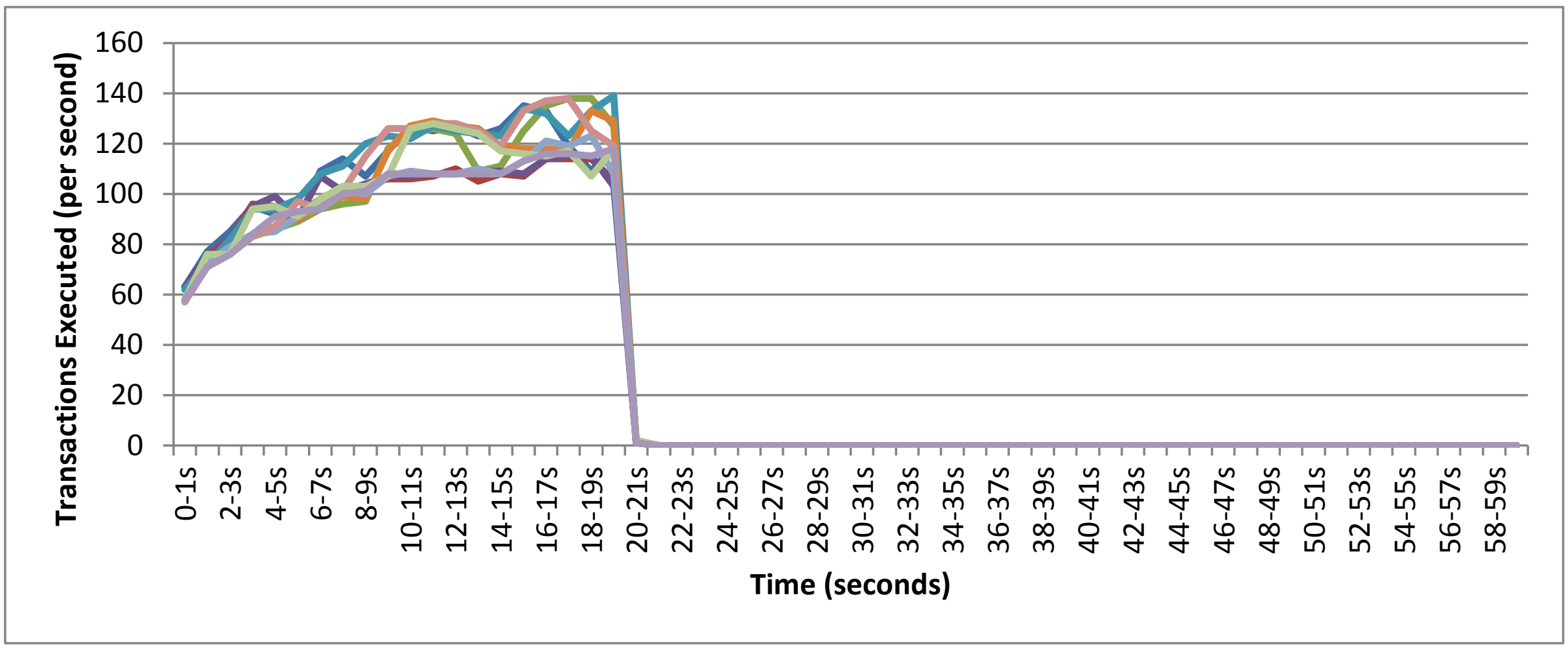

**Figure 71: Result of experiment 2, test 1E**

There are no replicas of test available once the database instance is killed off after 20 seconds, so no more queries succeed after this point. This is the expected behaviour of the DDBMS.

### ***Test 2A***

This test evaluates how the system responds to the failure of the machine holding a replica of a table and its Table Manager.

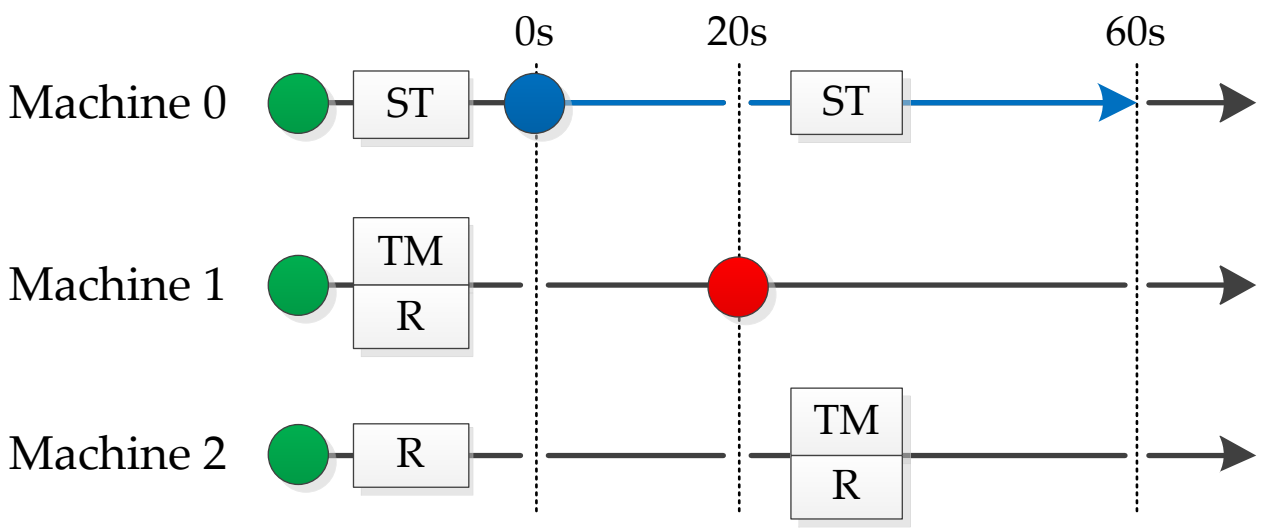

**Figure 72: Execution of 2A over time.**

### *Script:*

```
[Global Parameters: System-wide replication factor = 3]
{start_machine id="0"}
{start_machine id="1"}
{start_machine id="2"}
{create_table id="1" name="workloadTable" prepopulate_with="300"}
{0} {execute_workload="short.workload" duration="60000"}
{sleep="20000"}
```

```
{terminate_machine id="1"}
```

***Summary***

After twenty seconds this script terminates *machine 1*, which holds the Table Manager of *workloadTable* and a replica of *workloadTable*. No other machines are available to create new replicas, but the Table Manager is recreated on *machine 2*.

***Results***

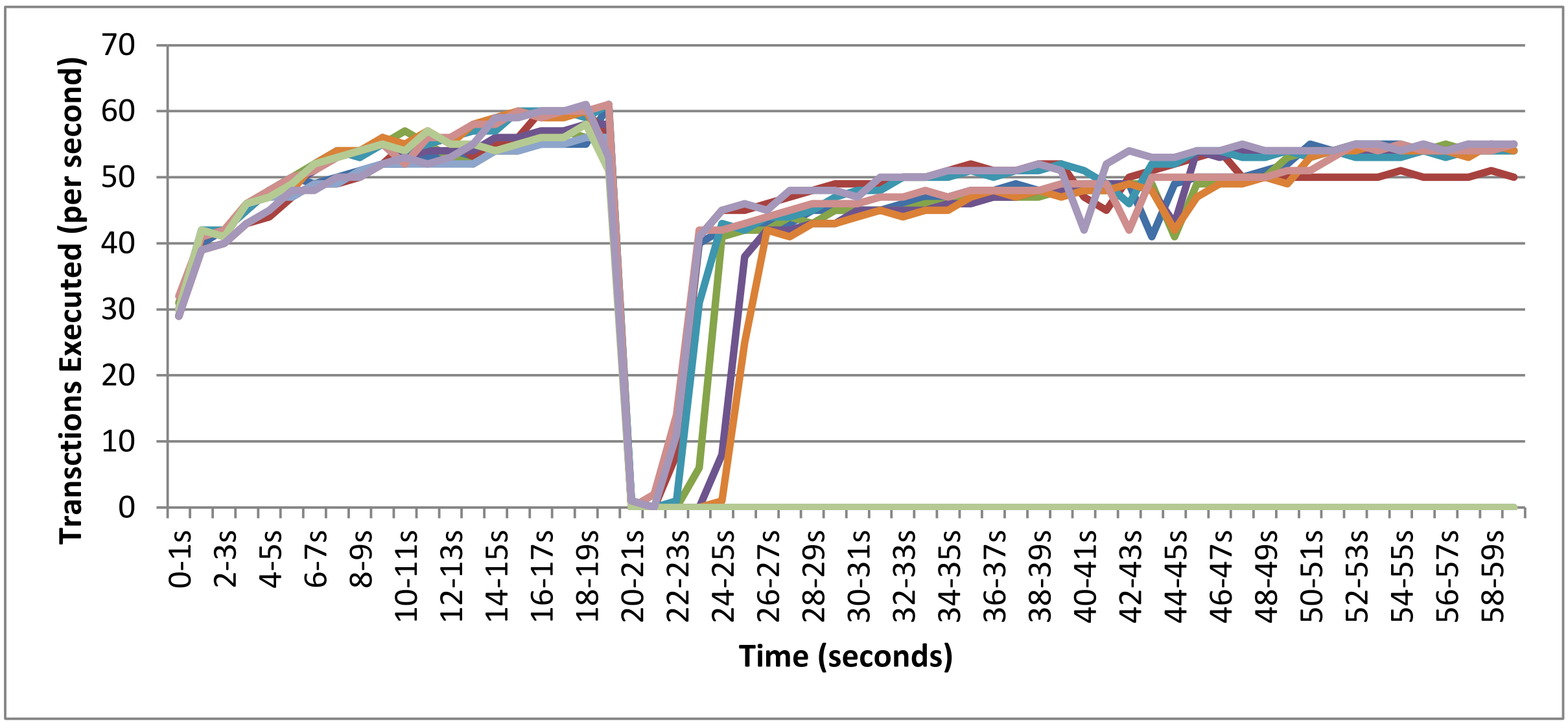

**Figure 73: Results of experiment 2, test 2A.**

This test is similar to *1A*, but in this case the instance that is killed after 20 seconds holds both a replica and the active Table Manager, so the Table Manager must be recreated before any queries proceed.

The system takes on average 3.6 seconds to recover from the failure of this machine, excluding two test runs that did not recover from the failure. This delay is a necessary part of the recovery process in H2O, unlike the delay in *1A* which could be masked by use of asynchronous updates.

The two test runs that did not recover are attributed to bugs in H2O that were not found at the time of evaluation. Each of the scripts used in this evaluation includes a number of checks to ensure that the replication factor of the database is as expected at various points in the script's execution. The failing test runs had the correct replication factor at all points, so their failure is not thought to be the result of a fundamental inability to recover from the failure of a machine, but the result of bugs in the software.

***Test 2B***

This test evaluates the ability of the system to recover from the failure of a Table Manager and then immediately re-replicate data to a new instance.

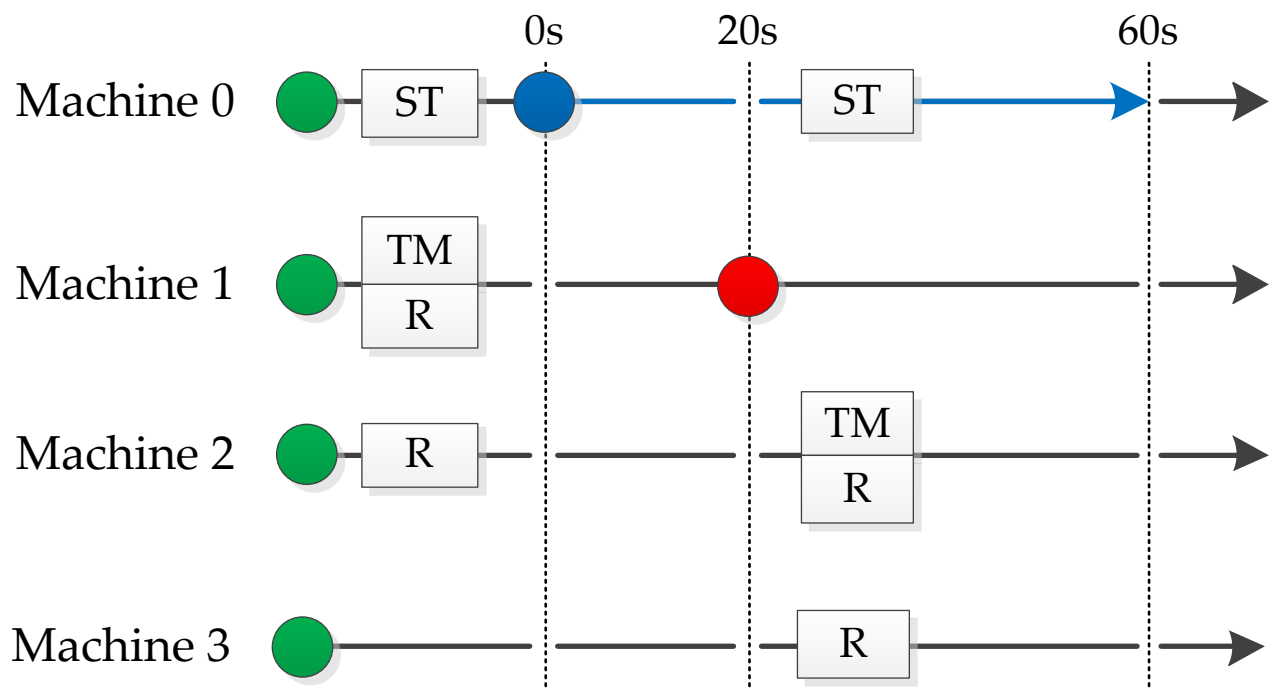

**Figure 74: Execution of 2B over time.**

***Script:***

```
[Global Parameters: System-wide replication factor = 2]
{start_machine id="0"}
{start_machine id="1"}
{start_machine id="2"}
{start_machine id="3"}
{create_table id="1" name="workloadTable" prepopulate_with="300"}
{0} {execute_workload="short.workload" duration="60000"}
{sleep="20000"}
{terminate_machine id="1"}
```

***Summary***

After twenty seconds this script terminates *machine 1*, which holds the Table Manager of *workloadTable* and a replica of *workloadTable*. The Table Manager is recreated and a new replica is created on *machine 3* to increase the replication factor back to two.

***Results***

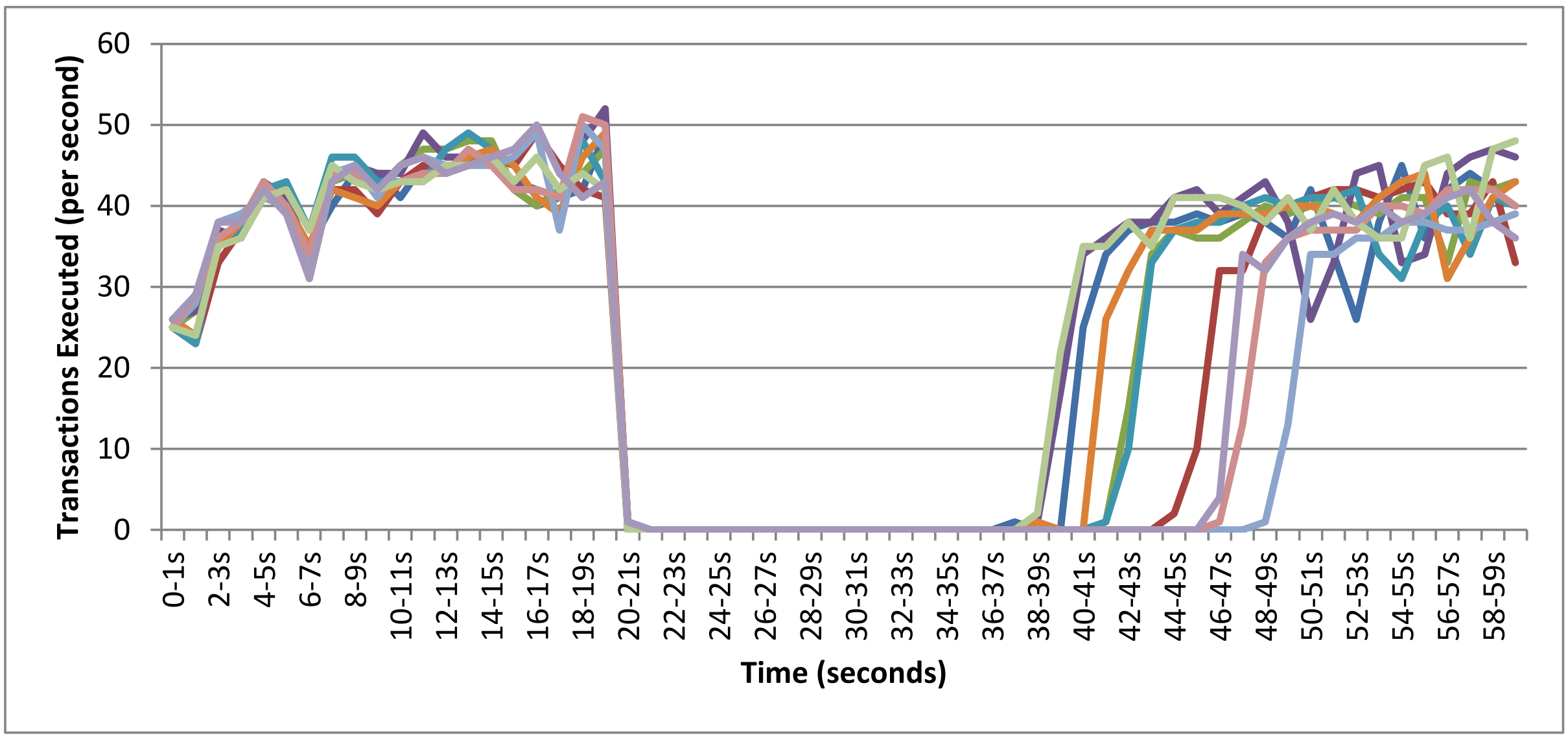

**Figure 75: Result of experiment 2, test 2B.**

This test is the same as 1B, but the machine being terminated holds both the Table Manager and a replica of *workloadTable*.

On average, it takes 23 seconds to recover from the failure of *machine 1*, much longer than in previous tests. In this test H2O immediately attempts to create a new replica upon creation of the Table Manager. This feature is designed to ensure that the replication factor is actively maintained at a desirable level, but these results show that it renders the system unavailable for a greater period of time than a typical failure (compared against previous tests).

***<u>Test 2C</u>***

This test evaluates the ability of the system to recreate the Table Manager after failure, then re-replicate data to a new machine which starts 12 seconds later.

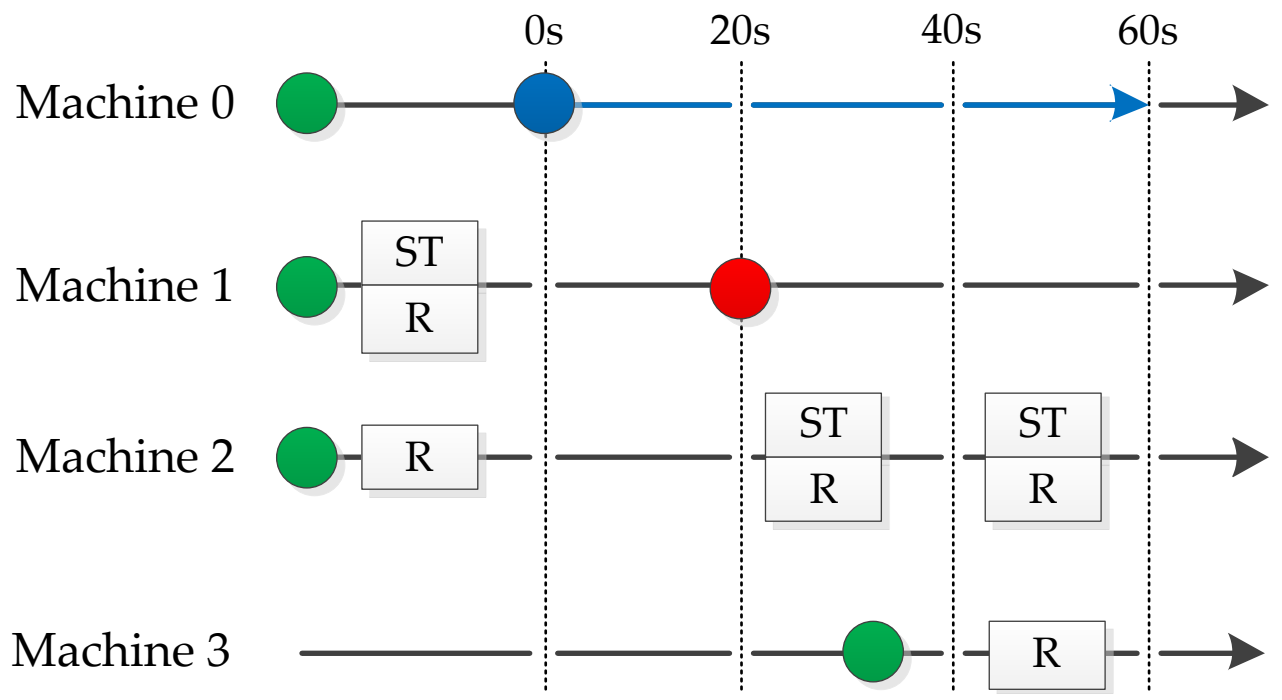

**Figure 76: Execution of 2C over time.**

***Script:***

```
[Global Parameters: System-wide replication factor = 3]
```

```
{start_machine id="0"}
{start_machine id="1"}
{start_machine id="2"}
{create_table id="1" name="workloadTable" prepopulate_with="300"}
{0} {execute_workload="short.workload" duration="60000"}
{sleep="20000"}
{terminate_machine id="1"}
{sleep="12000"}
{start_machine id="3" block-workloads="true"}
```

***Summary***

After twenty seconds this script terminates *machine 1*, which holds the Table Manager of *workloadTable* and a replica of *workloadTable*. After another twelve seconds a new machine is started and a new replica is created onto this machine to increase the replication factor back to two.

***Results***

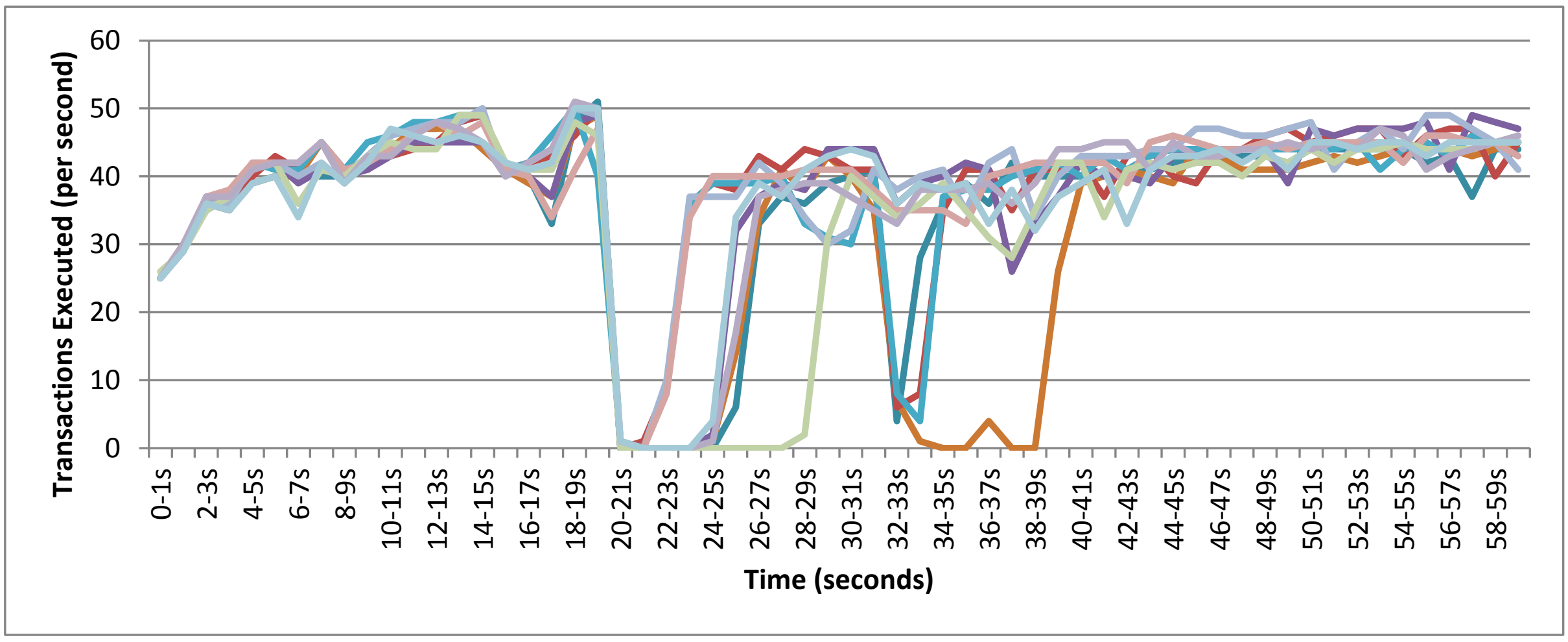

**Figure 77: Result of experiment 2, test 2C.**

This test is the same as *2A* up until a new machine is started after 32 seconds. It takes on average 3.7 seconds to recover from the failure of *machine 1*. When the new machine is started most test runs show no noticeable drop in throughput, while some show a short delay (on average, 1.4 seconds) before throughput returns to its previous level.

***<u>Test 2D</u>***

This test evaluates the ability of the system to recover from the failure of a Table Manager after failure, and to re-replicate to the failed machine when it is later restarted.

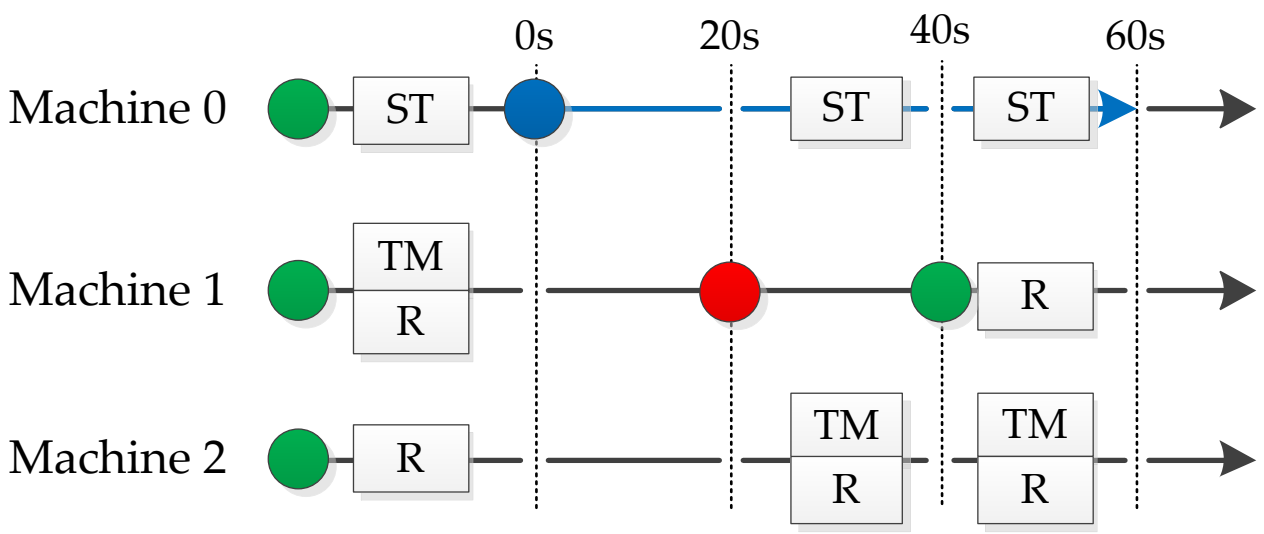

**Figure 78: Execution of 2D over time.**

***Script:***

```
[Global Parameters: System-wide replication factor = 3]
{start_machine id="0"}
{start_machine id="1"}
{start_machine id="2"}
{create_table id="1" name="workloadTable" prepopulate_with="300"}
{0} {execute_workload="short.workload" duration="60000"}
{sleep="20000"}
{terminate_machine id="1"}
{sleep="20000"}
{start_machine id="1" block-workloads="true"}
```

***Summary***

After twenty seconds this script terminates *machine 1*, which holds the Table Manager of *workloadTable* and a replica of *workloadTable*. After another 20 seconds *machine 1* is restarted and a new replica is created on this machine to increase the replication factor back to two.

***Results***

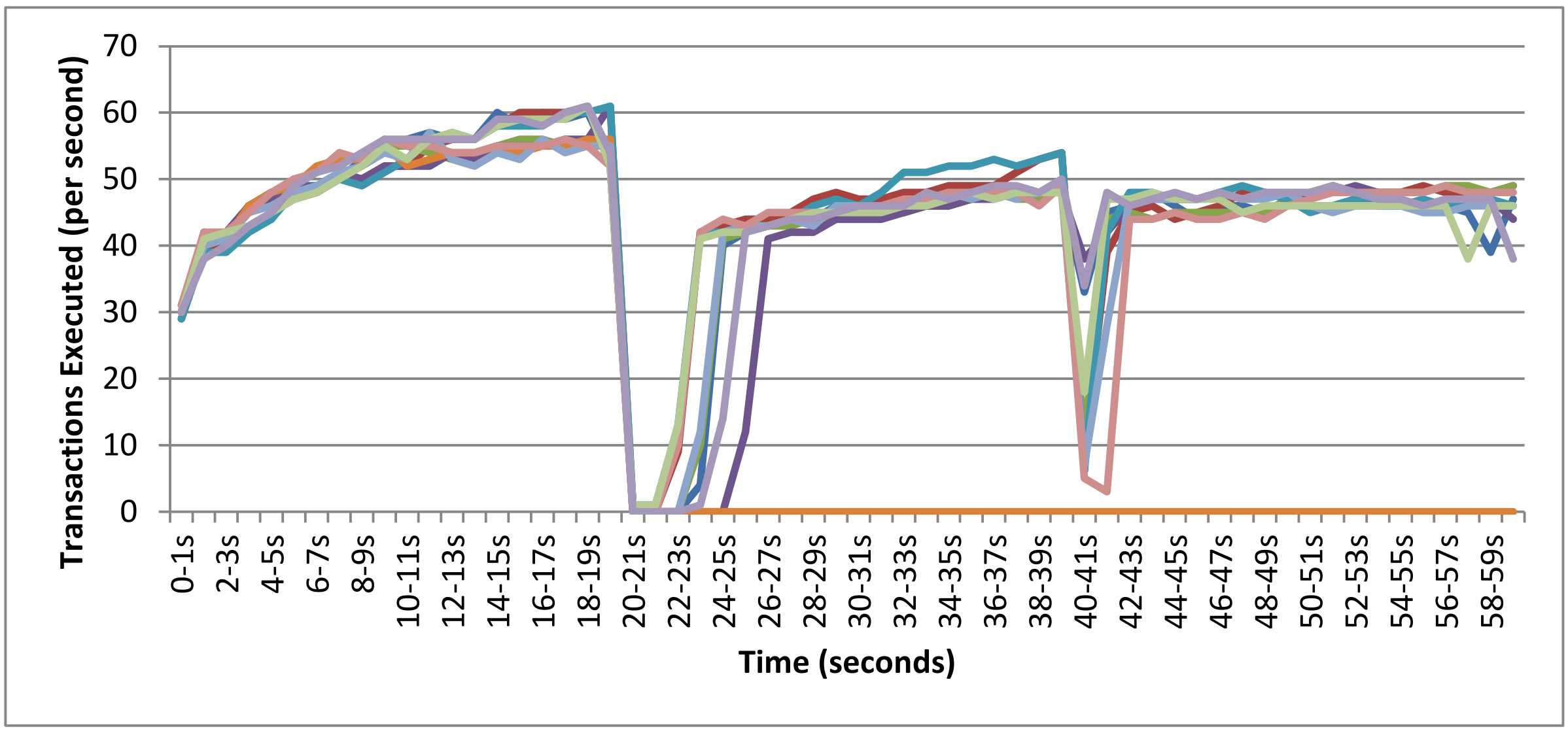

**Figure 79: Results of experiment 2, test 2D.**

This test is the same as *1D*, but in this case the Table Manager is on the machine that is killed off after 20 seconds. When this machine restarts there is a short drop in throughput (1.2 seconds, on average) as the system creates a new replica of *workloadTable*.

In one test run the database stops executing queries completely after machine 1 is killed off. As with the failure in 2A, this is the thought to be the result of a bug in the database software.

***<u>Test 2E</u>***

This script evaluates how the system responds to the failure of a machine which holds the only replica of a table and its Table Manager.

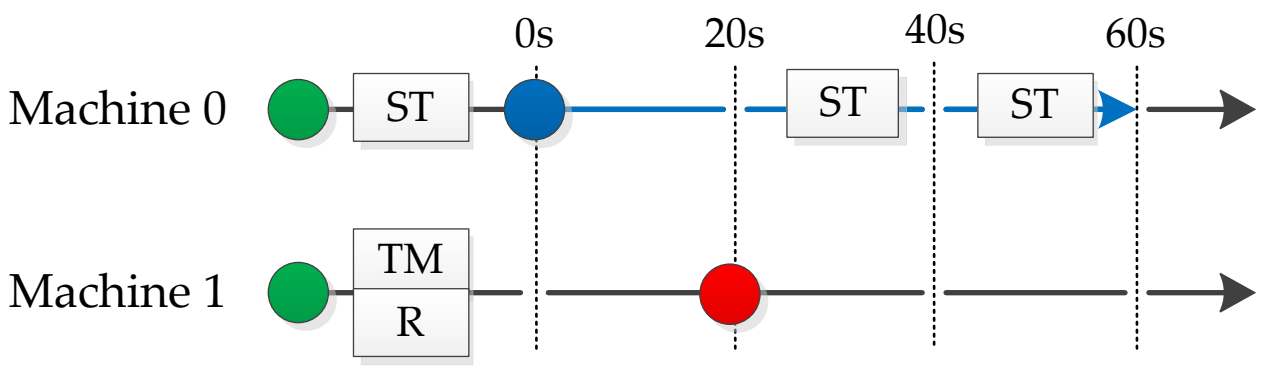

**Figure 80: Execution of 2E over time.**

***Script:***

```
[Global Parameters: System-wide replication factor = 3]
{start_machine id="0"}
{sleep="3000"}
{start_machine id="1"}
{create_table id="1" name="workloadTable" prepopulate_with="300"}
{0} {execute_workload="short.workload" duration="60000"}
{sleep="20000"}
{terminate_machine id="1"}
```

***Summary***

After twenty seconds this script terminates *machine 1*, which holds the Table Manager of *workloadTable* and a replica of *workloadTable*. There are no other replicas available so queries can no longer be executed.

***Results***

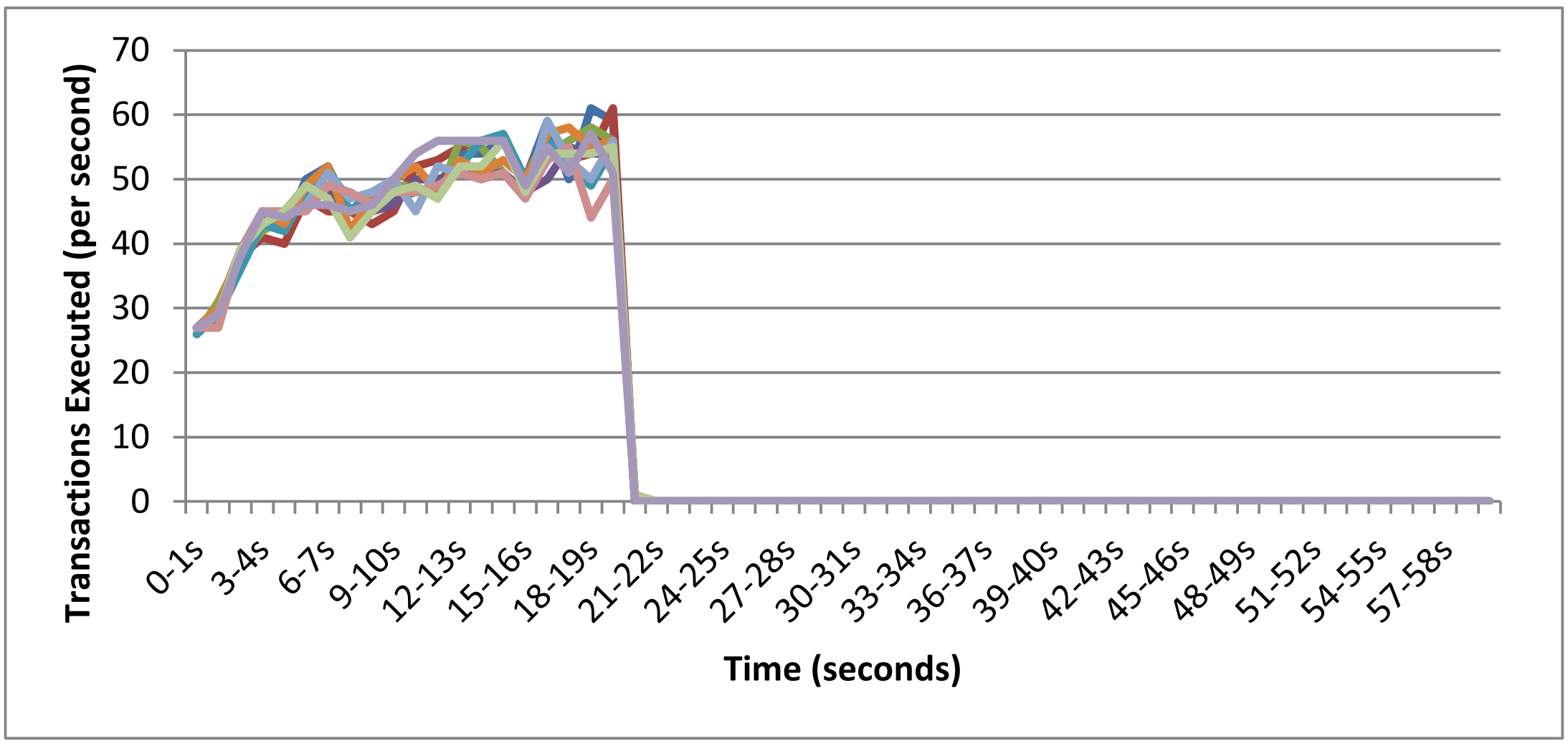

**Figure 81: Result of experiment 2, test 2E.**

When the machine is killed after 20 seconds there are no more replicas of the Table Manager or the table data of *workloadTable*, so no more transactions can be executed.

***<u>Test 3A</u>***

This test evaluates how the system responds to the failure of the machine holding the System Table and one of its two replicas. The replicas shown in Figure 82 (and the corresponding figures in 3B-E) refer to replicas of System Table state.

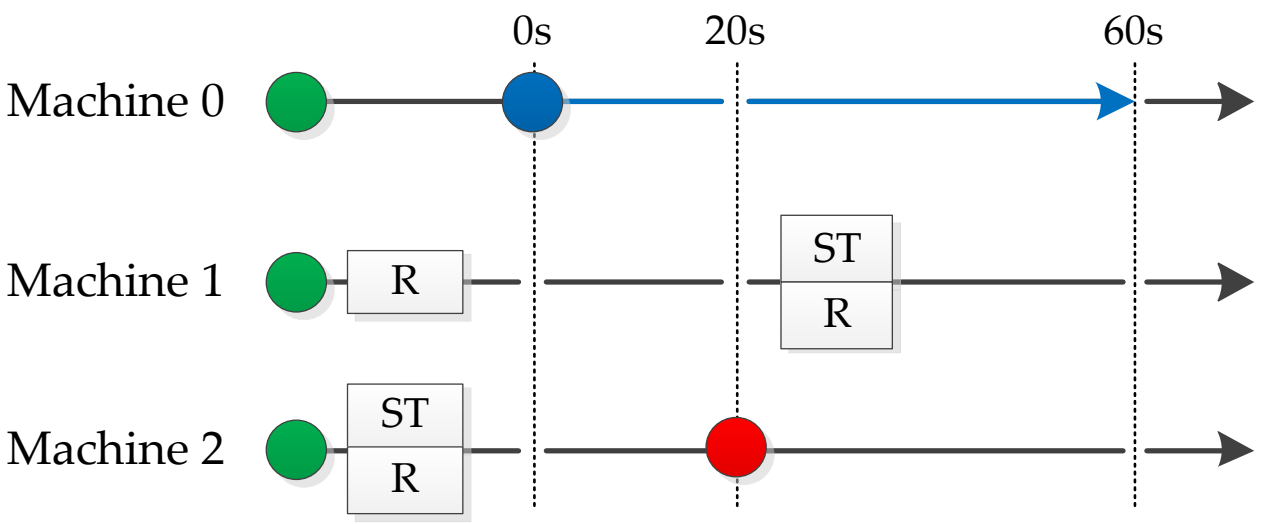

**Figure 82: Execution of 3A over time.**

***Script:***

```
[Global Parameters: System-wide replication factor = 3]
{start_machine id="0"}
{start_machine id="1"}
{start_machine id="2"}
{2} MIGRATE SYSTEMTABLE NO_REPLICATE
{0} {execute_workload="st.workload" duration="60000"}
{sleep="20000"}
{terminate_machine id="2"}
```

***Summary***

After twenty seconds this script terminates *machine 2*, which holds the active System Table and a replica of its state. No other machines exist on which to replicate data but another machine is able to re-instantiate the System Table.

***Results***

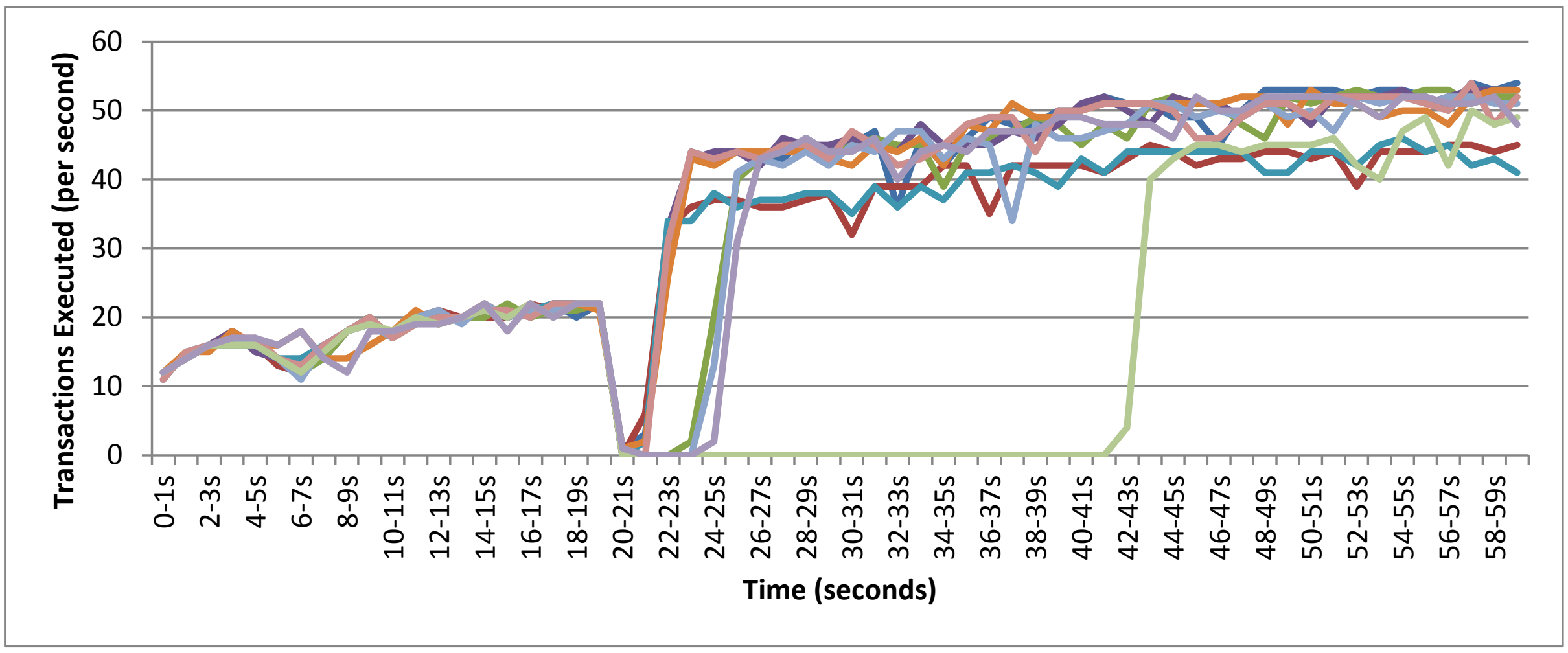

**Figure 83: Result of experiment 2, test 3A.**

This test runs a workload of *CREATE* and *DROP* table queries, rather than the previous tests which use a workload of *SELECT* and *UPDATE* queries against a table.

When the System Table is killed off after 20 seconds, it takes on average 3.8 seconds to recover (median, 2 seconds) and begin committing schema updates.

After the system has recovered transaction throughput almost doubles from an average of 20.4 per second (between 10-20 seconds) to 38.3 per second (between 25-35 seconds). The effect of a lower replication factor is more evident in this test compared to the previous tests, because the System Table workload consists entirely of updates, whereas the workload used in tests 1A-2E is a mix of reads and updates. Read requests are not sent to every instance, so they involve fewer cross-network messages[52].

It is unclear why some of the test runs take significantly longer to recover from the failure of *machine 2*.

[52] The speed of a read is mostly dependant on the required computation at a database instance and the volume of data returned. In these tests, reads are executed faster than writes.

***Test 3B***

This test evaluates the ability of the system to recover from the failure of the System Table and then immediately re-replicate System Table state to a new instance.

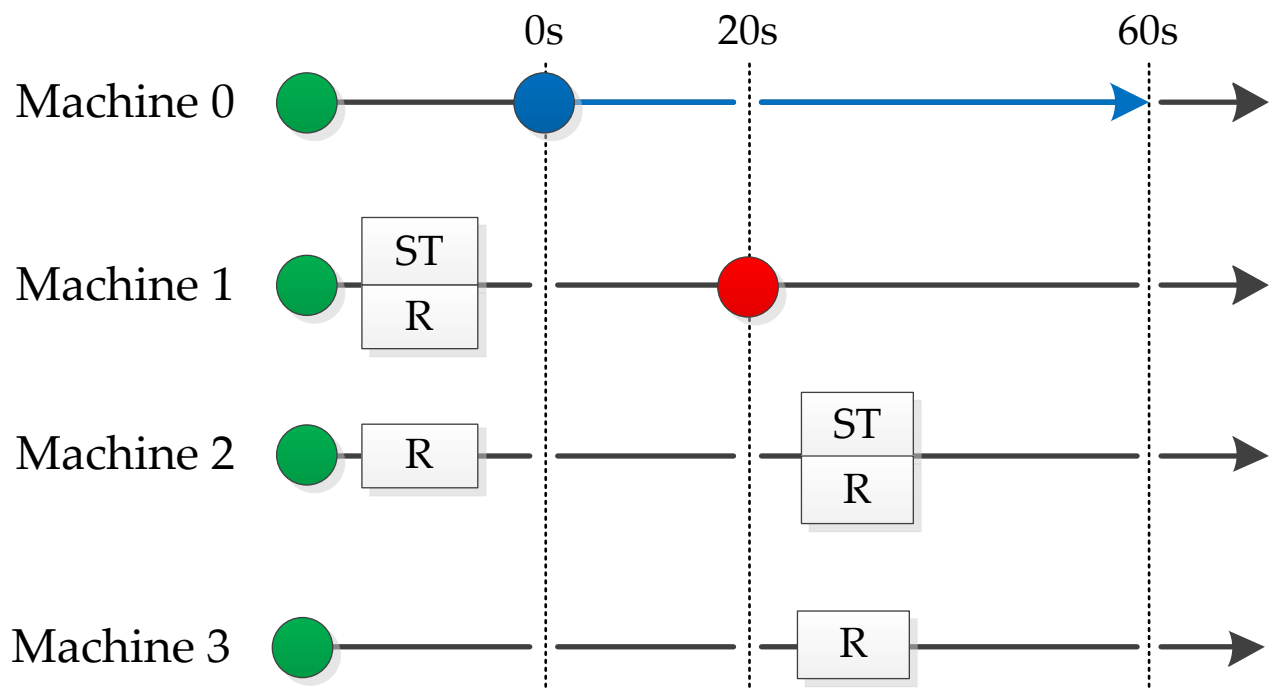

**Figure 84: Execution of 3B over time.**

***Script:***

```
[Global Parameters: System-wide replication factor = 2]
{start_machine id="0"}
{start_machine id="1"}
{start_machine id="2"}
{start_machine id="3"}
{1} MIGRATE SYSTEMTABLE NO_REPLICATE
{0} {execute_workload="st.workload" duration="60000"}
{sleep="20000"}
{terminate_machine id="1"}
```

***Summary***

After twenty seconds this script terminates *machine 1*, which holds the active System Table and a replica of its state. After recreating the System Table, the system replicates data and meta-data to *machine 3* to increase replication factor back to two.

***Results***

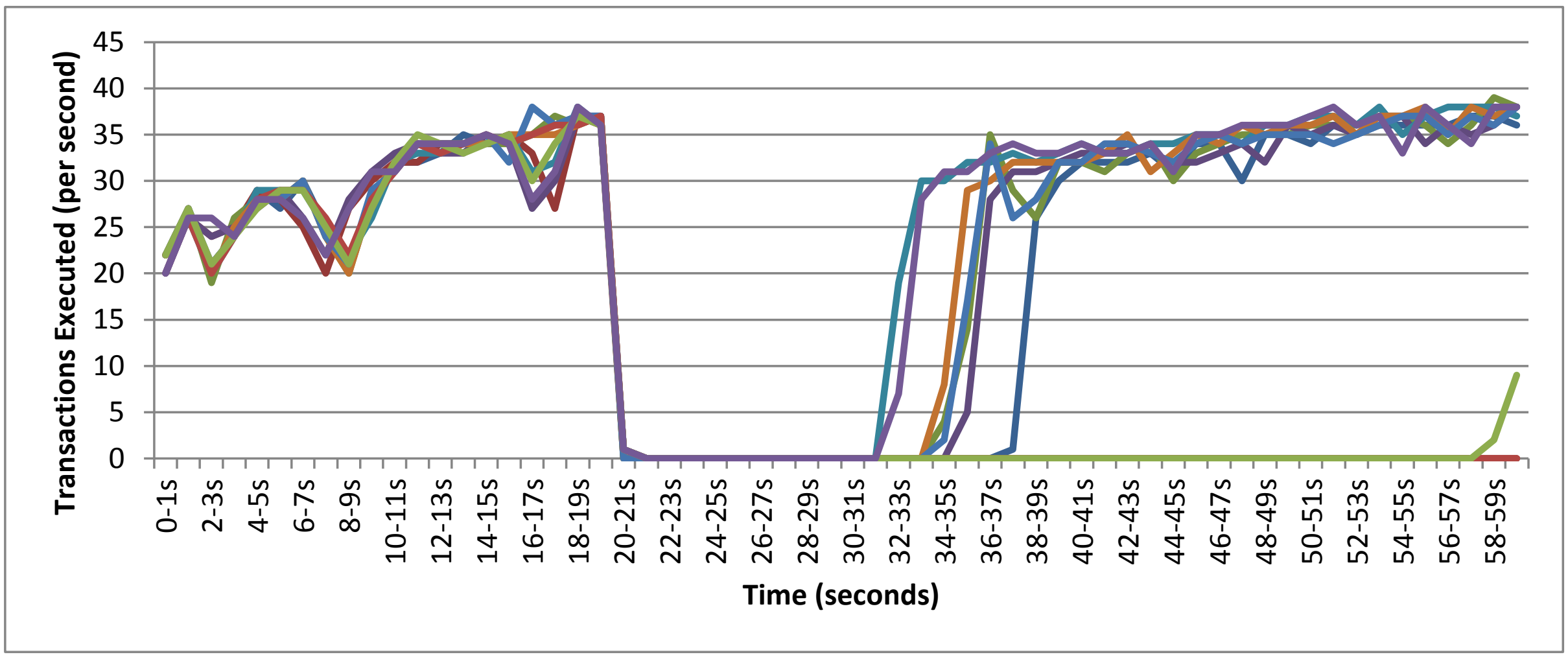

**Figure 85: Result of experiment 2, test 3B.**

When the System Table machine fails, it takes on average 17 seconds to recover from this failure (median, 13 seconds), excluding two test runs which did not recover at all. As with the results in 2B, this test shows that the system takes a significantly greater amount of time to recover from failure when a new replica is created immediately on recovery.

***Test 3C***

This test evaluates the ability of the system to recreate the System Table after failure, then re-replicate System Table state to a new machine which starts 20 seconds later.

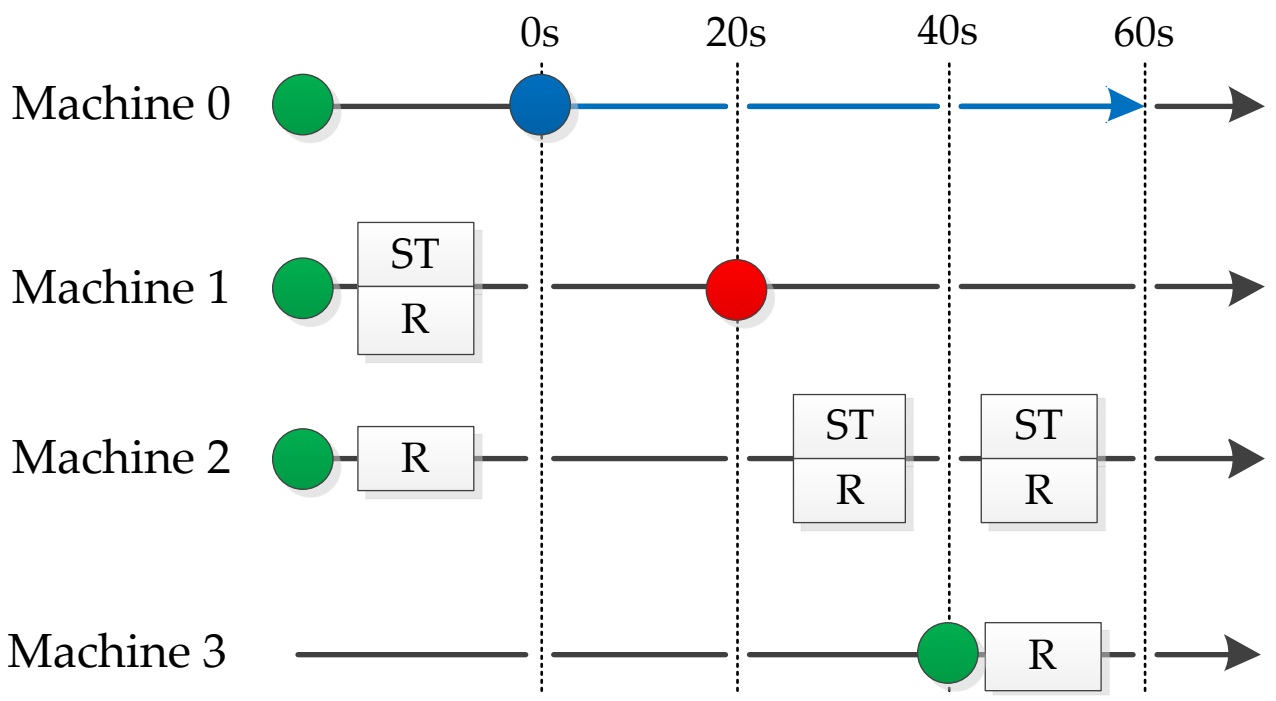

**Figure 86: Execution of 3C over time.**

***Script:***

```
[Global Parameters: System-wide replication factor = 3]
{start_machine id="0"}
{start_machine id="1"}
{start_machine id="2"}
{1} MIGRATE SYSTEMTABLE NO_REPLICATE
{0} {execute_workload="st.workload" duration="60000"}
{sleep="20000"}
```

```
{terminate_machine id="1"}
{sleep="20000"}
{start_machine id="3" block-workloads="true"}
```

***Summary***

After twenty seconds this script terminates *machine 1*, which holds the System Table and a replica of the System Table. After another twenty seconds a new machine, *machine 3*, is started and System Table state is replicated there.

***Results***

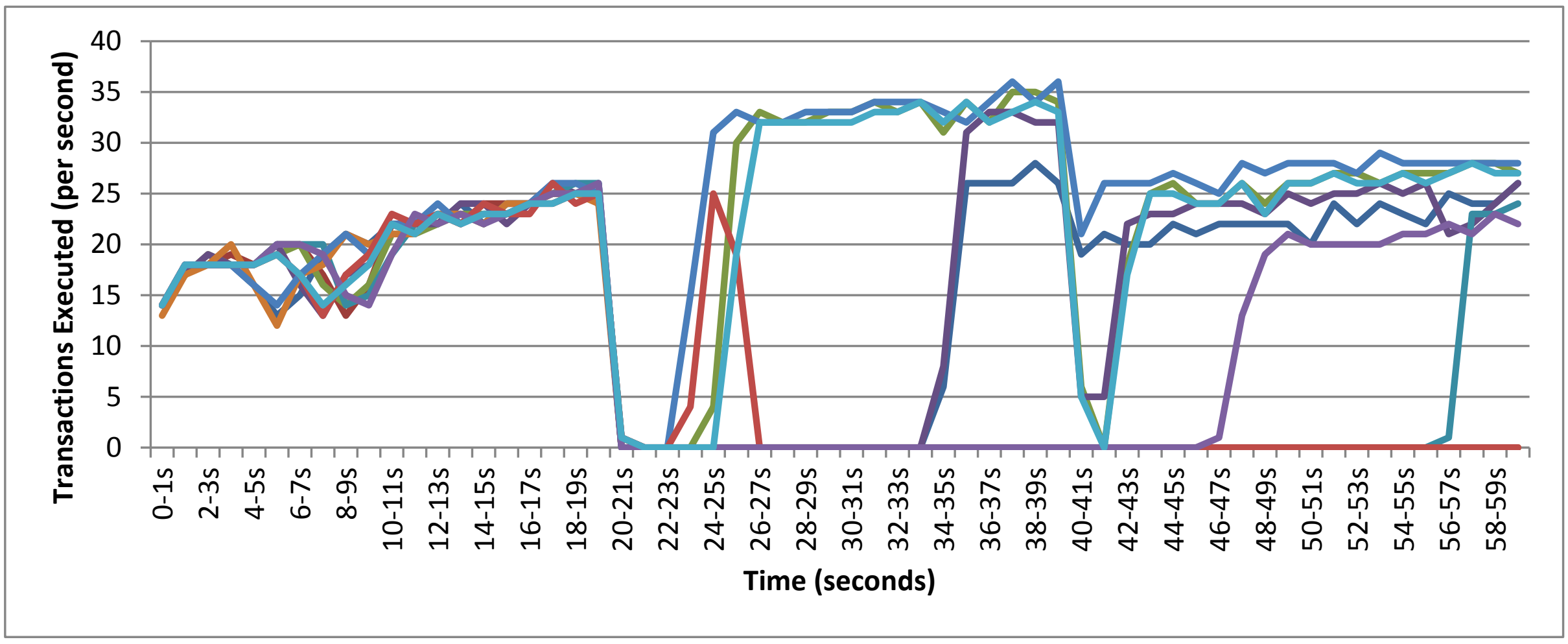

**Figure 87: Results of experiment 2, test 3C.**

The results of this experiment are much more variable than other tests, with 3 test runs failing to recover from the initial failure of *machine 1*. As with other tests which have failing test runs, this is not an indicator of insufficient replication factor, but rather bugs in H2O.

The test runs that succeed take on average 10 seconds to recover from the initial failure of *machine 1* (median, 5 seconds).

When *machine 3* is started after 40 seconds, throughput decreases from an average of 18.7 transactions per second, to 15.9 transactions per second. This is the result of an additional replica being created on *machine 3*.

***<u>Test 3D</u>***

This test evaluates the ability of the system recover from the failure of the System Table after failure, and to re-replicate System Table state to the failed machine when it is later restarted.

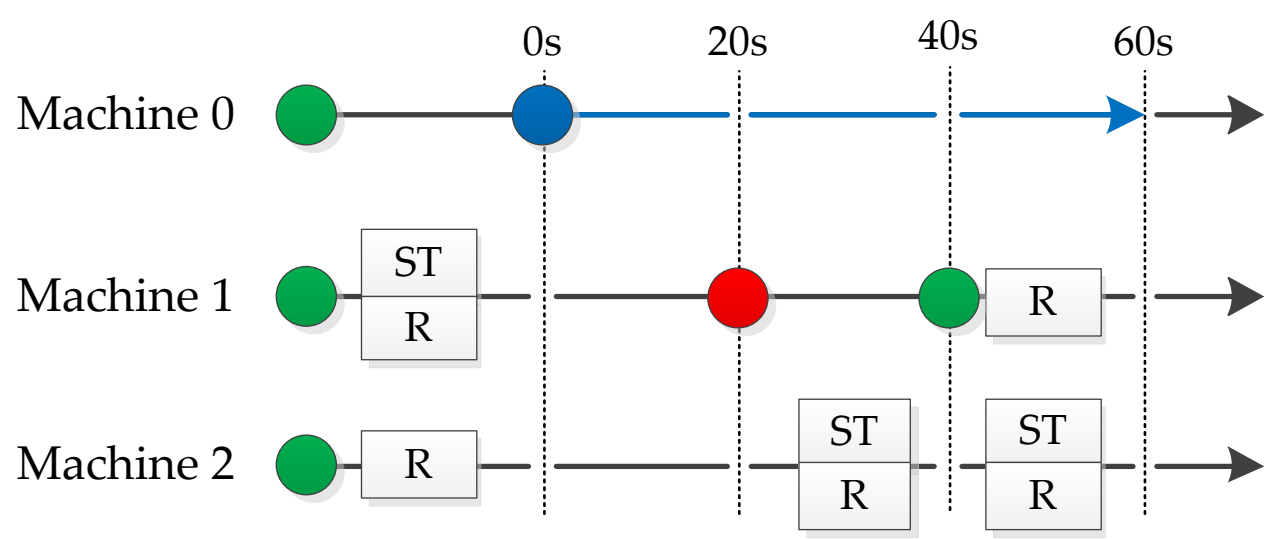

**Figure 88: Execution of 3D over time.**

### *Script:*

```
[Global Parameters: System-wide replication factor = 3]
{start_machine id="0"}
{start_machine id="1"}
{start_machine id="2"}
{1} MIGRATE SYSTEMTABLE NO_REPLICATE
{0} {execute_workload="st.workload" duration="60000"}
{sleep="20000"}
{terminate_machine id="1"}
{sleep="20000"}
{start_machine id="1" block-workloads="true"}
```

### *Summary*

After twenty seconds this script terminates *machine 1*, which holds the System Table and a replica of the System Table. After another twenty seconds *machine 1* is restarted and System Table state is replicated there again.

### *Results*

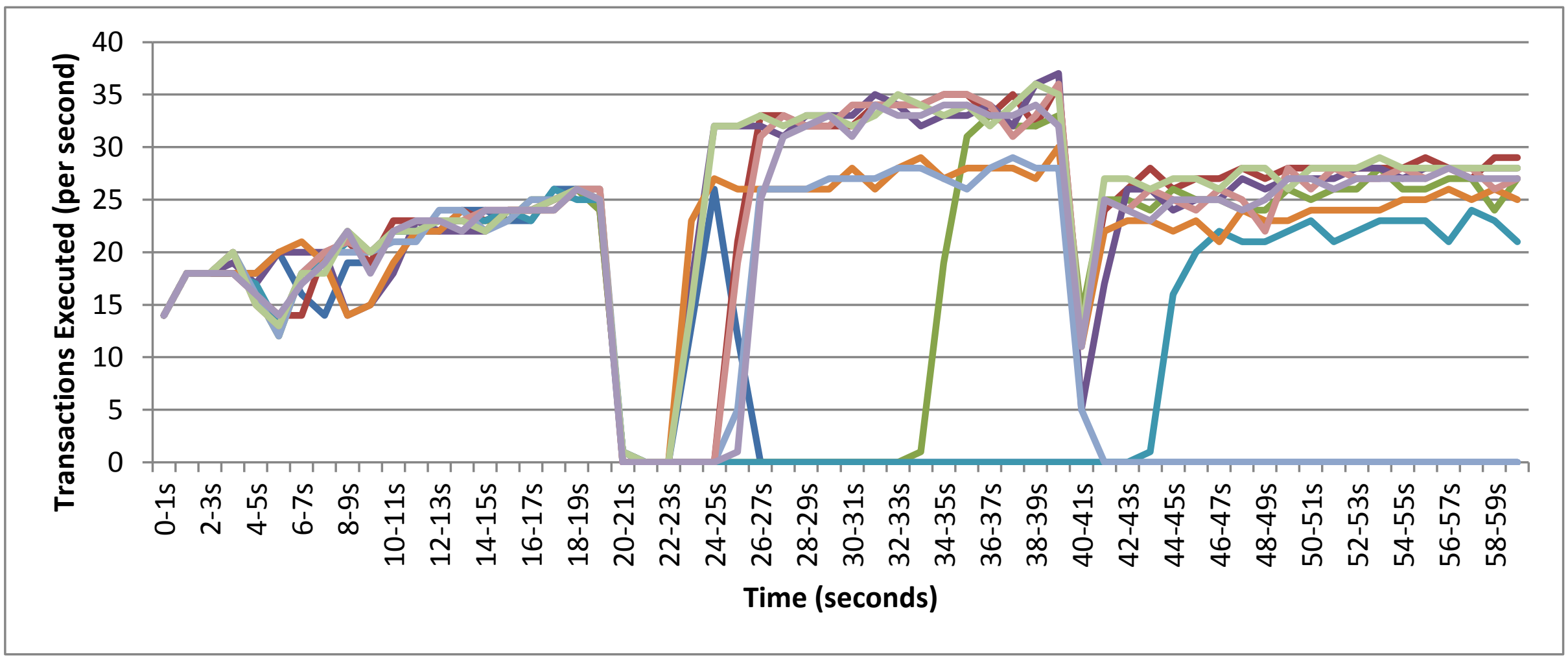

**Figure 89: Results of experiment 2, test 3D.**

Test *3D* is identical to *3C* but the machine being started is the same machine that was killed off after 20 seconds. The results for each of these tests are similar, but the temporary drop-off in throughput at 40 seconds is more consistent in this test. As with *1D* and *2D*, this drop-off is a result of H2O's design, whereby out-of-date replicas must be dropped before a new replica can be created in its place.

***Test 3E***

This script evaluates how the system responds to the failure of a machine which holds the System Table and the only replica of its state.

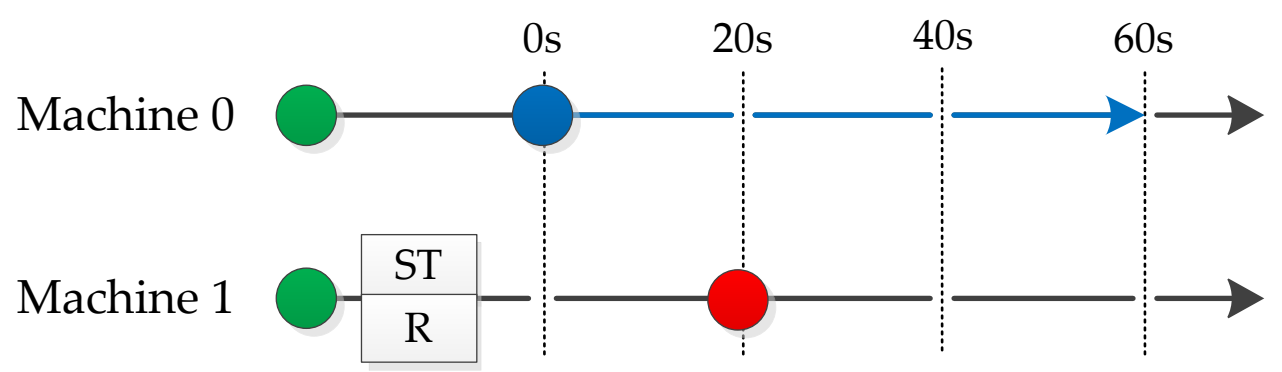

**Figure 90: Execution of 3E over time.**

***Script:***

```
[Global Parameters: System-wide replication factor = 3]
{start_machine id="0"}
{start_machine id="1"}
{1} MIGRATE SYSTEMTABLE NO_REPLICATE
{0} {execute_workload="st.workload" duration="60000"}
{sleep="20000"}
{terminate_machine id="1"}
```

***Summary***

After twenty seconds this script terminates *machine 1*, which holds the System Table and a replica of the System Table. No other machines hold System Table replicas, so no System Table operations can commit.

***Results***

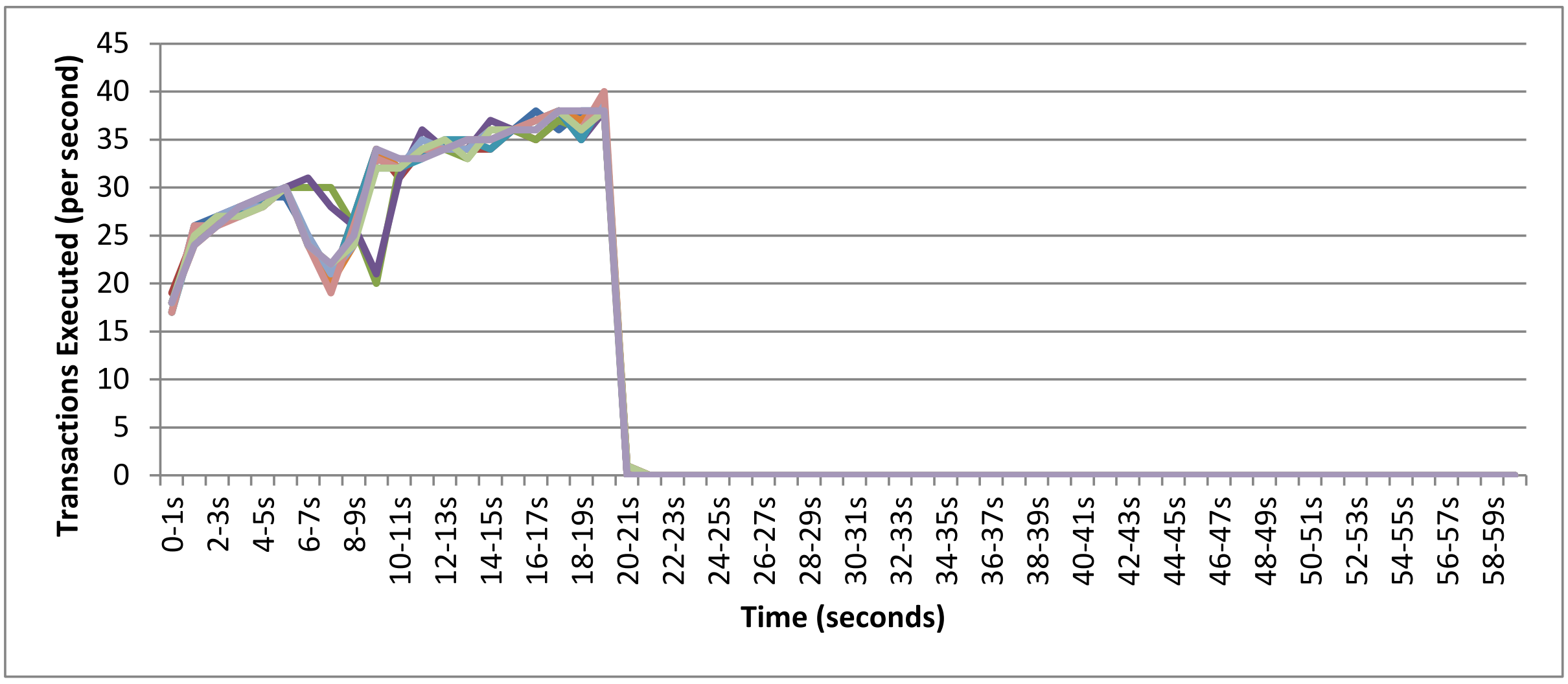

**Figure 91: Results of experiment 2, test 3E.**

When the machine is killed after 20 seconds there are no more replicas of the System Table available, so no more transactions can be executed.

***Test 4A***

This test evaluates how the system responds to the failure of the machine holding the System Table, a Table Manager, and a table replica. For each of the replicas shown in Figure 92 (and the corresponding figures in *4B-E*) there are replicas of the System Table, the Table Manager and the table on the given machine.

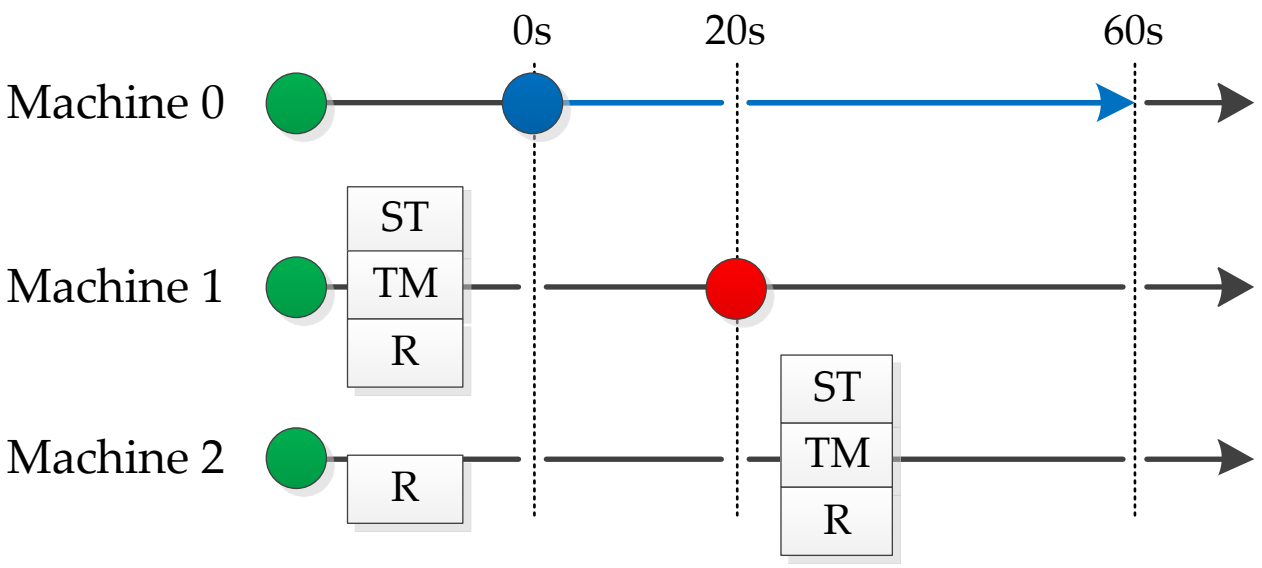

**Figure 92: Execution of 4A over time.**

***Script:***

```
[Global Parameters: System-wide replication factor = 3]
{start_machine id="0"}
{start_machine id="1"}
{start_machine id="2"}
{1} MIGRATE SYSTEMTABLE NO_REPLICATE
{create_table id="1" name="workloadTable" prepopulate_with="300"}
{0} {execute_workload="short.workload" duration="60000"}
```

```
{sleep="20000"}
{terminate_machine id="1"}
```

***Summary***

After twenty seconds this script terminates *machine 1*, which holds the System Table, the Table Manager, and a replica of the *workloadTable* table. No other machines exist on which to replicate data but another machine is able to re-instantiate the System Table and the Table Manager.

***Results***

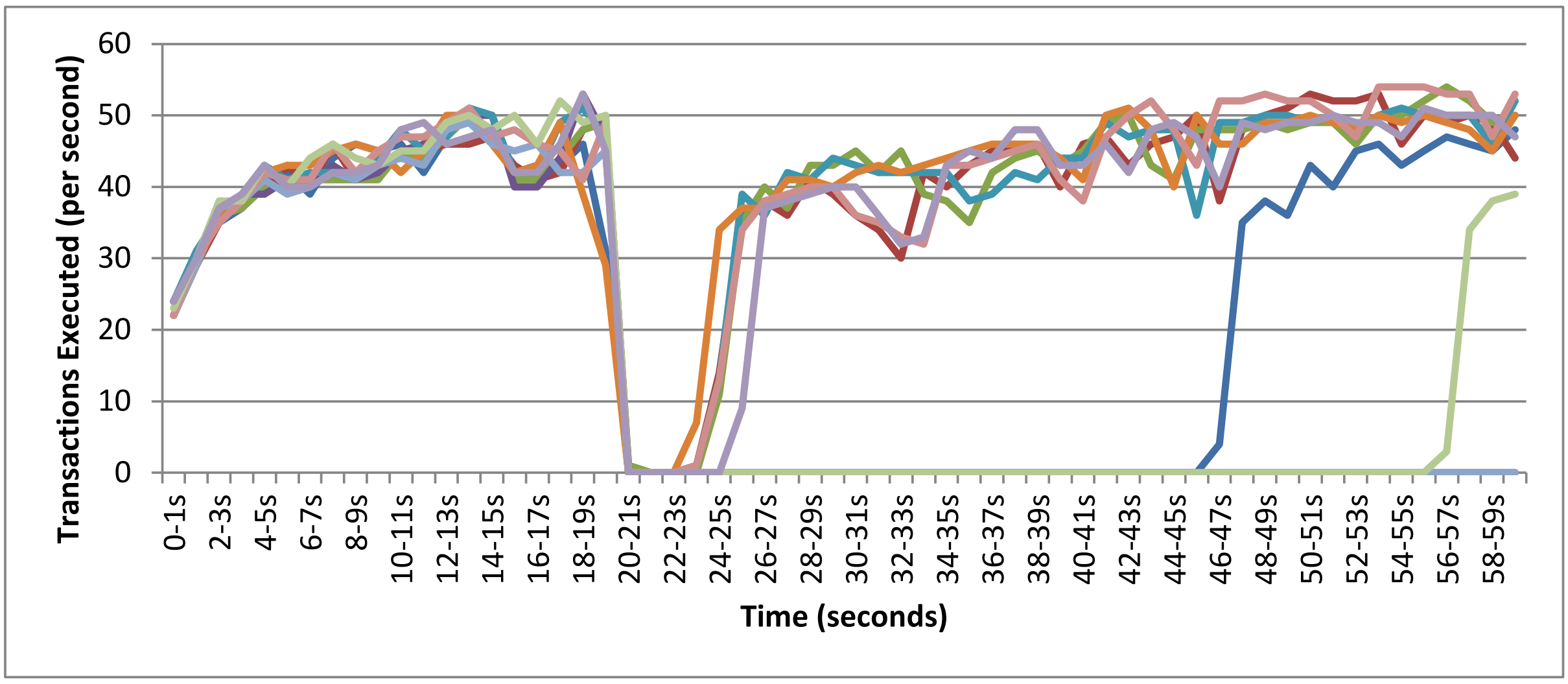

**Figure 93: Results of experiment 2, test 4A.**

This test is effectively a combination of 2A and 3A, as the machine being killed is running the System Table and Table Manager of *workloadTable*, as well as storing a replica of *workloadTable*.

When *machine 1* is killed, both the System Table and the Table Manager fail, so they must be recreated in sequence. This takes on average 10 seconds, and a median of 4 seconds due to a number of late recovering test runs. In addition, two test runs never recover from the failure of *machine 1*.

In contrast, in test *2A* (where only the Table Manager failed) it took on average 3.6 seconds to recover, and in test *3A* (where only the System Table failed) it took on average 3.8 seconds to recover.

***Test 4B***

This test evaluates the ability of the system to recover from the failure of the System Table and the Table Manager, and then immediately re-replicate System Table and Table Manager state to a new instance.

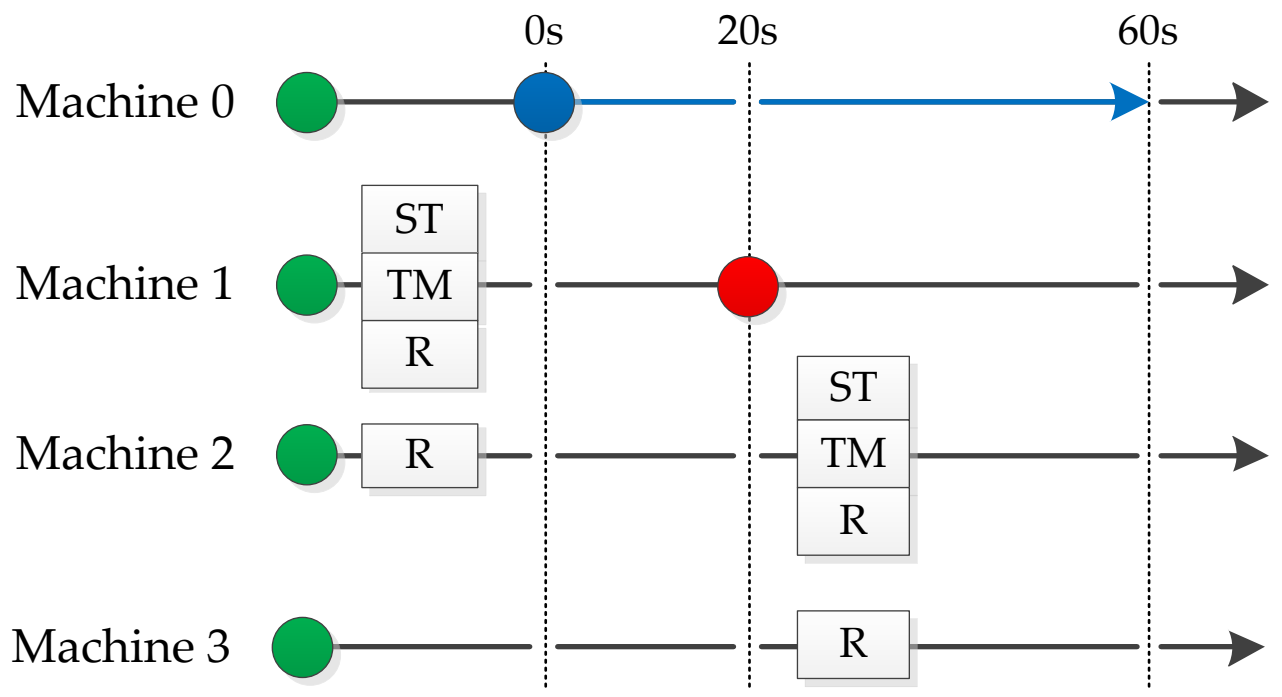

**Figure 94: Execution of 4B over time.**

***Script:***

```
[Global Parameters: System-wide replication factor = 2]
{start_machine id="0"}
{start_machine id="1"}
{start_machine id="2"}
{start_machine id="3"}
{1} MIGRATE SYSTEMTABLE NO_REPLICATE
{create_table id="1" name="workloadTable" prepopulate_with="300"}
{0} {execute_workload="short.workload" duration="60000"}
{sleep="20000"}
{terminate_machine id="1"}
```

***Summary***

After twenty seconds this script terminates *machine 1*, which holds the System Table, the Table Manager, and a replica of *workloadTable*. There is another machine available on which to replicate data, so the replication factor eventually increases again.

***Results***

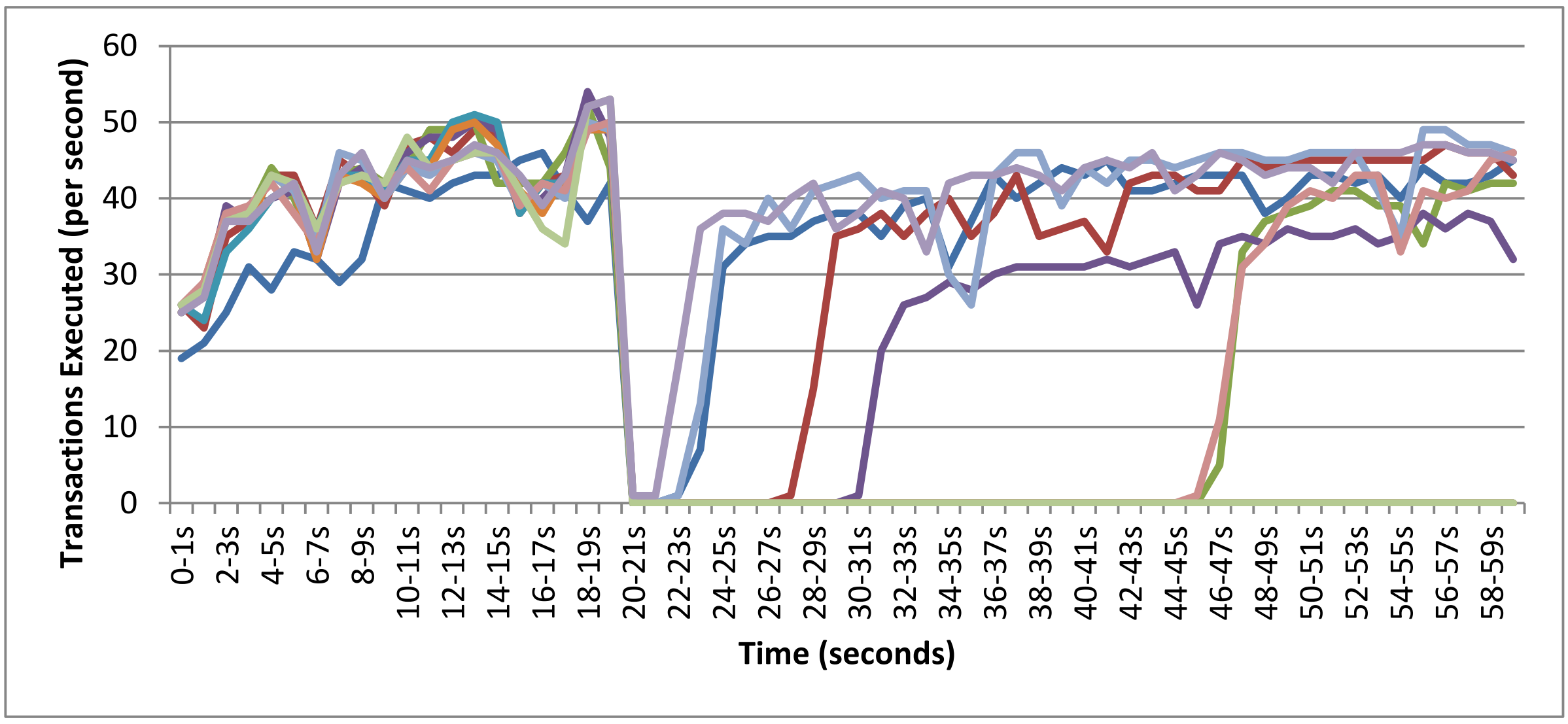

**Figure 95: Results of experiment 2, test 4B.**

This test is effectively a combination of 2B and 3B — the machine running the Table Manager and System Table is killed off, and when they are recreated they immediately increase the replication factor on another available machine.

It takes on average 11.4 seconds to recover from the failure of *machine 1*, though recovery time varies between 3 and 27 seconds. Three test runs did not recover at all after the failure.

***Test 4C***

This test evaluates the ability of the system to recreate the System Table and Table Manager after failure, and then re-replicate System Table and Table Manager state to a new machine which starts 20 seconds later.

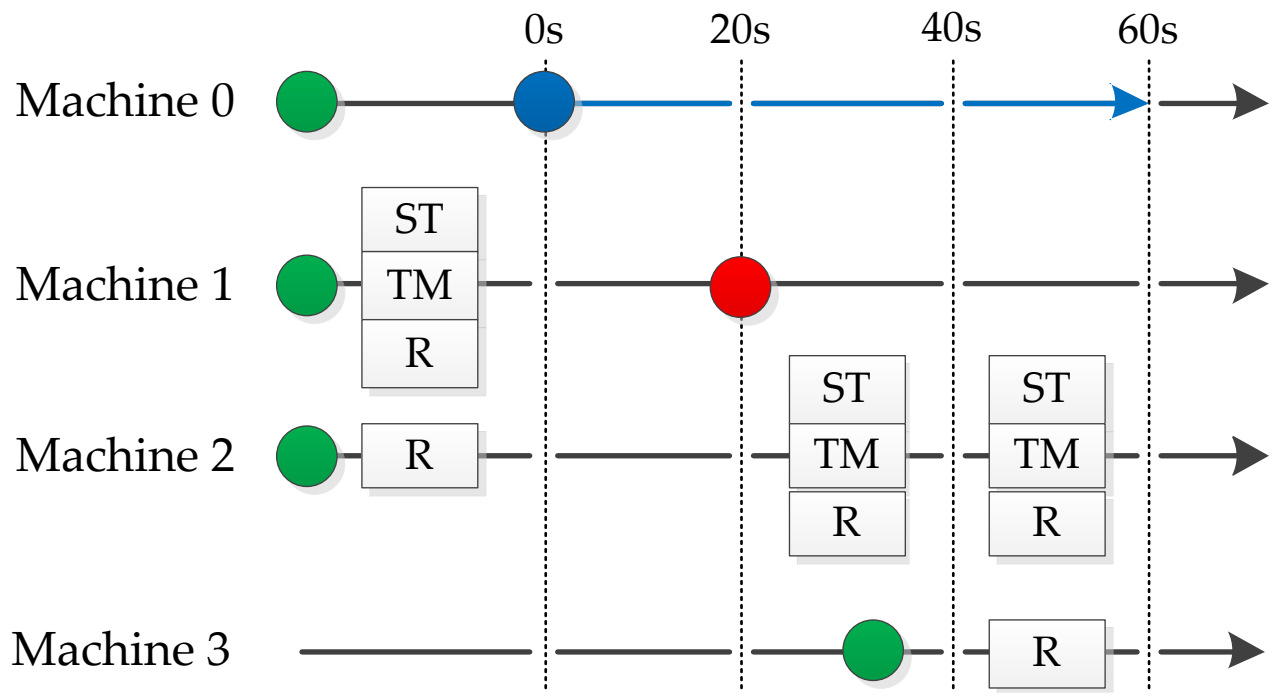

**Figure 96: Execution of 4C over time.**

***Script:***

```
[Global Parameters: System-wide replication factor = 3]
```

```
{start_machine id="0"}
{start_machine id="1"}
{start_machine id="2"}
{1} MIGRATE SYSTEMTABLE NO_REPLICATE
{create_table id="1" name="workloadTable" prepopulate_with="300"}
{0} {execute_workload="short.workload" duration="60000"}
{sleep="20000"}
{terminate_machine id="1"}
{sleep="20000"}
{start_machine id="3" block-workloads="true"}
```

***Summary***

After twenty seconds this script terminates *machine 1*, which holds the System Table, the Table Manager, and a replica of the *workloadTable* table. After another twenty seconds a new machine starts and data is replicated onto this machine.

***Results***

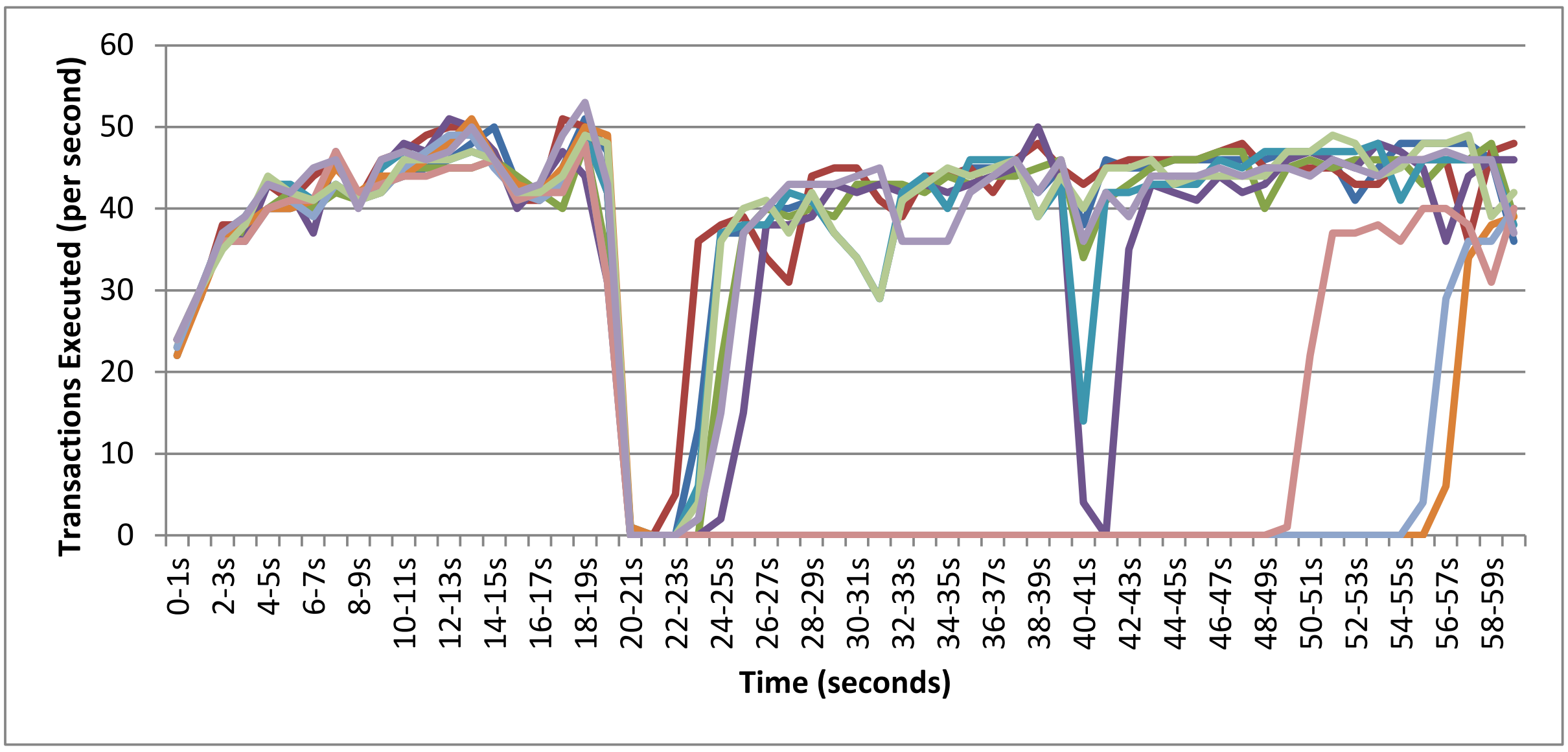

**Figure 97: Results of experiment 2, test 4C.**

This test is effectively a combination of 2C and 3C.

It takes the system an average of 12 seconds to recover from the failure of machine 1, with recovery time ranging between 3 and 37 seconds. The test runs that have recovered prior to 40 seconds show a slight drop in throughput at 40 seconds, which is the result of data being replicated to *machine 3*.

The increased variability in the results of tests *4A-4C* compared to those preceding them is likely due to the increased complexity of recovering in sequence from the failure of both the System Table and Table Manager, two distinct components in H2O. The complexity of this recovery sequence must be recognised in any evaluation of H2O's design.

***<u>Test 4D</u>***

This test evaluates the ability of the system to recreate the System Table and Table Manager after failure, and then re-replicate System Table and Table Manager state to the same machine when it is restarted 20 seconds later.

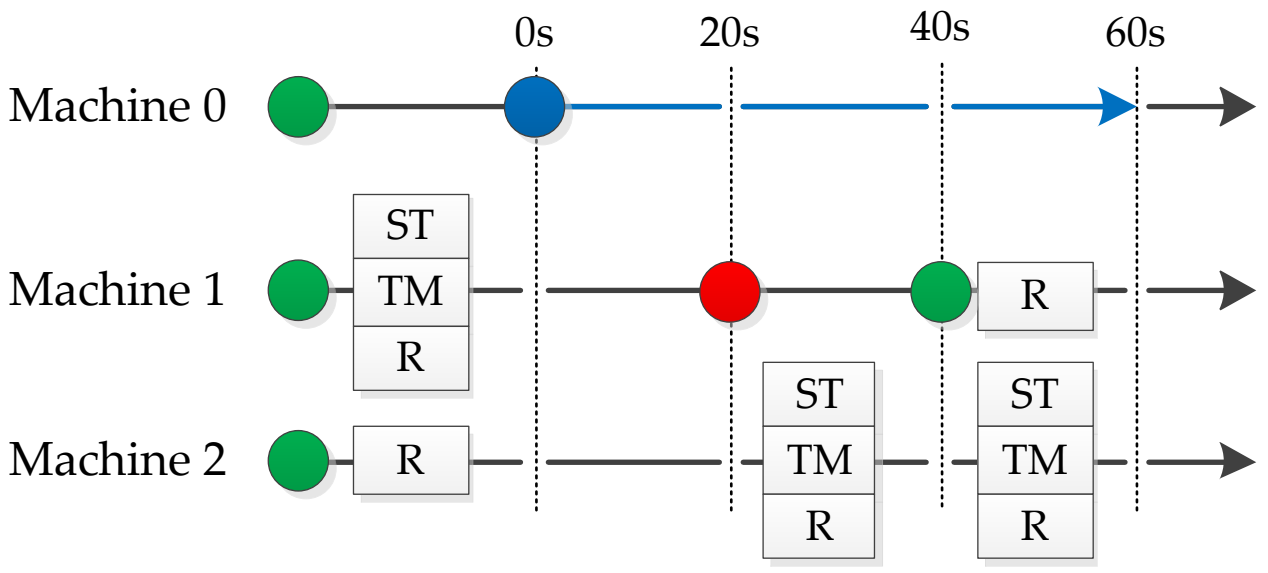

**Figure 98: Execution of 4D over time.**

***Script:***

```
[Global Parameters: System-wide replication factor = 3]
{start_machine id="0"}
{start_machine id="1"}
{start_machine id="2"}
{1} MIGRATE SYSTEMTABLE NO_REPLICATE
{create_table id="1" name="workloadTable" prepopulate_with="300"}
{0} {execute_workload="short.workload" duration="60000"}
{sleep="20000"}
{terminate_machine id="1"}
{sleep="20000"}
{start_machine id="1" block-workloads="true"}
```

***Summary***

After twenty seconds this script terminates *machine 1*, which holds the System Table, the Table Manager, and a replica of the *workloadTable* table. After another twenty seconds *machine 1* is restarted and data is replicated onto this machine.

***Results***

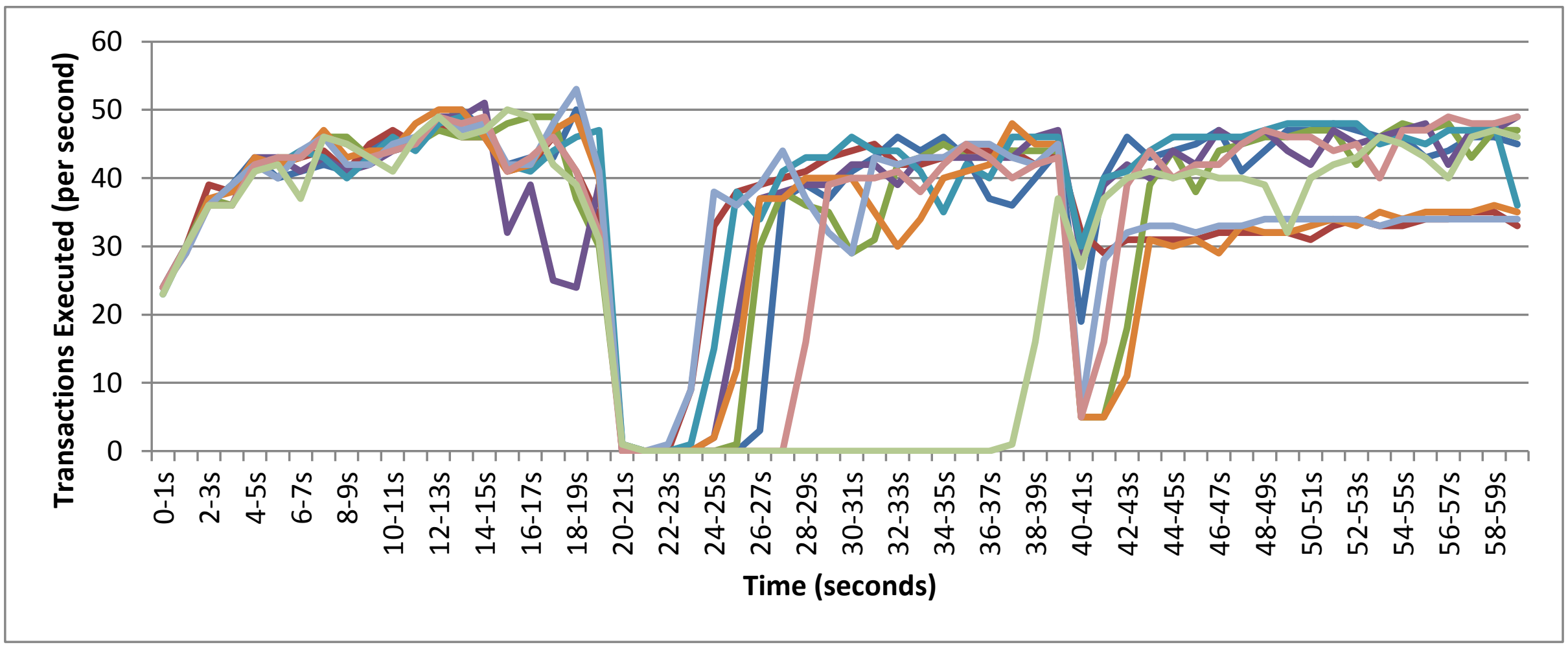

**Figure 99: Result of experiment 2, test 4D.**

This test is effectively a combination of 2D and 3D.

It takes the system 6.1 seconds on average to recover from the failure of *machine 1*, which is substantially different than the average of 12 seconds in *test 4C* (both test scripts are identical up to the 40 second mark). This difference is due to a number of outliers in *4C* greatly increasing the average recovery time; median recovery time for *4C* is 3.5 seconds, compared to 4.5 seconds in *4D*.

When *machine 1* is restarted after 40 seconds, throughput decreases while a new replica is created on *machine 1*. This follows the same pattern that is shown in both *2D* and *3D*.

***<u>Test 4E</u>***

This script evaluates how the system responds to the failure of a machine which holds the System Table, a Table Manager and the only replica of its table.

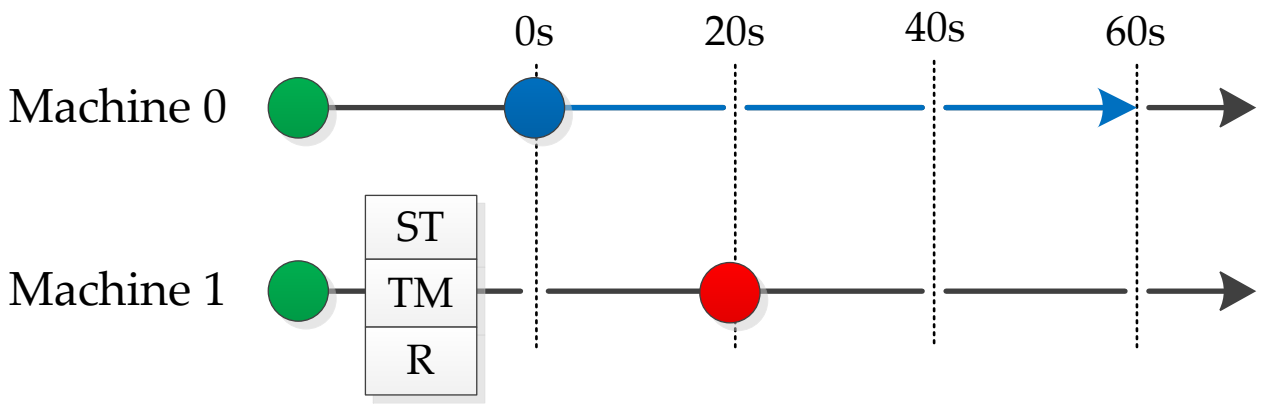

**Figure 100: Execution of 4E over time.**

***Script:***

```
[Global Parameters: System-wide replication factor = 3]
{start_machine id="0"}
```

```
{sleep="3000"}
{start_machine id="1"}
{1} MIGRATE SYSTEMTABLE NO_REPLICATE
{create_table id="1" name="workloadTable" prepopulate_with="300"}
{0} {execute_workload="short.workload" duration="60000"}
{sleep="20000"}
{terminate_machine id="1"}
```

***Summary***

After twenty seconds this script terminates *machine 1*, which holds the System Table, the Table Manager, and a replica of the *workloadTable* table. No other replicas are available, so no queries can be executed.

***Results***

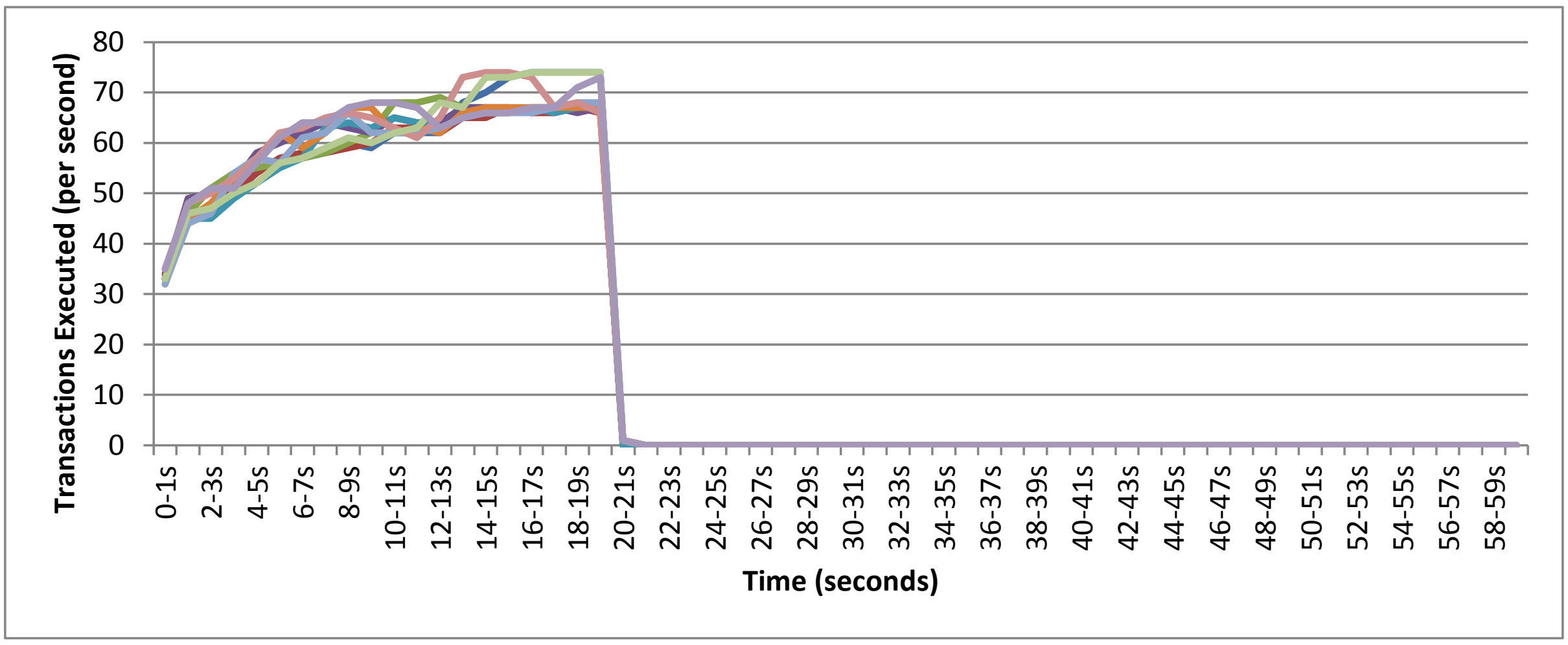

**Figure 101: Result of experiment 2, test 4E.**

When the machine is killed off after 20 seconds there are no available replicas of the System Table or the Table Manager for *workloadTable*, meaning no transactions can be executed.

### 7.3.7 Conclusion

Experiment 2 shows how H2O responds to the failure of machines while transactions are executing.

In numerous test runs in experiment 2 the database does not recover from the failure of a node. As is discussed in the previous section, these failures are thought to be due to flaws in the implementation of H2O that are not inherent in its design (though this theory was not proven in these evaluations).

Some of the delays in recovering from failure could be removed by optimizing H2O's current implementation. In the tests in *1A-1D* there is a period where no queries can be executed after the failure of a machine, even though that machine does not contain the Table Manager or System Table. This delay, due to an update to the failed machine timing out, could be removed by using partially asynchronous query execution[53]. Similarly, in *2B* there is a large delay while the Table Manager is restarted because this operation also synchronously creates a new replica on another machine. If the replica was created asynchronously the delay in recovering from the machine failure would be smaller (closer to that of *2A*), but there would be a lengthier period where the replication factor of the table was lower than desirable.

Tests *4A- 4C* show that the time it takes the system to recover from the failure of both the System Table and Table Manager is highly variable (the recovery time in *4C* ranges from 3 to 37 seconds). This indicates that the complexity of the system's design, where the System Table and a Table Manager must be recovered in sequence, makes recovery time unpredictable. To reduce this complexity the database could combine the roles of the System Table and Table Managers, but there are various reasons against this discussed in Chapter 5. It is also possible that the delays in recovery are a facet of H2O's implementation and not a result of its design, but this cannot be proven with the results available.

---

[53] This refers to the concept of allowing a transaction to commit when $n$ of $m$ replicas have committed, where $n$ is considered a sufficient replication factor less than $m$.

H2O is able to recover from most failures in this experiment in less than 10 seconds. This length of delay may be too great for some systems, which may limit potential applications of a workstation-based DDBMS. For comparison, *Amazon Relational Database Service* guarantees 99.5% uptime [93], which allows for 50 minutes and 24 seconds of downtime per week — if it took H2O 10 seconds to recover from every node failure, this would allow for up to 302 failures per week or 43 failures per day while still providing 99.5% uptime.

# Chapter 8: Conclusion

# 8 CONCLUSION

This thesis is motivated by the observation that workstation machines within organizations are often underutilized, and existing software is not capable of fully harnessing this capacity.

Existing workstation-based systems make use of the spare computational capacity of workstation machines, but they typically do not make full use of available storage capacity. Databases are designed to manage the structured storage of data, but existing systems are not designed to run on workstation machines, which are less reliable than server clusters and primarily used for other user-centric tasks.

DDBMSs are the focus of this thesis as they represent a particular challenge in testing the viability of workstation-based systems. They must be able to store large volumes of up-to-date, highly-structured data, and they are judged on the speed of their response to queries.

These are the motivations behind the following goal, originally stated in *chapter 1*:

*This thesis investigates the viability of systems that use the un-utilized capacity of workstations to provide services, such as databases, that are typically run in server clusters.*

The following sections summarise the process by which this goal was investigated, and present a conclusion to this investigation.

## 8.1 THESIS SUMMARY

*Chapter 2* presents an overview of ***background work*** in both database and workstation-based systems, to identify the areas of existing work that are relevant to this thesis.

It identifies that DDBMSs can be categorized by the locale of their intended use, as the trade-offs made explicit in the CAP theorem determine when a system will become unavailable and what type of transactional guarantees it can provide.

The ***related work*** discussed in *chapter 3* builds on this background by showing how existing solutions are designed, and identifying what problems they are aiming to solve.

It shows that most clustered DDBMSs assume that network partitions are rare, so they typically guarantee ACID transactions, but require manual intervention in the event of failure. Systems that are able to automatically recover from failure are typically designed to run over larger or less reliable sets of machines, but they provide fewer transactional guarantees and often only eventual consistency.

This background and related work is used in *chapter 4* to form a set of ***requirements*** which explicitly identify the needs of a workstation-based DDBMS, and workstation-based systems more generally.

A workstation-based system must be designed with the expectation that machines often fail or become unavailable, and the recognition that the set of machines that is available may gradually shift over time.

The ***design*** of D2O, the database introduced in *chapter 5*, is developed from the aforementioned requirements. It is motivated by these requirements and by the design of the systems presented earlier in *chapter 3*.

D2O is designed to provide the same ACID transactional guarantees as many clustered DDBMSs, but it is also able to automatically recover from the failure of database instances. To achieve this it uses synchronous replication, which allows it to recover from the failure of an individual instance without losing data. It uses a locator server architecture to ensure that active database instances can be found by an instance, even in cases where the set of active instances has changed completely since the last time the instance started.

A subset of the functionality described in D2O is implemented in H2O for the purpose of evaluation in *chapter 7*. H2O is evaluated to determine the performance of the architecture in comparison to existing DDBMSs, and to determine how this architecture responds to the failure of individual nodes. These results are used below to determine the viability of workstation-based systems.

## 8.2 CONCLUSIONS

This thesis specifically evaluates the viability of workstation-based systems in terms of the performance of H2O and its response to failure.

In addition, the design of D2O looks at how the requirements of a workstation-based system (stated in chapter 4) can be met. This design meets all of the requirements of chapter 4, but for *self-containment* — the requirement that the system should be able to run entirely over workstation machines. The reason for this, discussed in 5.7, highlights a limitation of workstation-based systems — namely, it is not possible to have a system with dynamic membership that uses majority consensus to select a co-ordinator (in this case, the System Table). This potentially limits the applications of workstation-based systems, as it requires that some set of machines must be used to run the majority consensus protocol. These machines could be workstation machines, but their failure would affect the availability of the entire system more than the failure of a regular workstation machine.

*Chapter 7* describes two experiments which are used to determine H2O's viability as a workstation-based system. *Experiment 1* evaluates its transaction throughput compared to two other DDBMSs, and *experiment 2* evaluates its ability to recover from the failure of database instances.

The results of *experiment 1* show that while H2O is slower than a comparable asynchronous replication DDBMS (*MySQL*), it has performance equal to another synchronous replication DDBMS (*PGCluster*). These results show that performing synchronous replication and flushing transactions to disk immediately (with no write delay) substantially reduces throughput, though both features are necessary for a workstation-based system.

The performance of H2O's architecture is validated by the closeness of its performance to *PGCluster*, the other synchronous replication system being tested. This result shows that H2O has comparable performance to similar clustered DDBMSs, which makes it a viable alternative to these systems.

The results of *experiment 2* show that it typically takes H2O less than 10 seconds to recover from a machine failure, though it does take up to 40 seconds in some cases and in other cases the current H2O implementation failed completely. These recovery times are short enough that H2O could endure roughly 300 failures per week and still meet the 99.5% uptime guarantee that *Amazon* give to users of *Amazon Relational Database Service*.

There is insufficient data available to categorically state whether a modern workstation environment typically suffers this many failures, but these results indicate that a workstation-based system can be viable in at least some environments.

The primary limitation of a workstation-based system is its vulnerability to power failures and other factors outwith the control of an organization. Clustered systems can limit the damage of such events by using uninterruptible power sources (and in more extreme cases, generators), but workstation machines typically rely on mains power. It would be difficult to guarantee uptime for a workstation-based system as a result of these external factors, not because of the downtime caused by intermittent machine failure.

Despite this, lower uptime guarantees are probably acceptable to many database applications, including non-production database deployments and small scale deployments such as that of the school administrator discussed in the introduction to this work.

### 8.2.1 Limitations of Evaluations

Both *experiment 1* and *experiment 2* indicate that a workstation-based system can be viable, both in terms of speed and availability, but there are a number of factors not evaluated by these experiments that prevent this statement being made conclusively.

Neither experiment tests the effectiveness of H2O on an actual workstation machine, so it is unclear what effect a user's activities would have on transaction throughput and on the availability of the system.

Neither experiment tests data placement, as the current implementation of H2O places the Table Manager on the machine that executed the *CREATE TABLE* request, rather than using resource monitoring information to determine optimal placement. An evaluation of the most effective method of data placement is left as future work, and is discussed in 8.4.

None of the tests in experiment 2 execute workloads with more than one table, which make it simpler to recover from machine failure. Further testing is needed to establish the typical recovery time when a machine running multiple Table Managers fails.

Finally, the autonomic and resource-aware aspects of D2O's design are not evaluated in this thesis. Without evaluation of these features it is not possible to determine whether a workstation-based system can run effectively without disrupting the activities of a user(s) of the machines in the system. D2O's autonomic functionality also needs to be evaluated to determine whether a workstation-based system can be run with minimal administrator intervention.

## 8.3 REVIEW OF CONTRIBUTIONS

The primary contribution of this thesis is the knowledge gained in the design of D2O, and the implementation and evaluation of H2O.

This work extends the state of the art in workstation-based computing by explicitly identifying the requirements of a software system designed for such an environment. It introduces a new database architecture, containing several novel components, which is designed to operate over dynamically changing sets of workstation machines.

*Chapters 2 and 3* clearly describe the problem space of database system design with respect to distribution and locale, and illustrate the challenges associated with workstation-based computing. These chapters contribute an improved understanding of this area, and a survey of modern distributed database architectures.

*Chapter 4* identifies a general set of requirements for workstation-based systems based on the challenges identified in previous chapters.

The design of *D2O* in *chapter 5*, and the knowledge gained from this design, is the primary contribution of this thesis. There are three key aspects to this design, which distinguish it from existing work and allow it to run over sets of workstation machines:

*D2O* uses table-level locking and replication, which gives it greater flexibility in re-partitioning data across available resources. This design allows it to operate over dynamically changing sets of machines with short downtime, in contrast to existing DDBMSs which focus on more static sets of machines and require manual recovery.

Unlike existing DDBMSs, *D2O* is designed to be resource-aware. It is designed to make use of resource monitoring information to determine data placement and to move data and processes when resources become scarce on some machines. This ability is discussed in the context of an autonomic, resource-aware architecture.

A discovery mechanism involving so-called *locator servers* allows database instances to connect to an existing database system without prior knowledge of other database instances, and to operate correctly in the presence of a network partition. The importance of this approach is discussed in relation to the CAP theorem and its effect on all distributed systems.

Finally, *chapter 5* introduces a mechanism for giving identities to database instances and maintaining these identities when the databases change location.

The implementation of D2O, *H2O,* is evaluated in two experiments which are used to determine the effectiveness of a workstation-based database system in terms of performance and fault tolerance. These experiments contribute to existing work by comparing the costs of synchronous and asynchronous replication, both by showing the cost of immediately committing transactions to disk, and by showing the effect of node failure and replication factor on the availability of a replicated database system. These results are used to establish the viability of a workstation-based DDBMS.

## 8.4 FUTURE WORK

This section discusses possible future work on this research topic.

***An Autonomic, Resource-Aware Database***

H2O makes some use of resource monitoring data, but it does not use this data to move replicas, or move the System Table and Table Managers, as resource availability changes. More research is needed to determine the best architecture for this requirement.

***Generic Ad-hoc Cloud Architecture***

This thesis discusses the requirements of a workstation-based system with particular focus on database systems, in contrast to the ad hoc cloud computing proposal in [46], which describes a generic framework for workstation-based computing. There are a number of possible future directions in research in this area, including the development of an ad hoc cloud infrastructure which provides common functionality such as resource monitoring and failure detection to applications.

This work would investigate the challenges in developing such a system, particularly the extent to which common cloud components such as resource monitoring can be separated from application-specific components, and the work involved in adapting existing applications for workstation-based computing. It would be evaluated on the ease of deploying such solutions in comparison to existing work, and on the extent to which autonomic management can be used for tasks that typically require manual administration.

Further work is also needed to establish whether significant cost savings can be produced through the exploitation of underused resources.

***Evaluating Data Placement***

An evaluation of approaches to data placement is a promising avenue of future research. The databases discussed in this work use either a hashing or heuristic approach to data placement, but these alternate solutions were not evaluated in the design of H2O.

An extension of this work is an evaluation of these approaches in a resource-aware system such as H2O, where the effectiveness of each approach is compared over time, in a dynamically changing database system.

***Evaluating Atomic Commit Protocols***

H2O uses two-phase commit to provide atomicity in distributed updates, but D2O's design uses three-phase commit. It is unclear which is the most appropriate atomic commit protocol, for a system which expects a higher rate of machine failure than clustered systems.

The experiment in 7.3 evaluates the cost of failure in relation to query execution time. This experiment could be extended to analyse the effect of the atomic commit protocol on failure recovery time and on transaction throughput, to establish the situations where a given protocol is most appropriate.

***Evaluating Failure Detection Mechanisms***

H2O uses *Chord* as a primitive failure detector, but a number of alternative approaches could be more effective. For example, machines in the heartbeat-based systems discussed in *chapter 3* send periodic messages to a set of nodes to detect failure. If this set is randomly chosen each time a heartbeat message is sent, each node will eventually check the liveliness of every other node. This may be more effective than *Chord*, which may not detect the failure of a machine if a number of successive nodes in the ring fail at once (as discussed in *chapter 5*).

The effectiveness of these methods of failure detection could be evaluated in the context of workstation-based systems, where failure is likely to be more common than the systems for which the failure detectors discussed in *chapter 3* are designed.

## 8.5 Concluding Remarks

In summary, this thesis introduces the area of interactive, workstation-based computing, and presents and evaluates the design of a workstation-based database system.

It contributes to existing understanding on the design of distributed database systems, and illustrates the trade-offs that must be made in the design of a workstation-based system.

It shows that, to have ACID transactions, a workstation-based database system must sacrifice some performance by requiring that transactions are synchronously replicated and immediately flushed to disk. A workstation-based system is typically less reliable than machines in a cluster, and local user processes may prevent a database instance from executing queries, so workstation-based machines typically sacrifice control over availability.

For a workstation-based system to be partition tolerant it must still use an external mechanism, such as the locator servers introduced in this work, to ensure that a partition does not exist. This makes it difficult to provide a workstation-based system that uses no external services.

These restrictions impact the performance of a workstation-based system, but the experimental results presented in this thesis indicate that such a system is still viable and could be used for many applications.

These results show the potential of interactive, workstation-based systems, and lay a foundation for future work in this field.

# Appendices

# APPENDIX

## 1 EXPERIMENT 1 CONFIGURATION SETTINGS

### 1.1 MYSQL CONFIGURATIONS

The following configuration is used by the MySQL master instance in each of the benchmarks in 0.

```
[mysqld]
server-id=1

log-bin=mysql-bin
binlog-do-db=benchmarksql

datadir=/var/lib/mysql
socket=/var/lib/mysql/mysql.sock
user=mysql

old_passwords=1

[mysqld_safe]
log-error=/var/log/mysqld.log
pid-file=/var/run/mysqld/mysqld.pid
```

The following is representative of a slave instance configuration file, used by the benchmarks in 0:

```
[mysqld]
server-id=2
master-host=compute-0-1.local
master-user=repl
master-password=iamaslave
master-connect-retry=60
replicate-do-db=benchmarksql

datadir=/var/lib/mysql
socket=/var/lib/mysql/mysql.sock
user=mysql

old_passwords=1
```

```
[mysqld_safe]
log-error=/var/log/mysqld.log
pid-file=/var/run/mysqld/mysqld.pid
```

## 1.2 PGCLUSTER CONFIGURATIONS

The following is the configuration used in *PGCluster* for various config files.

***postgresql.conf (the configuration used on all instances):***

```
listen_addresses = '*'
port = 5432
```

***pgreplicate.conf (the configuration for the replication node):***

```
<Cluster_Server_Info>
  <Host_Name>compute-0-4</Host_Name>
  <Port>5432</Port>
  <Recovery_Port>7001</Recovery_Port>
</Cluster_Server_Info>
<Cluster_Server_Info>
  <Host_Name>compute-0-1.local</Host_Name>
  <Port>5432</Port>
  <Recovery_Port>7001</Recovery_Port>
</Cluster_Server_Info>
<LoadBalance_Server_Info>
  <Host_Name>compute-0-2</Host_Name>
  <Recovery_Port>6001</Recovery_Port>
</LoadBalance_Server_Info>
<Host_Name>compute-0-3</Host_Name>
<Replication_Port>8001</Replication_Port>
<Recovery_Port>8101</Recovery_Port>
<RLOG_Port>8301</RLOG_Port>
<Response_Mode>normal</Response_Mode>
<Use_Replication_Log>no</Use_Replication_Log>
<Replication_Timeout>1min</Replication_Timeout>
<LifeCheck_Timeout>3s</LifeCheck_Timeout>
<LifeCheck_Interval>15s</LifeCheck_Interval>
<Log_File_Info>
  <File_Name>/tmp/pgreplicate.log</File_Name>
  <File_Size>1M </File_Size>
  <Rotate>3</Rotate>
</Log_File_Info>
```

***pglb.conf (the configuration used on the load balancer):***

```
<Cluster_Server_Info>
  <Host_Name>compute-0-4</Host_Name>
  <Port>5432</Port>
  <Max_Connect>32</Max_Connect>
</Cluster_Server_Info>
<Cluster_Server_Info>
  <Host_Name>compute-0-1.local</Host_Name>
  <Port>5432</Port>
  <Max_Connect>32</Max_Connect>
</Cluster_Server_Info>
<Host_Name>compute-0-2.local</Host_Name>
<Backend_Socket_Dir>/tmp</Backend_Socket_Dir>
<Receive_Port>5432</Receive_Port>
<Recovery_Port>6001</Recovery_Port>
<Max_Cluster_Num>128</Max_Cluster_Num>
<Use_Connection_Pooling>no</Use_Connection_Pooling>
<LifeCheck_Timeout>3s</LifeCheck_Timeout>
<LifeCheck_Interval>15s</LifeCheck_Interval>
  <Log_File_Info>
  <File_Name>/tmp/pglb.log</File_Name>
  <File_Size>1M</File_Size>
  <Rotate>3</Rotate>
</Log_File_Info>
```

***pg_hba.conf (firewall configuration used on all instances):***

```
local all all trust
host all all 10.1.255.249/32 trust
host all all 10.1.255.250/32 trust
host all all 10.1.255.251/32 trust
host all all 10.1.255.252/32 trust
host all all 10.1.255.253/32 trust
```

# 2 EXPERIMENT 2 CO-ORDINATION SCRIPTS

This section lists the scripts used by the H2O evaluation co-ordinator in the experiments described in section 7.3.

## 2.1 WORKLOAD SAMPLES

These workloads represent include a single transaction which is executed repeatedly for the length of an evaluation.

### 2.1.1 Insert / Select Workload (Table Manager Workload)

```
  SET AUTOCOMMIT OFF;
  INSERT INTO workloadTable VALUES (<loop-counter/>, <generated-string/>,
<generated-long/>);
  SELECT * FROM workloadTable WHERE int_a > 679153090560;
  DELETE FROM workloadTable WHERE id=<last-loop-counter/>
  COMMIT;
  SET AUTOCOMMIT ON;
```

### 2.1.2 Create / Drop Workload (System Table Workload)

```
  CREATE TABLE IF NOT EXISTS workloadTable (id int);
  <sleep>5</sleep>
  DROP TABLE IF EXISTS workloadTable;
```

## 2.2 WORKLOADS

### 2.2.1 Test 1A

```
  {start_machine id="0"}
  {sleep="3000"}
  {start_machine id="1"}
  {start_machine id="2"}
  {create_table id="1" name="workloadTable" schema="id int, str_a varchar(40),
int_a BIGINT" prepopulate_with="300"}
  {check_repl_factor name="workloadTable" expected="2"}
  {sleep="5000"}
  {check_meta_repl_factor expected="3"}
  {check_meta_repl_factor name="workloadTable" expected="2"}
  {0} {execute_workload="short.workload" duration="60000"}
  {sleep="20000"}
  {terminate_machine id="2"}
  {sleep="5000"}
  {check_repl_factor name="workloadTable" expected="1"}
```

### 2.2.2 Test 1B

```
{start_machine id="0"}
{sleep="3000"}
{start_machine id="1"}
{start_machine id="2"}
{start_machine id="3"}
{create_table id="1" name="workloadTable" schema="id int, str_a varchar(40),
int_a BIGINT" prepopulate_with="300"}
{check_repl_factor name="workloadTable" expected="2"}
{sleep="5000"}
{check_meta_repl_factor expected="2"}
{check_meta_repl_factor name="workloadTable" expected="2"}
{0} {execute_workload="short.workload" duration="60000"}
{sleep="20000"}
{terminate_machine id="2"}
{sleep="5000"}
{check_repl_factor name="workloadTable" expected="2"}
```

### 2.2.3 Test 1C

```
{start_machine id="0"}
{sleep="3000"}
{start_machine id="1"}
{start_machine id="2"}
{create_table id="1" name="workloadTable" schema="id int, str_a varchar(40),
int_a BIGINT" prepopulate_with="300"}
{check_repl_factor name="workloadTable" expected="2"}
{sleep="5000"}
{check_meta_repl_factor expected="3"}
{check_meta_repl_factor name="workloadTable" expected="2"}
{0} {execute_workload="short.workload" duration="60000"}
{sleep="20000"}
{terminate_machine id="2"}
{sleep="5000"}
{check_repl_factor name="workloadTable" expected="1"}
{sleep="15000"}
{start_machine id="3" block-workloads="true"}
{sleep="20000"}
{check_repl_factor name="workloadTable" expected="2"}
```

### 2.2.4 Test 1D

```
{start_machine id="0"}
{sleep="3000"}
{start_machine id="1"}
{start_machine id="2"}
```

```
{create_table id="1" name="workloadTable" schema="id int, str_a varchar(40),
int_a BIGINT" prepopulate_with="300"}
{check_repl_factor name="workloadTable" expected="2"}
{sleep="8000"}
{check_meta_repl_factor expected="3"}
{check_meta_repl_factor name="workloadTable" expected="2"}
{0} {execute_workload="short.workload" duration="60000"}
{sleep="20000"}
{terminate_machine id="2"}
{sleep="5000"}
{check_repl_factor name="workloadTable" expected="1"}
{sleep="15000"}
{start_machine id="2" block-workloads="true"}
{sleep="20000"}
{check_repl_factor name="workloadTable" expected="2"}
```

### 2.2.5 Test 1E

```
{start_machine id="0"}
{sleep="3000"}
{start_machine id="1"}
{create_table id="1" name="workloadTable" schema="id int, str_a
varchar(40), int_a BIGINT" prepopulate_with="300"}
{check_repl_factor name="workloadTable" expected="1"}
{sleep="5000"}
{check_meta_repl_factor expected="2"}
{check_meta_repl_factor name="workloadTable" expected="1"}
{0} MIGRATE TABLEMANAGER workloadTable;
{0} {execute_workload="short.workload" duration="60000"}
{sleep="20000"}
{terminate_machine id="1"}
{sleep="15000"}
{check_repl_factor name="workloadTable" expected="0"}
```

### 2.2.6 Test 2A

```
{start_machine id="0"}
{sleep="3000"}
{start_machine id="1"}
{start_machine id="2"}
{create_table id="1" name="workloadTable" schema="id int, str_a varchar(40),
int_a BIGINT" prepopulate_with="300"}
{sleep="20000"}
{check_meta_repl_factor expected="3"}
{check_repl_factor name="workloadTable" expected="2"}
{0} {execute_workload="short.workload" duration="60000"}
```

```
{sleep="20000"}
{terminate_machine id="1"}
{sleep="5000"}
{check_repl_factor name="workloadTable" expected="1"}
```

### 2.2.7 Test 2B

```
{start_machine id="0"}
{sleep="3000"}
{start_machine id="1"}
{start_machine id="2"}
{start_machine id="3"}
{create_table id="1" name="workloadTable" schema="id int, str_a varchar(40),
int_a BIGINT" prepopulate_with="300"}
{sleep="20000"}
{check_meta_repl_factor expected="2"}
{check_repl_factor name="workloadTable" expected="2"}
{0} {execute_workload="short.workload" duration="60000"}
{sleep="20000"}
{terminate_machine id="1"}
{sleep="20000"}
{check_repl_factor name="workloadTable" expected="2"}
```

### 2.2.8 Test 2C

```
{start_machine id="0"}
{sleep="3000"}
{start_machine id="1"}
{start_machine id="2"}
{create_table id="1" name="workloadTable" schema="id int, str_a varchar(40),
int_a BIGINT" prepopulate_with="300"}
{sleep="20000"}
{check_meta_repl_factor expected="3"}
{check_repl_factor name="workloadTable" expected="2"}
{0} {execute_workload="short.workload" duration="60000"}
{sleep="20000"}
{terminate_machine id="1"}
{sleep="10000"}
{check_repl_factor name="workloadTable" expected="1"}
{sleep="2000"}
{start_machine id="3" block-workloads="true"}
{sleep="10000"}
{check_repl_factor name="workloadTable" expected="2"}
```

### 2.2.9 Test 2D

```
{start_machine id="0"}
{sleep="3000"}
```

```
{start_machine id="1"}
{start_machine id="2"}
{create_table id="1" name="workloadTable" schema="id int, str_a varchar(40),
int_a BIGINT" prepopulate_with="300"}
{sleep="20000"}
{check_meta_repl_factor expected="3"}
{check_repl_factor name="workloadTable" expected="2"}
{0} {execute_workload="short.workload" duration="60000"}
{sleep="20000"}
{terminate_machine id="1"}
{sleep="5000"}
{check_repl_factor name="workloadTable" expected="1"}
{sleep="15000"}
{start_machine id="1" block-workloads="true"}
{sleep="5000"}
{check_repl_factor name="workloadTable" expected="2"}
```

### 2.2.10 Test 2E

```
{start_machine id="0"}
{sleep="3000"}
{start_machine id="1"}
{create_table id="1" name="workloadTable" schema="id int, str_a varchar(40),
int_a BIGINT" prepopulate_with="300"}
{sleep="10000"}
{check_meta_repl_factor expected="2"}
{check_repl_factor name="workloadTable" expected="1"}
{0} {execute_workload="short.workload" duration="60000"}
{sleep="20000"}
{terminate_machine id="1"}
{sleep="20000"}
{check_meta_repl_factor name="workloadTable" expected="0"}
```

### 2.2.11 Test 3A

```
{start_machine id="0"}
{sleep="3000"}
{start_machine id="1"}
{start_machine id="2"}
{sleep="10000"}
{2} MIGRATE SYSTEMTABLE NO_REPLICATE
{sleep="10000"}
{check_meta_repl_factor expected="2"}
{0} {execute_workload="st.workload" duration="60000"}
{sleep="20000"}
{terminate_machine id="2"}
{sleep="30000"}
```

```
{check_meta_repl_factor expected="1"}
```

### 2.2.12 Test 3B

```
{start_machine id="0"}
{sleep="3000"}
{start_machine id="1"}
{start_machine id="2"}
{start_machine id="3"}
{1} MIGRATE SYSTEMTABLE NO_REPLICATE
{sleep="30000"}
{check_meta_repl_factor expected="2"}
{0} {execute_workload="st.workload" duration="60000"}
{sleep="20000"}
{terminate_machine id="1"}
{sleep="10000"}
{check_meta_repl_factor expected="2"}
```

### 2.2.13 Test 3C

```
{start_machine id="0"}
{sleep="3000"}
{start_machine id="1"}
{start_machine id="2"}
{1} MIGRATE SYSTEMTABLE NO_REPLICATE
{sleep="10000"}
{check_meta_repl_factor expected="2"}
{0} {execute_workload="st.workload" duration="60000"}
{sleep="20000"}
{terminate_machine id="1"}
{sleep="20000"}
{start_machine id="3" block-workloads="true"}
{sleep="10000"}
{check_meta_repl_factor expected="2"}
```

### 2.2.14 Test 3D

```
{start_machine id="0"}
{sleep="3000"}
{start_machine id="1"}
{start_machine id="2"}
{1} MIGRATE SYSTEMTABLE NO_REPLICATE
{sleep="10000"}
{check_meta_repl_factor expected="2"}
{0} {execute_workload="st.workload" duration="60000"}
{sleep="20000"}
{terminate_machine id="1"}
{sleep="20000"}
```

```
{start_machine id="1" block-workloads="true"}
{sleep="10000"}
{check_meta_repl_factor expected="2"}
```

### 2.2.15 Test 3E

```
{start_machine id="0"}
{sleep="3000"}
{start_machine id="1"}
{1} MIGRATE SYSTEMTABLE NO_REPLICATE
{sleep="10000"}
{check_meta_repl_factor expected="1"}
{0} {execute_workload="st.workload" duration="60000"}
{sleep="20000"}
{terminate_machine id="1"}
{sleep="10000"}
{check_meta_repl_factor expected="0"}
```

### 2.2.16 Test 4A

```
{start_machine id="0"}
{sleep="3000"}
{start_machine id="1"}
{start_machine id="2"}
{1} MIGRATE SYSTEMTABLE NO_REPLICATE
{create_table id="1" name="workloadTable" schema="id int, str_a varchar(40), int_a BIGINT" prepopulate_with="300"}
{sleep="10000"}
{check_meta_repl_factor expected="2"}
{0} {execute_workload="short.workload" duration="60000"}
{sleep="20000"}
{sleep="20000"}
{terminate_machine id="1"}
{sleep="10000"}
{check_meta_repl_factor expected="1"}
```

### 2.2.17 Test 4B

```
{start_machine id="0"}
{sleep="3000"}
{start_machine id="1"}
{start_machine id="2"}
{start_machine id="3"}
{1} MIGRATE SYSTEMTABLE NO_REPLICATE
{create_table id="1" name="workloadTable" schema="id int, str_a varchar(40), int_a BIGINT" prepopulate_with="300"}
{check_repl_factor name="workloadTable" expected="2"}
{sleep="10000"}
```

```
{check_meta_repl_factor expected="2"}
{0} {execute_workload="short.workload" duration="60000"}
{sleep="20000"}
{terminate_machine id="1"}
{sleep="5000"}
{check_repl_factor name="workloadTable" expected="2"}
```

### 2.2.18 Test 4C

```
{start_machine id="0"}
{sleep="3000"}
{start_machine id="1"}
{start_machine id="2"}
{1} MIGRATE SYSTEMTABLE NO_REPLICATE
{create_table id="1" name="workloadTable" schema="id int, str_a varchar(40),
int_a BIGINT" prepopulate_with="300"}
{check_repl_factor name="workloadTable" expected="2"}
{sleep="10000"}
{check_meta_repl_factor expected="2"}
{0} {execute_workload="short.workload" duration="60000"}
{sleep="20000"}
{terminate_machine id="1"}
{sleep="20000"}
{start_machine id="3" block-workloads="true"}
{sleep="30000"}
{check_meta_repl_factor expected="2"}
```

### 2.2.19 Test 4D

```
{start_machine id="0"}
{sleep="3000"}
{start_machine id="1"}
{start_machine id="2"}
{1} MIGRATE SYSTEMTABLE NO_REPLICATE
{create_table id="1" name="workloadTable" schema="id int, str_a varchar(40),
int_a BIGINT" prepopulate_with="300"}
{check_repl_factor name="workloadTable" expected="2"}
{sleep="10000"}
{check_meta_repl_factor expected="2"}
{0} {execute_workload="short.workload" duration="60000"}
{sleep="20000"}
{terminate_machine id="1"}
{sleep="20000"}
{start_machine id="1" block-workloads="true"}
{sleep="5000"}
{check_repl_factor name="workloadTable" expected="2"}
{sleep="20000"}
```

```
{check_meta_repl_factor expected="2"}
```

### 2.2.20 Test 4E

```
{start_machine id="0"}
{sleep="3000"}
{start_machine id="1"}
{1} MIGRATE SYSTEMTABLE NO_REPLICATE
{create_table id="1" name="workloadTable" schema="id int, str_a varchar(40),
int_a BIGINT" prepopulate_with="300"}
{check_repl_factor name="workloadTable" expected="1"}
{sleep="10000"}
{check_meta_repl_factor expected="1"}
{0} {execute_workload="short.workload" duration="60000"}
{sleep="20000"}
{terminate_machine id="1"}
{sleep="20000"}
{check_repl_factor name="workloadTable" expected="0"}
{check_meta_repl_factor expected="0"}
```

# 3 Source Code

All of the code used in the evaluations in this thesis can be found at the following location:

http://archive.cs.st-andrews.ac.uk/h2o

## BIBLIOGRAPHY

[1] MySQL, **"A Guide to Database High Availability",** 2006 [White Paper, available via http://www.computerworlduk.com/white-paper/open-source/5656/a-guide-to-database-high-availability/, 26-Feb-2012].

[2] PostgreSQL, **"PostgreSQL",** 2011 [Computer Program, available via http://www.postgresql.org/, 03-May-2011].

[3] Apache Foundation, **"CouchDB",** 2011 [Computer Program, available via http://couchdb.apache.org/, 01-May-2011].

[4] Greenplum, **"Greenplum Database 4.0: Critical Mass Innovation",** 2010 [White Paper, available at http://www.emc.com/collateral/hardware/white-papers/h8072-greenplum-database-wp.pdf, 26-Feb-2012].

[5] GenieDB, **"Technology Overview: Solving SQL Scale-out",** 2010 [White Paper, available at http://www.geniedb.com/wp-content/uploads/2011/04/sql-overview-revised.pdf, 26-Feb-2012].

[6] DeCandia, G., Hastorun, D., Jampani, M., Kakulapati, G., Lakshman, A., Pilchin, A., Sivasubramanian, S., Vosshall, P., Vogels, W., **"Dynamo: amazon's highly available key-value store",** ACM SIGOPS Operating Systems Review, vol. 41, no. 6, p. 220, 2007.

[7] M. Stonebraker, **"The case for shared nothing"** in Database Engineering Bulletin, vol. 9, no. 1, pp. 4–9, 1986.

[8] M. Hogan (ScaleDB), **"Shared-Disk vs Shared-Nothing."** [White Paper, available at http://www.scaledb.com/pdfs/WP_SDvSN.pdf, 26-Feb-2012].

[9] M. Michael, J. E. Moreira, D. Shiloach, and R. W. Wisniewski, **"Scale-up x scale-out: A case study using nutch/lucene"**, in 2007 IEEE International Parallel and Distributed Processing Symposium, 2007, pp. 1–8.

[10] Xeround, **"Database Scalability and Availability in the Cloud"**, 2010 [White Paper, available vis http://xeround.com/developers/mysql-cloud-db-technical-whitepaper/, 20-Nov-2011]

[11] Caurino, C., Jones, E.P.C., Popa, R.A., Malviya, N., Wu, Eugene, Madden, S., Balakrishnan, H., Zeldovich, N., **"Relational Cloud: A Database-as-a-Service for the Cloud"**, in Proceedings of the Conference on Innovative Data Systems Research 2011.

[12] Amazon, **"Amazon EC2"** [Webpage: http://aws.amazon.com/ec2/, 29-Apr-2011].

[13] D. Terry, M. Theimer, K. Peterson, A. Demers, and M. Spreitzer, **"Managing a Update Conflicts in Bayou , a Weakly Connected Replicated Storage System"**, in Proceedings of the Fifteenth ACM Symposium on Operating Systems Principles, 1995.

[14] D. Skeen and M. Stonebraker, **"A Formal Model of Crash Recovery in a Distributed System"** in IEEE Transactions on Software Engineering, vol. 9, no. 3, pp. 219-228, 1983.

[15] T. Connolly and C. Begg, **"Database Systems: A Practical Approach to Design, Implementation, and Management"**, Addison Wesley, 2002.

[16] L. Lamport, **"Paxos Made Simple"** in ACM SIGACT News, vol. 32, no. 4, pp. 18–25, 2001.

[17] Baker, J., Bond, C., Corbett, J. C., Furman, J., Khorlin, A., Larson, J., Léon, J. M., **"Megastore: Providing Scalable, Highly Available Storage for Interactive Services"** in Proceedings of the Conference on Innovative Data Systems Research, pp. 223-234, 2011.

[18] T. D. Chandra, R. Griesemer, and J. Redstone, **"Paxos made live: an engineering perspective"**, in Proceedings of the twenty-sixth annual ACM Symposium on Principles of Distributed Computing, 2007, pp. 398–407.

[19] J. Gray and L. Lamport, **"Consensus on transaction commit"** in ACM Transactions on Database Systems, vol. 31, no. 1, pp. 133–160, 2006.

[20] E. A. Brewer, **"Towards robust distributed systems"** in Proceedings of the Annual ACM Symposium on Principles of Distributed Computing, vol. 19, pp. 7–10, 2000.

[21] S. Gilbert and N. Lynch, **"Brewer's conjecture and the feasibility of consistent, available, partition-tolerant web services"** in ACM SIGACT News, vol. 33, no. 2, p. 59, 2002.

[22] D. Abadi, **"CAP, PACELC, and Determinism"**, 2011 [Presentation, available via http://www.slideshare.net/abadid/cap-pacelc-and-determinism, 26-Feb-2012].

[23] M. Stonebraker, **"Urban Myths about SQL"**, 2010 [Presentation, available at http://voltdb.com/_pdf/VoltDB-MikeStonebraker-SQLMythsWebinar-060310.pdf, 26-Feb-2012].

[24] P. Kopietz, P. Scharf, M. Skaf, and S. Chakravarty, **"Consistent Hashing and Random Trees: Distributed Caching Protocols for Relieving Hot Spots on the World Wide Web"** in Proceedings of the twenty-ninth annual ACM symposium on Theory of Computing, 1997.

[25] E. Petrank and D. Rawitz, **"The hardness of cache conscious data placement"** in Proceedings of the 29th ACM SIGPLAN-SIGACT symposium on Principles of programming languages, pp. 101–112, 2002.

[26] S. Gribble, A. Halevy, Z. Ives, M. Rodrig, and D. Suciu, **"What can databases do for peer-to-peer"** in WebDB Workshop on Databases and the Web, 2001.

[27] Stonebraker, M., Aoki, P. M., Litwin, W., Pfeffer, A., Sah, A., Sidell, J., Staelin, C., et al., **"Mariposa: a wide-area distributed database system"** in the International Journal on Very Large Data Bases, vol. 5, no. 1, pp. 48-63, 1996.

[28] B. Fitzpatrick, **"Memcached"** [Computer Program, available via http://www.memcached.org/, 19-Jan 2011].

[29] M. Özsu and P. Valduriez, **"Principles of Distributed Database Systems"**, 2nd ed. Prentice Hall, 1999.

[30] Stonebraker, M., Abadi, D. J., Batkin, A., Chen, X., Cherniack, M., Ferreira, M., Lau, E., et al., **"C-store: a column-oriented DBMS"** in Proceedings of the 31st international conference on Very Large Data Bases, 2005.

[31] VoltDB, **"VoltDB Technical Overview"**, 2010 [White Paper, available at http://voltdb.com/_pdf/VoltDBTechnicalOverviewWhitePaper.pdf, 26-Feb-2012].

[32] TPC, **"Transaction Processing Performance Council (TPC)."** [Webpage: http://www.tpc.org, 19-Apr-2011].

[33] E. P. C. Jones, D. J. Abadi, and S. Madden, **"Low overhead concurrency control for partitioned main memory databases"** in Proceedings of the 2010 International Conference on Management of Data, pp. 603–614, 2010.

[34] I. Gupta, T. D. Chandra, and G. S. Goldszmidt, **"On scalable and efficient distributed failure detectors"** in Proceedings of the Twentieth Annual ACM ymposium on Principles of Distributed Computing, pp. 170–179, 2001

[35] T. D. Chandra and S. Toueg, **"Unreliable failure detectors for reliable distributed systems"** in Journal of the ACM, vol. 43, no. 2, pp. 225-267, 1996.

[36] X. Défago, P. Urbán, N. Hayashibara, and T. Katayama, ., **"On accrual failure detectors"**, 2004 [Technical Report, available via https://dspace.jaist.ac.jp/dspace/bitstream/10119/4785/1/IS-RR-2004-011.pdf, 26-Feb-2012]

[37] W. Chen, S. Toueg, and M. K. Aguilera, **"On the quality of service of failure detectors"** in IEEE Transactions on computers, vol. 51, no. 1, pp. 561–580, 2002.

[38] X. Défago, P. Urbán, and N. Hayashibara, **"The $\varphi$ accrual failure detector",** 2004 [Technical Report, available via http://ddg.jaist.ac.jp/pub/HDY+04.pdf, 26-Feb-2012]

[39] A. Schiper, **"Failure detection vs group membership in fault-tolerant distributed systems: Hidden trade-offs"** in Proceedings of the Second Joint International Workshop on Process Algebra and Probabilistic Methods, Performance Modeling and Verification, pp. 1-15, 2002.

[40] M. Massie, B. Chun, and D. Culler, **"The ganglia distributed monitoring system: design, implementation, and experience"** in Parallel Computing, vol. 30, no. 7, pp. 817-840, 2004.

[41] Tierney, B., Aydt, R., Gunter, D., Smith, W., Swany, M., Taylor, V., & Wolski, R., **"A grid monitoring architecture"**, 2002 [Technical Report, available via http://www-didc.lbl.gov/GGF-PERF/GMA-WG/papers/GWD-GP-16-1.pdf, 26-Feb-2012]

[42] Zanikolas and R. Sakellariou, **"A taxonomy of grid monitoring systems"** in Future Generation Computer Systems, vol. 21, no. 1, pp. 163-188, 2005.

[43] R. L. Ribler, J. S. Vetter, H. Simitci, and D. A. Reed, **"Autopilot: Adaptive Control of Distributed Applications"** in the Seventh International Symposium on High Performance Distributed Computing, pp. 172-179, 1998.

[44] S. Fisher, **"R-GMA: Relational Grid Monitoring Architecture"** ,2001 [Technical Report, accessible at http://www.r-gma.org/pub/ah03_148.pdf, Feb-26-2012].

[45] J. O. Kephart and D. M. Chess, **"The vision of autonomic computing"** in IEEE Computer, vol. 36, no. 1, pp. 41-50, 2003.

[46] G. N. C. Kirby, A. Dearle, A. Macdonald, and A. Fernandes, **"An Approach to Ad Hoc Cloud Computing"**, 2010 [Technical Report, available via http://arxiv.org/abs/1002.4738, 26-Feb-2012]

[47] D. P. Anderson, **"BOINC: A System for Public-Resource Computing and Storage"** in the Fifth IEEE/ACM International Workshop on Grid Computing, pp. 4-10, 2005.

[48] M. Litzkow, M. Livny, and M. Mutka, **"Condor-a hunter of idle workstations"** in the Proceedings of the 8th International Conference of Distributed Computing Systems, vol. 43, 1988.

[49] A. L. Beberg, D. L. Ensign, G. Jayachandran, S. Khaliq, and V. S. Pande, **"Folding@ home: Lessons from eight years of volunteer distributed computing"** in 2009 IEEE International Symposium on Parallel & Distributed Processing, 2009.

[50] PGCluster, **"PGCluster"**, 2005 [Computer Program, available via http://pgcluster.projects.postgresql.org, 03-May-2011].

[51] A. Mitani, **"PGCluster-II"** at PGCon2007, 2007 [Presentation, available via, http://www.pgcon.org/2007/schedule/attachments/26-pgcluster2pgcon.pdf, 26-Feb-2012]. [52] Oracle, "MySQL Cluster 7.0 & 7.1: Architecture and New Features." pp. 1-29, 2010.

[53] Oracle, **"MySQL Cluster 7.0 & 7.1: Architecture and New Features.",** 2010 [White Paper, available via http://sun.systemnews.com/articles/146/3/MySQL/23055 26-Feb-2012]

[54] GenieDB, **"Beating the CAP Theorem."**, 2010 [White Paper, available via http://www.geniedb.com/wp-content/uploads/2011/03/cap-theorem.pdf, 26-Feb-2012].

[55] Clustrix, **"A New Approach: Clustrix Sierra Database Engine"**, 2010 [White Paper, available at http://www.clustrix.com/Default.aspx?app=LeadgenDownload&shortpath=docs%2fClustrix_A_New_Approach.pdf, 26-Feb-2012]

[56] S. Tsarev, **"MongoDB vs. Clustrix Comparison"**, 2011 [Webpage: http://sergeitsar.blogspot.com/2011/02/mongodb-vs-clustrix-comparison-part-2.html, 04-Feb-2011].

[57] Stonebraker, M., Madden, S., Abadi, D. J., Harizopoulos, S., Hachem, N., & Helland, P., **"The end of an architectural era (it's time for a complete rewrite)"** in Proceedings of the 33rd international conference on Very Large Data Bases, pp. 1150–1160, 2007.

[58] Kallman, R., Kimura, H., Natkins, J., Pavlo, A., Rasin, A., Zdonik, S., Jones, E. P. C., **"H-Store: a high-performance, distributed main memory transaction processing system"** in Proceedings of the VLDB Endowment, vol. 1, no. 2, pp. 1496–1499, 2008.

[59] Carey, M. J., DeWitt, D. J., Franklin, M. J., Hall, N. E., McAuliffe, M. L., Naughton, J. F., Schuh, D. T., **"Shoring up persistent applications"**, 1994, pp. 383-394.

[60] S. Harizopoulos, D. J. Abadi, S. Madden, and M. Stonebraker, **"OLTP through the looking glass, and what we found there"** in Proceedings of the 2008 ACM SIGMOD International Conference on Management of Data, pp. 981–992, 2008.

[61] VoltDB, **"VoltDB Stored Procedures"**, 2010 [Webpage: http://voltdb.com/blog/voltdb-stored-procedures, 07-Apr-2011].

[62] VoltDB, **"VoltDB: The Lifecycle of a Transaction"**, 2010 [Webpage: http://voltdb.com/blog/lifecycle-transaction, 07-Apr-2011].

[63] Vertica, **"The Vertica ® Analytic Database Technical Overview"**, 2010 [White Paper, accessible at http://www.vertica.com/wp-content/uploads/2011/01/VerticaArchitectureWhitePaper.pdf, 26-Feb-2012].

[64] B. R. Iyer and D. Wilhite, **"Data Compression Support in Databases"** in Proceedings of the 20th International Conference on Very Large Data Bases, pp. 695-704, 1994.

[65] D. J. Abadi, S. Madden, and M. Ferreira, **"Integrating compression and execution in column-oriented database systems"** in SIGMOD international conference on Management of data, vol. pages, pp. 671-682, 2006.

[66] G. Graefe and L. D. Shapiro, G. Graefe and L. D. Shapiro, **"Data Compression and Database Performance"** in Proceedings of the 1991 Symposium on Applied Computing, 1991.

[67] S. Harizopoulos, V. Liang, D. J. Abadi, and S. Madden, **"Performance tradeoffs in read-optimized databases"** in Proceedings of the 32nd International Conference on Very Large Data Bases, 2006, p. 498.

[68] F. Yang and R. Yerneni, **"A Scalable Data Platform for a Large Number of Small Applications",** 2009 [White Paper, accessible via http://research.yahoo.com/files/ScalingSmallApps.pdf, 26-Feb-2012].

[69] Chang, F., Dean, J., Ghemawat, S., Hsieh, W. C., Wallach, D. A., Burrows, M., Chandra, T., **"Bigtable: A distributed storage system for structured data"** in Proceedings of the 7th USENIX Symposium on Operating Systems Design and Implementation (OSDI'06), 2006.

[70] S. Ghemawat, H. Gobioff, and S. Leung, **"The Google file system"** in Proceedings of the Nineteenth ACM Symposium on Operating Systems Principles, vol. 19, p. 22, 2003.

[71] R. Burgess, **"Bigtable and Boxwood"**, 2009 [Presentation, available via http://www.slideshare.net/Eweaver/bigtable-and-boxwood, 26-Feb-2012].

[72] M. Burrows, **"The Chubby lock service for loosely-coupled distributed systems"** in Seventh Symposium on Operating System Design and Implementation, 2006.

[73] A. Lakshman, **"Cassandra: a decentralized structured storage system"** in ACM SIGOPS Operating Systems Review, 2010.

[74] P. Hunt, M. Konar, F. P. Junqueira, and B. Reed, **"ZooKeeper: wait-free coordination for internet-scale systems"** in Proceedings of the 2010 USENIX annual technical conference, pp. 11–11, 2010.

[75] J. C. Anderson, J. Lehnardt, and N. Slater, **"CouchDB: The Definitive Guide"**. O'Reilly, ISBN 978-0-596-15589-6, 2010.

[76] P. Helland, **"Life beyond distributed transactions: an apostate's opinion"** in Proceedings of the Conference on Innovative Data Systems Research, pp. 132-141, 2007.

[77] VoltDB, "Transaction Ordering and Replication", 2011. Available: http://voltdb.com/blog/transaction-ordering-and-replication. [Accessed: 17-Feb-2011].

[78] J. MacCormick, N. Murphy, M. Najork, C. A. Thekkath, and L. Zhou, **"Boxwood: Abstractions as the foundation for storage infrastructure"** in Symposium on Operating System Design and Implementation (OSDI), 2004.

[79] I. Stoica, R. Morris, D. Karger, M. F. Kaashoek, and H. Balakrishnan, **"Chord: A scalable peer-to-peer lookup service for internet applications"** in Proceedings of the 2001 conference on Applications, technologies, architectures, and protocols for computer communications, 2001.

[80] B. Lowekamp, N. Miller, R. Karrer, T. Gross, and P. Steenkiste, **"Design, implementation, and evaluation of the REMOS network monitoring system"** in the Journal of Grid Computing, pp. 1-34, 2003.

[81] W. lodzimierz Funika, M. Smetek, and M. Bubak, **"Integrating of the OCM-G Monitoring System into the GRID superscalar"** Grid Systems, Tools and Environments, p. 66, 2006.

[82] W. Lang and J. M. Patel, **"Towards eco-friendly database management systems"** in Proceedings of the Conference on Innovative Data Systems Research, 2009.

[83] W. Lang, J. M. Patel, and J. F. Naughton, **"On energy management, load balancing and replication"** in ACM SIGMOD Record, vol. 38, no. 4, p. 35, 2010.

[84] H. Hsiao and D. J. DeWitt, **"Chained declustering: a new availability strategy for multiprocessor database machines"** in Proceedings of the sixth international Conference on Data Engineering, pp. 456-465, 1990.

[85] G. Kirby, A. Macdonald, and A. Dearle, **"NUMONIC"**, 2010 [Computer Program, available via http://blogs.cs.st-andrews.ac.uk/numonic/ , 03-Mar-2011].

[86] Stonebraker, M., Aoki, P. M., Litwin, W., Pfeffer, A., Sah, A., Sidell, J., Staelin, C., **"Mariposa: A Wide-Area Distributed Database System"** in Very Large Data Bases Journal, vol. 5, no. 1, pp. 48-63, 1996.

[87] T. Mueller, **"H2 Database"**, 2009 [Computer Program, available via http://www.h2database.com, 26 Feb-2012].

[88] T. Mueller, **"H2 Documentation"**, 2011 [Documentation, available at http://www.h2database.com/h2.pdf, 03-Jun-2011].

[89] D. Crockford, **"JSON (JavaScript Object Notation)",** 2006 [Webpage: http://www.json.org/, 11-Sep-2011].

[90] Hyperic**, "SIGAR"**, 2011 [Computer Program, available via http://www.hyperic.com/products/sigar, 10-Jun-2011].

[91] D. Lussier and S. Martin, **"BenchmarkSQL",** 2011 [Computer Program, available via http://sourceforge.net/projects/benchmarksql/, 26-Feb-2012].

[92] C. Levine, **"Standard Benchmarks for Database Systems"** in Proceedings of the Special Interest Group on Management of Data (SIGMOD), 1997.

[93] Amazon, **"Amazon EC2 SLA."** [Webpage: http://aws.amazon.com/ec2-sla/, 12-Feb-2012].

[94] L. A. Barroso and U. Hölzle, **"The Case for Energy-Proportional Computing"** in IEEE Computer, vol. 40, no. 12, pp. 33-37, 2007.

[95] B. Kemme and G. Alonso, **"Don't be lazy, be consistent: Postgres-R, a new way to implement database replication"** in Proceedings of the 26th International Conference on Very Large Databases, pp. 134–143, 2000.

[96] M. J. Fischer, N. A. Lynch, and M. S. Paterson, **"Impossibility of distributed consensus with one faulty process"** in Journal of the ACM (JACM), vol. 32, no. 2, pp. 374–382, 1985.

[97] E. Krishnamurthy, **"On the design and administration of secure database transactions"** in ACM SIGSAC Review, pp. 63-70, 1992.

[98] MySQL, **"Statements That Cause an Implicit Commit"** [Webpage: http://dev.mysql.com/doc/refman/5.1/en/implicit-commit.html, 08-Mar-2011].